\documentclass[]{iopart}

\usepackage{iopams}
\expandafter\let\csname equation*\endcsname\relax 
\expandafter\let\csname endequation*\endcsname\relax 
\usepackage{amsmath}
\usepackage{amssymb}
\usepackage{amsfonts}				

\usepackage{graphicx}
\usepackage{dcolumn}
\usepackage{bm}
\usepackage{dsfont}
\usepackage{color}
\usepackage{enumerate}

\usepackage{url}
\usepackage[colorlinks=true ,citecolor=blue,urlcolor=blue]{hyperref}
\usepackage[square,sort&compress,comma,numbers]{natbib}

\definecolor{Nathanblue}{rgb}{0.7,0.14,0.91}

\newcommand{\onlinecite}[1]{\nocite{#1}\citenum{#1}}

\def\nm{{\ {\rm nm}}}                       
\def\micron{{\ \mu{\rm m}}}                 
\def\g{{\ {\rm g}}}                          
\def\P{{\ {\rm P}}}                         
\def\Hz{{\ {\rm Hz}}}                       
\def\MHz{{\ {\rm MHz}}}                     
\def\GHz{{\ {\rm GHz}}}                     
\def\THz{{\ {\rm THz}}}                     
\def\ms{{\ {\rm ms}}}                       
\def\Schrodinger{{Schr\"odinger\ }}

\def\Er{E_R}                            
\def\El{E_L}                            
\def\kr{k_R}                            
\def\kl{k_L}                            
\def\Cs133{^{133}\rm{Cs}}					
\def\Rb87{^{87}{\rm Rb}}                     
\def\Na23{^{23}{\rm Na}}                     
\def\K40{^{40}\rm{K}}					
\def\K41{^{41}{\rm K}}                     
\def\Li6{^{6}{\rm Li}}                       
\def\Li7{^{7}{\rm Li}}                       
\def\ex{{\mathbf e}_x}                            
\def\ey{{\mathbf e}_y}                            
\def\ez{{\mathbf e}_z}                            
\def\shorttimes{\!\times\!}                            
\DeclareMathAlphabet\mathbfcal{OMS}{cmsy}{b}{n}
\def\bra#1{\mathinner{\langle{#1}|}}
\def\ket#1{\mathinner{|{#1}\rangle}}

{\catcode`\|=\active
  \gdef\Braket#1{\left<\mathcode`\|"8000\let|\BraVert {#1}\right>}}
\def\BraVert{\egroup\,\mid@vertical\,\bgroup}

\def\bs#1{\boldsymbol{#1}}

\def\be{\begin{equation}}
\def\ee{\end{equation}}
\def\bea{\begin{eqnarray}}
\def\eea{\end{eqnarray}}

\begin{document}

\article[Light-induced gauge fields for ultracold atoms]{Review}{Light-induced gauge fields for ultracold atoms}

\author{N. Goldman}
\ead{nathan.goldman@lkb.ens.fr and ngoldman@ulb.ac.be}
\address{Laboratoire Kastler Brossel, CNRS, UPMC, ENS, Coll\`ege de France, 11 place Marcelin Berthelot, 75005, Paris, France}
\address{Center for Nonlinear Phenomena and Complex Systems, Universit\'e Libre de Bruxelles, CP 231, Campus Plaine, B-1050 Brussels, Belgium}

\author{G. Juzeli\={u}nas}
\ead{gediminas.juzeliunas@tfai.vu.lt}
\address{Institute of Theoretical Physics and Astronomy, Vilnius University, \\
A.~Go\v{s}tauto 12, LT-01108 Vilnius, Lithuania}

\author{P. \"Ohberg}
\ead{p.ohberg@hw.ac.uk}
\address{SUPA, Institute of Photonics and Quantum Sciences, Heriot-Watt University, Edinburgh EH14 4AS, United Kingdom}

\author{I.~B.~Spielman}
\ead{ian.spielman@nist.gov}
\address{Joint Quantum Institute, National Institute of Standards and Technology, and University of Maryland, Gaithersburg, Maryland, 20899, USA}



\begin{abstract}
Gauge fields are central in our modern understanding of physics at all scales.  At the highest energy scales known, the microscopic universe is governed by particles interacting with each other through the exchange of gauge bosons.  At the largest length scales, our universe is ruled by gravity, whose gauge structure suggests the existence of a particle -- the graviton-- that mediates the gravitational force.  At the mesoscopic scale, solid-state systems are subjected to gauge fields of different nature: materials can be immersed in external electromagnetic fields, but they can also feature emerging gauge fields in their low-energy description.  \\
In this review, we focus on another kind of gauge field: those {\it engineered} in systems of ultracold neutral atoms.  In these setups, atoms are suitably coupled to laser fields that generate effective gauge potentials in their description. Neutral atoms ``feeling" laser-induced gauge potentials can potentially mimic the behavior of an electron gas subjected to a magnetic field, but also, the interaction of elementary particles with non-Abelian gauge fields. Here, we review different realized and proposed techniques for creating gauge potentials -- both Abelian and non-Abelian -- in atomic systems and discuss their implication in the context of quantum simulation. While most of these setups concern the realization of background and classical gauge potentials, we  conclude with more exotic proposals where these synthetic fields might be made dynamical, in view of simulating interacting gauge theories with cold atoms.
\end{abstract}

\tableofcontents

\maketitle

\section{Introduction}\label{sect:introduction}



The laboratory realization of ultracold neutral atomic gases such as Bose-Einstein condensates~\cite{Anderson1995,Davis1995a} and degenerate Fermi gases~\cite{DeMarco1999a} -- quantum gases -- delivered remarkably versatile experimental systems that can realize physical effects with analogues throughout physics.  The coherence properties of Bose-Einstein condensates (BECs) allow them to address concepts from optics and nonlinear optics: classical and quantum atom optics~\cite{Ketterle1996,Jeltes2007}.  Quantum gases have shed light on many effects predicted in the context of traditional condensed matter systems such as the bosonic superfluid to Mott transition in optical lattices~\cite{Fisher1989,Jaksch1998,Greiner2002}, and the Bardeen-Cooper-Schrieffer crossover in degenerate Fermi gases~\cite{Greiner2003a,Bourdel2004,Bartenstein2004,Zwierlein2004,Kinast2004}.  Even phenomena commonplace in high energy physics can occur in ultracold settings, where Higgs modes have been observed~\cite{Bissbort2011,Endres2012}, unconventional ``color'' superfluidity~\cite{Williams2009} is possible, and where confinement mechanisms  \cite{Kapit2011,Banerjee2013,Edmonds2013} and axion electrodynamics ~\cite{Bermudez:2010} have been predicted.

Atomic quantum gases are charge neutral, and therefore, they are not affected by external electromagnetic fields the way electrons are.  However, atom-light coupling allows for the creation of versatile gauge potentials that effectively emerge in the atoms dynamics, allowing experimental access to a panoply of new phenomena at the quantum level. Using this technology, atoms can be subjected to static Abelian gauge fields, offering a framework where synthetic electric and magnetic fields can be experimentally tuned with lasers (see the first experimental works at NIST \onlinecite{Lin2009a,Lin2009b,Lin2011a}). These setups can also be extended to generate versatile non-Abelian gauge potentials \cite{Wilczek:1984}. These static non-Abelian gauge fields could be tailored so as to reproduce the effects of Rashba-type spin-orbit couplings, but also, to mimic a variety of properties encountered in the context of high-energy physics. The first experimental steps towards the realization of a two-dimensional spin-orbit-coupled atomic gas have been reported in Refs. \cite{Lin2011,Wang2012,Cheuk2012,Zhang2012PRL,Fu2013,Zhang2013PRA,Qu2013,LeBlanc2013}, where the spin-orbit coupling acts along a single spatial dimension.
Mimicking magnetic and spin-orbit effects in cold atom laboratories enables the assembly of quantum simulators of new kinds of exotic quantum matter~\cite{Lewenstein2007,Bloch2008a,Lewenstein2012}. Indeed, cold atomic gases are ideally suited for quantum simulation, as numerous physical parameters governing the systems dynamics are experimentally tunable: particle density, confining potentials, effective dimensionality~\cite{Gorlitz2001}, and even the collisional properties~\cite{Chin2008} can be easily controlled in the same laboratory.  Taken together, this greatly enlarges the range of systems that can realize Richard Feynman's vision~\cite{Feynman1982} for constructing physical quantum emulators of systems or situations that are computationally or analytically intractable.

Gauge theories, with their associated gauge potentials, are central in our understanding of the interactions between elementary particles.  Electromagnetism is the simplest example, where the scalar and vector potentials together describe the coupling between charged matter and electromagnetic fields.  In the standard model, interactions are mediated by more complex gauge fields which often are of a non-Abelian character.  Also, the idea of emergent gauge fields, where the low energy sector of a more complicated system is described by an effective gauge theory, is not new. Mead and Truhlar~\cite{Mead:1979} and Berry~\cite{Berry:1984} noted that the adiabatic motion of quantum particles with internal structure can be described in terms of an effective ``geometric'' gauge potential.  This property was first studied in molecular physics, where the Jahn-Teller effect revealed the geometric phases and corresponding vector potentials~\cite{Mead:1979,Berry:1984,Mead1992,Shapere1989,Bohm2003}.   

The adiabatic and Born-Oppenheimer approximations are closely linked to geometric gauge fields in atomic systems. These geometric vector potentials appear when each atom's external motion is described separately from it's internal dynamics, yet the Hamiltonian governing the internal dynamics parametrically depends on the atomic position (for example, via the light-matter interaction).  In this context, the possibility of emergent vector potentials was first noted by Taichenachev et al~\cite{Taichenachev1992}, Dum and Olshanii~\cite{Dum1996}, as well as by Visser and Nienhuis~\cite{Visser1998}.  Refs.~\onlinecite{Juzeliunas2004,Juzeliunas2006,Zhu2006,Spielman2009,Gunter2009} proposed setups for systematically engineering vector potentials which provide a non-zero artificial (synthetic) magnetic field for quantum degenerate gases.  These synthetic magnetic fields were recently experimentally realized~\cite{Lin2009b,Lin2009a}, whereas the effect of geometric scalar potentials in optical lattices was experimentally observed a decade earlier~\cite{Dutta1999}.  When the local atomic internal states ``dressed'' by the laser fields have degeneracies, effective non-Abelian gauge potentials  can be formed~\cite{Unanyan99PRA,Ruseckas2005,Liu2009}, often manifesting as a spin-orbit coupling would in material systems. These artificial spin-orbit couplings lead to the spin Hall effect in atomic systems \cite{Liu2007,Zhu2006}, as recently demonstrated experimentally \cite{Beeler:2013}.  Artificial gauge fields are therefore a highly versatile tool for creating exotic condensed matter analogs in atomic gases~\cite{Lin2011,Wang2012,Cheuk2012,Zhang2012PRL,Zhang2012PRL,Fu2013,Zhang2013PRA,Qu2013}.

These approaches can be extended in a powerful way by adding optical lattice potentials~\cite{Eckardt:2005,Lewenstein2007,Bloch2008a,Lewenstein2012,Cooper2008}, where the link to quantum simulation of condensed matter phenomena is particularly evident. Here, the artificial magnetic field can be understood as resulting from a laser induced tunneling between the lattice sites \cite{Jaksch:2003,Gerbier:2010}, or lattice shaking \cite{Aidelsburger:2013,Ketterle:2013}.  Numerous theoretical proposals for simulating  condensed matter models and realizing strongly correlated systems have been put forth, and recently, artificial gauge potentials corresponding to staggered ~\cite{Aidelsburger:2011,Jimenez-Garcia2012,Struck2012}  and uniform \cite{Aidelsburger:2013,Ketterle:2013,Aidelsburger:2014} magnetic fluxes have been produced in optical lattices.

All of these schemes create {\it static} gauge fields, in the sense that they are described by additional terms in the atomic Hamiltonian (although the gauge fields can still have an externally imposed time dependence, leading to effective electric fields~\cite{Lin2011a}).  Dynamical gauge fields (which are described by their own Hamiltonian and are not just imposed) are important in many areas of physics, from particle physics where the gauge fields are the fundamental force carriers (e.g., Ref.~\onlinecite{Cheng1991}), to many-body condensed-matter physics where they appear in effective field theories~\cite{Levin2005}.  As such, a number of proposals exist for creating dynamic gauge fields with ultracold atoms~\cite{Wiese:2013uh,Kapit2011,Banerjee2012,Tagliacozzo2011,Banerjee2013,Zohar2013,Tagliacozzo2013}, but to date their complexity has stymied experimental realization. 

By emulating a fully dynamical field theory, which includes gauge fields, it is certainly tempting to envisage a quantum simulator that can address open questions from the Standard Model \cite{Wiese:2013uh}. Mapping out the complete QCD phase diagram is a formidable task. It is also an NP hard problem, and therefore highly intractable using classical computation. A special purpose quantum computer able to emulate the corresponding machinery from the Standard Model  would significantly contribute to our understanding of the fundamental forces and processes in Nature.

Although the fundamental nature of being able to create gauge fields for charge neutral ultracold quantum gases, with clear links to particle physics and the forces of Nature, is a compelling argument for pursuing such an endeavor, there are also other, more practical, motivations to create these gauge fields. Magnetic fields and spin-orbit coupling, in particular, appear to provide a route towards the preparation of topological states of matter \cite{Qi2011,Hasan2010}, with some quite remarkable properties and promises for future applications \cite{Nayak2008}. Ultracold gases subject to artificial gauge fields provide an alternative route for reaching such exotic states of matter, with some added benefits from their unique probes and unprecedented flexibility in controlling many experimental parameters in these systems \cite{Bloch2008a,Lewenstein2007}. Topology is the branch of mathematics that deals with properties of geometric objects that do not change under smooth deformations \cite{nakahara}. The great interest in topological states of matter relies on the fact that such states are robust against external perturbations (e.g. finite temperature, noise, or in general experimental imperfections). There are many intriguing phenomena associated with topological matter. The most striking is the existence of metallic edge states in a material that is insulating in the bulk  \cite{Qi2011,Hasan2010}. In the integer quantum Hall effect \cite{vonKlitzing:1986,Hatsugai:1993}, these edge modes carry precisely one quantum of conductance, which  leads to the quantization of the Hall conductivity. Due to the bulk-edge correspondence \cite{Hatsugai:1993}, much of the properties and information of topological systems can be extracted from the edge states. Depending on the details of the particular physical setup (e.g. the lattice geometry, the interactions, the number of magnetic flux per particle, the presence of spin-orbit coupling), these edge modes may turn out to have very exotic properties \cite{Cooper2008,Wen:1995fg,Nayak2008,Qi2011}. For instance, some topological edge states do not satisfy the traditional statistics of fermions or bosons. These enigmatic \emph{anyons} \cite{Nayak2008,Qi2011}, have not been identified in Nature, but are expected to live as excitations in quantum Hall liquids \cite{Nayak2008} and topological superfluids \cite{Qi2011}. The unusual braiding properties associated with the so-called non-Abelian anyons \cite{Nayak2008}, together with their robustness against imperfections and noise, makes topological matter a promising candidate for building an error-free quantum computer \cite{Nayak2008}, which has all the potential to revolutionize modern technology. Generating and probing anyonic excitations in cold-atom setups, using artificial gauge fields, is certainly one of the most important goals in the field. One of the challenges and open questions would be how to develop schemes which allow for the creation and stability of such states, e.g. against various decay and heating processes that are generally present in experiments, and by doing so manipulate anyonic excitations in a well-controlled environment.

In this Review, we summarize different techniques for creating artificial gauge potentials in cold atom systems (both implemented and proposed), pedagogically describing the main physical mechanisms behind each.  We then illustrate the gauge potential's role in a number of applications and highlight the connections between these engineered gauge potentials and other branches of physics.  Since the publication of a shorter Review of Modern Physics Colloquium on artificial gauge fields for ultracold atoms  \cite{Dalibard2011}, there has been a great deal of theoretical and experimental activities in the area, which are reflected in the present Review.   

The manuscript is organized as follows. In Sect.~\ref{sect:rotation}, we review the initial technique where rotating gases experienced effective uniform magnetic fields.  In Sect.~\ref{sect:gaugefields} we present a general framework for light-induced gauge potentials.  In Sect.~\ref{sect:lightmatter} we study the basic interaction between laser fields and atoms, providing guidelines for designing realistic artificial gauge potentials in alkali systems. We then show in Sect.~\ref{sect:schemes} how light-matter interactions can optically induce Abelian gauge potentials for ultracold atoms.  Sect.~\ref{sect:soc} considers the schemes for generating non-Abelian gauge potentials and spin-orbit coupling for ultra cold atoms. In Sect.~\ref{sect:interactions}, we study how collisions are altered by the light-matter coupling, illustrating their role first for ground-state BEC's, and for the pairing mechanisms in interacting Fermi gases. Sect.~\ref{sect:lattices} adds optical lattices, and describe how synthetic magnetic fluxes and spin-orbit couplings can be engineered in a lattice environment. Then, in Sect.~\ref{sect:simulation}, we discuss several quantum simulators based on the gauge potential concept, and comment on experimental techniques for detecting the gauge field's effects. In Sect.~\ref{sect:dynamical}, we then briefly discuss the concept of interacting gauge theories with their unconventional current nonlinearities, and also recent proposals for emulating fully dynamical gauge fields in optical lattices.  We conclude by a summary of the current techniques available for creating artificial gauge potentials and indicate potential applications of these ideas.


\section{Non-inertial frame}\label{sect:non-inertial_frame} 



Artificial gauge fields result from spatially and/or temporally inhomogeneous
Hamiltonians. In this Section we consider two situations where one can
eliminate the time-dependence of a trapping potential by going
to a non-inertial frame of reference.

\subsection{Rotation\label{sect:rotation}}

Both conceptually, and experimentally, the most simple example of an artificial gauge field appears in a spatially rotating frame \cite{Cooper2008,Fetter2009}. 
This exploits the familiar equivalence between the Coriolis force in a
rotating system and the Lorentz force acting on a charged particle
in a uniform magnetic field.

\begin{figure}
\begin{centering}
\includegraphics[width=4in]{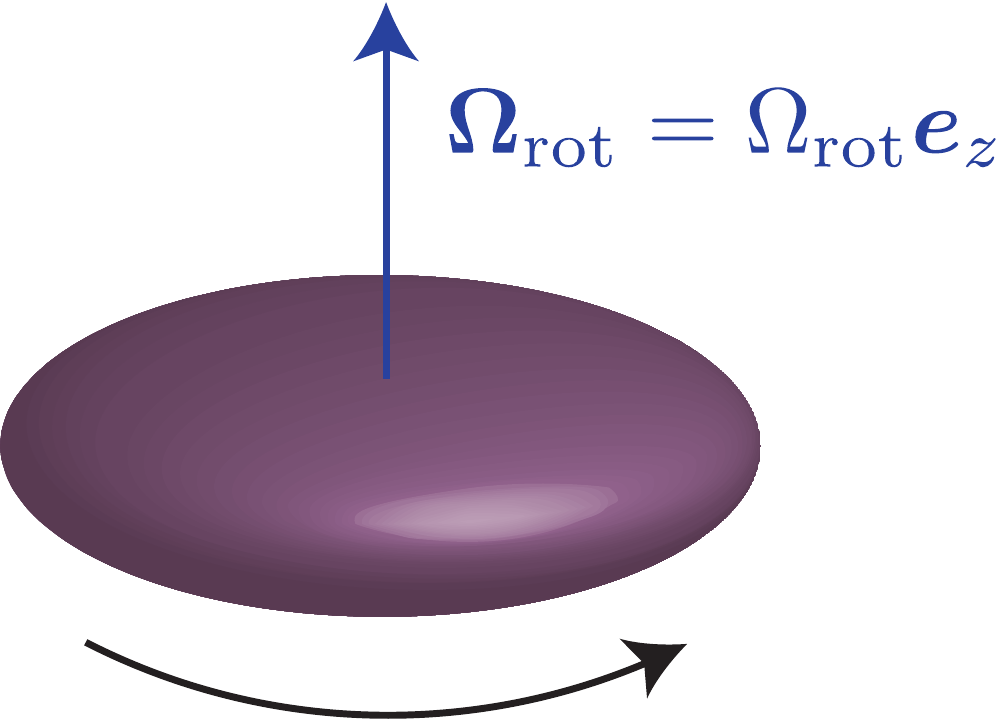} 
\par\end{centering}

\caption[Rotating atomic cloud]{The atomic cloud rotating with an angular frequency $\Omega_{\mathrm{rot}}$
around the $\mathbf{e}_{z}$ axis. }

\label{fig:rotating_cloud} 
\end{figure}

Let us consider in detail the quantum dynamics of an electrically
neutral atom in a trap rotating with an angular frequency $\Omega_{\mathrm{rot}}$
around the $\mathbf{e}_{z}$ axis, as depicted in Fig.~\ref{fig:rotating_cloud}. Recalling that a spatial rotation
by an angle $\theta=\Omega_{\mathrm{rot}}t$ around the rotation vector
$\boldsymbol{\Omega}_{\mathrm{rot}}=\Omega_{\mathrm{rot}}\mathbf{e}_{z}$
is described by the transformation $R_{z}(t)=\exp(-it\boldsymbol{\Omega}_{\mathrm{rot}}\cdot\mathbf{L}/\hbar)$
involing an orbital angular momentum operator $\mathbf{L}=\mathbf{r}\times\mathbf{p}$,
the atomic Hamiltonian is
\begin{equation}
H(t)=\frac{{\bf p}^{2}}{2m}+V\left(\mathbf{r}^{\prime}\right)\,,\quad\mathbf{r}^{\prime}=R_{z}(t)\mathbf{r}R_{z}^{\dagger}(t)\label{eq:H-initial}
\end{equation}
where $\mathbf{r}=x\mathbf{e}_{x}+y\mathbf{e}_{y}+z\mathbf{e}_{z}$, and $\mathbf{p}=p_x\mathbf{e}_{x}+p_y\mathbf{e}_{y}+p_z\mathbf{e}_{z}$ are respectively the position and momentum vectors in the inertial frame of reference, and $\mathbf{r}^{\prime}=x^{\prime}\mathbf{e}_{x}+y^{\prime}\mathbf{e}_{y}+z\mathbf{e}_{z}$ is the position vector in the rotating frame, with $x^{\prime}=x\cos\theta+y\sin\theta$ and $y^{\prime}=-x\sin\theta+y\cos\theta$.  The usual canonical commutation relations between the Cartesian components of the position and momentum vectors $\left[r_l,p_j\right] = i \hbar \delta_{l,j}$ allows us to represent the momentum vector $\mathbf{p}=-i\hbar\left(\mathbf{e}_{x}\partial_{x}+\mathbf{e}_{y}\partial_{y}+\mathbf{e}_{z}\partial_{z}\right)\equiv-i\hbar\boldsymbol{\nabla}$ for problems explicitly expressed in the coordinate representation.  Thus any time dependence of the trapping potential $V\left(\mathbf{r}^{\prime}\right)$ emerges exclusively through the temporal dependence of the rotating vector
$\mathbf{r}^{\prime}$. 

Because the magnitude of the momentum is unchanged by rotations,
$R_{z}(t)\mathbf{p}^{2}R_{z}^{\dagger}(t)=\mathbf{p}^{2}$, the Hamiltonian $H(t)$ is related to its time-independent counterpart via the unitary transformation $R_{z}(t)$ 
\begin{equation*}
H(t)=R_{z}(t)\left[\frac{{\bf p}^{2}}{2m}+V\left(\mathbf{r}\right)\right]R_{z}^{\dagger}(t)\,.\label{eq:H-initial-1}
\end{equation*}
The time-dependent \Schrodinger equation (TDSE) $i\hbar\partial_{t}\ket{\psi}=H(t)\ket{\psi}$
governs the system's dynamics.  Inserting the transformed wavefunction
$\ket{\psi}=R_{z}(t)\ket{\psi^{\prime}}$ into the TDSE yields a rotating frame TDSE
\begin{equation*}
i\hbar\partial_{t}\ket{\psi^{\prime}}=H^{\prime}\ket{\psi^{\prime}}\,,\label{eq:TDSE-rot-frame}
\end{equation*}
with the time-independent Hamiltonian 
\begin{equation}
H^{\prime}={\bf p}^{2}/2m+V\left(\mathbf{r}\right)-\boldsymbol{\Omega}_{\mathrm{rot}}\cdot\mathbf{L}\,,\label{eq:H-rot-frame}
\end{equation}
where the term $\boldsymbol{\Omega}_{\mathrm{rot}}\cdot\mathbf{L}=i\hbar R_{z}^{\dagger}\partial_{t} R_{z}$
results from the temporal dependence of $R_{z}(t)$.  Using $\boldsymbol{\Omega}_{\mathrm{rot}}\cdot\mathbf{L}=\left(\boldsymbol{\Omega}_{\mathrm{rot}}\times\mathbf{r}\right)\cdot\mathbf{p}$,
the Hamiltonian $H^{\prime}$ can be represented as 
\begin{equation}
H^{\prime}=\frac{\left({\bf p}-{{\mathbfcal A}}\right)^{2}}{2m}+V\left(\mathbf{r}\right)+W_{\mathrm{rot}}({\bf r})\,.\label{eq:RotHamiltonian}
\end{equation}
The emerging symmetric-gauge vector potential 
\begin{equation*}
{{\mathbfcal A}}=m\boldsymbol{\Omega}_{\mathrm{rot}}\times\mathbf{r}=m\Omega_{\mathrm{rot}}\left(x\mathbf{e}_{y}-y\mathbf{e}_{x}\right)\,,\label{eq:A-rot}
\end{equation*}
describes the cyclotron motion of the atom in the $\mathbf{e}_{x}-\mathbf{e}_{y}$
plane. An additional anti-trapping (centrifugal) potential 
\begin{equation*}
W_{\mathrm{rot}}({\bf r})=-\frac{{{\mathbfcal A}}^{2}}{2m}=-\frac{1}{2}m\Omega_{\mathrm{rot}}^{2}\left(x^{2}+y^{2}\right)\,,\label{eq:W-rot}
\end{equation*}
repels the atom away from the rotation axis $\mathbf{e}_{z}$. The
Hamiltonian (\ref{eq:RotHamiltonian}) has the same form as that for
a particle with a unit charge moving in a uniform magnetic field \cite{Landau:1987} ${\mathbfcal B}=\boldsymbol{\nabla}\times{{\mathbfcal A}}=2m\Omega_{\mathrm{rot}}\mathbf{e}_{z}$.
The above analysis does not involve any assumption concerning a specific
form of the trapping potential $V(\mathbf{r})$. Thus the
creation of an artificial magnetic flux via rotation can be applied
not only to the usual trapping potentials ~\cite{Matthews1999a,Madison2000,Abo-Shaeer2001}
and also to other structures, such as rotating optical lattices
\cite{Tung2006,Williams2010} or superfluid atom circuits with a rotating
weak link \cite{Wright13PRL}.  The centrifugal potential can compensate for harmonic trapping potentials
\begin{equation*}
V\left(\mathbf{r}\right)=\frac{1}{2}m\left(\omega_{x}^{2}x^{2}+\omega_{y}^{2}y^{2}\right)\label{eq:V-Harmonic},
\end{equation*}
when the rotation frequency approaches the trap frequencies $\Omega_{\mathrm{rot}}\rightarrow\omega_{x}$ and $\Omega_{\mathrm{rot}}\rightarrow\omega_{y}$.  In this limit, the
problem reduces to that of an unconfined free particle in the constant magnetic field $\mathbf{B}=2m\Omega\mathbf{e}_{z}$. Interestingly, the associated cyclotron frequency $\Omega_{\mathrm{c}}=B_{\mathrm{rot}}/m=2\Omega_{\mathrm{rot}}$ is twice the rotation frequency. 

Having seen how the single-particle Hamiltonian transforms into the
rotating frame, we now turn to the question of interactions. For now,
consider an arbitrary pairwise interaction $V\left(|{\bf r}_{1}-{\bf r}_{2}|\right)$
which is a function only of the separation between particles. Under
the transformation to the rotating frame the potential 
\begin{eqnarray*}
R_{z}(t)V\left(|{\bf r}_{1}-{\bf r}_{2}|\right)R_{z}^{\dagger}(t) & =V\left(|{\bf r}_{1}-{\bf r}_{2}|\right)
\end{eqnarray*}
 is unchanged. This follows from the identities $R_{z}(t){\bf r}^{2}R_{z}^{\dagger}(t)={\bf r}^{2}$, and $R_{z}(t){\bf r}_{1}\cdot{\bf r}_{2}R_{z}^{\dagger}(t)={\bf r}_{1}\cdot{\bf r}_{2}$
which simply state that relative geometry is not changed by rotations.
The potential $V\left(|{\bf r}_{1}-{\bf r}_{2}|\right)$
remains the same, but the two body problem does change due to the
emerging vector and centrifugal potentials.

\begin{figure}[tb]
\begin{center}
\includegraphics[width=4.5in]{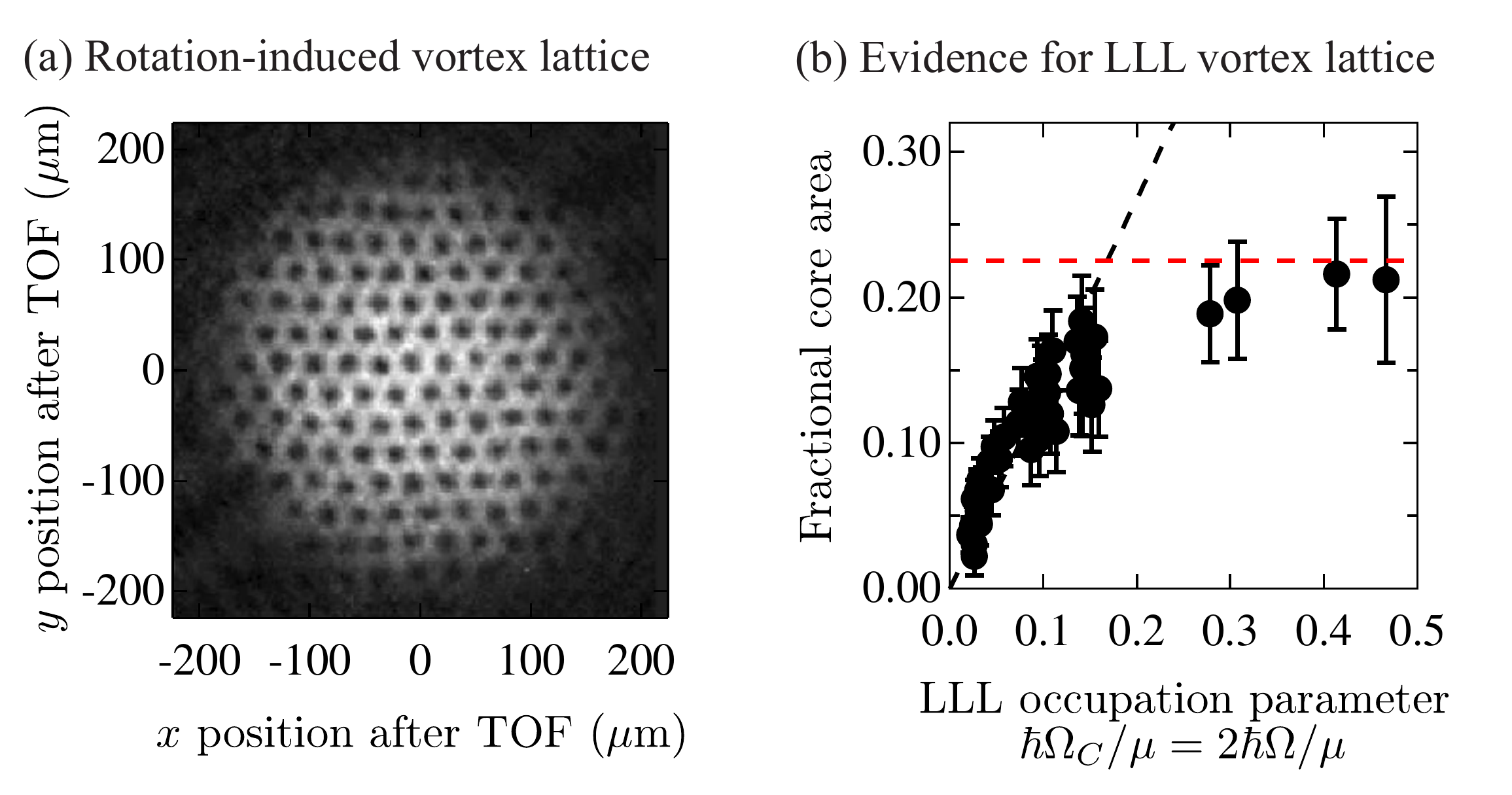}
\end{center}
\caption[Rapidly rotating Bose-Einstein condensates]{Rapidly rotating Bose-Einstein condensates.  (a) Representative image of a symmetrically trapped rapidly rotating BEC, with $\Omega/\omega_{x,y}\approx0.95$, showing a well ordered vortex lattice.  (b) Rotating systems entering the lowest Landau level (LLL).  For slowly rotating systems, vortices occupy only a small fraction of the system (black dashed line), whereas in the LLL the vortex density is constant (red-dashed line).  As the chemical potential falls below the effective cyclotron frequency, the BEC can be well described by a wavefunction projected into the LLL.  Figures are included by permission of Eric Cornell, and first appeared in [P. Engels, {\it et al.} PRL {\bf 90}, 170405 (2003), and V. Schweikhard, {\it et al.} PRL {\bf 92}, 040404 (2004)], Refs.~\cite{Engels2003} and \cite{Schweikhard2004}.
}
\label{fig_rotation}
\end{figure}

Ultracold atoms have been rotated to large angular frequency with
spectacular success (see Fig.~\ref{fig_rotation}a) by several groups,
lead by pioneering experiments at JILA~\cite{Matthews1999a}, ENS~\cite{Madison2000},
and MIT~\cite{Abo-Shaeer2001} in conventional harmonic traps.
These experiments addressed several important technical questions:
(1) how to start an ultracold atomic gas rotating; (2) how to keep
it rotating; and (3) how to detect rotation.

Questions (1) and (2) are related.  As is evident in Eq.~\ref{eq:RotHamiltonian}, the desired rotating frame Hamiltonian should have no remnant time-dependance.   The trapping potential must be asymmetric in the $\ex$-$\ey$ plane in order to induce rotation into an initially non-rotating system.  Generally this is achieved either by rotating an initially deformed harmonic trap, or by stirring with focused ``tweezer'' lasers. (The initial JILA experiment used an ingenious technique involving transitions between internal atomic states~\cite{Matthews1999a}, but adopted the deformed trap method to great success as described below.)  For experiments featuring the most rapid rotation -- equivalently the largest effective magnetic fields -- the trap potential is generally returned to near-perfect axial symmetry.  Any non-rotating component of the potential (in the lab frame) transforms to an unwanted rotating contribution (in the rotating frame) that can frictionally heat, or slow rapidly rotating clouds (the edges of which can easily exceed the critical velocity for superfluid flow in the lab frame).  With sufficient effort, it is possible to achieve nearly perfect axial symmetry, and in the later JILA experiments, there was no discernible decrease in angular frequency for the lifetime of their atomic ensembles~\cite{Engels2003,Schweikhard2004}.

For superfluid Bose gases, there are three primary techniques for detecting rotation.  Firstly, the presence of vortices in a BEC directly indicates the existence of an effective magnetic field, or rotation.  These vortices result in a change in the atomic density over a small region, with a length scale set by the condensate's healing length $\xi = \left(\hbar^2/2 m \mu\right)^{1/2}$, where $\mu$ is the chemical potential.  Typically $\xi$ is between $0.3\micron$ and $1\micron$: below the usual imaging resolution.  As a result, vortices have not been measured directly in the trap, and are revealed by removing the confining potential and allowing the BEC and vortices to expand before imaging.  Secondly, the evolution of collective modes directly reveals rotation ~\cite{Chevy2000,Haljan2001}.  Lastly, rotation can be inferred by observing the weakening of the trap from the centripetal anti-confinement~\cite{Schweikhard2004}.

In spite of the success creating large vortex lattices, rotating systems have not entered the strongly correlated regime where the filling factor $\nu \lesssim 10$.  This results from a technical limitation: at some point it becomes difficult to increase the angular momentum per particle and remain in equilibrium.  The most rapidly rotating BEC just barely entered the lowest Landau level (LLL) regime, where the vast majority of the particles reside in the LLL (unlike electron systems, this can occur at quite large $\nu$).  A BEC will enter the LLL when the occupation parameter $\hbar\Omega_c / \mu$ becomes large; the onset of this crossover was  observed by the JILA group (see Fig.~\ref{fig_rotation}b) using a series of ingenious evaporation tricks to increase the mean angular momentum per particle.  A secondary technical challenge for studying strongly correlated systems is to confine the atoms into 2D planes. The confinement along $\ez$ should be large compared to the radial trapping.  This is generally achieved using one-dimensional optical lattices, but technically it is difficult to create lattices with the required axial symmetry for rapid rotation experiments.  

Most experiments with rotating systems worked with large atomic gases in a single trapping potential, however, a recent experiment created an optical lattice with $\sim10$ atoms per site, where each lattice site was separately rotated~\cite{Gemelke2010} (as opposed to rotating the optical lattice in its entirety~\cite{Tung2006,Williams2010}).  In this regime, very rapid rotation is possible owing to the very strong confinement in individual lattice sites, and the authors argue they have entered the few-atom strongly correlated regime.

\subsection{Shaking\label{sub:Shaking}}

Besides rotation, shaking optical lattices is another
widely used experimental technique \cite{Madison1998,Struck2011,Struck2012,Arimondo2012,Struck:2013,Windpassinger2013RPP,Dalibard_time_dep},
which can trigger non-trivial topological effects and gauge structures in a rather direct manner \cite{Hauke:2012,Goldman:2014uz,Zhai2014-Floquet,Baur:2014ux,Eckardt-2014-unpubl}. Driven-induced gauge fields, using modulated cold-atom systems, is deeply related to the concept of ``Floquet topological states" \cite{Kitagawa2010,Galitski2011NP,Cayssol:2013gk} and strain-induced magnetic fields \cite{Levy:2010hl}, which are currently explored in solid-state laboratories. In particular, a shaken atomic system can be tailored so as to reproduce the dynamics of electronic systems subjected to circularly polarized light [see below and Ref. \cite{Zhai2014-Floquet}]; such an observation is instructive since light-driven electronic systems were predicted to produce Floquet topological states in materials such as graphene \cite{Oka2009,Galitski2011NP,Cayssol:2013gk}. 

To give a general description of the phenomenon,
let us consider atoms in an arbitrary trapping potential (e.g. 
a periodic optical lattice or a harmonic potential), which is affected by shaking. The Hamiltonian
$H(t)$ then has the form of Eq.~(\ref{eq:H-initial}), in which the
trapping potential $V\left(\mathbf{r}^{\prime}\right)$ depends on
the position vector  $\mathbf{r}^{\prime}\equiv\mathbf{r}^{\prime}(t)$ defined by
\begin{equation}
\mathbf{r}^{\prime}(t)=\mathbf{r}-\mathbf{r}_{0}\left(t\right)=R_{{\mathbf{r}}_{0}}(t)\mathbf{r}R_{{\mathbf{r}}_{0}}^{\dagger}(t)\,,\quad\mathrm{with}\quad R_{{\mathbf{r}}_{0}}(t)=\exp\left[-i\mathbf{p}\cdot {\mathbf{r}}_{0}\left(t\right)/\hbar\right]\,.\label{eq:shifted_frame}
\end{equation}
Namely, $\mathbf{r}^{\prime}(t)$ contains a time-dependent
shift $\mathbf{r}_{0}\left(t\right)$ with respect to the inertial frame vector $\mathbf{r}$, which depends on the shaking protocol \cite{Dalibard_time_dep}. 

In contrast to the case of rotation, the orientation of the coordinate system
is not altered. Only the origin $\mathbf{r}_{0}\left(t\right)$
changes in time. If the motion was chosen to be linear in time, ${\mathbf{r}}_{0}\left(t\right)=\mathbf{v}_{0}t$,
the transformation would imply a simple transition to another inertial frame.
Here, we are interested in another scenario, where the trapping potential
is modulated in a periodic manner, $\mathbf{r}_{0}\left(t+T\right)=\mathbf{r}_{0}\left(t\right)$,
with $T=2\pi/\omega$ being the period of the shaking. For instance, considering a single harmonic,
the linear and circular driving are respectively described by 
\begin{equation}
\mathbf{r}_{0}\left(t\right)=\kappa \mathbf{e}_{x}\sin\left(\omega t\right)\quad\mathrm{and}\quad\mathbf{r}_{0}\left(t\right)=\kappa \mathbf{e}_{x}\sin\left(\omega t\right)- \kappa \mathbf{e}_{y}\cos\left(\omega t\right)\,,\label{eq:linear-circular}
\end{equation}
where  $\kappa$ denotes the driving amplitude.

The transformation to the non-inertial frame does not alter the momentum,
$R_{{\mathbf{r}}_{0}}(t)\mathbf{p}R_{{\mathbf{r}}_{0}}^{\dagger}(t)=\mathbf{p}$,
so that the transformed Hamiltonian
reads 
\begin{align}
H^{\prime}&=R_{{\mathbf{r}}_{0}}^{\dagger}(t)H(t)R_{{\mathbf{r}}_{0}}(t)-i\hbar R_{{\mathbf{r}}_{0}}^{\dagger}(t)\partial_{t}R_{{\mathbf{r}}_{0}}(t) \notag \\
&={\bf p}^{2}/2m+V\left(\mathbf{r}\right)-\mathbf{p}\cdot\dot{\mathbf{r}}_{0}\left(t\right)\,,\label{eq:H-shaken-frame}\\
&=\frac{\left({\bf p}-{\mathbfcal A}\right)^{2}}{2m}+V\left(\mathbf{r}\right)\,,\quad\mathrm{with}\quad{\mathbfcal A}=m\dot{\mathbf{r}}_{0}\left(t\right),\label{eq:H-shaken-frame-1}
\end{align}
where the vector potential ${\mathbfcal A}\equiv{\mathbfcal A}\left(t\right)\,$
represents the momentum of a particle moving with the frame velocity
$\dot{\mathbf{r}}_{0}\left(t\right)$.  In the last relation \eqref{eq:H-shaken-frame-1}, we have neglected the uniform
term $-{\mathbfcal A}^2/2m$,
which can be eliminated by including a position-independent phase
factor to the atomic wave-function. 

The time-dependent vector potential yields a spatially uniform force
${\mathbfcal F}=-\dot{{\mathbfcal A}}\left(t\right)=-m\ddot{\mathbf{r}}_{0}\left(t\right)\,$
acting on atoms due to acceleration of the non-inertial frame.
In contrast to the transformation to a rotating frame considered in the previous Sect. \ref{sect:rotation},
the vector potential ${\mathbfcal A}\left(t\right)$ does not have any spatial dependence, but instead, 
it is time-dependent \cite{Dalibard_time_dep}. Note also that the vector
potential ${\mathbfcal A}\left(t\right)\,$ associated with a linear (resp. circular)
harmonic shaking given by Eq.~(\ref{eq:linear-circular}) is equivalent to the one emerging for a charged particle
in a linear (resp. circular) polarized electric field \cite{Cohen-Tannoudji1989,Jackson:book,Oka2009}. As mentioned above, this analogy with electronic systems emphasizes the possibility to generate driven-induced (Floquet) topological phases \cite{Cayssol:2013gk} with shaken optical lattices \cite{Hauke:2012,Goldman:2014uz,Zhai2014-Floquet,Baur:2014ux,Eckardt-2014-unpubl}. The effective gauge structures and topological phases emanating from lattice modulations will be further addressed in Section \ref{sec:shaking lattice}.


\section{Geometric gauge potentials}\label{sect:gaugefields}


\subsection{Formulation\label{sub:Formulation-gauge-potentials}}

Geometric gauge potentials arise throughout physics \cite{Mead:1979,Berry:1984,Wilczek:1984,Moody1986,Zygelman1987,Zee1988,Jackiw1988,Shapere1989,Mead1992,Bohm2003,Xiao2010,Zygelman2012}.
One place they can emerge in cold atom systems is when the atomic center of mass motion
is coupled to its internal (``spin'') degrees of freedom \cite{Dum1996,Visser1998,Dutta1999,Grynberg2001,Dudarev2004,Juzeliunas2004,Ruseckas2005,Juzeliunas2006,Zhu2006,Gunter2009,Spielman2009,Zhang:2005EPJD,Liu:2008review}.
To understand these gauge fields, we begin with the full (and in general
time-dependent) atomic Hamiltonian 
\begin{equation}
\hat{\tilde{H}}\equiv\hat{\tilde{H}}\left(\mathbf{r},t\right)=\left(\frac{{\bf p}^{2}}{2m}+V\right)\hat{1}+\hat{M}\,,\label{eq:H-full}
\end{equation}
where $V\equiv V\left(\mathbf{r},t\right)$ is the state-independent
trapping potential, $\mathbf{r}$ and $\mathbf{p}$ are the atomic
center of mass coordinate and momentum, and $\hat{1}$ is the unit
operator acting on the internal atomic degrees of freedom. The operator
$\hat{M}\equiv\hat{M}\left(\mathbf{r},t\right)$ includes the Hamiltonian
for the atomic internal motion, as well as the atom-light coupling
term. As a result, $\hat{M}$ explicitly depends on $\mathbf{r}$,
and in general time. Hats over operators (like $\hat{M}$ and $\hat{1}$)
signify that they act on the internal atomic degrees of freedom; center
of mass operators such as $\mathbf{r}$ and $\mathbf{p}$, will be
hatless. In this section, we adopt the coordinate representation for
the center of mass motion, so $\mathbf{r}$ simply denotes
the atomic position, and $\mathbf{p}=-i\hbar\boldsymbol{\nabla}$
is the associated momentum operator. In subsequent sections, we will
turn to the momentum representation to view the gauge potentials from
another angle~\cite{Spielman2009}.

The operator $\hat{M}\left(\mathbf{r},t\right)$ can be cast in terms
of the atomic bare internal states $\left|m\right\rangle $ 
\begin{equation}
\hat{M}\left(\mathbf{r},t\right)=\sum_{n,m=1}^{N}\left|n\right\rangle M_{nm}\left(\mathbf{r},t\right)\left\langle m\right|\,,\label{eq:state-vector-full-1}
\end{equation}
so the position and time dependence of $\hat{M}\left(\mathbf{r},t\right)$ comes exclusively from the matrix elements $M_{nm}\left(\mathbf{r},t\right)$. Here
 $N$ is the number of atomic internal states involved. 

\begin{figure}
\begin{centering}
\includegraphics[width=4in]{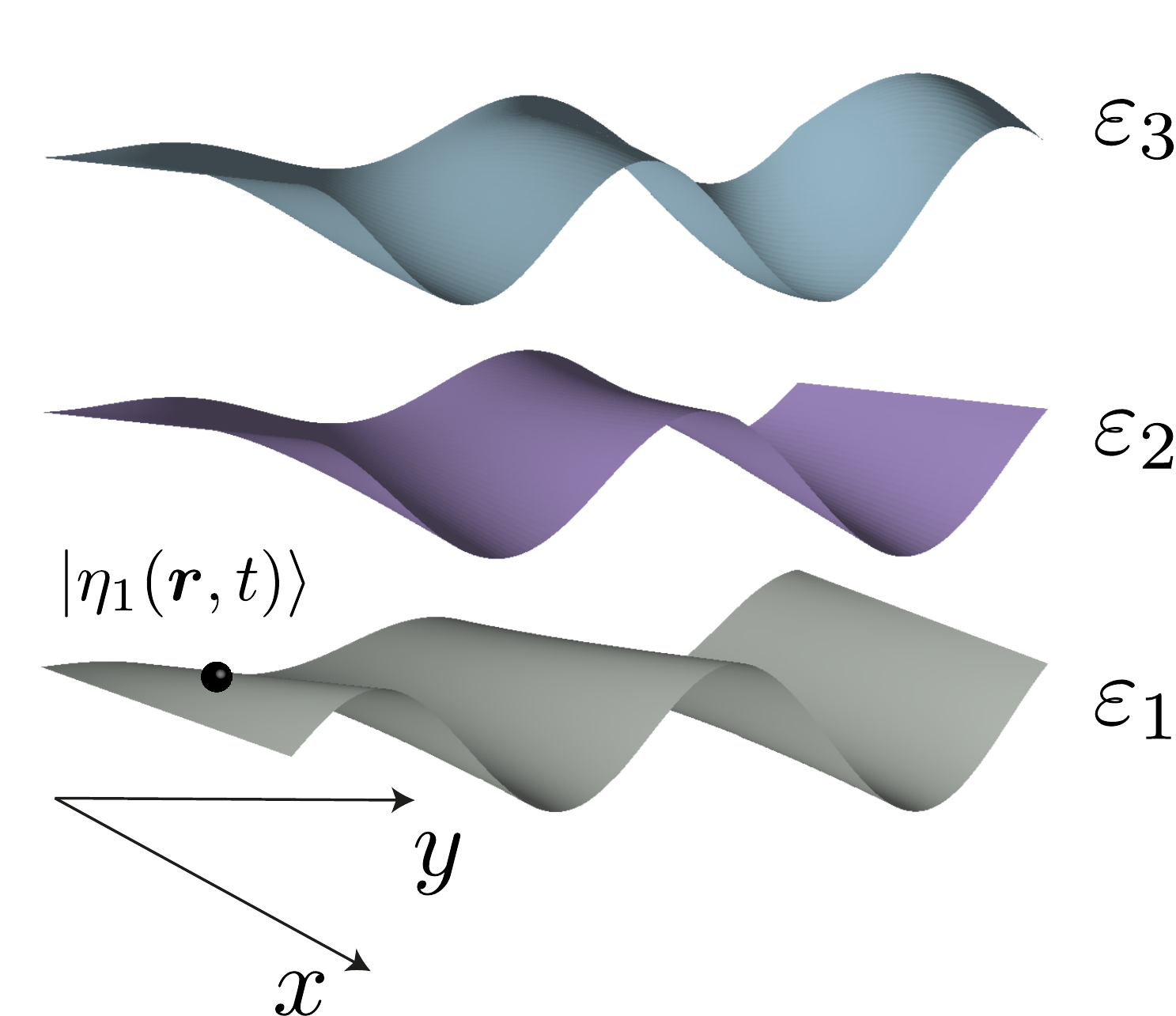} 
\par\end{centering}

\caption[Adiabatic surfaces]{Illustration of the position-dependence of the eigenenergies 
$\varepsilon_{m}\equiv\varepsilon_{m}\left(\mathbf{r},t\right)$  
of the atomic dressed states  
$\left|\eta_{m}\right\rangle \equiv\left|\eta_{m}\left(\mathbf{r},t\right)\right\rangle $. }

\label{fig:fig_adiabatic} 
\end{figure}

The diagonalisation of the operator $\hat{M}\left(\mathbf{r},t\right)$
provides a set of eigenstates $\left|\eta_{m}\right\rangle \equiv\left|\eta_{m}\left(\mathbf{r},t\right)\right\rangle $ ($m=1,\,2,\,\ldots\, N$),
known as the atomic dressed states, with eigenenergies $\varepsilon_{m}\equiv\varepsilon_{m}\left(\mathbf{r},t\right)$
that depend on the atomic position $\mathbf{r}$, as depicted in Fig.~\ref{fig:fig_adiabatic}. Any atomic state-vector can be expanded in this position-dependent basis 
\begin{equation}
\tilde{\ket{\psi}} =\sum_{ m=1}^{N}\psi_{m}\left(\mathbf{r},t\right)\left|\eta_{m}\left(\mathbf{r},t\right)\right\rangle \,,\label{eq:state-vector-full}
\end{equation}
where $\psi_{m}\left(\mathbf{r},t\right)\equiv\psi_{m}$ is a wave-function
for the centre of mass motion of the atom in the $m$-th internal
dressed state.

The atomic bare and dressed basis states are connected via a position-dependent
unitary transformation $\left|\eta_{m}\left(\mathbf{r},t\right)\right\rangle =\hat{R}\left|m\right\rangle $
which makes the operator $\hat{M}\left(\mathbf{r},t\right)$
diagonal:

\begin{equation}
\hat{R}^{\dagger}\hat{M}\left(\mathbf{r},t\right)\hat{R}=\hat{\varepsilon}\,,\quad\textrm{with}\quad\hat{\varepsilon}=\sum_{m=1}^{N}\left|m\right\rangle \varepsilon_{m}\left(\mathbf{r},t\right)\left\langle m\right|.\label{eq:M-transformed-R}
\end{equation}
Replacing the original state vector $\tilde{\ket{\psi}}$,
Eq.~(\ref{eq:state-vector-full}), by the transformed one
\begin{equation}
\ket{\psi} =\hat{R}^{\dagger}\tilde{\ket{\psi}} =\sum_{n=1}^{N} \psi_{n}  \left(\mathbf{r},t\right)\left|n\right\rangle  \,,\label{eq:state-vector-full-transformed}
\end{equation}
the Hamiltonian governing the
transformed TDSE $i\hbar\partial_{t} \tilde{\ket{\psi}} =\hat{H}\left(\mathbf{r},t\right) \tilde{\ket{\psi}} $
reads $\hat{H}=\hat{R}^{\dagger}\hat{\tilde{H}}\left(\mathbf{r},t\right)\hat{R}-i\hbar\hat{R}^{\dagger}\partial_{t}\hat{R}\,$,
giving 
\begin{equation}
\hat{H}=\frac{\left(\mathbf{p}-\hat{{\mathbfcal A}}\right)^{2}}{2m}+V+\hat{\varepsilon}+\hat{\Phi}\,.\label{eq:H-transformed-R}
\end{equation}
The vector operator 
\begin{equation}
\hat{{\mathbfcal A}}=i\hbar\hat{R}^{\dagger}\nabla\hat{R}=\sum_{n,m=1}^{N}\left|n\right\rangle {\mathbfcal A}_{nm}\left\langle m\right|,\quad{\mathbfcal A}_{nm}=i\hbar\left\langle \eta_{n}\right|\nabla\left|\eta_{m}\right\rangle \,\label{eq:A-nm}
\end{equation}
emerges due to the spatial dependence of the atomic dressed states, whereas
the additional scalar operator 
\begin{equation}
\hat{\Phi}=-i\hbar\hat{R}^{\dagger}\partial_{t}\hat{R}=\sum_{n,m=1}^{N}\left|n\right\rangle \Phi_{nm}\left\langle m\right|\,,\quad\Phi_{nm}=-i\hbar\left\langle \eta_{n}\right|\partial_{t}\left|\eta_{m}\right\rangle \label{eq:Phi_nm}
\end{equation}
arises because of their temporal dependence. For the sake of simplicity,
we have omitted the unit operator $\hat{1}$ multiplying the momentum
operator $\mathbf{p}$ and the state-independent potential $V$ in
the transformed Hamiltonian $\hat{H}$, Eq.~(\ref{eq:H-transformed-R}).

\subsection{Adiabatic approximation\label{sub:Adiabatic-approximation}}

If a subset containing $q\le N$ dressed states is well separated
in energy from the remaining ones, it is appropriate to make an adiabatic
(Born-Oppenheimer) approximation by projecting 
the system's dynamics onto the truncated space of the internal states, using the operator
$\hat{P}^{\left(q\right)}=\sum_{m=1}^{q}\left|m\right\rangle \left\langle m\right|$.
The Hamiltonian $\hat{H}^{\left(q\right)}=\hat{P}^{\left(q\right)}\hat{H}\hat{P}^{\left(q\right)}$
describing such a reduced dynamics reads 
\begin{equation}
\hat{H}^{\left(q\right)}=\frac{\left(\mathbf{p}-\hat{{\mathbfcal A}}^{\left(q\right)}\right)^{2}}{2m}+\hat{V}_{\mathrm{tot}}^{\left(q\right)}\,,\qquad\hat{V}_{\mathrm{tot}}^{\left(q\right)}=V+\hat{\varepsilon}^{\left(q\right)}+\hat{\Phi}^{\left(q\right)}+\hat{W}^{\left(q\right)},\label{eq:H-matrix-reduced}
\end{equation}
where $\hat{\varepsilon}^{\left(q\right)}$, $\hat{{\mathbfcal A}}^{\left(q\right)}$
and $\hat{\Phi}^{\left(q\right)}$ are the projections onto the reduced
subspace of the corresponding operators featured in the full Hamiltonian
$\hat{H}$, Eq.~(\ref{eq:H-transformed-R}). Here we have omitted the projector $\hat{P}^{\left(q\right)}$ multiplying $\mathbf{p}$
and the state-independent potential $V$. 

An extra operator
\begin{equation}
\hat{W}^{\left(q\right)}=\frac{1}{2m}\hat{P}^{\left(q\right)}\hat{{\mathbfcal A}}\left(\hat{1}-\hat{P}_{q}\right)\hat{{\mathbfcal A}}\hat{P}^{\left(q\right)}\label{eq:W-q-general}
\end{equation}
 with the matrix elements 
\begin{equation}
W_{nm}=\frac{1}{2m}\sum_{l=q+1}^{N}{\mathbfcal A}_{nl}\cdot{\mathbfcal A}_{lm}\,,\quad n\,,m=1,\,\ldots\,,q\,\label{eq:W-nm}
\end{equation}
results from projecting $\hat{{\mathbfcal A}}^{2}$ onto the selected
subspace of the internal dressed states. The potential $\hat{W}^{\left(q\right)}$
can be interpreted as a kinetic energy of the atomic micro-trembling
due to off-resonance non-adiabatic transitions to the omitted dressed
states with $m>q$ \cite{Aharonov1992,Cheneau2008}. 

In this way, the geometric vector potential $\hat{{\mathbfcal A}}^{\left(q\right)}$ and scalar potential $\hat{W}^{\left(q\right)}$ emerge from the atomic dressed states'
spatial dependence. The potential $\hat{\Phi}^{\left(q\right)}$ on the other hand, stems
from their time dependence. The latter $\hat{\Phi}^{\left(q\right)}$
describes the population transfer between the atomic levels due to
the temporal dependence of the external fields \cite{Unanyan98OC,Unanyan99PRA,Juzeliunas207LJP}.

When the truncated space includes a single dressed state ($q=1$)
well separated from the others, as is the case in Fig.~\ref{fig:fig_adiabatic}, the vector and scalar potentials reduce
to the ordinary commuting vector and scalar fields, 
\begin{equation}
\hat{{\mathbfcal A}}^{\left(q\right)}\rightarrow{\mathbfcal A}=i\hbar\left\langle \eta_{1}\right|\nabla\left|\eta_{1}\right\rangle \quad\textrm{and}\quad\hat{V}_{\mathrm{tot}}^{\left(q\right)}\rightarrow V_{\mathrm{tot}}=\frac{1}{2m}\sum_{l=2}^{N}{\mathbfcal A}_{nl}\cdot{\mathbfcal A}_{lm}\,.\label{eq:A-V--q-eq-1}
\end{equation}
In this case, the resulting artificial electric and magnetic fields
are the standard ${\mathbfcal E}=-\boldsymbol{\nabla}V_{\mathrm{tot}}-\partial_{t}{\mathbfcal A}$
and ${\mathbfcal B}=\boldsymbol{\nabla}\times{\mathbfcal A}$. However,
for $q>1$ the more general relations discussed in the following are
required.

\subsection{Artificial magnetic and electric fields: Abelian and non-Abelian cases}
\label{abvsnonab}

The vector and scalar potentials $\hat{{\mathbfcal A}}^{\left(q\right)}$
and $\hat{W}^{\left(q\right)}$ featured in the projected Hamiltonian
$\hat{H}^{\left(q\right)}$, Eq.~(\ref{eq:H-matrix-reduced}), can
be related to the operator generalizations of conventional magnetic
and electric fields. The situation is more complicated than for classical
electromagnetism, because the scalar potential and the Cartesian components
of the vector potential are now operators which do not necessarily
commute.

To elucidate the problem, we turn to the Heisenberg equations of motion
governing the projected dynamics. The velocity operator defined via
the Heisenberg equation is 
\begin{equation}
\hat{\mathbf{v}}=-\frac{i}{\hbar}[\mathbf{r},\hat{H}^{\left(q\right)}]=\frac{1}{m}(\mathbf{p}-\hat{{\mathbfcal A}^{\left(q\right)}})\,.\label{eq:v-explicit}
\end{equation}
 The acceleration is defined via the Heisenberg equation for the velocity
operator $\dot{\hat{\mathbf{v}}}=\partial_{t}\hat{\mathbf{v}}-i[\hat{\mathbf{v}},\hat{H}^{\left(q\right)}]/\hbar$.
Since the only explicit time dependence of $\hat{\mathbf{v}}$ resides
in $\hat{{\mathbfcal A}}^{\left(q\right)}$, then $\partial_{t}\hat{\mathbf{v}}=\partial_{t}\hat{{\mathbfcal A}}^{\left(q\right)}/m$,
giving 
\begin{equation}
\dot{\hat{\mathbf{v}}}=-\frac{1}{m}\partial_{t}\hat{{\mathbfcal A}}^{\left(q\right)}-\frac{im}{2\hbar}[\hat{\mathbf{v}},\hat{v}^{2}]-\frac{i}{\hbar}[\hat{\mathbf{v}},\hat{V}_{\mathrm{tot}}^{\left(q\right)}].\label{eq:v-derivative}
\end{equation}
 This equivalently gives the Cartesian components of the acceleration
\begin{equation}
\dot{\hat{v}}_{k}=-\frac{1}{m}\partial_{t}\hat{{\mathcal{A}}}_{k}^{\left(q\right)}-\frac{im}{2\hbar}(\hat{v}_{l}[\hat{v}_{k},\hat{v}_{l}]+[\hat{v}_{k},\hat{v}_{l}]\hat{v}_{l})-\frac{i}{\hbar}[\hat{v}_{k},\hat{V}_{\mathrm{tot}}^{\left(q\right)}],\label{eq:v-derivative-components}
\end{equation}
 where a summation over the repeated Cartesian indices is again implied.
Expressing the velocity commutators $m^{2}[\hat{v}_{k},\hat{v}_{l}]=i\hbar\hat{{\mathcal{F}}}_{kl}$
in terms of the antisymmetric tensor 
\begin{equation}
\hat{{\mathcal{F}}}_{kl}=\partial_{k}{\hat{\mathcal{A}}}_{l}^{\left(q\right)}-\partial_{l}{\hat{\mathcal{A}}}_{k}^{\left(q\right)}-\frac{i}{\hbar}[{\hat{\mathcal{A}}}_{k}^{\left(q\right)},{\hat{\mathcal{A}}}_{l}^{\left(q\right)}]\,,\label{eq:v-k--v-l-commutators}
\end{equation}
 one arrives at the equation of motion 
\begin{equation}
m\dot{\mathbf{\hat{v}}}=\frac{1}{2}(\hat{\mathbf{v}}\times\hat{{\mathbfcal B}}^{\left(q\right)}-\hat{{\mathbfcal B}}^{\left(q\right)}\times\hat{\mathbf{v}})+\hat{{\mathbfcal E}}^{\left(q\right)}\,.\label{eq:Heisenberg-eq-v-result}
\end{equation}
 The vector operator 
\begin{equation}
\hat{{\mathbfcal B}}^{\left(q\right)}=\nabla\times\hat{{\mathbfcal A}}^{\left(q\right)}-\frac{i}{\hbar}\hat{{\mathbfcal A}}^{\left(q\right)}\times\hat{{\mathbfcal A}}^{\left(q\right)}\,\label{eq:B}
\end{equation}
with components $\hat{{\mathcal{B}}}_{j}^{\left(q\right)}=\frac{1}{2}\epsilon_{jkl}\,\hat{{\mathcal{F}}}_{kl}$
is the artificial magnetic field (Berry curvature) providing the Lorentz
force. Using Eq. (\ref{eq:A-nm}) for ${\mathbfcal A}_{nm}$ together
with the completeness relation, the matrix elements of the curvature
can be represented as a sum over the eliminated states 
\begin{equation}
{\mathbfcal B}_{nm}=-\frac{i}{\hbar}\sum_{l=q+1}^{N}{\mathbfcal A}_{nl}\times{\mathbfcal A}_{lm}\,,\quad n,m=1,\,\ldots\,,q\,.\label{eq:B_nm}
\end{equation}
Hence, the Berry curvature is non-zero only for the reduced atomic
dynamics $(q<N)$ when some of the atomic states are eliminated. The
same applies to the geometric scalar potential $W_{nm}$ given by
Eq.(\ref{eq:W-nm}).

Additionally the atom is affected by an effective electric field 
\begin{equation}
\hat{{\mathbfcal E}}^{\left(q\right)}=-\partial_{t}\hat{{\mathbfcal A}}^{\left(q\right)}-\nabla\hat{V}_{\mathrm{tot}}+\frac{i}{\hbar}[\hat{{\mathbfcal A}}^{\left(q\right)},\hat{V}_{\mathrm{tot}}^{\left(q\right)}]\,\label{eq:E-Prime}
\end{equation}
 that contains a commutator $[\hat{{\mathbfcal A}}^{\left(q\right)},\hat{V}_{\mathrm{tot}}^{\left(q\right)}]$
together with the usual gradient $\nabla\hat{V}_{\mathrm{tot}}^{\left(q\right)}$
and induced $\partial_{t}\hat{{\mathbfcal A}}^{\left(q\right)}$ contributions.

 When all the Cartesian components of $\hat{{\mathbfcal A}}^{\left(q\right)}$
commute with each other, the vector potential is said to be Abelian (this is always
the case when $q=1$). Otherwise, when some components do not commute, $\left [ \hat{{\mathcal A}}^{\left(q\right)}_j , \hat{{\mathcal A}}^{\left(q\right)}_l \right ] \ne 0 $ for some $j$ and $l$, we will state that the system exhibits a non-Abelian gauge potential $\hat{{\mathbfcal A}}^{\left(q\right)}$ \footnote{We note that genuine non-Abelian properties are captured by the non-Abelian character of the field strength  \cite{Mead1992},  $[\hat{\mathcal{F}}_{kl} (\bs r), \hat{\mathcal{F}}_{k^{\prime} l^{\prime}} (\bs r^{\prime})] \ne 0$, or by the non-commutativity of successive loop operations, see e.g. \cite{Zhang:2008na}. However, in this Review, a vector potential satisfying the criterium $[A_j,A_k] \ne 0$ will generally be referred to as a ``non-Abelian gauge potential". This issue will be further addressed in the Section \ref{non-Abelian_Wilson}.}.

For Abelian vector
potentials, we recover ${\mathbfcal B}^{\left(q\right)}=\nabla\times{\mathbfcal A}^{\left(q\right)}$
and ${\mathbfcal E}^{\left(q\right)}=-\partial_{t}{\mathbfcal A}^{\left(q\right)}-\nabla V_{\mathrm{tot}}^{\left(q\right)}$.
An Abelian geometric vector potential $\hat{{\mathbfcal A}}^{\left(q\right)}$
has  already been engineered for ultracold atoms in the cases 
$q=1$ \cite{Lin2009a,Lin2009b,Lin2011a} and $q=2$ \cite{Lin2011,Wang2012,Cheuk2012,Zhang2012PRL,Fu2013PRA,Fu2013,Zhang2013PRA,Qu2013,LeBlanc2013}.
For a single adiabatic state ($q=1$) the spatial dependence of $\hat{{\mathbfcal A}}^{\left(q\right)}$
yield an artificial magnetic field~\cite{Lin2009a,Lin2009b} and
its temporal dependence generate an effective electric field~\cite{Lin2011a}. 
We shall return to the latter issue in the Sec. \ref{sec:Electric field} on experimentally creating the artificial electric field.

\subsection{Geometric gauge potentials and rotation\label{sub:Geometric-gauge-potentials}}

Now we turn to an interplay between geometric gauge potentials and
rotation, motivated by a possibility for simultaneously generating
an Abelian vector potential (via rotation) and a light-induced non-Abelian
geometric vector potential (equivalent to spin-orbit coupling). Here
we outline a general framework for adding an Abelian gauge potential
via rotation to a non-Abelian geometric gauge potential~\cite{Lu2007,Burrello:2010,Radic2011PRA,Zhou-Zhoua-Wu2011PRA,Xu2011PRL,Liu-Liu2012PRA,Zhang:2013PLA}.
This scenario has several associated subtle issues: (1) In proper rotation
experiments, the whole Hamiltonian should be static in the rotating
frame. In this context, both the center of mass and coupling terms
should be rotating, implying that the whole laser system should rotate.
(2) Generally, techniques for creating artificial gauge fields involve
Raman-resonant transitions, and in such cases the {}``orientation''
of Zeeman terms results from a relative phase (for the $\ex$ and
$\ey$ components) and detuning from resonance (for the $\ez$ component),
not geometry. The following discussion focuses on the broad picture
and will not treat these points.

Let us suppose that the system
is described by the Hamiltonian (\ref{eq:H-full}), in which both
the center of mass and internal contributions rotate with the frequency
$\Omega_{\mathrm{rot}}$ around $\mathbf{e}_{z}$. Both the state-independent
and state-dependent potentials are assumed to be time-independent
in the rotating frame of reference, so that in the lab frame the time-dependence
of $V\left(\mathbf{r},t\right)=V\left(\mathbf{r}^{\prime}\right)$
and $\hat{M}\left(\mathbf{r},t\right)=\hat{M}\left(\mathbf{r}^{\prime}\right)$
come exclusively from the rotating radius vector $\mathbf{r}^{\prime}\equiv\mathbf{r}^{\prime}\left(\mathbf{r},t\right)$.
This means the trapping potential as well as all the lasers producing
the atom-light coupling $\hat{M}$ should rotate with the same frequency
$\Omega_{\mathrm{rot}}$ around the rotation vector $\boldsymbol{\Omega}_{\mathrm{rot}}$. 
Reference~\onlinecite{Radic2011PRA} includes
a comparison with an experimentally more feasible situation where the center of mass
degrees of freedom are set into rotation, but the lasers (and hence
internal degrees of freedom) are not. 

\begin{figure}
\begin{centering}
\includegraphics[width=4in]{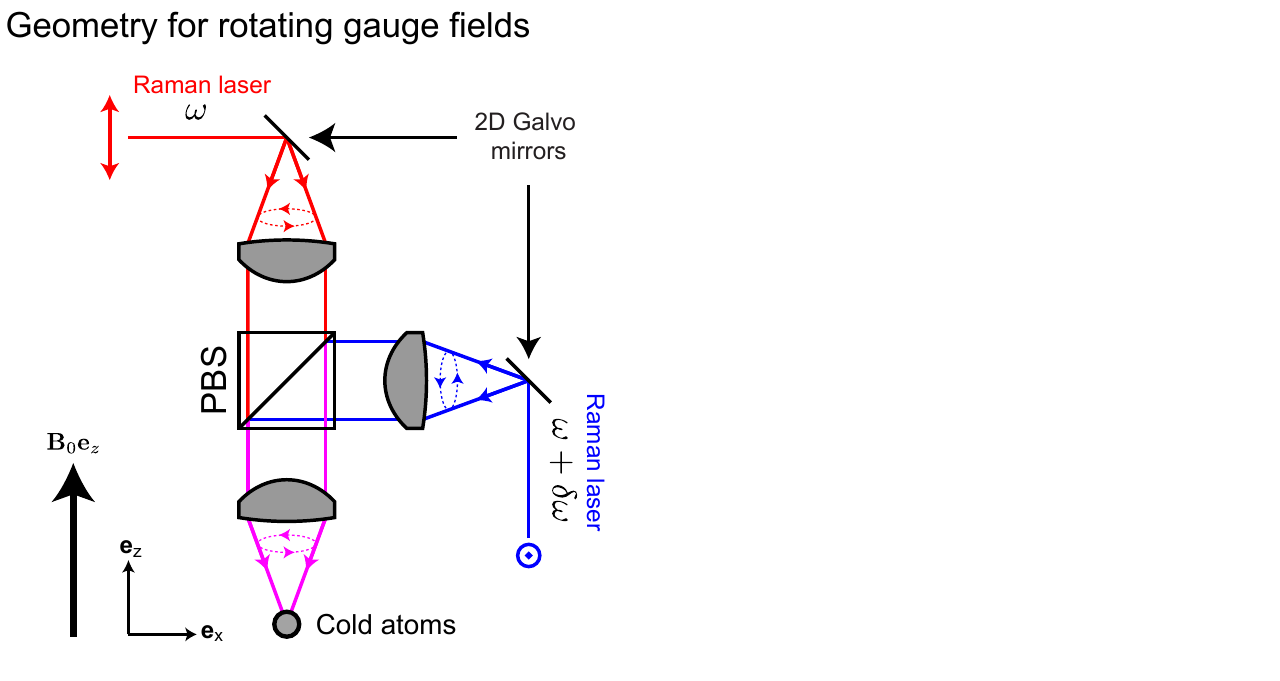} 
\par\end{centering}

\caption[Rotating Raman lasers]{Laser geometry for creating rotating artificial gauge fields.  ``NIST-style'' artificial gauge fields~\cite{Lin2009a,Lin2009b} and spin orbit coupling~\cite{Lin2011} can be induced in the rotating frame using a pair of two-dimensional galvanometer-based scanning mirrors (``galvo's'') that steer the Raman beams on circular trajectories in an otherwise circularly symmetric confining potential.  In this example the lasers are combined on a polarizing beam splitter (PBS), producing the polarizations required to drive Raman transitions.  Because of the rotating orientation of the lasers' polarization vectors, there are in principle additional polarization present in this diagram, however, these lead only to changes in the overall scalar potential.  This contribution can be tuned to zero by suitable choice of wavelength.}
\label{Fig_RotatingGaugeFields} 
\end{figure}

Note that realizing the rotating laser configuration is a challenging task, but  this geometry can be realized by extending techniques for creating dynamical~\cite{Huckans2009a} and rotating~\cite{Williams2010} optical lattices.  Figure~\ref{Fig_RotatingGaugeFields} depicts the required experimental geometry.  This setup uses a pair of scanning galvanometer-based mirrors that direct the Raman lasers to intersect at the depicted cloud of atoms with wave-vectors ${\bf k}_1$ and ${\bf k}_2$, with difference $\boldsymbol\delta {\bf k} = {\bf k}_1 - {\bf k}_2$ that rotates sinusoidally in the $\ex-\ey$ plane. 

Assuming that the internal degrees of freedom are characterized by
the total angular momentum $\hat{{\bf F}}$, the transformation
to the rotating frame of the original Hamiltonian $\hat{\tilde{H}}$
{[}Eq.~(\ref{eq:H-full}){]} is described by the unitary operator
$\hat{R}_{z}^{\mathrm{full}}(t)\equiv\exp\left[-it\boldsymbol{\Omega}_{\mathrm{rot}}\cdot(\mathbf{L}+\hat{\mathbf{F}})/\hbar\right]$
involving the sum of the center of mass angular momentum $\mathbf{L}=\mathbf{r}\times\mathbf{p}$
and the internal angular momentum $\hat{\mathbf{F}}$. Transforming to the rotating frame
the full atomic state-vector $\tilde{\ket{\psi}} =\hat{R}_{z}^{{\rm full}}(t)\tilde{\ket{\psi^{\prime}}}$,
one arrives at a TDSE for $\tilde{\ket{\psi^{\prime}}}$
described by the Hamiltonian $\hat{\tilde{H}}^{\prime}$. The latter is connected to the laboratory frame Hamiltonian $\hat{\tilde{H}}$ as 
$\hat{\tilde{H}}^{\prime}=\hat{\tilde{H}}-\boldsymbol{\Omega}_{\mathrm{rot}}\cdot(\mathbf{L}+\hat{\mathbf{F}})$ and hence 
\begin{equation}
\hat{\tilde{H}}^{\prime}=\left[\frac{\left(\mathbf{p}-{\mathbfcal A}_{\mathrm{rot}}\right)^{2}}{2m}+V\left(\mathbf{r}\right)+W_{\mathrm{rot}}\left(\mathbf{r}\right)\right]\hat{I}+\hat{M}\left(\mathbf{r}\right)-\boldsymbol{\Omega}_{\mathrm{rot}}\cdot\hat{\mathbf{\mathbf{F}}}\,,\label{eq:H-full-rotating-frame}
\end{equation}
where the rotation vector potential ${\mathbfcal A}_{\mathrm{rot}}=m\boldsymbol{\Omega}_{\mathrm{rot}}\times\mathbf{r}$
and the antitrapping (centrifugal) potential $W_{\mathrm{rot}}=-{\mathcal{A}}_{\mathrm{rot}}^{2}/2m$
are as in Eq. (\ref{eq:A-rot}). The additional term $-\boldsymbol{\Omega}_{\mathrm{rot}}\cdot\hat{\mathbf{\mathbf{F}}}$
is analogous to the Hamiltonian describing the atomic internal spin
in a magnetic field proportional to the rotation vector $\boldsymbol{\Omega}_{\mathrm{rot}}$.
This is not surprising, because the rotation manifests itself like
the magnetic field not only for the centre of mass motion, but also
for the internal dynamics of the atom. Consequently the operator $-\boldsymbol{\Omega}_{\mathrm{rot}}\cdot\hat{\mathbf{F}}$
introduces a $\hbar\Omega_{\mathrm{rot}}$ shift of the atomic spin
states. For typical rotation experiments, the shift $\hbar\Omega_{\mathrm{rot}}\lesssim h\times100\Hz$
is small compared to the other frequencies characterizing the atomic
internal dynamics.

The state-dependent operator $\hat{M}\left(\mathbf{r}\right)-\boldsymbol{\Omega}_{\mathrm{rot}}\cdot\hat{\mathbf{\mathbf{F}}}$
featured in $\hat{\tilde{H}}^{\prime}$ has a set of time-independent
eigenstates $\left|\eta_{m}\right\rangle \equiv\left|\eta_{m}\left(\mathbf{r}\right)\right\rangle $
with eigenenergies $\varepsilon_{m}\equiv\varepsilon_{m}\left(\mathbf{r}\right)$.
Subsequently, like in the Sec. \ref{sub:Formulation-gauge-potentials},
the (rotating frame) atomic state-vector $\tilde{\ket{\psi^{\prime}}}$
is transformed to $\tilde{\ket{\psi^{\prime}}}=\hat{R}^{\dagger} \ket{\psi^{\prime}}$
by means of the unitary transformation $\hat{R}\equiv\hat{R}\left(\mathbf{r},t\right)=\sum_{m=1}^{N}\left|\eta_{m}\left(\mathbf{r}\right)\right\rangle \left\langle m\right|$
converting the atomic bare states into the dressed ones. The transformed
Hamiltonian $\hat{H}^{\prime}=\hat{R}^{\dagger}\hat{\tilde{H}}^{\prime}\left(\mathbf{r},t\right)\hat{R}-i\hbar\hat{R}^{\dagger}\partial_{t}\hat{R}\,$
has the same form as Eq.(\ref{eq:H-transformed-R}) with the geometric
vector potential $\hat{{\mathbfcal A}}$ replaced by $\hat{{\mathbfcal A}}+{\mathbfcal A}_{\mathrm{rot}}$
and the centrifugal potential $W_{\mathrm{rot}}$ added: 
\begin{equation}
\hat{H}^{\prime}=\frac{\left(\mathbf{p}-{\mathbfcal A}_{\mathrm{rot}}-\hat{{\mathbfcal A}}\right)^{2}}{2m}+V+W_{\mathrm{rot}}+\hat{\varepsilon}\,,\label{eq:H-transformed-R-1}
\end{equation}
where we put $\Phi=0$ because the dressed states are time-independent
in the rotating frame. 

The adiabatic approximation outlined in Sect.~\ref{sub:Adiabatic-approximation}
gives the effective Hamiltonian for the atomic dynamics in the reduced
internal space
\begin{equation}
\hat{H}^{\prime\left(q\right)}=\frac{\left(\mathbf{p}-{\mathbfcal A}_{\mathrm{rot}}-\hat{{\mathbfcal A}}^{\left(q\right)}\right)^{2}}{2m}+\hat{V}_{\mathrm{tot}}^{\left(q\right)}\,,\qquad\hat{V}_{\mathrm{tot}}^{\left(q\right)}=V+W_{\mathrm{rot}}+\hat{\varepsilon}^{\left(q\right)}+\hat{W}^{\left(q\right)}\,,\label{eq:H-matrix-reduced-1}
\end{equation}
where the geometric scalar potential $\hat{W}^{\left(q\right)}$ is
as in Eqs.(\ref{eq:W-q-general})-(\ref{eq:W-nm}) . Thus, rotation
adds an Abelian potential ${\mathbfcal A}_{\mathrm{rot}}$ to the
geometric vector potential $\hat{{\mathbfcal A}}^{\left(q\right)}$, and also introduces
the centrifugal potential $W_{\mathrm{rot}}=-{\mathcal{A}}_{\mathrm{rot}}^{2}/2m$.


\section{Light matter interaction}\label{sect:lightmatter}





In this Section, we study the basic interaction between laser-light
and atoms, before explicitly analyzing light-induced gauge potentials.
Specifically, we consider the far off-resonant 
coupling~\cite{Happer1967,Cohen-Tannoudji1972,Deutsch1998,Grimm:2000}
between an alkali atom in its electronic ground state manifold and
an oscillatory optical field, providing guidelines for designing realistic
artificial gauge fields in alkali systems. The same basic line of
reasoning is generally valid in other atomic systems, but the ground
and excited states will of course differ, and specific conclusions
may not cross over~\cite{Cui2013}.

\subsection{Light-matter coupling} We focus on the largest light-matter coupling term -- the electric
dipole (see Ref.~\cite{Cohen-Tannoudji1998}, for example) -- that
links electronic motion to the optical electric field ${\bf E(t)}$
with vector components $E_{j}$ ($j=1,2,3$) and frequency $\omega$ 
\begin{equation}
\hat{H}_{{\rm dip}}=\hat{{\bf d}}\cdot{\bf E(t)}=\hat{d}_{j}E_{j}\cos(\phi_{j}-\omega t)\,,\label{eq:H-dip}
\end{equation}
where the summation over repeated Cartesian indices is assumed. Here,
$\hat{{\bf d}}=-e\sum_{\alpha}\hat{{\bf r}}_{\alpha}$ is the electric
dipole operator, $e$ is the electron's charge, and $\hat{{\bf r}}_{\alpha}$
is the position of the $\alpha$'th electron within the atom. We will
study the impact of this coupling term on the atomic ground state
manifold to second order in perturbation theory, in a step-by-step
manner including different effects in order of decreasing importance.

\begin{figure}
\begin{centering}
\includegraphics[width=5in]{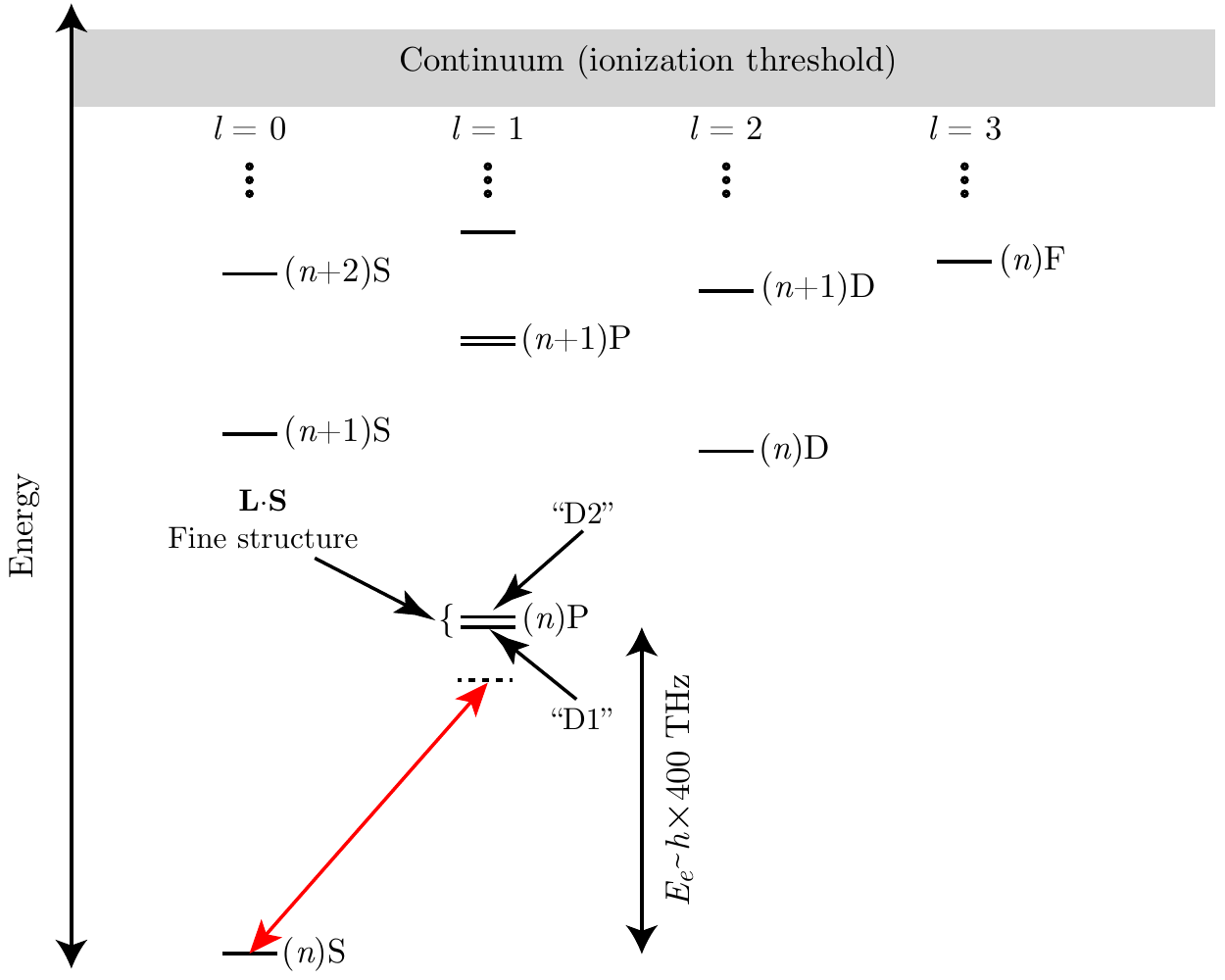} 
\par\end{centering}

\caption[Alkali atom level structure]{Typical alkali atom level structure. This figure illustrates the
typical level structure of the alkali atoms with the primary $n{\rm S}$
to $n{\rm P}$ transition (with their typical $\approx h\times400\THz$ transition energies, and labeled by the archaic D1 and D2 designations for the two fine-structure resolved group of states) required for the discussed herein.  This transition is driven by the
far-detuned coupling laser indicated in red. Also pictured is the
much smaller $\Delta_{{\rm FS}}$ fine-structure splitting (ranging
from $\approx10\GHz$ in $^{6}{\rm Li}$ to $\approx17\THz$ in $\Cs133$)
that enables state-dependent ``vector'' light shift. }

\label{fig:AlkaliLevelStructure} 
\end{figure}

We analyze the lowest energy electric dipole transition in an alkali
atom, between the ground $(n){\rm S}$ electron orbital and the excited
$(n){\rm P}$ orbital with excitation energy $E_{e}$, as pictured
in Fig.~\ref{fig:AlkaliLevelStructure}. The atom is then described
by the Hamiltonian 
\begin{align}
\hat{H}_{{\rm at}} & =E_{e}\hat{P}_{e}+\frac{A_{{\rm FS}}}{\hbar^{2}}\hat{{\bf L}}\cdot\hat{{\bf S}},\label{eq:AtomicFineStructure}
\end{align}
 where $A_{{\rm FS}}$ is the fine-structure coupling constant; $\hat{P}_{g,e}$
are the projectors onto the space of ground or excited states; and
$\hat{{\bf L}}$ and $\hat{{\bf S}}$ are the total electronic orbital
angular momentum and the spin, respectively. (The electronic orbital
momentum $\hat{{\bf L}}$ should not be confused with the orbital
momentum of the atomic center of mass motion ${\bf L}$ appearing
in the analysis of the rotating systems in the Sects.~\ref{sect:rotation}
and \ref{sect:gaugefields}.)

The orbital angular momentum $\hat{{\bf L}}$ has eigenstates $\{\ket{l=0,m_{L}=0}$,
$\ \ket{l=1,m_{L}=0,\pm1}\}$, with eigenvalues $\hbar^{2}l(l+1)$
and $\hbar m_{L}$ for $\hat{{\bf L}}^{2}$ and $\hat{L}_{z}$, respectively;
the electron spin $\hat{{\bf S}}$ along the quantization axis $\ez$
has the eigenstates $\left\{ \ket{s=1/2,m_{S}=\pm1/2}\right\} $.
In a similar manner, the atomic nuclear spin $\hat{{\bf I}}$ provides
the eigenstates $\left\{ \ket{i,m_{I}=-i,\cdots,i}\right\} $. Because
the electronic degree of freedom {[}$(n)$ or $(n+1)${]} is uniquely
defined by the total electronic angular momentum quantum number, the
operator $\hat{P}_{g}=\hat{1}-\hat{L}^{2}/2\hbar^{2}$ projects onto
the space of ground electronic states {[}$(n){\rm S}$, with $l=0${]}
and $\hat{P}_{e}=\hat{L}^{2}/2\hbar^{2}$ projects onto the set of
electronic excited states {[}$(n){\rm P}$, with $l=1${]}, as long
as only these two sets of states are involved.

In the alkali atoms, there are three excited-state energy scales: 
 the ground-excited splitting $E_{e}$, the fine
structure splitting $\Delta_{{\rm FS}}$ (with eigenstates of the combined angular momentum
$\hat{{\bf J}}=\hat{{\bf L}}+\hat{{\bf S}}$), and 
the typical scale of hyperfine splittings $\Delta_{{\rm HFS}}$ (with eigenstates of the
total angular momentum $\hat{{\bf F}}=\hat{{\bf L}}+\hat{{\bf S}}+\hat{{\bf I}}$).
In any given alkali atom, these are ordered $E_{e}\gg\Delta_{{\rm FS}}\gg\Delta_{{\rm HFS}}$.
 In an experiment, the detuning from the excited states is generally large
compared to $\Delta_{{\rm HFS}}$. Consequently, to analyze the effect
of far off-resonant light fields on ground state atoms, it suffices
to include just the electronic excited level structure as corrected
by the $\hat{{\bf L}}\cdot\hat{{\bf S}}$ fine-structure term. The
hyperfine-structure of the excited level enables further transitions,
but these can only be resolved when the lasers are tuned so near to the atomic resonance
that the rate of spontaneous emission prohibits any practical use for
engineering artificial gauge fields, or quantum simulation in general.

\subsection{Rotating wave approximation}

We focus on optical fields not too far from resonance, $|E_{e}-\hbar\omega|\ll E_{e}$.
In that case it is convenient to transform the atomic excited state
vectors $\ket{e}$ into the rotating frame $\ket{e^{\prime}}=\exp(i\omega t)\ket{e}$
via the unitary transformation 
\begin{equation}
\hat{U}_{{\rm rot}}(t)=\exp(-i\omega t\hat{P}_{e})\equiv\hat{P}_{g}+\hat{P}_{e}\exp(-i\omega t),\label{eq:U}
\end{equation}
 where the transformed state vector is $\ket{\psi^{\prime}}=\hat{U}_{{\rm rot}}^{\dagger}(t)\ket{\psi}$,
and the projection operators $\hat{P}_{g}$ and $\hat{P}_{e}$ are
as defined above. In this temporally rotating frame, the dipole Hamiltonian
{[}Eq.~(\ref{eq:H-dip}){]} is 
\begin{equation}
\hat{H}_{{\rm dip^{\prime}}}=\hat{U}_{{\rm rot}}^{\dagger}(t)\hat{H}_{{\rm dip}}\hat{U}_{{\rm rot}}(t)\approx\frac{1}{2}\left[\tilde{E}_{j}^{*}\hat{P}_{g}\hat{d}_{j}\hat{P}_{e}+\tilde{E}_{j}\hat{P}_{e}\hat{d}_{j}\hat{P}_{g}\right],\label{eq:H-dip-rot}
\end{equation}
 where we introduced the complex electric field $\tilde{E}_{j}=E_{j}\exp(i\phi_{j})$
and applied the rotating wave approximation (RWA) by removing the
terms oscillating at the frequencies $\omega$ and $2\omega$ in the
second relation of Eq.~(\ref{eq:H-dip-rot}). In the rotating frame
and after the RWA, the full atomic and coupling Hamiltonian is 
\begin{align}
\hat{H}_{{\rm full}} & =\hat{H}_{{\rm at}^{\prime}}+\hat{H}_{{\rm dip^{\prime}}},\label{eq:AtomicFineStructureRWA}
\end{align}
 where the rotating frame atomic Hamiltonian $\hat{H}_{{\rm at}^{\prime}}=\hat{H}_{{\rm at}}-\hbar\omega\hat{P}_{e}$
acquires an extra term $-\hbar\omega\hat{P}_{e}\equiv-i\hbar\hat{U}_{{\rm rot}}^{\dagger}(t)\left[\partial_{t}\hat{U}_{{\rm rot}}(t)\right]$
from the the unitary transformation's temporal dependence. Hence,
the excited state energy $E_{e}$ is supplanted by the detuning $\Delta_{e}=E_{e}-\hbar\omega$
in the transformed Hamiltonian 
\begin{align}
\hat{H}_{{\rm at}^{\prime}} & =\Delta_{e}\hat{P}_{e}+\frac{A_{{\rm FS}}}{\hbar^{2}}\hat{{\bf L}}\cdot\hat{{\bf S}}\equiv\Delta_{e}\hat{P}_{e}+\frac{A_{{\rm FS}}}{2\hbar^{2}}\left(\hat{J}^{2}-\hat{L}^{2}-\hat{S}^{2}\right).\label{eq:H_0}
\end{align}

\subsection{Effective atomic ground-state Hamiltonian}

\subsubsection{General}

Within the ground state manifold (adiabatically eliminating the excited
states), the effect of the RWA electric dipole term $\hat{H}_{{\rm dip^{\prime}}}$
is described to second-order by the effective atomic Hamiltonian 
\begin{align}
\hat{H}_{{\rm eff}} & =-\hat{P}_{g}\hat{H}_{{\rm dip^{\prime}}}\hat{H}_{{\rm at}^{\prime}}^{-1}\hat{H}_{{\rm dip^{\prime}}}\hat{P}_{g}\,,
\label{eq:H_eff-def}
\end{align}
which can be represented as
\begin{align}
\hat{H}_{{\rm eff}} & =-\frac{1}{4}\tilde{E}_{i}^{*}\hat{D}_{i,j}\tilde{E}_{j}\,.\label{eq:H_eff}
\end{align}
Here we have introduced the rank-2 Cartesian tensor operator 
\begin{align}
\hat{D}_{i,j} & =\hat{P}_{g}\hat{d}_{i}\hat{P}_{e}\hat{H}_{{\rm at}^{\prime}}^{-1}\hat{P}_{e}\hat{d}_{j}\hat{P}_{g}\equiv\hat{P}_{g}\hat{d}_{i}\hat{H}_{{\rm at}^{\prime}}^{-1}\hat{d}_{j}\hat{P}_{g}\label{eq:D_ij}
\end{align}
 acting on the ground state manifold. In the second relation of Eq.~\eqref{eq:D_ij}
we have omitted the excited state projectors $\hat{P}_{e}$, because
the atomic ground states do not have a permanent electric dipole moment:
$\hat{P}_{g}\hat{d}_{j}\hat{P}_{g}=0$.

Operators of this form can be expressed as the sum 
\begin{align}
\hat{D}_{i,j} & =\hat{D}_{i,j}^{(0)}+\hat{D}_{i,j}^{(1)}+\hat{D}_{i,j}^{(2)}\label{eq:D_ij-alternative}
\end{align}
 of irreducible tensor operators of rank-0 (transforms as a scalar
under rotation) 
\begin{align}
\hat{D}_{i,j}^{(0)} & =\frac{1}{3}\delta_{i,j}{\rm Tr}\hat{D}\equiv\frac{1}{3}\hat{D}_{l,l}\delta_{i,j}\,,\label{eq:D_ij-0}
\end{align}
 rank-1 (transforms as a vector under rotation) 
\begin{align}
\hat{D}_{i,j}^{(1)} & =\frac{1}{2}\left(\hat{D}_{i,j}-\hat{D}_{j,i}\right)\equiv\frac{1}{2}\epsilon_{i,j,k}\epsilon_{i^{\prime},j^{\prime},k}\hat{D}_{i^{\prime},j^{\prime}}\,,\label{eq:D_ij-1}
\end{align}
 and rank-2 
\begin{align}
\hat{D}_{i,j}^{(2)} & =\frac{1}{2}\left(\hat{D}_{i,j}+\hat{D}_{j,i}\right)-\frac{1}{3}\hat{D}_{i,j}^{(0)}\,,\label{eq:D_ij-2}
\end{align}
 where the summation over the repeated Cartesian indices is again
implied. This separation allows one to classify the light-matter interaction
in a powerful way.

\subsubsection{No fine structure}

Let us begin by considering the simple case when $A_{{\rm FS}}=0$,
implying that the excited states are degenerate, and allowing us to
completely neglect electronic and nuclear spin. In this case, we replace
the atomic Hamiltonian $\hat{H}_{{\rm at}^{\prime}}$ with $\Delta_{e}\hat{P}_{e}$,
and the tensor $\hat{D}_{i,j}$ in Eq.~\eqref{eq:D_ij} takes the
particularly simple form 
\begin{align}
\hat{D}_{i,j} & =\Delta_{e}^{-1}\hat{P}_{g}\hat{d}_{i}\hat{d}_{j}\hat{P}_{g}.\label{eq:LightShiftNoFS}
\end{align}
 In the present case, let us study how Eq.~\eqref{eq:LightShiftNoFS}
transforms under rotations as effected by the unitary operator $\hat{R}({\boldsymbol{\theta}})=\exp(-i\hat{{\bf L}}\cdot{\boldsymbol{\theta}}/\hbar)$.
Because the ground S state has $l=0$, it is clear that $\hat{L}_{x,y,z}\hat{P}_{g}=0$,
therefore 
\begin{align}
\hat{R}({\boldsymbol{\theta}})\hat{D}_{i,j}\hat{R}^{\dagger}({\boldsymbol{\theta}}) & =\hat{D}_{i,j}.
\end{align}
 Thus $\hat{D}_{i,j}$ transforms like a scalar under rotation and
must be proportional to the unit tensor. Consequently only the zero-rank
contribution $\hat{D}_{i,j}^{(0)}$ is non-zero 
\begin{align}
\hat{D}_{i,j}^{(0)} & =\hat{P}_{g}\frac{\hat{d}_{l}\hat{d}_{l}}{3\Delta_{e}}\delta_{ij}\hat{P}_{g}=-4u_{s}\delta_{ij}\hat{P}_{g},\label{eq:D^0_ij}
\end{align}
 where $u_{s}$ is proportional to the atoms' ac polarizability (without
including the fine-structure corrections) 
\begin{align}
u_{s} & =-\frac{\left|\bra{l=0}|{\bf d}|\ket{l^{\prime}=1}\right|^{2}}{12\left(E_{e}-\hbar\omega\right)}.\label{eq:u_s}
\end{align}
 Here $\left|\bra{l=0}|{\bf d}|\ket{l^{\prime}=1}\right|^{2}\equiv\sum_{m_{L}^{\prime}=0,\pm1}\left|\bra{l=0,m_{L}=0}{\bf d}\ket{l^{\prime}=1,m_{L}^{\prime}}\right|^{2}\approx4 e^2 a_{0}^{2}$
is the commonly used reduced matrix element and $a_{0}$ is the Bohr
radius (the reduced matrix element is a single number, even though
the conventional notation makes it look like it should have three
vector components).

Equations \eqref{eq:H_eff} and \eqref{eq:D^0_ij} yield the \textit{scalar
light shift} 
\begin{align}
\hat{H}_{{\rm eff}}^{(0)} & =u_{s}|\tilde{{\bf E}}|^{2}\hat{P}_{g},\label{eq:H_eff^0}
\end{align}
 Alternatively \textit{$\hat{H}_{{\rm eff}}^{(0)}$} can be expressed
as 
\begin{align}
\hat{H}_{{\rm eff}}^{(0)} & =-\frac{3\pi\hbar^{3}c^{2}I}{2E_{e}^{3}}\frac{\hbar\Gamma_{1\rightarrow0}}{\Delta_{e}},\label{eq:H_eff^0-alt}
\end{align}
where we defined the intensity $I=\epsilon_{0}c|\tilde{{\bf E}}|^{2}/2$, used
the speed of light $c$, the transition's natural linewidth $\Gamma_{1\rightarrow0}$,
the electric constant $\epsilon_{0}$, and the expression 
\begin{align}
\hbar\Gamma_{l^{\prime}\rightarrow l} & =\frac{E_{e}^{3}}{3\pi\epsilon_{0}\hbar^{3}c^{3}}\frac{2l+1}{2l^{\prime}+1}\left|\bra{l}|{\bf d}|\ket{l^{\prime}}\right|^{2}\label{eq:Gamma}
\end{align}
 for the linewidth~\cite{Loudon2000}. This result is valid whenever
it is safe to ignore the excited state fine structure; a good zero-order
approximation when the $\Delta_{e}=E_{e}-\hbar\omega$ detuning from
the optical transition is much larger than $\Delta_{{\rm FS}}$. This
result reminds us of several important facts: the scalar light shift
is a potential that is independent of optical polarization, scales
like $1/\Delta_{e}$, and for red-detuned beams ($\Delta_{e}>0$)
is attractive.

\subsubsection{Including fine structure\label{sub:Including-fine-structure}}

Let us now turn to the effects of the fine structure by recalling
that the excited eigenstates $\left|j,m_{J}\right\rangle $ of the
Hamiltonian $\hat{H}_{{\rm at}^{\prime}}$ Eq.~\eqref{eq:H_0} corresponding
to $l=1$ are characterised by the quantum numbers $j$ and $m_{J}$
of the combined angular momentum $\hat{{\bf J}}=\hat{{\bf L}}+\hat{{\bf S}}$,
with $j=1/2$ and $j=3/2$. Specifically we have $\hat{H}_{{\rm at}^{\prime}}\left|j,m_{J}\right\rangle =\Delta_{j}\left|j,m_{J}\right\rangle $,
where $\Delta_{1/2}=\Delta_{e}-A_{FS}$ and $\Delta_{3/2}=\Delta_{e}+A_{FS}/2$
are the detunings for the corresponding fine structure transitions.

Next we shall determine the tensor $\hat{D}_{i,j}=\hat{P}_{g}\hat{d}_{i}\hat{H}_{{\rm at}^{\prime}}^{-1}\hat{d}_{j}\hat{P}_{g}$
describing the light-induced coupling between the atomic ground states,
without explicitly involving the excited fine-structure states $\left|j,m_{J}\right\rangle $.
For this we use Eq.~\eqref{eq:H_0} for $\hat{H}_{{\rm at}^{\prime}}$
and write down a Dyson-type equation for the inverse atomic Hamiltonian
$\hat{H}_{{\rm at}^{\prime}}^{-1}$ projected onto the excited-state
manifold, 
\begin{align}
\hat{H}_{{\rm at}^{\prime}}^{-1} & =\frac{1}{\Delta_{e}}\hat{P}_{e}-\frac{\alpha}{\Delta_{e}}\hat{{\bf L}}\cdot\hat{{\boldsymbol{\sigma}}}\hat{H}_{{\rm at}^{\prime}}^{-1},\label{eq:AtomicFineStructureSolved-expansion}
\end{align}
with $\alpha=A_{{\rm FS}}/2\hbar\Delta_{e}$ and $\hat{{\bf S}}=\hbar\hat{{\boldsymbol{\sigma}}}/2$.


Calling on the communtation relations $\left[\hat{L}_{i},\hat{d}_{j}\right]=i\hbar\epsilon_{ijk}\hat{d}_{k}$
\cite{Landau:1987} together with $\hat{{\bf L}}\hat{P}_{g}=0$, we replace $\hat{P}_{g}\hat{d}_{i}\hat{{\bf L}}\cdot\hat{{\boldsymbol{\sigma}}}$
with the commutator $-\hat{P}_{g}\left[\hat{{\bf L}}\cdot\hat{{\boldsymbol{\sigma}}}\,,\hat{d}_{i}\right]=i\hbar\epsilon_{i\ell m}\hat{P}_{g}\hat{d}_{m}\hat{\sigma}_{l}$.
Using the relation $\hat{P}_{g}\hat{d}_{i}\hat{d}_{j}\hat{P}_{g}=\left|\langle|\!|{\bf d}|\!|\rangle\right|^{2}\hat{P}_{g}\delta_{ij}/3$,
Eqs.~\eqref{eq:D_ij} and \eqref{eq:AtomicFineStructureSolved-expansion}
yield 
\begin{align}
\hat{D}_{ij} & =\frac{\left|\langle|\!|{\bf d}|\!|\rangle\right|^{2}}{3\Delta_{e}}\delta_{ij}\hat{P}_{g}-i\alpha\hbar\epsilon_{i\ell m}\hat{\sigma}_{\ell}\hat{D}_{mj}\,,\label{eq:D_ij-Dyson}
\end{align}
 where the reduced matrix element $\left|\langle|\!|{\bf d}|\!|\rangle\right|^{2}\equiv\left|\bra{l=0}\!|{\bf d}|\!\ket{l^{\prime}=1}\right|^{2}$
has already been featured in the previous Subsection. As Eq.~\eqref{eq:D_ij-Dyson}
contains operators that act only within the electronic ground state,
we may now omit the projectors $\hat{P}_{g}$. Furthermore, since
the only operators acting in the electronic ground state are the Pauli
matrices, $\hat{D}_{ij}$ must take the form 
\begin{equation}
\hat{D}_{ij}=D^{(0)}\delta_{ij}+iD^{(1)}\epsilon_{ijk}\sigma_{k}\,,\label{eq:D-^ij-Ansatz}
\end{equation}
 where we also used the fact that $\hat{D}_{ij}$ can be decomposed
into a scalar and vector component (no rank-2 component is possible
with Pauli matrices alone).

Replacing the proposed solution \eqref{eq:D-^ij-Ansatz}, permuting
one Levi-Civita symbol and exploiting the identity $\hat{\sigma}_{i}\hat{\sigma}_{j}=\delta_{ij}+i\epsilon_{ijk}\sigma_{k}$,
Eq.~\eqref{eq:D_ij-Dyson} gives 
\begin{align*}
\left(D^{(0)}-2\alpha\hbar D^{(1)}\right)\delta_{ij}+i\epsilon_{ijk}\left(D^{(1)}-\alpha\hbar D^{(0)}-\alpha\hbar D^{(1)}\right)\hat{\sigma}_{k} & =\frac{\left|\langle|\!|{\bf d}|\!|\rangle\right|^{2}}{3\Delta_{e}}\delta_{ij}\,.
\end{align*}
 The resulting pair of linear equations has solutions 
\begin{align*}
D^{(0)} & =\frac{1}{3}\left(\frac{2}{\hbar\alpha+1}-\frac{1}{2\hbar\alpha-1}\right)\frac{\left|\langle|\!|{\bf d}|\!|\rangle\right|^{2}}{3\Delta_{e}} & {\rm and} &  & D^{(1)} & =\frac{\alpha\hbar}{1-\hbar\alpha}D^{(0)}.
\end{align*}
 In the initial notation with $\hbar\alpha=A_{{\rm FS}}/2\Delta_{e}$
and $\Delta_{e}=E_{e}-\hbar\omega$, we associate the denominators
appearing in the scalar coefficient $D^{(0)}$ with the detuning from
the excited-state fine-structure split levels $E_{{\rm D1}}=E_{e}-A_{{\rm FS}}$
and $E_{{\rm D2}}=E_{e}+A_{{\rm FS}}/2$, giving 
\begin{align*}
D^{(0)} & =\frac{\left|\langle|\!|{\bf d}|\!|\rangle\right|^{2}}{9}\left(\frac{2}{E_{D2}-\hbar\omega}+\frac{1}{E_{D1}-\hbar\omega}\right).
\end{align*}
 Additionally introducing an average $\bar{E}=(2E_{{\rm D1}}+E_{{\rm D2}})/3$
and the fine-structure splitting $\Delta_{{\rm FS}}=3A_{{\rm FS}}/2$,
the complete tensor operator takes the form 
\begin{align}
\hat{D}_{ij} & =D^{(0)}\left[\delta_{ij}+i\epsilon_{ijk}\frac{2}{3\hbar}\frac{\Delta_{{\rm FS}}}{\bar{E}-\hbar\omega}\hat{J}_{k}\right]\,,
\end{align}
 where we used $\hat{J}_{i}=\hat{S}_{i}$ as is suitable in the $l=0$
electronic ground state. This leads to the light shift 
\begin{align}
\hat{H}_{{\rm eff}} & =-\frac{1}{4}\tilde{E}_{i}^{*}\hat{D}_{i,j}\tilde{E}_{j}=\left[u_{{\rm s}}\left(\tilde{{\bf E}}^{*}\cdot\tilde{{\bf E}}\right)+\frac{iu_{{\rm v}}\left(\tilde{{\bf E}}^{*}\times\tilde{{\bf E}}\right)}{\hbar}\cdot\hat{{\bf J}}\right]\hat{P}_{g}\label{eq:H_eff-result}
\end{align}
 in terms of the scalar and vector polarizabilities 
\begin{align}
u_{{\rm s}} & =-\frac{\left|\langle|\!|{\bf d}|\!|\rangle\right|^{2}}{36}\left(\frac{2}{E_{D2}-\hbar\omega}+\frac{1}{E_{D1}-\hbar\omega}\right), & {\rm and} &  & u_{{\rm v}} & =\frac{2u_{{\rm s}}\Delta_{{\rm FS}}}{3\left(\bar{E}-\hbar\omega\right)}\,.\label{eq:u_s,v}
\end{align}
 When $E_{D2}=E_{D1}$, we recover our previous result, Eqs.~\eqref{eq:u_s}-\eqref{eq:H_eff^0}.

As an illustration, consider an atom illuminated with circularly $\sigma^{+}$
(along $\ez$) polarized light, $\tilde{{\bf E}}=-|\tilde{{\bf E}}|^{2}\left(\ex+i\ey\right)/\sqrt{2}$.
For this field $\tilde{{\bf E}}^{*}\times\tilde{{\bf E}}=-i|\tilde{{\bf E}}|^{2}\ez$,
so the vector light shift is described by the operator $\hat{J}_{z}$
acting on the ground state manifold. Consequently $\hat{H}_{{\rm eff}}^{(1)}$
provides opposite light shifts for the spin-up and spin-down atomic
ground states.

\subsubsection{Complete electronic ground state\label{sub:Complete-electronic-ground}}

In the presence of an external magnetic field, an alkali atom in its
electronic ground state manifold is described by the Hamiltonian 
\begin{align}
\hat{H}_{{\mathbf{B}}}= & A_{{\rm hf}}\hat{\mathbf{I}}\!\cdot\!\hat{\mathbf{J}}+\frac{\mu_{B}}{\hbar}{\mathbf{B}}\!\cdot\!\left(g_{J}\hat{\mathbf{J}}+g_{I}\hat{\mathbf{I}}\right),\label{eq:H_B}
\end{align}
 where $\mu_{B}$ is the Bohr magneton (henceforth the projector $\hat{P}_{g}$
is kept implicit in the ground state operators). The first term in
Eq.~\eqref{eq:H_B} takes into account the coupling between the electron
and nuclear spins of the atom, and is described by the magnetic dipole
hyperfine coefficient $A_{{\rm hf}}$. The second (Zeeman) term includes
separate contributions from the electronic spin $\hat{\mathbf{J}}=\hat{\mathbf{L}}+\hat{\mathbf{S}}$
and the nuclear angular momentum $\hat{\mathbf{I}}$, along with their
respective Land\'e factors $g_{J}$ and $g_{I}$. For the alkali atoms
$|g_{I}/g_{J}|\simeq0.0005$, so we will safely neglect the nuclear
contribution $\mu_{B}g_{I}{\mathbf{B}}_{{\rm }}\cdot\hat{\mathbf{I}}/\hbar$
to the Zeeman term.

The previous Subsection showed that laser fields induce the
scalar and vector light shifts featured in the effective ground-state
Hamiltonian $\hat{H}_{{\rm eff}}$, Eq.~\eqref{eq:H_eff-result}.
The two terms can be independently controlled by choosing the laser
frequency $\omega$ and polarization. Evidently, the vector light
shift is a contribution to the total Hamiltonian acting like an effective
magnetic field \cite{Happer1967,Deutsch1998,Dudarev2004,Sebby-Strabley2006,Juz-Spielm2012NJP}
\begin{align}
{\mathbf{B}}_{{\rm eff}} & =\frac{iu_{v}(\tilde{\mathbf{E}}^{*}\!\times\!\tilde{\mathbf{E}})}{\mu_{B}\g_{J}}\label{eq:B_eff-electr}
\end{align}
 that affects $\hat{\mathbf{J}}$ but not the nuclear spin $\hat{\mathbf{I}}$.
Thus the complete effective Hamiltonian for the ground state atoms
affected by the magnetic field and light reads 
\begin{align}
\hat{H}_{{\mathbf{B}}\&{\rm {\mathbf{E}}}}=\ u_{s}(\tilde{\mathbf{E}}^{*}\!\cdot\!\tilde{\mathbf{E}}) & +\frac{\mu_{B}g_{J}}{\hbar}\left({\mathbf{B}}+{\mathbf{B}}_{{\rm eff}}\right)\!\cdot\!\hat{\mathbf{J}}+A_{{\rm hf}}\hat{\mathbf{I}}\!\cdot\!\hat{\mathbf{J}},\label{eq:H_B-E}
\end{align}
 where ${\mathbf{B}}_{{\rm eff}}$ acts as a true magnetic field and
adds vectorially with ${\mathbf{B}}$. Instead of using the full Breit-Rabi
equation~\cite{Breit1931,Budker2004} for the Zeeman energies, we
assume that the Zeeman shifts are small in comparison with the hyperfine
splitting corresponding to the linear, or anomalous, Zeeman regime.
In this case, the effective Hamiltonian for a single manifold of total
angular momentum $\hat{\mathbf{F}}=\hat{\mathbf{J}}+\hat{\mathbf{I}}$
states is obtained by replacing $g_{J}\hat{\mathbf{J}} \rightarrow g_{F}\hat{\mathbf{F}}$
in the magnetic field interaction term \cite{Budker2004}, giving

\begin{align}
\hat{H}_{{\mathbf{B}}\&{\rm {\mathbf{E}}}}^{\left(f\right)}= & \ u_{s}(\tilde{\mathbf{E}}^{*}\!\cdot\!\tilde{\mathbf{E}})+\frac{\mu_{B}g_{F}}{\hbar}\left({\mathbf{B}}+{\mathbf{B}}_{{\rm eff}}\right)\!\cdot\!\hat{\mathbf{F}},\label{eq:H_B-E-result}
\end{align}
 where we have introduced the hyperfine Land\'e g-factor 
\begin{align*}
g_{F} & =g_{J}\frac{f(f+1)-i(i+1)+j(j+1)}{2f(f+1)}
\end{align*}
 and omitted the last term $A_{{\rm hf}}\hat{\mathbf{I}}\!\cdot\!\hat{\mathbf{J}}=A_{{\rm hf}}\left(\hat{\mathbf{F}}^{2}-\hat{\mathbf{J}}^{2}-\hat{\mathbf{I}}^{2}\right)/2$
which is constant for the single hyperfine manifold with fixed $f$.
For example, in $\Rb87$'s for which $i=3/2$ and $j=1/2$ in the
electronic ground state, the lowest energy hyperfine manifold with $f=1$
corresponds to $g_{F}=-g_{J}/4\approx-1/2$.

In this way, at second order of perturbation the light shift contains
the term $u_{s}(\tilde{\mathbf{E}}^{*}\!\cdot\!\tilde{\mathbf{E}})$
described as a rank-0 (scalar) tensor operator resulting directly
from the electric dipole transition, as well as the term $\mu_{B}g_{F}{\mathbf{B}}_{{\rm eff}}\!\cdot\!\hat{\mathbf{F}}/\hbar$
described by the rank-1 (vector) tensor operator emerging from adding
excited state fine structure. Since the latter operator is linear
with respect to the total momentum $\hat{\mathbf{F}}$, it can induce
only transitions which change $m_{F}$ by $\pm1$.

Adding effects due to the excited state hyperfine coupling
leads to an additional rank-2 tensor contribution to the light-induced
coupling within the atomic ground-state manifold. Because the scale of
this term is set by the $\MHz$ to $\GHz$ scale excited-state hyperfine
couplings, the smallness makes these light shifts unsuitable for
most equilibrium-system applications in quantum gases.

\subsection{Bichromatic light field\label{sub:Bichromatic-light-field}}

\subsubsection{General analysis \label{sec:TwoRaman-general}}

We now turn to a situation frequently encountered in the current experiments
on artificial gauge potentials~\cite{Lin2009a,Lin2009b,Lin2011a}
where an ensemble of ultracold atoms is subjected to a magnetic field
${\mathbf{B}}=B_{0}\ez$ and is simultaneously illuminated by possibly
several laser beams with two frequencies $\omega$ and $\omega+\delta\omega$.
The frequency separation $\delta\omega=g_{F}\mu_{B}B_{0}/\hbar+\delta$
differs by a small detuning $\delta$ (with $|\delta/\delta\omega|\ll1$)
from the linear Zeeman shift between $m_{F}$ states. The effective
magnetic field induced by the complex electric field ${\mathbf{E}}={\mathbf{E}}_{\omega_{-}}\exp(-i\omega t)+{\mathbf{E}}_{\omega_{+}}\exp\left[-i(\omega+\delta\omega)t\right]$
can be cast into its frequency components using Eq.~\eqref{eq:B_eff-electr}
\begin{align}
{\mathbf{B}}_{{\rm eff}}= & {\mathbf{B}}_{{\rm eff\,0}}+{\mathbf{B}}_{{\rm eff\,-}}e^{-i\delta\omega t}+{\mathbf{B}}_{{\rm eff\,+}}e^{i\delta\omega t},\label{eq:B_eff-bichromatic}
\end{align}
 where 
\begin{align}
{\mathbf{B}}_{{\rm eff\,0}}= & \frac{iu_{v}}{\mu_{{\rm B}}g_{J}}\left({\mathbf{E}}_{\omega_{-}}^{*}\shorttimes{\mathbf{E}}_{\omega_{-}}+{\mathbf{E}}_{\omega_{+}}^{*}\shorttimes{\mathbf{E}}_{\omega_{+}}\right)\,,\quad{\mathbf{B}}_{{\rm eff\,\mp}}=\frac{iu_{v}}{\mu_{{\rm B}}g_{J}}{\mathbf{E}}_{\omega_{\mp}}^{*}\shorttimes{\mathbf{E}}_{\omega_{\pm}}\label{eq:B_eff-components}
\end{align}
 are the corresponding amplitudes. The first term ${\mathbf{B}}_{{\rm eff\,0}}$
adds to the static bias field $B_{0}\ez$, and the remaining time-dependent
terms ${\mathbf{B}}_{{\rm eff\,\pm}}e^{\pm i\delta\omega t}$ can drive
transitions between different $m_{F}$ levels.

The real magnetic field is assumed to be much greater than the effective
one $B_{0}\gg\left|{\mathbf{B}}_{{\rm eff}}\right|$, and the frequency
difference $\delta\omega$ is taken to be large compared to the kinetic
energy scales. In this case, the time-dependence of the Hamiltonian
$\hat{H}_{{\mathbf{B}}\&{\rm {\mathbf{E}}}}^{\left(f\right)}$ in Eq.~\eqref{eq:H_B-E-result}
can be eliminated via the unitary transformation $\hat{S}=\exp\left(-i\delta\omega t\hat{F}_{z}\right)$
and the subsequent application of the rotating wave approximation
(RWA) by removing the terms oscillating at the frequencies $\delta\omega$
and $2\delta\omega$ in the transformed Hamiltonian $\hat{S}^{\dagger}\hat{H}_{{\mathbf{B}}\&{\rm {\mathbf{E}}}}^{\left(f\right)}\hat{S}-i\hbar S^{\dagger}\partial_{t}S$.
The transformation does not alter $\hat{F}_{z}$, whereas the raising
and lowering operators $\hat{F}_{\pm}=\hat{F}_{x}\pm i\hat{F}_{y}$
change to $\hat{F}_{\pm}\exp\left(\pm i\delta\omega t\right)$.

The resulting Hamiltonian can be represented as 
\begin{eqnarray}
\hat{H}_{{\rm RWA}} & =V({\mathbf{r}})\hat{1}+\hat{M}({\bf r}).\label{RWA_Hamiltonian}
\end{eqnarray}
The state-independent (scalar) potential is given by
\begin{equation}
V({\mathbf{r}})=u_{s}\left({\mathbf{E}}_{\omega_{-}}^{*}\!\cdot\!{\mathbf{E}}_{\omega_{-}}+{\mathbf{E}}_{\omega_{+}}^{*}\!\cdot\!{\mathbf{E}}_{\omega_{+}}\right)\label{eq:V-scalar-shift}
\end{equation}
and
\begin{equation}
\hat{M}({\bf r})={\boldsymbol{\Omega}}\!\cdot\!\hat{\mathbf{F}}=\Omega_{z}\hat{F}_{z}+\Omega_{-}\hat{F}_{+}+\Omega_{+}\hat{F}_{-}\label{eq:M-r}
\end{equation}
where 
\begin{equation}
\Omega_{z}=\delta+\frac{\mu_{B}g_{F}}{\hbar}{\mathbf{B}}_{{\rm eff\,0}}\!\cdot\!\ez\,\label{eq:Omega_z_Explicit}
\end{equation}
 and 
\begin{equation}
\Omega_{\pm}\equiv\frac{\Omega_{x}\pm i\Omega_{y}}{2}=\frac{\mu_{B}g_{F}}{2\hbar}\left[{\mathbf{B}}_{{\rm eff\,\pm}}\!\cdot\!\left(\ex\pm i\ey\right)\right]\,\label{eq:Omega_pm_Explicit}
\end{equation}
 are the circular components of the RWA effective Zeeman magnetic
field ${\boldsymbol{\Omega}}$. The latter Zeeman field
is explicitly
\begin{eqnarray}
{\boldsymbol{\Omega}}= & \left[\delta+i\frac{u_{v}g_{F}}{\hbar g_{J}}\left({\mathbf{E}}_{\omega_{-}}^{*}\shorttimes{\mathbf{E}}_{\omega_{-}}+{\mathbf{E}}_{\omega_{+}}^{*}\shorttimes{\mathbf{E}}_{\omega_{+}}\right)\!\cdot\!\ez\right]\ez\nonumber \\
 & -\frac{u_{v}g_{F}}{\hbar g_{J}}{\rm Im}\left[\left({\mathbf{E}}_{\omega_{-}}^{*}\shorttimes{\mathbf{E}}_{\omega_{+}}\right)\!\cdot\!\left(\ex-i\ey\right)\right]\ex\label{eq:Omega-final-general}\\
 & -\frac{u_{v}g_{F}}{\hbar g_{J}}{\rm Re}\left[\left({\mathbf{E}}_{\omega_{-}}^{*}\shorttimes{\mathbf{E}}_{\omega_{+}}\right)\!\cdot\!\left(\ex-i\ey\right)\right]\ey.\nonumber 
\end{eqnarray}
The form of the effective coupling operator $\hat{M}({\bf r})$ shows
that, while it is related to the initial vector light shifts, the
RWA effective Zeeman magnetic field ${\mathbf{\Omega}}$ is composed
of both static and resonant couplings in a way that goes beyond the
restrictive ${\mathbf{B}}_{{\rm eff}}\propto i{\mathbf{E}}^{*}\shorttimes{\mathbf{E}}$
form.

In this way, the operator $\hat{F}_{z}$ featured
in the vector coupling $\hat{M}({\bf r})$, Eq.~\eqref{eq:M-r} induces
the opposites light shifts for the spin-up and spin-down atomic ground
states, whereas the operators $\hat{F}_{\pm}$ describe the Raman
transitions which change $m_{F}$ by $\pm1$. If the detuning from
atomic resonance $\Delta_{e}$ is large compared to the excited state
fine structure splitting $\Delta_{{\rm FS}}$, the magnitude of this
vector coupling proportional to $u_{v}$ drops off as $\Delta_{{\rm FS}}/\Delta_{e}^{2}$,
not like the scalar shift $V({\mathbf{r}})$ which drops off as $1/\Delta_{e}$.
Since the amount of the off-resonant light scattering is also proportional
to $1/\Delta_{e}^{2}$, the balance between the off-resonant scattering
and the vector coupling is bounded, and cannot be improved by large
detuning. While this is a modest problem for rubidium ($15\nm$ fine
structure splitting), it turns to be a serious obstacle for atoms
with smaller fine structure splitting such as potassium ($\approx4\nm$)
and lithium ($\approx0.02\nm$).

\subsubsection{Two Raman beams \label{sec:TwoRaman}}

\begin{figure}[t!]
\begin{centering}
\includegraphics[width=4.5in]{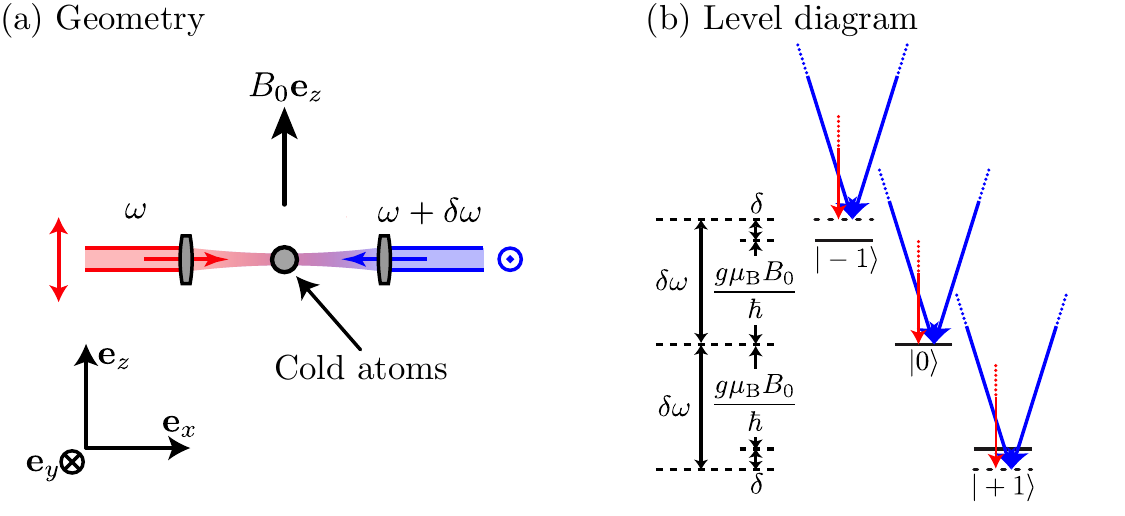} 
\par\end{centering}

\caption[Generic Geometry]{Typical two beam experimental geometry. (a) Laser geometry for creating
abelian artificial gauge fields showing two counter propagating linearly
polarized beams in the $\ex$-$\ey$ plane with frequency $\omega$
and $\omega+\Delta$. (b) Physical level diagram for three-level total
angular momentum $f=1$ case, as is applicable for the common alkali
atoms $^{7}$Li, $^{23}$Na, $^{39}{\rm K}$, $^{41}$K, and $^{87}$Rb
(all of which have $g_{F}=-g_{J}/4\approx-1/2)$. The generally small
quadratic Zeeman effect that makes the $\ket{-1}$-$\ket{0}$ energy
difference different from the $\ket{0}$-$\ket{+1}$ splitting, is
not included in this diagram.}

\label{fig_layout} 
\end{figure}

Within this formalism, let us consider a straightforward example of
two counter propagating Raman beams depicted in Fig.~\ref{fig_layout}. Such
a setup is often used in current experiments on synthetic gauge
fields~\cite{Lin2009a,LeBlanc2012,Jimenez-Garcia2012} and  spin-orbit
coupling\cite{Wang2012,LeBlanc2013} for ultracold atoms. In this
case 
\begin{align}
{\mathbf{E}}_{\omega_{-}} & = Ee^{i\kr x}\ey & {\rm and} &&
{\mathbf{E}}_{\omega_{+}} & = Ee^{-i\kr x}\ez\label{eq:InitialFields-1}
\end{align}
 describe the electric field of two lasers counterpropogating along
$\ex$ with equal intensities, crossed linear polarization and
wave-number $\kr=2\pi/\lambda$, giving ${\mathbf{B}}_{{\rm eff\,0}}=0$
and ${\mathbf{B}}_{{\rm eff\,\pm}}=\mp\frac{iu_{v}E^{2}}{\mu_{{\rm B}}g_{J}}e^{\pm i2\kr x}\ex$.
Consequently $\Omega_{z}=\delta$ and 
\begin{equation}
2\Omega_{\pm}=\Omega_{x}\pm i\Omega_{y}=\Omega_{R}e^{\pm i\left(2\kr x-\pi/2\right)}\,,\label{eq:2omega_pm}
\end{equation}
 where $\Omega_{R}=\left(g_{F}/g_{J}\right)u_{v}E^{2}/\hbar$ is the
Rabi frequency of the Raman coupling. The resulting effective Zeeman
field ${\boldsymbol{\Omega}}$ entering the vector light shift is
\begin{equation}
{\boldsymbol{\Omega}}=\delta\ez+\Omega_{R}\left[\sin\left(2\kr x\right)\ex-\cos\left(2\kr x\right)\ey\right].\label{eq:onedimensioneffectivefield}
\end{equation}
Together, $V({\mathbf{r}})=2u_{s}E^{2}$ and ${\boldsymbol{\Omega}}({\mathbf{r}})$
describe a constant scalar light shift along with a spatially rotating Zeeman field. 

It is evident from Eqs.~\eqref{eq:B_eff-electr} and \eqref{eq:Omega_pm_Explicit} that for this spatially rotating Zeeman field to be non-zero, the effective Zeeman field ${\mathbf{B}}_{\rm eff}$ must not be parallel to the quantizing axis defined by the applied magnetic field ${\mathbf{B}}=B_{0}\ez$.  In the present case, this implies that the counter propagating beams cannot be aligned along $\ez$.  Physically, any projection of ${\mathbf{B}}_{\rm eff}$ along $\ez$ yields a spin-dependent optical lattice \cite{Deutsch1998,Grimm:2000,McKay2010NJP} rather than a spatially rotating Zeeman field.   But since the Raman beams have different frequencies, the spin-dependent lattice is moving fast and vanishes after performing the RWA.

Note also that in order to produce the vector light shifts like the one given by Eq.~(\ref{eq:onedimensioneffectivefield}), there
is no need for the two Raman beams to counter-propagate. When the beams intersect with opening angle $\alpha$, the momentum exchange is changed from $2\kr$ to $2\kl = 2\kr\cos\left[\left(\pi-\alpha\right)/2\right]$.
Since this altered wave-vector simply changes the energy and length scales of the problem, we shall concentrate on the counter propagating
setup, Eqs.~\eqref{eq:InitialFields-1}.
To obtain an exact correspondence with the notations used
in most of the experimental papers on the artificial gauge fields~\cite{Lin2009a,Lin2009b,Lin2011a,Williams2012,LeBlanc2012,Jimenez-Garcia2012}
and the spin-orbit coupling~\cite{Lin2011,Zhang2012PRL,Wang2012,Cheuk2012,LeBlanc2013},
one needs to interchange the $\ez$ and $\ey$ axes  in our analysis.

It is instructive to note that the $z$ component of the effective Zeeman field
${\boldsymbol{\Omega}}$, Eq.~\eqref{eq:onedimensioneffectivefield}, does not depend on the
intensity of the Raman beams. The situation is not universal. For example, consider
a pair of Raman laser beams with complex electric fields  
\begin{align}
{\mathbf{E}}_{\omega_{-}} & = E_{-}e^{i\kr z}\left(\ex-i\ey\right)/\sqrt{2}, & {\rm and} &&
{\mathbf{E}}_{\omega_{+}} & = E_{+}e^{i\kr x}\ez\label{eq:InitialFields-2-1},
\end{align}
intersecting with angle $\alpha=\pi/2$, and where the $\omega_-$ beam is circularly polarized~\cite{Gunter2009}.  In this case ${\mathbf{E}}_{\omega_{-}}^*\times {\mathbf{E}}_{\omega_{-}} = -i\left|E_{-}\right|^2\ez$,  ${\mathbf{E}}_{\omega_{+}}^*\times {\mathbf{E}}_{\omega_{+}} = 0$, and  ${\mathbf{E}}_{\omega_{-}}^*\times {\mathbf{E}}_{\omega_{+}} = i E_{-}^* E_+ (\ex + i\ey)/\sqrt{2}$ giving
\begin{equation}
\hbar\Omega_{z}=\hbar\delta+\left(u_{v}\frac{g_{F}}{g_{J}}\right)\left|E_{-}\right|^{2}\,,\quad\hbar\Omega_{\pm}=-\left(u_{v}\frac{g_{F}}{g_{J}}\right)\frac{E_{-}E_{+}}{\sqrt{2}}e^{\pm i\kr\left(z-x\right)}\,,\label{eq:Omega_z--Omega_pm}
\end{equation}
so $\Omega_{z}$ becomes intensity-dependent, providing an extra control
of the light-induced gauge potential.

Finally, more sophisticated spatial dependence of the Zeeman field ${\boldsymbol{\Omega}}$ can be created using additional Raman beams with specially chosen polarizations, for example leading to optical flux lattices \cite{Cooper2011,Juz-Spielm2012NJP}, to be considered in the Section \ref{sub:Flux-latt}.


\section{Schemes for creating Abelian gauge potentials}\label{sect:schemes}




In this Section, we describe how the atom-light coupling can be exploited to generate Abelian gauge potentials in cold-atom systems. We present useful schemes based on the adiabatic motion of atomic dressed states, and we eventually describe the specific experimental setup used at NIST to generate synthetic gauge fields. 

\subsection{The $\Lambda$ setup}
\label{subsect:Lambda}

\begin{figure}
\begin{centering}
\includegraphics[width=8cm]{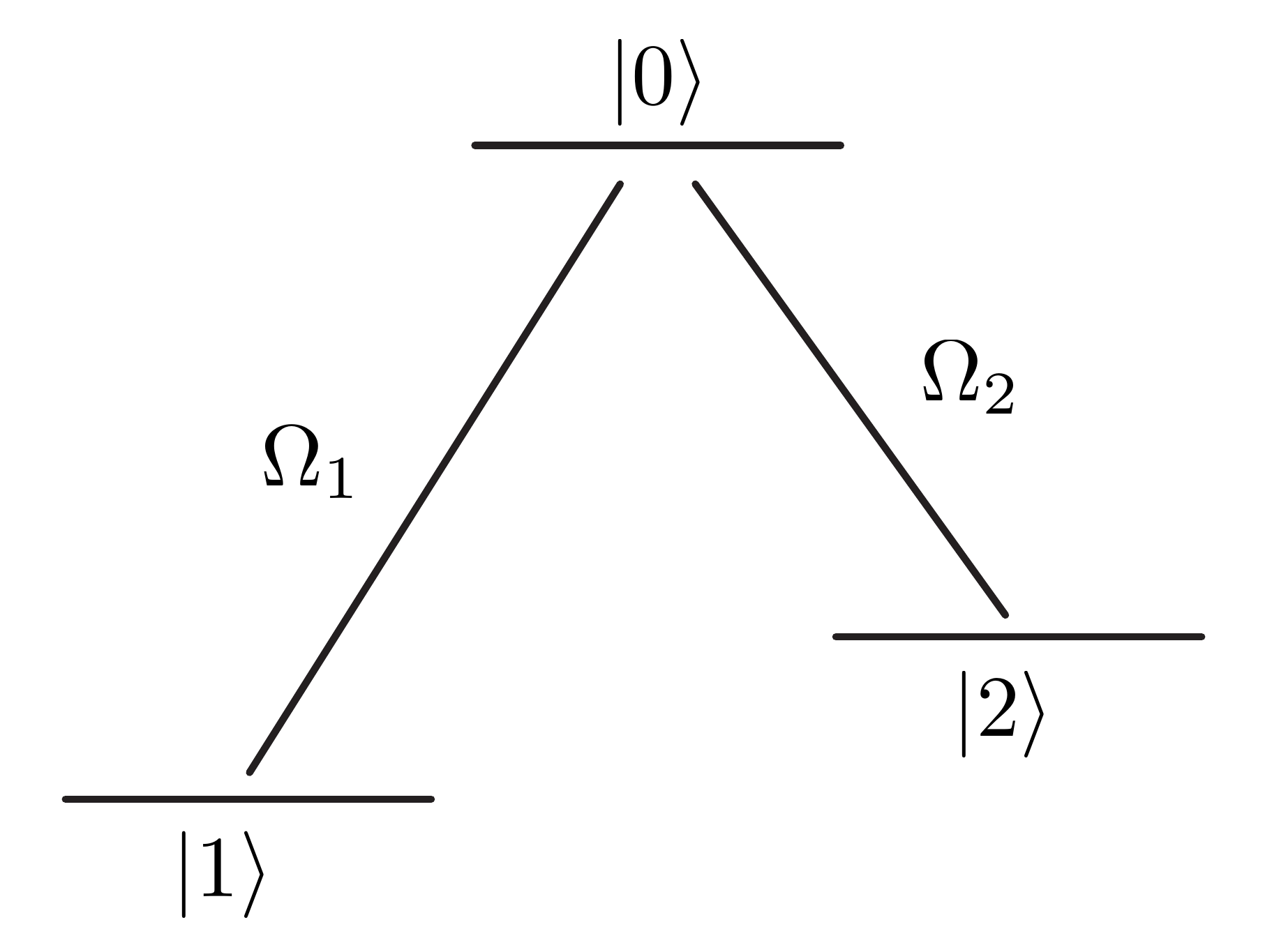} 
\par\end{centering}
\caption{The $\Lambda$ scheme of the atom-light coupling in which the laser
beams induce atomic transitions $|1\rangle\rightarrow|0\rangle$ and
$|2\rangle\rightarrow|0\rangle$ characterized by the Rabi frequencies
$\Omega_{1}$ and $\Omega_{2}$. }
\centering{}\label{Lambda} 
\end{figure}

Let us begin our analysis of Abelian gauge potentials with the illustrative case of
atoms characterized by the $\Lambda$ type level structure, where the
laser beams couple two atomic internal states $|1\rangle$ and $|2\rangle$
with a third one $|0\rangle$, see Figure \ref{Lambda}. The atom light coupling operator
$\hat{M}$ featured in the full atomic Hamiltonian (\ref{eq:H-full})
has then the following form: 
\begin{equation}
\hat{M}(\mathbf{r})=\hbar\delta\left(|2\rangle\langle2|-|1\rangle\langle1|\right)+\hbar\Delta|2\rangle\langle2|+\hbar(\Omega_{1}|0\rangle\langle1|+\Omega_{2}|0\rangle\langle2|+\mathrm{H.c.})/2,\label{eq:M-Lambda}
\end{equation}
 where the frequencies $\delta$ and $\Delta$ characterize the detuning
between the atomic states. The Rabi frequencies $\Omega_{1}\equiv\Omega_{1}({\bf r})$
and $\Omega_{2}\equiv\Omega_{2}({\bf r})$ are generally complex and
position-dependent.

In the context of the light-induced gauge potential for ultracold
atoms, the $\Lambda$ scheme was first explored in the late 90's \cite{Dum1996,Visser1998,Dutta1999}.
Subsequently it was shown that the scheme can yield a non-zero artificial
magnetic field (inducing the Lorentz force for electrically neutral
atoms) using non-trivial arrangements of laser fields \cite{Juzeliunas2004,Juzeliunas2005,Juzeliunas2005b,Zhang:2005EPJD,Juzeliunas2006,Zhu2006,Cheneau2008,Song2008,Gunter2009,Cooper2010}
or position-dependent detuning \cite{Spielman2009,Lin2009b}. In most
of these treatments the state $|0\rangle$ is assumed to be the atomic
excited state coupled resonantly to the ground states $|1\rangle$
and $|2\rangle$ by laser beams, as shown in fig. \ref{Lambda}. The
involvement of the excited states is inevitably associated with a
substantial dissipation due to the spontaneous emission.

To avoid such losses, one can choose $|0\rangle$ to be an atomic
ground state coupled to the other two ground states $|1\rangle$ and
$|2\rangle$ via the Raman transitions \cite{Spielman2009,Cooper2010}.
In that case $\Omega_{1}$ and $\Omega_{2}$ featured in the coupling
operator (\ref{eq:M-Lambda}) represent the Raman Rabi frequencies.
This forms a Ladder scheme which is equivalent to the $\Lambda$ sheme.
In particular, if $\Omega_{1}=\Omega_{2}^{*}=\Omega_{+}\sqrt{2}$ and $\Delta=0$,
the Hamiltonian (\ref{eq:M-Lambda}) reduces to the Hamiltonian given
by Eq.~\eqref{eq:M-r} for an atom in the $F=1$ ground state manifold
affected by the RWA effective Zeeman field ${\boldsymbol{\Omega}}=\left({\rm Re}\Omega_{+},\,{\rm Im}\Omega_{+},\,\delta\right)$.
The latter ${\boldsymbol{\Omega}}$ is induced by the Raman transitions
and the real magnetic field. In this case the states $|1\rangle$,
$|0\rangle$ and $|2\rangle$ corresponds to the magnetic sublevels
with $m=-1$, $m=0$ and $m=1$, respectively. In the next Subsection
we shall analyse in detail the gauge potentials resulting
from such a coupling between the magnetic sublevels.

If $\Omega_{1}\ne\Omega_{2}^{*}$ the atom-light coupling operator
(\ref{eq:M-Lambda}) can no longer be represented as a Hamiltonian
for the atomic spin (or quasispin) interacting with the effective
magnetic field. Assuming exact resonance between the atomic levels
$1$ and $2$ ($\delta=0$), the interaction operator $\hat{M}(\mathbf{r})$
given by Eq. (\ref{eq:M-Lambda}) has a single dressed eigenstate $|D\rangle$
known as the dark or uncoupled state:
\begin{equation}
|D\rangle=\frac{|1\rangle-\zeta|2\rangle}{1+\zeta^{2}}\,,\quad\zeta=\frac{\Omega_{1}}{\Omega_{2}}=\left|\zeta\right|e^{iS}\,,\label{eq:D-state}
\end{equation}
where $S$ is a relative phase between the two Rabi frequencies.
The state $|D\rangle$ contains no
contribution from the excited state $|0\rangle$ and is characterised by
a zero eigenvalue $\varepsilon_{D}=0$.
Dark states are frequently encountered in quantum optics. They play
an important role in applications such as the electromagnetically
induced transparency (EIT) \cite{Arimondo1996,Harris1997,Lukin2003,Fleischhauer2005}
and the Stimulated Raman Adiabatic Passage (STIRAP) \cite{Bergmann1998,Vitanov2001,Shapiro2007},
relying on the fact that the excited level $|0\rangle$ is not populated
for the dark state atoms and spontaneous decay is therefore suppressed.

As we have seen in Section \ref{sub:Formulation-gauge-potentials},
if the atom is in a selected internal dressed state well separated
from the remaining dressed states, an Abelian geometric vector
potential emerges for the centre of mass motion. For atoms in the internal
dark state the adiabatic motion takes place if the total Rabi frequency
$\sqrt{\left|\Omega_{1}\right|^{2}+\left|\Omega_{2}\right|^{2}}$
exceeds the characteristic kinetic energy of the atomic motion. The
corresponding vector potential ${\mathbfcal A}=i\hbar\left\langle D\right|\nabla\left|D\right\rangle $
and the associated magnetic field ${\mathbfcal B}=\boldsymbol{\nabla}\times{\mathbfcal A}$
are 
\begin{equation}
{\mathbfcal A}=-\hbar\frac{|\zeta|^{2}}{1+|\zeta|^{2}}\nabla S\,,\label{eq:vector-2}
\end{equation}
 
\begin{equation}
{\mathbfcal B}=\hbar\frac{\nabla S\times\nabla|\zeta|^{2}}{(1+|\zeta|^{2})^{2}}\,.\label{eq:B-eff-2}
\end{equation}
The geometric scalar potential for the dark state atoms is given by
\begin{equation}
W=\frac{\hbar^{2}}{2m}\frac{(\nabla|\zeta|)^{2}+|\zeta|^{2}(\nabla S)^{2}}{(1+|\zeta|^{2})^{2}}.\label{eq:phi-2}
\end{equation}
 One easily recognizes that the vector gauge potential yields
a non-vanishing artificial magnetic ${\mathbfcal B}$ only if the
gradients of the relative intensity and the relative phase are both
non-zero and not parallel to each other. Therefore the synthetic magnetic
field cannot be created for the dark state atoms of the $\Lambda$ scheme using the plane-waves
driving the transitions $\left|1\right>\rightarrow \left|0\right>$ and $\left|2\right>\rightarrow \left|0\right>$
 \cite{Dum1996,Visser1998}. However, plane
waves can indeed be used in more complex tripod \cite{Ruseckas2005,Stanescu2007a,Jacob2007,Juzeliunas2008PRA,Stanescu2008,Vaishnav08PRL,Juzeliunas2010}
or closed loop \cite{Campbell2011,Galitski2013} setups to generate non-Abelian gauge fields for a pair of degenerate internal dressed states,
as we shall see in the Section \ref{sect:soc}.

\begin{figure}
\begin{centering}
\includegraphics[width=0.45\textwidth]{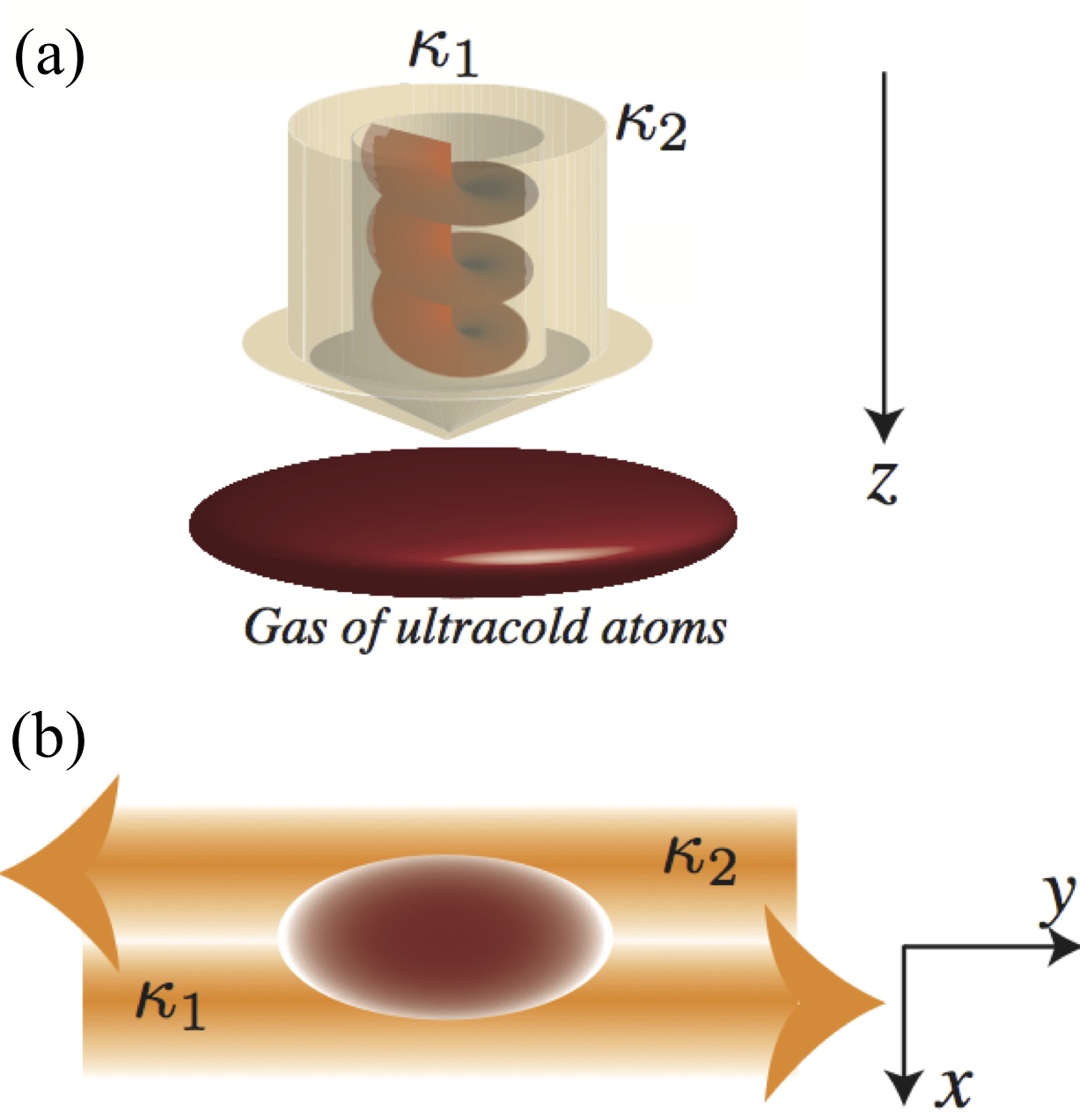} 
\par\end{centering}

\caption{(a) The light beams carrying optical vortices \cite{Juzeliunas2004,Juzeliunas2005,Juzeliunas2005b,Zhang:2005EPJD,Song2008,Cooper2010}
or (b) the counterpropagating beams with spatially shifted profiles
\cite{Juzeliunas2006,Zhu2006,Sun2007,Cheneau2008} provide a non-zero
artificial magnetic field for the atoms in the laser-dressed states
of the $\Lambda$ scheme shown in fig. \ref{Lambda}.}
\label{Lambda-1} 
\end{figure}

Equation (\ref{eq:B-eff-2}) has a very intuitive interpretation \cite{Juzeliunas2006}.
The vector $\nabla[|\zeta|^{2}/(1+|\zeta|^{2})]$ connects the ``center
of mass'' of the two light beams and $\nabla S$ is proportional
to the vector of the relative momentum of the two light beams. Thus
a nonvanishing ${\mathbfcal B}$ requires a \textit{relative orbital
angular momentum} of the two light beams. This is the case for light
beams carrying optical vortices \cite{Juzeliunas2004,Juzeliunas2005,Juzeliunas2005b,Zhang:2005EPJD,Song2008}
or if one uses two counterpropagating light beams of finite diameter
with an axis offset \cite{Juzeliunas2006,Cheneau2008}, as depicted
in Fig. \ref{Lambda-1}. A more detailed analysis of these schemes is
available in the cited references and the previous reviews \cite{Juzeliunas2008-Chapter,Dalibard2011}.

\subsection{Spins in effective Zeeman fields \label{sub:Spin-in-Magn-field}}

\subsubsection{General treatment \label{sub:General-spin-in-Zeeman-field}}

Let us now turn to a situation where the atom-light Hamiltonian can
be represented as an interaction of a spin $ $$\hat{\mathbf{F}}$
with an effective Zeeman field 
\begin{equation}
\hat{M}({\bf r})={\boldsymbol{\Omega}}\!\cdot\!\hat{\mathbf{F}}=\Omega\left(\cos\theta\hat{F}_{z}+\sin\theta\cos\phi\hat{F}_{x}+2\sin\theta\sin\phi\hat{F}_{y}\right)\label{eq:M-r-spin-interaction-general}
\end{equation}
 where we have parametrized the Zeeman vector $\boldsymbol{\Omega}=(\Omega_{x},\Omega_{y},\Omega_{z})$
in terms of the spherical angles $\theta$ and $\phi$ shown in Fig. 9.
\begin{align}
\tan\phi & =\frac{\Omega_{y}}{\Omega_{x}}, &  & {\rm and} & \qquad\cos\theta & =\frac{\Omega_{z}}{\Omega}\label{eq:tan-phi--theta}
\end{align}
where $\Omega$ is the length of the Zeeman vector. Such a coupling
can be produced using bichromatic laser beams considered in the previous
Section, leading to Eq.~\eqref{eq:M-r} which has the same form as
Eq.(\ref{eq:M-r-spin-interaction-general}). For the spin $1$ case, the
Hamiltonian (\ref{eq:M-r-spin-interaction-general}) corresponds to the
$\Lambda$ (ladder) scheme described by the Hamiltonian (\ref{eq:M-Lambda})
as long as $\Omega_{1}=\Omega_{2}^{*}$ and $\Delta=0$. Note that
the operator $\hat{\mathbf{F}}$ does not necessarily represent the
true atomic spin operator; it can be other atomic vector operators
with Cartesian components obeying the angular momentum algebra.

\begin{figure}
\begin{centering}
\includegraphics[width=2.0in]{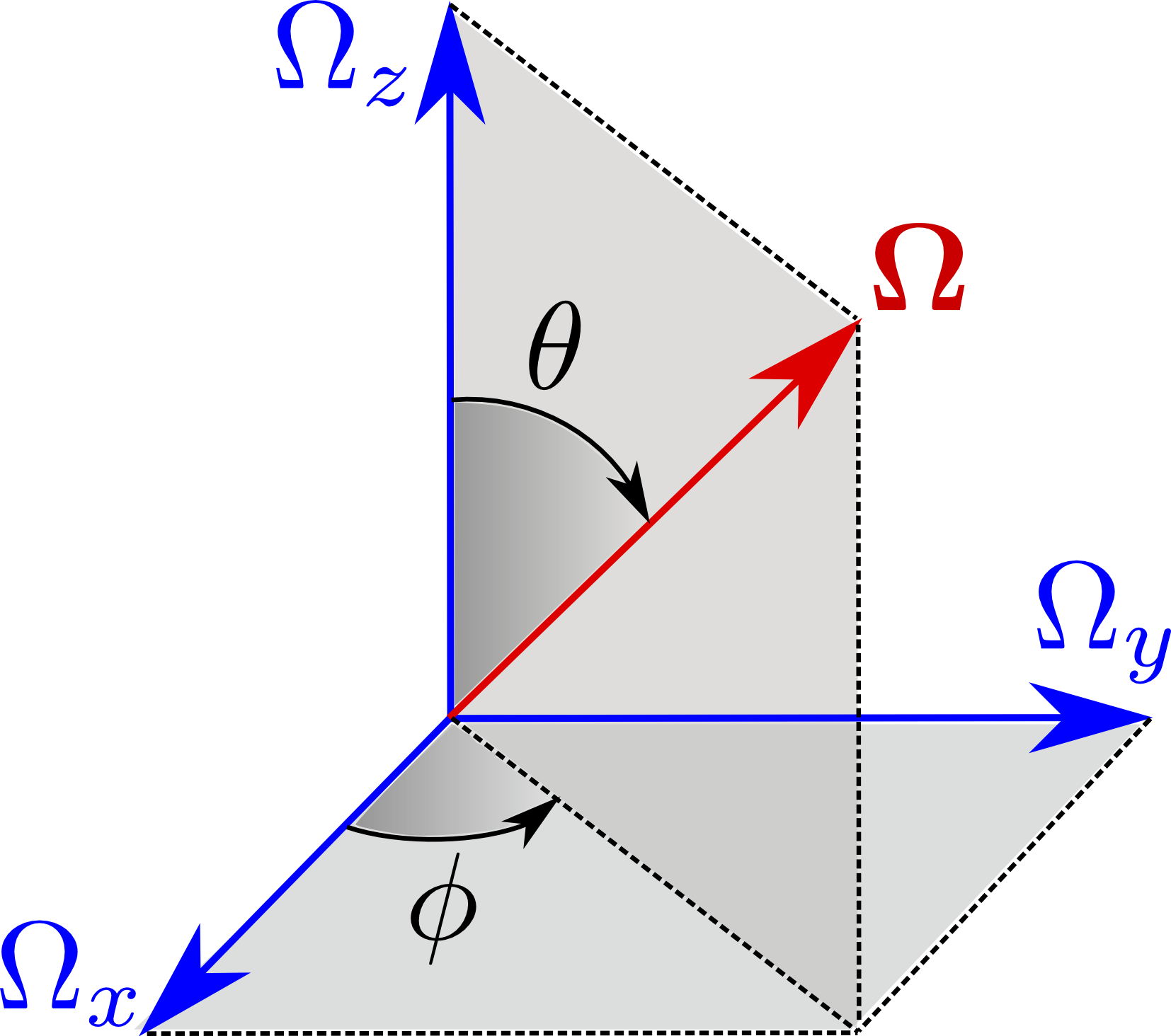} 
\caption{Representation of the coupling vector $\boldsymbol{\Omega}=(\Omega_{x},\Omega_{y},\Omega_{z})$
in terms of the spherical angles $\theta$ and $\phi$ .}
\end{centering}
\label{fig:Omega-spherical} 
\end{figure}

The coupling Hamiltonian $\hat{M}$ can be diagonalised via a unitary
transformation 
\begin{equation}
\hat{R}=e^{-i\hat{F}_{z}\phi/\hbar}e^{-i\hat{F}_{y}\theta/\hbar}e^{i\hat{F}_{z}\phi/\hbar}.\label{eq:S-Omega}
\end{equation}
 The first operator ${e^{-i\hat{F}_{z}\phi/\hbar}}$ rotates
the spin around the $z$-axes by the angle $\phi$ eliminating the
phase $\phi$ in $\hat{M}({\bf r})$, Eq.(\ref{eq:M-r-spin-interaction-general}).
The subsequent transformation using the second operator $e^{-i\hat{F}_{y}\theta/\hbar}$
rotates the spin around the $y$-axes by the angle $\theta$ making
the operator $\hat{M}({\bf r})$ proportional to $\hat{F}_{z}$. Although
the third operator $e^{i\hat{F}_{z}\phi/\hbar}$ no longer affects
the transformed operator $\hat{M}({\bf r})$, its inclusion ensures
that the whole transformation $\hat{R}$ reduces to the unit operator
for zero polar angle $\theta=0$ and arbitrary azimuthal angle $\phi$.
Thus one has 
\begin{equation}
\hat{R}^{\dagger}\hat{M}\hat{R}=\Omega\hat{F}_{z}\,.\label{eq:F-Omega}
\end{equation}

The transformed Hamiltonian $\Omega\hat{F}_{z}$ has eigenstates $\left|f,\, m\right\rangle \equiv\left|m\right\rangle $
characterized by the total angular momentum $f$ and its $\ez$ projection
$m$. In the following, we use both notations $m$ and $m_F$ to denote the projection of the total angular momentum. Therefore the eigenstates $\left|m,{\boldsymbol{\Omega}}\right\rangle \equiv\left|m,\theta,\phi\right\rangle$  of the original atom-light Hamiltonian
$\hat{M}$ read 
\begin{equation}
\left|m,{\boldsymbol{\Omega}}\right\rangle =  \hat{R} \left|m \right\rangle =e^{i\left(m-\hat{F}_{z}/\hbar\right)\phi}e^{-i\hat{F}_{y}\theta/\hbar}\left|m\right\rangle ,\label{eq:eigenstates-m_F-explicit}
\end{equation}
the corresponding eigenenergies being 
\begin{align}
\varepsilon_{m} & =\hbar m\Omega\,,\quad m=-f,\,\ldots\,,f\,,\label{eq:eigenstates-F_z-tilde-1}
\end{align}
where both the eigenenergies $\varepsilon_{m}(\mathbf{r})$ and
also the local eigenstates $\left|m,{\boldsymbol{\Omega}}(\mathbf{r})\right\rangle $
are position-dependent through the position-dependence of the
effective Zeeman vector ${\boldsymbol{\Omega}}(\mathbf{r})$. It is to be noted that
 similar kinds of eigenstates give rise to artificial gauge potentials
describing the rotation of diatomic molecules~\cite{Moody1986,Bohm92JMP}
and the physics of atomic collisions~\cite{Zygelman1987,Zygelman:1990}.

If an atom is prepared in a dressed state $\left|m,{\boldsymbol{\Omega}}\right\rangle $
and if its characteristic kinetic energy is small compared to the
energy difference between adjacent spin states $\Delta E=\hbar\Omega$,
the internal state of the atom will adiabatically follow the dressed
state $\left|m,{\boldsymbol{\Omega}}\right\rangle $ as the atom moves,
and contributions due to other states with $m_{F}^{\prime}\ne m_{F}$
can be neglected. Projecting the atomic dynamics onto the selected
internal eigenstate $\left|m_{F},{\boldsymbol{\Omega}}\right\rangle $
yields a reduced \Schrodinger equation for the atomic center of mass
motion affected by the geometric vector potential ${\mathbfcal A}\equiv{\mathbfcal A}_{m}(\mathbf{r})=i\hbar\bra{m,{\boldsymbol{\Omega}}}\boldsymbol{\nabla}\ket{m,{\boldsymbol{\Omega}}}$
and the scalar potential $W\equiv W_{n}(\mathbf{r})$ which are generally defined
by Eq.~(\ref{eq:A-V--q-eq-1}). In the current situation these potentials explicitly read
\cite{Juz-Spielm2012NJP} 
\begin{equation}
{\mathbfcal A}(\mathbf{r})=\hbar m\left(\cos\theta-1\right)\boldsymbol{\nabla}\phi\label{eq:vect-spin}
\end{equation}
 and 
\begin{align}
W_{m}(\mathbf{r}) & =\frac{\hbar^{2}}{4m}\left[f(f+1)-m^{2}\right]\left[\sin^{2}\theta\left(\nabla\phi\right)^{2}+\left(\nabla\theta\right)^{2}\right],\label{eq:W-spin}
\end{align}

The geometric vector potential ${\mathbfcal A}(\mathbf{r})$ yields
the artificial magnetic field 
\begin{equation}
{\mathbfcal B}(\mathbf{r})=\boldsymbol{\nabla}\times{\mathbfcal A}(\mathbf{r})=\hbar m_{F}\boldsymbol{\nabla}\left(\cos\theta\right)\times\boldsymbol{\nabla}\phi , \label{eq:B-spin}
\end{equation}
 which is non-zero if $\boldsymbol{\nabla}\phi$ and $\boldsymbol{\nabla}\left(\cos\theta\right)$
are not parallel to each other. Both ${\mathbfcal A}(\mathbf{r})$
and ${\mathbfcal B}(\mathbf{r})$ have the largest magnitude for maximum
absolute values of the spin projection $m$ and are zero for $m=0$.
On the other hand, the geometric scalar potential $W(\mathbf{r})$
is maximum for $m=0$ and reduces with increasing $\left|m\right|$.

The atomic motion is affected by three distinct scalar potentials:
(a) the state independent potential $V(\mathbf{r})$ representing
the ``scalar light shift'' given by Eq.~\eqref{eq:V-scalar-shift}
for bichromatic fields, (b) the ``adiabatic scalar potential''
$\varepsilon_{m}(\mathbf{r})\equiv\varepsilon_{m}$ arising from spatial
variations in the magnitude of the Zeeman vector $\boldsymbol{\Omega}(\mathbf{r})$,
and (c) the ``geometric scalar potential'' $W(\mathbf{r})$ described
above. All three contribute to the potential energy of atoms in the
dressed state basis.

\subsubsection{A pair of Raman beams \label{sub:A-pair-of}}

For two counterpropagating Raman beams, the effective Zeeman field
${\boldsymbol{\Omega}}$, Eq.~\eqref{eq:onedimensioneffectivefield},
defining the gauge fields, is characterised by the azimuthal angle
$\phi=2\kr x-\pi/2$ with the gradient 
\begin{equation}
\boldsymbol{\nabla}\phi=2\mathbf{\kr}\,,\quad\mathbf{\kr}=\kr\ex,\label{eq:grad-phi}
\end{equation}
 equal to the Raman recoil wave-vector $2\mathbf{\kr}$. On the other
hand, the polar angle $\theta$ and the length of the Zeeman vector
$\Omega$ are determined by the detuning $\delta$ and the Raman Rabi
frequency $\Omega_{R}$ with
\begin{equation}
\cos\theta=\frac{\delta}{\Omega}\,,\quad\Omega=\sqrt{\delta^{2}+\Omega_{R}^{2}} . \label{eq:cos-theta}
\end{equation}
 Typically experiments are performed in the lowest energy dressed
state, where $m=-f$ assumes its maximum absolute value. Equations~(\ref{eq:vect-spin})
and (\ref{eq:grad-phi})-(\ref{eq:cos-theta}) illustrate two important
points:  (1) the magnitude of the vector potential ${\mathbfcal A}_{m}({\bf r})$
is defined by the recoil momentum of the lasers as well as the spin
projection $\hbar m$, and (2) the vector potential is
strictly bounded between $\pm2\hbar\kr m_{F}$, where the maximum possible
value is the recoil momentum $2\hbar\kr$ times the spin $f$.

We note that employing an extra spatially uniform radio-frequency
magnetic field adds a constant term to the spatially oscillating $x$
component of effective Zeeman field ${\boldsymbol{\Omega}}$, Eq.~\eqref{eq:onedimensioneffectivefield}.
In that case both the adiabatic energy $\varepsilon_{m}$ and the
geometric scalar potential $W_{m}(\mathbf{r})$ become spatially oscillating
functions, thus creating a composite lattice potential in addition
to the vector potential ${\mathbfcal A}(\mathbf{r})$~\cite{Jimenez-Garcia2012}.
We shall discuss this issue in more detail in the Section \ref{sec:other schemes}.

Returning to Eqs.~(\ref{eq:B-spin})-(\ref{eq:cos-theta}),
we see that in order to have a non-zero ${\mathbfcal B}$, either the optical intensity
or the detuning must depend on $y$ or $z$, i.e. their gradient(s)
should not be parallel to the wave-vector $\mathbf{\kr}$.  The 
position-dependent detuning can be generated by the vector
light shift itself, which is absent for co-propagating light beams. This provides
an additional term analogous to $\delta$, but proportional to the
Raman Rabi frequency. This is the case if the laser beams propagate
at the angle $\alpha=\pi/2$, one of them being circularly polarised
\cite{Gunter2009}, as one can see in Eq.~\eqref{eq:Omega_z--Omega_pm}
of the previous Section. Therefore, by simply making $\delta=0$ and
having the intensity-dependent detuning (which
adds to $\delta$) depend on position in Eq.~\eqref{eq:Omega_z--Omega_pm},
it is possible to create artificial gauge potentials which are truly
geometric in nature, as the absolute light intensity then completely
vanishes from the azymuthal angle $\theta$.

The first experimental implementation of the synthetic magnetic field
\cite{Lin2009b} was based on a different insight, namely that 
with constant $\Omega_{R}$ a gradient of the detuning $\delta$ due to
a magnetic field gradient also generates a non-zero artificial magnetic
field ${\mathbfcal B}$ providing an artificial Lorentz force. The generated
field ${\mathbfcal B}$ is not purely geometric, and depends on the
relative strength of $\delta$ and $\Omega_{R}$ in addition to their
geometry.

It is noteworthy that the maximum magnetic flux produced by means
of two counterpropagating laser beams amounts to $2f\kr L$ Dirac flux
quanta and is thus proportional to the systems length $L$ \cite{Cooper2011}. This is
a drawback in creating very large magnetic fluxes necessary for reaching
 the fractional quantum Hall regimes. The use of the optical flux
lattices considered in the following Section overcomes this drawback
making the induced flux proportional to the area rather than the length
of the atomic cloud. 

Note also that instead of counterpropagating laser fields one can employ 
co-propagating beams carrying optical vortices with opposite vorticity 
\cite{Cooper2010}. For the first-order Laguerre-Gaussian beams the azimuthal  
angle $\phi$ entering the vector and scalar potentials is then twice the real space 
azimuthal angle. On the other hand, the Raman Rabi frequency  $\Omega_{R}$, entering in 
Eqs. (\ref{eq:grad-phi})-(\ref{eq:cos-theta})  for the polar angle, linearly depends on 
the cylindrical radius in  the vicinity of the vortex core. Non-zero artificial magnetic 
fields can therefore be generated without making the detuning position-dependent. Using vortex beams the number
of Dirac flux quanta imparted onto the atomic cloud is determined
by the vorticity of the beams \cite{Juzeliunas2004,Juzeliunas2005,Juzeliunas2005b,Cooper2010}
and is thus normally much smaller than that induced by counter-propagating
laser fields.

\subsubsection{Optical flux lattices \label{sub:Flux-latt}}

\begin{figure}
\begin{centering}
\includegraphics[width=0.4\textwidth]{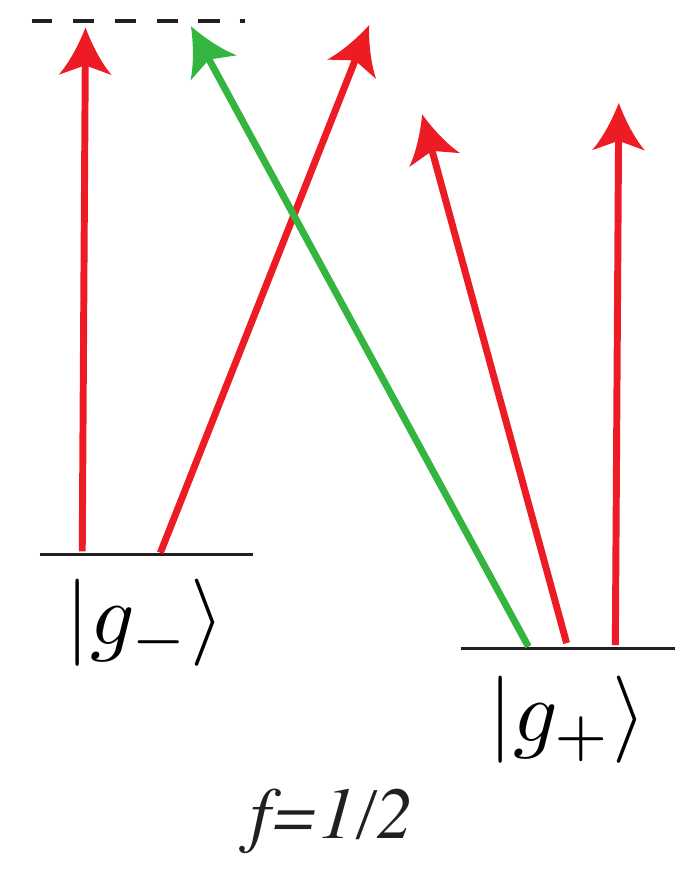} 
\end{centering}
\caption[Omega]{\label{fig:Raman-Scheme} Raman transitions providing the optical flux lattice for $f=1/2$}.
\end{figure}

Optical flux lattices were introduced by N. R. Cooper in Ref. \cite{Cooper2011a}. This concept is based on the observation that singularities in the synthetic vector potential could be created and exploited so as to produce periodic structures with non-trivial magnetic fluxes \cite{Cooper2011a,Juz-Spielm2012NJP}. Such optical flux lattices could be created using a bichromatic
laser field \cite{Cooper2011,Juz-Spielm2012NJP}, as considered in the
previous Section.  Following the proposal by Cooper and Dalibard \cite{Cooper2011},
we consider that the $f=1/2$ ground state magnetic sublevels are coupled via Raman
transitions (Fig.~\ref{fig:Raman-Scheme}) 
involving laser fields with two frequencies.
One frequency component ${\mathbf{E}}_{\omega_{+}}$ shown in 
green represents a circular $\sigma_{-}$ polarized plane wave propagating
along the $z$ axis. Another frequency field ${\mathbf{E}}_{\omega_{-}}$
shown in red has all three circular polarizations. Dalibard and
Cooper considered a situation where the field ${\mathbf{E}}_{\omega_{-}}$
represents a superposition of three linearly polarized plane waves
propagating in the $\ex$-$\ey$ plane and intersecting at 120 degrees,
with polarizations being tilted by some angle with respect to the
propagation plane \cite{Cooper2011}. In that case one can produce
triangular or hexagonal state-dependent lattices for the adiabatic
motion of laser-dressed atoms affected by a non-staggered magnetic
flux.

A square optical flux lattice is formed if the first frequency component
${\mathbf{E}}_{\omega_{+}}$ is again a circular $\sigma_{-}$ polarized
plane wave propagating along the $z$ axis, yet the second frequency
field ${\mathbf{E}}_{\omega_{-}}$ represents a sum of polarization-dependent
standing waves with a $\pi/2$ time-phase difference between the standing
waves oscillating in the $\ex$ and $\ey$ directions \cite{Juz-Spielman2012SPIE,Juz-Spielm2012NJP}%
\footnote{The standing waves with time-phase difference were previously exploited
in the context of the light forces \cite{Hemmerich1991,Hemmerich1992}.%
}. In that case one arrives at an atom-light coupling described by
the state-dependent Hamiltonian $\hat{M}\left(\mathbf{r}\right)=\boldsymbol{\Omega}\cdot\hat{\mathbf{F}}$
with a spatially periodic Zeeman field $\boldsymbol{\Omega}$: 
\begin{equation}
\Omega_{x}=b\Omega_{\parallel}\cos(x\pi/a)\,,\quad\Omega_{y}=b\Omega_{\parallel}\cos(y\pi/a)\,,\quad\Omega_{z}=\Omega_{\parallel}\sin(x\pi/a)\sin(y\pi/a)\,,\label{eq:Omega-x,y,z}
\end{equation}
 where a dimensionless ratio $b$ is controlled by changing the polarisations
of the standing waves \cite{Juz-Spielman2012SPIE,Juz-Spielm2012NJP}.

Atoms with a fixed spin projection $m$ (along the Zeeman field $\boldsymbol{\Omega}$)
then undergo an adiabatic motion in a square lattice potential $\varepsilon_{m}\left(\mathbf{r}\right)+W_{n}\left(\mathbf{r}\right)$ with a periodicity in the $\ex$ and $\ey$ directions twice smaller
than the periodicity $2a$ of the atom-light Hamiltonian $\hat{M}\left(\mathbf{r}\right)$.
The adiabatic motion is accompanied by a vector potential ${\mathbfcal A}_{m}({\bf r})$
containing two Aharonov-Bohm type singularities (per elementary cell)
corresponding to the points where $\Omega_{x}=\Omega_{y}=0$ and $\Omega_{z}<0$.
 In the vicinity of each singular point, the vector potential $\mathbf{{\mathbfcal A}}_{m}$
has a Dirac-string piercing the $\ex$-$\ey$ plane and carrying $-2m$
Dirac flux quanta (with $-f\le m\le f$) \cite{Juz-Spielm2012NJP} 
\begin{equation}
\mathbf{{\mathbfcal A}}_{m}\rightarrow\mathbf{{\mathbfcal A}}_{m}^{AB}=-2\hbar m\nabla\phi\,.\label{eq:A_m-AB}
\end{equation}
 Note that these singular fluxes are gauge-dependent and non-measurable.
One can shift each singular points to another location via a gauge
transformation \cite{Juz-Spielm2012NJP}, yet the singularities can
not be removed from the $\ex$-$\ey$ plane.

Due to the periodicity of the system, the total flux over the elementary
cell is equal to zero
\begin{align}
\frac{1}{\hbar}\oint_{{\rm cell}}\!{\mathbfcal A}_{m}\cdot d\mathbf{r} & =\frac{1}{\hbar}\iint_{{\rm cell}}\!{\mathbfcal B}_{{\rm m}}^{tot}\cdot d\mathbf{S}=0,\label{eq:flux-alpha}
\end{align}
 where ${\mathbfcal B}_{{\rm m}}^{tot}={\mathbfcal B}_{m}(\mathbf{r})+\sum{\mathbfcal B}_{{\rm m}}^{AB}(\mathbf{r})$
is the total magnetic flux density composed of the continuous (background)
magnetic flux density ${\mathbfcal B}_{m}(\mathbf{r})$ and a set
of above-mentioned gauge-dependent singular fluxes ${\mathbfcal B}_{{\rm m}}^{AB}(\mathbf{r})$
of the Aharonov-Bohm type corresponding to $\mathbf{{\mathbfcal A}}_{m}^{AB}$,
Eq.~(\ref{eq:A_m-AB}).

Thus it is strictly speaking impossible to produce a non-zero effective
magnetic flux over the elementary cell using the periodic atom-light
coupling. Yet this does not preclude having a non-staggered continuous
magnetic flux density ${\mathbfcal B}(\mathbf{r})$ over the elementary
cell, because of the existence of non-measurable Dirac strings.
Deducting the latter, the continuous physical flux over the elementary
cell is 
\begin{equation}
\frac{1}{\hbar}\iint_{{\rm cell}}\!{\mathbfcal B}_{m}\cdot d\mathbf{S}=-\frac{1}{\hbar}\sum\oint_{{\rm singul}}\!\mathbf{{\mathbfcal A}}_{m}\cdot d\mathbf{r}=4m\,.\label{eq:flux-alpha-prime}
\end{equation}
 where the summation is over the singular points of the vector potential
(emerging at $\cos\theta\rightarrow-1$ ) around which the contour
integration is carried out.

To summarize, the optical flux lattice contains a background non-staggered
magnetic field ${\mathbfcal B}$ plus an array of gauge-dependent
Dirac-string fluxes of the opposite sign as compared to the background.
The two types of fluxes compensate each other, so the total magnetic
flux over an elementary cell is zero, as it is required from the periodicity
of the Hamiltonian. However, the Dirac-string fluxes are non-measurable
and hence must be excluded from any physical consideration. As a result,
a non-staggered magnetic flux over the optical flux lattice appears,
and topologically non-trivial bands with non-zero Chern numbers can
be formed \cite{Cooper2011,Juz-Spielm2012NJP}. By choosing the proper
laser beams producing the optical flux lattice,  one can produce a configuration where
the lowest energy band is topologically non-trivial and nearly dispersionless, leading to a possible formation of fractional quantum Hall states \cite{CooperDalibard:2012}. Note also
that unlike other lattice schemes involving laser-assisted-tunneling methods, such as proposed in Refs. \cite{Jaksch:2003,Gerbier:2010} (see Section \ref{sect:lattices}), the concept of optical flux lattices is  based on the
adiabatic motion of the atoms. Finally, it is also convenient to reinterpret optical flux lattices as tight-binding models defined in reciprocal space \cite{CooperMoessner:2012}. This complementary picture offers an efficient method to establish the atom-light coupling configurations leading to non-trivial topological bands: following Ref. \cite{CooperMoessner:2012}, any tight-binding model exhibiting non-trivial Chern insulating phases, such as the Haldane model \cite{Haldane:1988}, could constitute the roots of an interesting optical flux lattice.

\subsection{Dressed states: explicit picture}

The preceding discussion followed our initial introduction to artificial
gauge fields (Sect.~\ref{sect:gaugefields}), where we first explicitly
solved the state-dependent part of the Hamiltonian $\hat{M}({\bf r})$,
thereby introducing a vector potential $\mathbf{{\mathbfcal A}}$
for the adiabatic motion of atoms in the dressed-state manifold. This
elegant description clearly illustrates the geometric origin of these
gauge fields, but the projection process is an approximation, and
in many cases,  relevant information can be obtained by directly 
solving the total Hamiltonian (including the kinetic energy part and the atom-light
coupling $\hat{M}$). This can be done explicitly if the atom-light
coupling is induced by plane waves ~\cite{Higbie2002,Spielman2009},
such as in the case of two counterpropagating laser beams considered
above.



 In order to establish a general framework that accommodates the effects of inter-particles collisions, we find convenient to use  the second-quantized formalism, where the Hamiltonian for a particle in a uniform time-independent magnetic field normal to a 2D plane is
\begin{eqnarray}\label{FieldHam}
\int d^2 {\bf x}\frac{\hbar^2}{2 m}\hat \psi^\dagger({\bf x}) \left\{\left[k_x - \frac{{\mathcal A}_x}{\hbar} \right]^2 + \left[k_y - \frac{{\mathcal A}_y}{\hbar} \right]^2\right\}\hat \psi({\bf x}),
\end{eqnarray}
and $\psi^\dagger({\bf x})$ is the field operator for the creation of a particle at position ${\bf x}$, with momentum represented by $\hbar k_{x,y}=-i\hbar\partial_{x,y}$.  We have in mind a situation which explicitly realizes Eq.~(\ref{FieldHam}) in a specific gauge. The actual gauge will depend on the details of the experimental setup, and will be in this case the Landau gauge with ${\mathbfcal A} = -{\mathcal B} y \ex$. The goal is to arrive at a Hamiltonian where the minimum of the energy-momentum dispersion relation $E({\bf k})$ becomes asymmetric~\cite{Higbie2002} and is displaced from zero momentum as a function of spatial position.  The dressed single-particle states are spin and momentum superpositions whose state decomposition depends on the effective vector potential ${\mathbfcal A}$. The canonical momentum associated with this vector potential can be observed by probing the internal state decomposition of the dressed states, as was done by Lin et al~\cite{Lin2009a}, see Sect. ~\ref{sec:ExImplementation}.

This approach relies on a collection of atoms with two or more electronic ground states which interact with two counter-propagating Raman coupling lasers aligned along $\ex$. If both Raman beams are far detuned from the ground to excited state transition there is negligible population in the excited state. The Raman beams then induce a coupling $\Omega_R\exp\left(\pm i \kr x\right)$ between ground states which can lead to synthetic magnetic fields.

\subsubsection{The two-level system}\label{subsect:twolevel}

We will first consider a coupled two level system with internal states $\ket{+}$ and $\ket{-}$, where exact solutions can be obtained ~\cite{Higbie2002}. Physically, these two states might be two $m_F$ levels in the ground state manifold of an alkali atom. For example, the $F=1$ manifold of Rb$^{87}$ at large enough field such that the quadratic Zeeman effect can resolve two out of the three Zeeman sublevels for the Raman or radio frequency transitions. In the frame rotating with angular frequency $\Delta_R/h$, the Raman fields are detuned $\delta= g\mu_B \Delta B/\hbar$ from resonance. The atom-light coupling term in the rotating wave approximation is then
\begin{eqnarray}
\hat H' =&\int d y \int\frac{dk_x}{2 \pi}\left\{\frac{\hbar\Omega_R}{2}\left[\hat \phi_+^\dagger(k_x-2\kr,y)\hat \phi_-(k_x,y) + {\rm h.c.}\right]\right. \\
& +\left.\frac{\hbar\delta}{2}\left[\hat\phi_+^\dagger(k_x,y)\hat \phi_+(k_x,y) - \hat\phi_-^\dagger(k_x,y)\phi_-(k_x,y)\right]\right\}
\end{eqnarray} 
The notation $\hat \phi_\sigma^\dagger(k_x,y)$ denotes the creation of a particle with wave vector $k_x$ along $\ex$ at position $y$, with $\sigma=\pm$.  Here $\hat H'$ also includes the Raman detuning terms.  In the following $\Omega_R$ and $\delta$ will be treated as spatially varying functions of $y$, but not $x$.  This is the explicit expansion of the ${\boldsymbol \Omega} \cdot \hat{\mathbf F}$ contribution to Eq.~(\ref{RWA_Hamiltonian}) with the effective magnetic field from Eq.~\eref{eq:onedimensioneffectivefield} in terms of field operators for the two-level case.

With no coupling, the Hamiltonian is given by $\hat {\mathcal H} =\hat H_x + \hat H_y + \hat U +\hat H_{\rm int}$,  which represent motion along $\ex$, motion along $\ey$, the external potential, and interparticle interactions respectively.  When expressed in terms of the real space field operators $\hat \psi_\sigma({\bf r})$we obtain
\begin{eqnarray}
\hat H_x &= \int d^2{\bf r}\sum_\sigma\hat\psi^\dagger_\sigma({\bf r})\left(-\frac{\hbar^2\partial_x^2}{2m}\right) \hat\psi_\sigma({\bf r})\\
\hat H_y &= \int d^2{\bf r}\sum_\sigma\hat\psi^\dagger_\sigma({\bf r})\left(-\frac{\hbar^2\partial_y^2}{2m}\right)\hat\psi_\sigma({\bf r})\\
\hat U &=  \int d^2{\bf r}\sum_\sigma\hat\psi^\dagger_\sigma({\bf r})U({\bf r})\hat\psi_\sigma({\bf r})\\
\hat H_{\rm int} &= \frac{g_{\rm 2D}}{2} \int d^2{\bf r}\sum_{\sigma,\sigma'} \hat\psi^\dagger_\sigma({\bf r}) \hat\psi^\dagger_{\sigma'}({\bf r}) \hat\psi_{\sigma'}({\bf r}) \hat\psi_{\sigma}({\bf r}).
\end{eqnarray}
The contact interaction for collisions between ultracold atoms in 3D is set by the 3D s-wave scattering length $a_s$, here assumed to be state independent.  Strong confinement in one direction results in an effective 2D coupling constant $g_{\rm 2D}=\sqrt{8\pi}\hbar^2 a_s/m l_{\rm HO}$ where $l_{\rm HO}$ is the harmonic oscillator length resulting from a strongly confining potential along $\ez$.  Finally, $U({\bf r})$ is an external trapping potential, which we assume also to be state-independent.

This problem can be solved exactly when considering free motion along $\ex$, i.e., treating only $\hat H_x$ and $\hat H'$, and going to the momentum representation for the atomic motion along the $x$ axis.  The second quantized Hamiltonian for these two contributions can be compactly expressed in terms of the operators $\hat{\varphi}^\dagger_\pm(q_x,y) = \hat\phi^\dagger_\pm(q_x\mp\kr,y)$.  Using this set, $\hat H\approx\hat H_x+\hat H'$ reduces to an integral over 2$\times$2 blocks 
\begin{eqnarray}
\hat H(q_x,y) &=
\left(
\begin{array}{cc}
\frac{\hbar^2(q_x - \kr)^2}{2m} + \frac{\hbar\delta}{2} & \frac{\hbar\Omega_R}{2} \\
\frac{\hbar\Omega_R}{2} & \frac{\hbar^2(q_x + \kr)^2}{2m} - \frac{\hbar\delta}{2} 
\end{array}\right)\label{TwoByTwo}
\end{eqnarray}
labeled by $q_x$ and $y$.  The expression in terms of $\hat\varphi$ operators instead of $\hat\phi$ is a gauge transformation which boosts the $\ket{+}$ and $\ket{-}$ states in opposite directions.  The dependence of the two-photon coupling $\Omega_R$ and detuning $\delta$ on $y$ has been suppressed for notational clarity.  The resulting Hamiltonian density for motion along $\ex$ at a fixed $y$ is then
\begin{eqnarray}
\hat H_x + \hat H' & = \int\frac{d q_x}{2\pi} \sum_{\sigma,\sigma'} \hat{\varphi}^\dagger_\sigma(q_x,y) \hat H_{\sigma, \sigma'}(q_x,y) \hat {\varphi}_{\sigma'}(q_x,y).
\end{eqnarray}
For each $q_x$, $\hat H(q_x,y)$ can be  diagonalized by the unitary transformation $\hat R(q_x,y) \hat H(q_x,y) \hat R^\dagger(q_x,y)$.  Unlike the approximate solutions discussed in the context of adiabatic gauge fields in Subsections~\ref{subsect:Lambda}-\ref{sub:Spin-in-Magn-field}, these represent the exact solution for motion along $\ex$.

The resulting eigenvalues of $\hat H(q_x,y)$,
\begin{eqnarray}
E_\pm(q_x,y)&=E_R\left[\left(\frac{q_x}{\kr}\right)^2+1\pm\frac{1}{2}\sqrt{\left(\frac{4 q_x}{\kr} - \frac{\hbar\delta}{\Er}\right)^2 + \left(\frac{\hbar\Omega_R}{\Er}\right)^2}\right]
\end{eqnarray}
give the effective dispersion relations in the dressed basis,  $\hat{\tilde{\varphi}}_\sigma(q)=\sum R_{\sigma,\sigma'}(q) \hat {\varphi}_{\sigma'}(q)$.  For each $q$ the eigenvectors of $H(q_x)$ form a family of states~\cite{Papoff1992}.  In terms of the associated real-space operators $\hat \psi'_\sigma({\bf r})$ these diagonalized terms of the initial Hamiltonian are
\begin{eqnarray}
\hat H_x + \hat H' &=\int d^2 {\bf r} \sum_{\sigma=\pm} \hat\psi'^\dagger_\sigma({\bf r}) E_\sigma\left(-i\hbar\frac{\partial}{\partial x},y\right)\hat\psi'_\sigma({\bf r}).\label{TwoLevel1D}
\end{eqnarray}
In analogy with the term ``crystal-momentum'' for particles in a lattice potential, we call the quantum number $q_x$ the ``quasi-momentum''.  Here $-i\hbar\partial_x$ is the real-space representation of the quasi-momentum $q_x$.  The symbol $E_\pm\left(-i\hbar\partial_x,y\right)$ is a differential operator describing the dispersion of the dressed eigenstates, just as the operator $E_x\left(-i\hbar\partial_x,y\right)=(-i\hbar\partial_x+eBy)^2/2m$ describes quadratic dispersion along $\ex$ of a charged particle moving in a magnetic field $B\ez$ in the Landau gauge.

To lowest order in $\Er/\hbar\Omega_R$ and second order in $q_x/\kr$, $E_\pm(q_x,y)$ can be expanded as
\begin{equation}
E_\pm  \approx \frac{\Er}{m^*}\left(\frac{q_x}{\kr} - \frac{\hbar\delta}{4\Er\pm\hbar\Omega_R}\right)^2 + \frac{2\Er\pm\hbar\Omega_R}{2}+ \frac{\hbar^2\delta^2(4\Er\pm\hbar\Omega_R)}{4(4\Er+\hbar\Omega_R)^2}.
\end{equation}
Atoms in the dressed potential are significantly changed in three ways: (1) The energies of the dressed state atoms are shifted by a scalar shift analogous to the sum of the adiabatic and geometric potentials $V({\bf r})$ and $W({\bf r})$. (2) Atoms acquire an effective mass $m^*/m = \hbar\Omega_R / (\hbar\Omega_R\pm4\Er)$. This change is absent in the adiabatic picture. (3) The centre of the dispersion relation is shifted to ${\mathcal A}_x/\hbar\kr = \hbar\delta / (\hbar\Omega_R\pm 4\Er)$.  Just as with the adiabatic case considered in the previous Subsections~\ref{subsect:Lambda}-\ref{sub:Spin-in-Magn-field}, ${\mathcal A}_x$ can depend on $y$ by a spatial dependence on $\Omega_R$, or, via $\delta(y)$ as described below.  In either case, the effective Hamiltonian is that of a charged particle in a magnetic field expressed in the Landau gauge.

\begin{figure}[tbp]
\begin{center}
\includegraphics[width=4.0in]{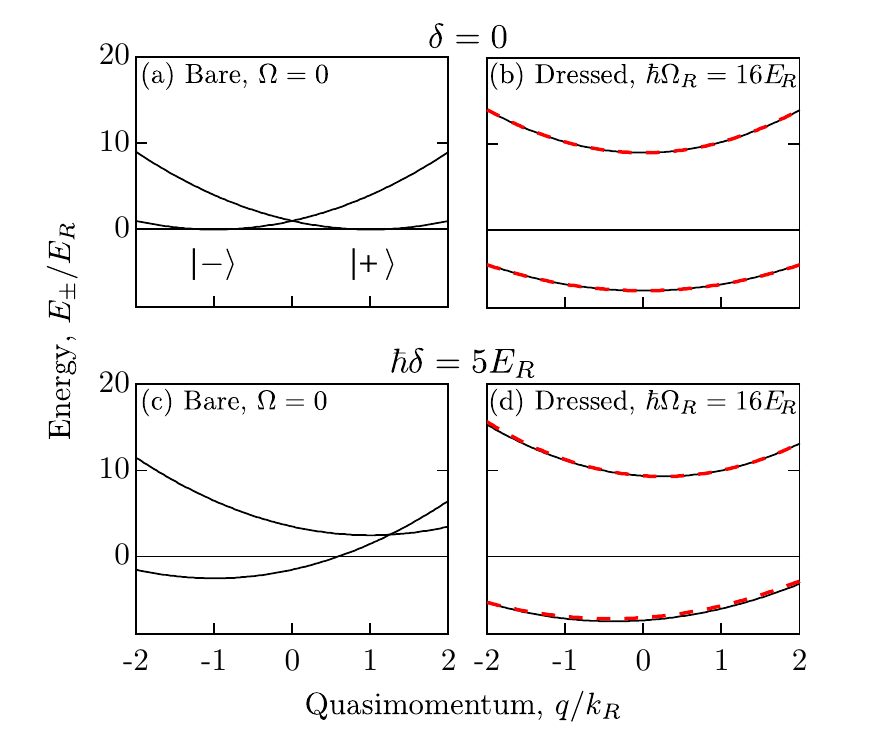}
\end{center}
\caption[Dressed state dispersion for the 2 level system]{Each panel illustrates the dressed state dispersion relations for two level dressed state atoms.  The horizontal axis is quasi-momentum $q_x$, and the vertical axis is the dressed state energy $E_\pm(q_x)$.  The black lines are the exact eigenvalues of Eq.~\eref{TwoByTwo}, and the red dashed lines are the analytic approximation.  (a) Bare potentials with Raman beams on resonance  $(\Omega_R=\delta=0\Er)$. (b) Dressed potentials with $\hbar\Omega_R=16\Er$, $\hbar\delta = 0\Er$. (c) Bare potentials with Raman beams off resonance  $(\Omega_R=0$ and $\hbar\delta = 5\Er$). (d) Dressed potentials with Raman beams off resonance ($\hbar\Omega_R=16\Er$, $\hbar\delta = 5\Er$).}

\label{2dressedstates}
\end{figure}

Figure \ref{2dressedstates} shows the dressed state dispersion relations in this model.  Panels (a) and (c) show the undressed case ($\Omega_R=0$) for detuning $\hbar\delta=0$ and $5\Er$ respectively.  Panels (b) and (d) depict the same detunings, for $\hbar\Omega_R=16\Er$, where the exact results (solid line) are displayed along with the approximate dispersion (red dashed line).  Fig.~\ref{2dressedstates}b shows the strongly dressed states for large $\Omega_R$, each of which is symmetric about $q=0$.  When detuned as in panel c, the dispersion is displaced from $q=0$. If space dependent this displacement leads to a non-trivial gauge potential.  In the limit of very small $\Omega_R$ the dressed curve $E_-(q_x,y)$ forms a double-well potential as a function of $k_x$.  In a related Raman-coupled system, Bose condensation in such double-well potentials were studied theoretically~\cite{Higbie2002,Montina2003,Higbie2004,Stanescu2008}, and has been explored in detail in the context of spin-orbit coupling~\cite{Lin2011,Zhang2012PRL,Wang2012,Cheuk2012}.
\subsubsection{Synthetic gauge fields}

\begin{figure}[tbp]
\begin{center}
\includegraphics[width=3.375in]{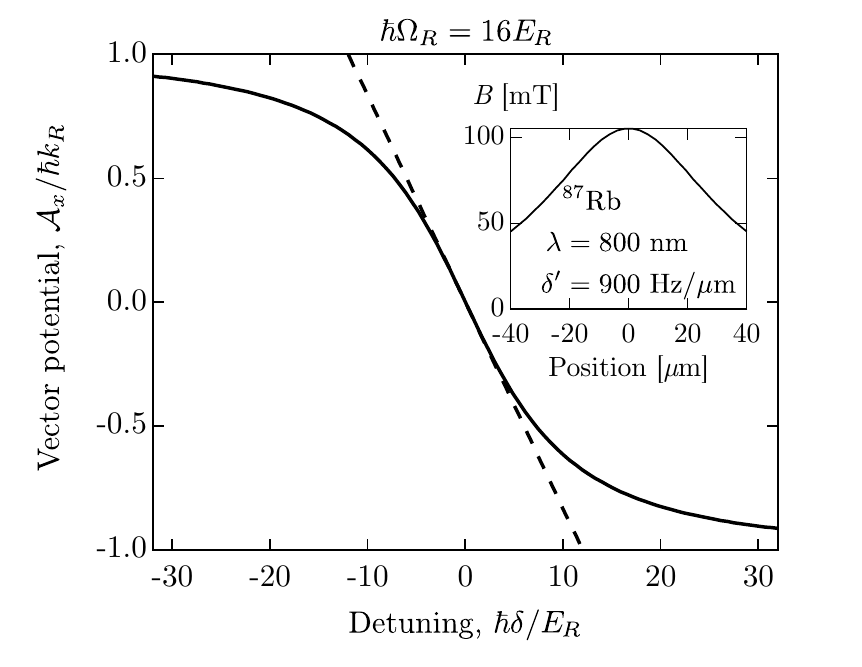}
\end{center}
\caption[Effective vector potential for the 2 level system]{
Effective vector potential ${\mathcal A}/\hbar\kr$ versus detuning $\hbar\delta/\Er$ for $\hbar\Omega_R = 16\Er$.  The solid line is the exact result and the dashed line is the lowest order expansion in $\delta$.  The inset shows the synthetic magnetic field $|{\mathcal B}|$, indicating the predicted degree of field inhomogeneity predicted quantities relevant to experiment, computed for Rb$^{87}$ with a detuning gradient $\hbar\delta'(y) = 900\Hz/\mu{\rm m}$, and $\lambda = 800\nm$ Raman lasers.
}
\label{Limitations}
\end{figure}

The  Raman coupling leads to a dressed dispersion along $\ex$, where motion along $\ey$ is largely unaffected.  When the detuning is made to vary linearly along $\ey$,  with $\delta(y)=\delta' y$, an effective single particle Hamiltonian contains a 2D effective vector potential ${\mathbfcal A}/\hbar \kr \approx \hbar\delta' y / (4\Er\pm \hbar\Omega_R) \ex$. The synthetic magnetic field is therefore ${\mathcal B}_z/\hbar\kr \approx -\hbar\delta' /(4\Er\pm\hbar\Omega_R)$.  Figure~\ref{Limitations} shows the vector potential as a function of detuning $\delta$.  As expected, the linear approximation discussed above (dashed line) is only valid for small $\delta$. The synthetic field therefore decreases from its peak value as $\delta$ increases (top inset).

This technique also modifies the trapping potential along $\ey$. In the adiabatic picture, this comprises the sum of the adiabatic and geometric potentials $V({\bf r})$ and $W({\bf r})$ which have no individual identity in this exact solution.  When the initial potential $U(x,y)$ is harmonic with trapping frequencies $\omega_x$ and $\omega_y$, the combined potential along $\ey$ becomes $U_\pm(y) = m [\omega_y^2+(\omega_\pm^*)^2] y^2/2$, where $m (\omega_\pm^*)^2/2 \approx \hbar^2\delta'^2 (4\Er\pm\hbar\Omega_R)/4\hbar\kr(4\Er+\hbar\Omega_R)^2$.

This contribution to the overall trapping potential is not unlike the centripetal term which appears in a rotating frame of reference, where a synthetic magnetic field ${\mathbfcal B}$ arises as well.  In the case of a frame rotating with angular frequency $\Omega$, the centripetal term gives rise to a repulsive harmonic term with frequency $\omega_{\rm rot}^2 = \left({\mathbfcal B} / 2 m\right)^2$.  In the present case the scalar trapping frequency can be rewritten in a similar form $\omega_\pm^{*2} = ({\mathcal B} / 2 m)^2\times\left|4\Er\pm\hbar\Omega_R\right|^3/\hbar\kr(4\Er+\hbar\Omega_R)^2$. The scalar potential may be attractive or repulsive, and it increases in relative importance with increasing $\Omega_R$.

We have so far omitted motion along $\ey$, interactions, and the effect of an external potential. The inclusion of these effects are discussed in detail in Ref.~\cite{Spielman2009}.  The outcome is that in the limit of large $\Omega_R$ these are unchanged by the transformation into the dressed basis.
Together these allow the construction of the real space Hamiltonian in the basis of localized spin-superposition states $\hat \psi'({\bf r})$.  The density distribution of these localized states is not perfectly localized, but are some fraction of an optical wavelength $\lambda$ in extent.  For interactions, this is important because dressed state atoms separated by this distance will in fact interact.
For particles starting in higher bands, transitions to lower bands are energetically allowed and a Fermi's Golden Rule argument thus gives rise to decay from all but the lowest energy dressed state~\cite{Spielman2006}.

\subsubsection{Limitations}

This technique is not without its limitations.  Foremost among them is the range of possible ${\mathcal A}_x/\hbar\kr$ shown in Fig.~\ref{Limitations} where $\hbar\Omega_R = 16\Er$.  While the linear expansion (dashed) is unbounded, the exact vector potential is bounded by $\pm\kr$.  For example, the hybridized combination of $\ket{+,q-\kr}$ and $\ket{-,q+\kr}$ cannot give rise to dressed states with minima more positive than $q = +\kr$ where the energy of the $\ket{+}$ states is minimized without dressing,  Fig.~\ref{2dressedstates} (c). The minima can also not be more negative than $q = -\kr$.

This limitation does not effect the maximum attainable field, only the spatial range over which this field exists.  Specifically, a linear gradient in $\delta(y)$ gives rise to the synthetic field ${\mathcal B}_z(y)$ which is subject to $\int_{-\infty}^{\infty}{\mathcal B}_z(y)dy=2\hbar\kr$.  This simply states that the vector potential -- bounded by $\pm \hbar\kr$ -- is the integral of the magnetic field.  Note however, that along $\ex$ the region of large ${\mathcal B}_z$ has no spatial bounds.

A second limitation of this technique is the assumption of strong Raman coupling between the Zeeman split states.   As already pointed out at the end of the Sec. \ref{sec:TwoRaman-general}, for alkali atoms, when the detuning from atomic resonance $\Delta_e$ is large compared to the excited state fine structure $\Delta_{{\rm FS}}$, the two-photon Raman coupling for $\Delta m_F=\pm 1$ transitions drops as $\Omega_R \propto \Delta_{{\rm FS}}/ \Delta_e^{2}$, not $\Delta_e^{-1}$ as for the AC Stark shift.   As a result, the ratio between the Raman coupling  $\Omega_R$ and off-resonant scattering cannot be increased by using larger detuning, and the only solution is to consider atoms with large fine structure splitting $\Delta_{{\rm FS}}$, such as rubidium.  

\subsubsection{Three-level system}\label{subsect:threelevel}

\begin{figure}[tbp]
\begin{center}
\includegraphics[width=3.375in]{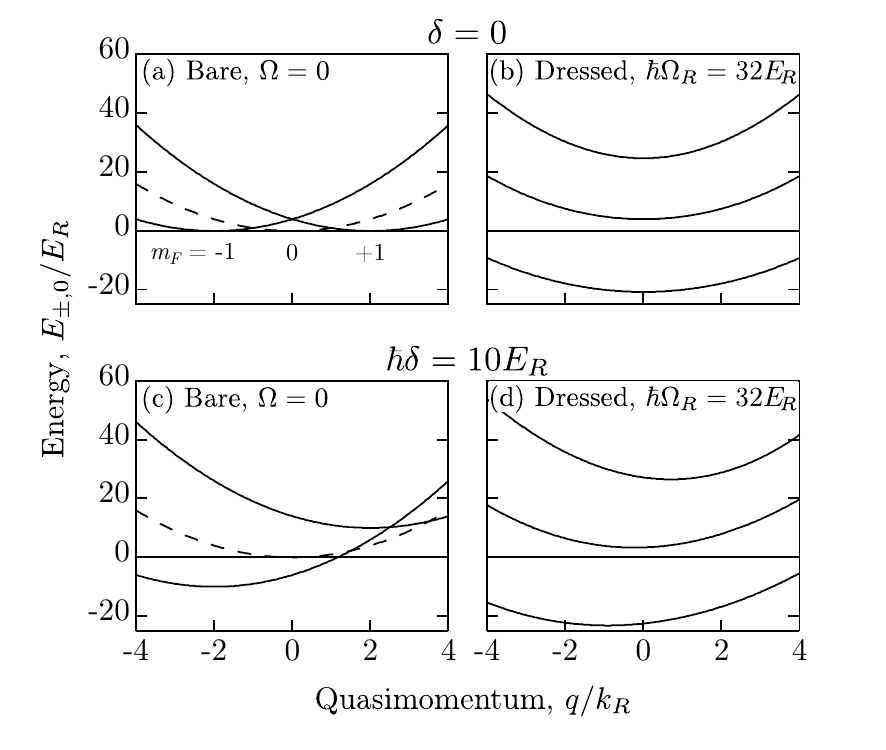}
\end{center}
\caption[Dressed state dispersion for the three-level system]{
Each panel denotes the dressed state dispersion relations $E(q_x)$ for three-level atoms dressed by counter-propagating Raman beams.  The horizontal axis is quasi-momentum $q_x$, and the vertical axis is energy.  (a) Bare potentials  (undressed) with Raman beams on resonance. (b) Dressed potentials ($\hbar\Omega_R=32\Er$, $\hbar\delta = 0\Er$). (c) Bare potentials (undressed) with Raman beams off resonance ($\hbar\delta = 5\Er$). (d) Dressed potentials with Raman beams off resonance ($\hbar\Omega_R=32\Er$, $\hbar\delta = 10\Er$).
}
\label{dressedstates}
\end{figure}

The range of possible effective vector potentials can be extended by coupling more states, for example the $m_F$ states of an $f>1/2$ manifold in the linear Zeeman regime.  We can follow the two-level example above, except there are no compact closed-form solutions.   In the adiabatic picture, Eqs~(\ref{eq:vect-spin})
and (\ref{eq:grad-phi})-(\ref{eq:cos-theta}), additional levels extended the range of the vector potential from $\pm 2 \hbar \kr$ (the two level case) to $\pm 2 \hbar f \kr$ for arbitrary $f$.

This can be illustrated by considering an optically-trapped system of Rb$^{87}$ atoms in the $f=1$ manifold in a small magnetic field which splits the three $m_F$ levels by $g\mu_B |{\bf B}|$. The $3\times3$ blocks $\hat H(q_x)$ describing the three internal states of the $f=1$ manifold are
\begin{equation}
\check H(q_x) = \left(
\begin{array}{ccc}
\frac{\hbar^2(q_x - 2\kr)^2}{2m} + \hbar\delta & \frac{\hbar\Omega_R}{2} & 0 \\
\frac{\hbar\Omega_R}{2} & {q_x}^2 + \epsilon & \frac{\hbar\Omega_R}{2} \\
0 & \frac{\hbar\Omega_R}{2} & \frac{\hbar^2(q_x + 2\kr)^2}{2m} - \hbar\delta
\end{array}\right).\label{ThreeLevelMatrix}
\end{equation}
In this expression, $\delta$ is the detuning of the two photon dressing transition from resonance, $\epsilon$ accounts for any quadratic Zeeman shift (not included in the Sect.~\ref{sub:Spin-in-Magn-field}), $\Omega_R$ is the two-photon transition matrix element, and $q_x$, in units of the recoil momentum $\kr$, is the atomic momentum displaced by a state-dependent term $q_x = k-2\kr$ for $m_F=-1$, $q_x = k$ for $m_F=0$, and $q_x = k+2\kr$ for $m_F=+1$.  Here, the three eigenvalues are denoted by $E_{\pm}$ and $E_0$.  As with the two-level case, the states associated with eigenvalues $E_\pm$ experience an effective vector potential which can be made position-dependent with a spatially varying detuning $\delta$. The $E_0$ also experiences a synthetic field, but at higher order in $\Omega_R^{-1}$.  A  magnetic field gradient along $\ey$ gives $\delta\propto y$, and generates a uniform synthetic magnetic field normal to the plane spanned by the dressing lasers and real magnetic field $\bf B$.

\subsection{Experimental implementation}\label{sec:ExImplementation}

The preceding sections described an overall procedure by which light-induced artificial gauge fields can be created. In this section, we look at three effects: the introduction of a spatially uniform vector potential ${\mathbfcal A}$, the use of a temporal gradient to induce an electric field  ${\mathbfcal E} = -\partial{\mathbfcal A}/\partial t$, and lastly the inclusion of a spatial gradient which gives rise to a magnetic field ${\mathbfcal B} = {\boldsymbol \nabla}\shorttimes{\mathbfcal A}$.  For more technical details we refer the reader to the original publications~\cite{Lin2009a,Lin2011a,Lin2009b}.

\subsubsection{The effective vector potential}
\begin{figure}[tb]
\begin{center}
\includegraphics[width=4.5in]{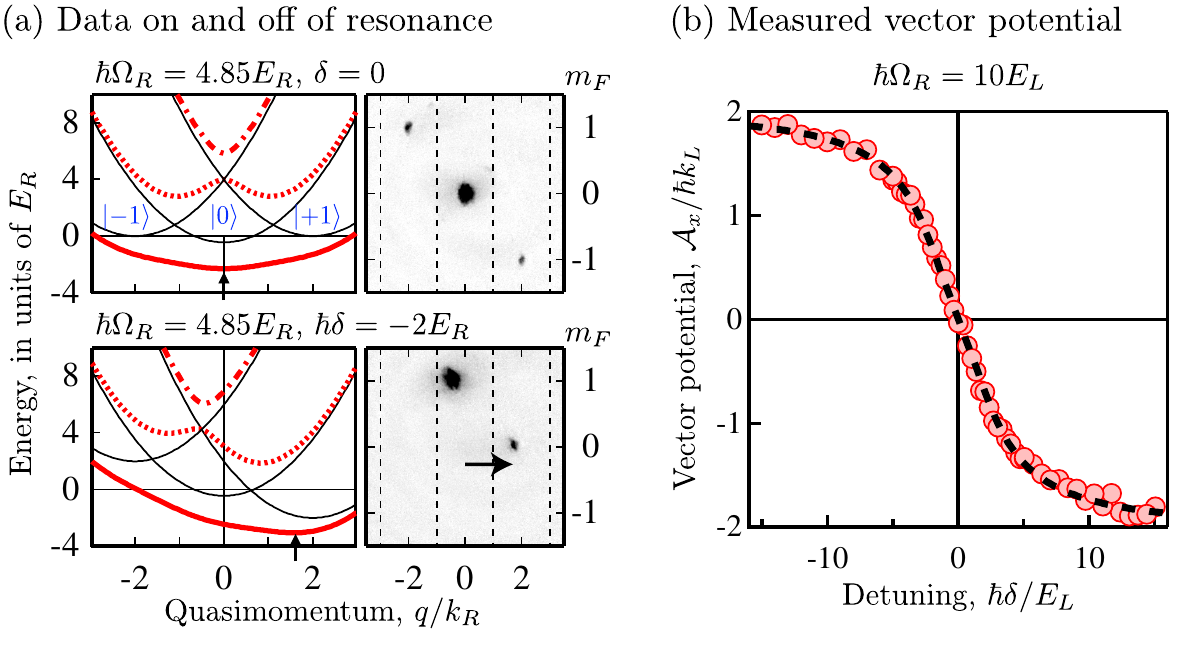}
\end{center}
\caption[Artificial vector potential]{The artificial vector potential.  (a) Left: computed dispersion relations, where the solid curves are the bare states and the dashed curves are the dressed states.  The vertical arrow locates the minimum of the lowest energy dressed curve.  Right: representative image of Raman-dressed atoms both on and off of resonance showing the three $m_F$ components differing in momentum by $\pm\hbar\kr$, displaced from zero $q$ when $\delta\neq0$.  The horizontal arrow depicts the displacement of the $m_F=0$ cloud from zero corresponding the the non-zero vector potential.  (b) Measured momentum of $m_F = 0$ component as a function of detuning $\delta$.  With the gauge choice ${\mathbfcal A}(\delta = 0) = 0$, this constitutes a direct measurement of the artificial vector potential ${\mathbfcal A}(\delta)$.  The dashed curve is the predicted vector potential for the experimental parameters.  Data first appeared in Y.-J. Lin, {\it et al.} PRL {\bf 102}, 130401 (2009), and Y.-J. Lin, {\it et al.} Nat. Phys. {\bf 7}, 531 (2011), Refs.~\cite{Lin2009a}, and \cite{Lin2011a}.
}
\label{fig_VectorPotential}
\end{figure}

Figure~\ref{fig_VectorPotential} shows the basic principle for creating artificial vector potentials.  In these experiments an optically trapped $\Rb87$ BEC was adiabatically transferred from the initial $\ket{f=-1,m_F = 1}$ state into the $\ket{q_x=0, -}$ dressed state, with quasimomentum $q=0$, which corresponds to the lowest energy dressed band at $\delta=0$, as identified by the black arrow in the top-left panel to Fig.~\ref{fig_VectorPotential}a. Once loaded into this state, the atoms were held for a brief equilibration time, at which point both the Raman lasers and the trapping lasers were suddenly turned off.  After the turn off, the atoms were allowed to travel ballistically along $\ex$, while a magnetic field gradient Stern-Gerlach separated the three $m_F$ components along $\ey$.  This process allowed the direct detection of the spin-momentum superpositions which comprise the Raman dressed states.  The top-right panel of Fig.~\ref{fig_VectorPotential}a shows that for $\delta=0$ the ground state BEC is located at $q=0$, and that the dressed state is made up of three different $m_F$ states each with momentum given by $k_{m_F} = q - 2 m_F \kr$.  

The lower two panels of Fig.~\ref{fig_VectorPotential}a depict an experiment where the BEC is loaded as described above.  The detuning is slowly ramped from $0$ to $\hbar\delta = -2\Er$.  During such a ramp, the BEC adiabatically remains at the minimum of $E_-(q)$.  The magnitude of the resulting shift is consequently associated with the artificial vector potential.  The results of many such experiments is shown in Fig.~\ref{fig_VectorPotential}b which compares the measured vector potential. In this case the Raman lasers intersected at $90$ degrees.  This has the effect of decreasing the effective recoil momentum to $\kl = \kr / \sqrt{2}$ and the recoil momentum to $\El = \Er/2$. The prediction of the model is obtained by numerically solving Eq.~\eref{ThreeLevelMatrix}  with no free parameters.

It is important to note that  the BEC was always at rest in these experiments. The uniform and static vector potential would therefore normally not be experimentally detectable. In this case, however, the vector potential's difference from the $\delta=0$ case is encoded into the spin-momentum superposition making up the dressed state, and can be accessed in experiment. See also Ref.~\cite{Lin2011a} for more details of this effect.

\subsubsection{Creating artificial electric fields}\label{sec:Electric field}

\begin{figure}[tb]
\begin{center}
\includegraphics[width=4.5in]{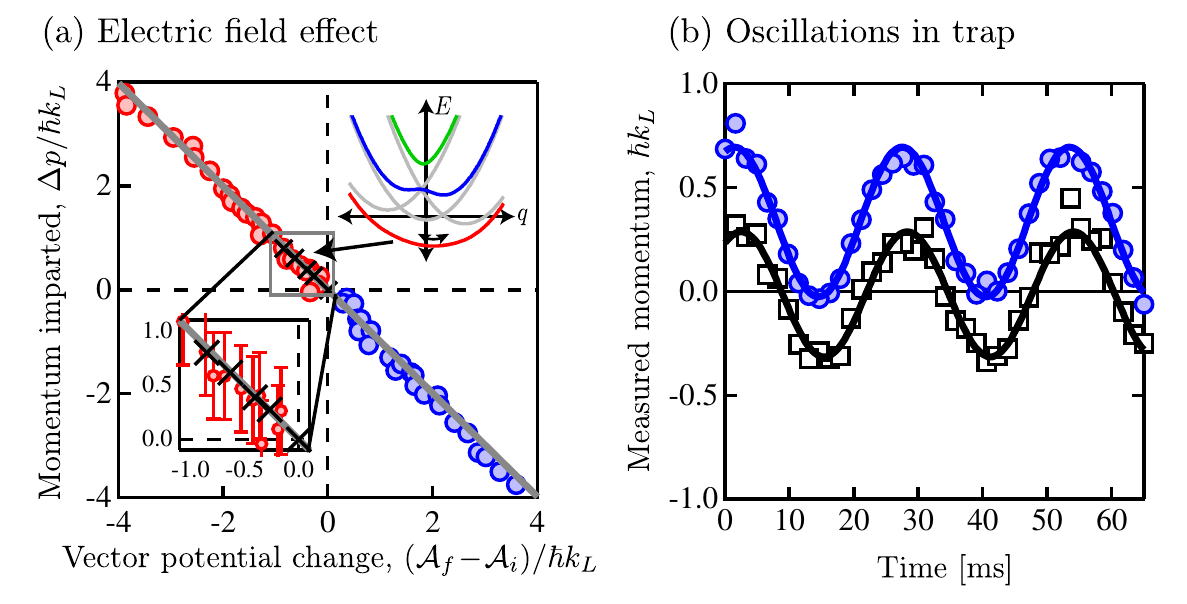}
\end{center}
\caption[Artificial electric field]{The artificial electric field.  (a)  Observed momentum kick resulting from a sudden change in ${\mathbfcal A}$.  The symbols indicate data in which the vector potential was changed from ${\mathbfcal A}_i\propto\ex$ to ${\mathbfcal A}_f = \pm 2\hbar\kl\ex$ ("+'' for red and "-" for blue), at which point the BEC's momentum was measured.  The black crosses (bottom inset) reflect data for which ${\mathbfcal A}_f=0$, where the BEC was allowed to oscillate in the harmonic trap. The amplitude in momentum of these oscillations is plotted.   The grey line is the expected outcome, with slope $-1$.  (b)  The in trap oscillation data, fitted to a sinusoidal model.  The blue circles depict data where ${\mathcal A}_f\neq0$.  As depicted by the top inset in (a), the measured canonical momentum oscillates around a non-zero value given by ${\mathcal A}_f$.  The black squares depict the average velocity of the three components measured after TOF multiplied by the effective mass $m^*\approx2.5m$.  Data first appeared in Y.-J. Lin, {\it et al.} Nat. Phys. {\bf 7}, 531 (2011), Ref.~\cite{Lin2011a}.
}
\label{fig_ElectricField}
\end{figure}


With access to an artificial vector potential it is natural to consider what kind of mechanical forces one can induce on the atoms with suitable gradients of this potential. Several possibilities are available. One can for instance change the vector potential from an initial ${\mathbfcal A}_i$ to a final ${\mathbfcal A}_f$ by suddenly changing $\delta$.  If ${\mathbfcal A}$ behaves as a real vector potential, then this change should be associated  with a force that accelerates the atoms, changing their {\it mechanical} momentum by an amount ${\boldsymbol \delta} {\bf p}_{\rm mech} = \left({\mathbfcal A}_i - {\mathbfcal A}_f\right)$.

There are currently two experiments by Lin et al., \cite{Lin2011a} which illustrate the effect of an artificial electric field. The atoms were first prepared at an initial detuning $\delta$ producing a non-zero ${\mathbfcal A}_i$, and then suddenly making $|\delta|\gg\Omega_R$. This results in ${\mathbfcal A}_f = \pm2\hbar\kl\ex$ which depends on the sign of $\delta$, see Fig.~\ref{fig_VectorPotential}b. Combined with free expansion of the cloud the final momentum can be measured, shown in Fig.~\ref{fig_ElectricField}a.


Alternatively, ${\mathbfcal A}$ can be changed only slightly, as illustrated in Fig.~\ref{fig_ElectricField}b. By taking the average velocity of all three spin-momentum components the quasimomentum and also the group velocity can be monitored.  The subsequent evolution in the harmonic confining potential is then monitored. Fig.~\ref{fig_ElectricField}b depicts the mechanical momentum $p_{\rm mech}$, related to the group velocity,  by multiplying the average velocity by the effective mass $m^*\approx 2.5 m$.  This quantity has the same amplitude as the canonical momentum oscillations, but is centered on zero.   

\subsubsection{Inclusion of a magnetic field} \label{ian:vortex}

\begin{figure}[h!]
\begin{center}
\includegraphics[width=4.5in]{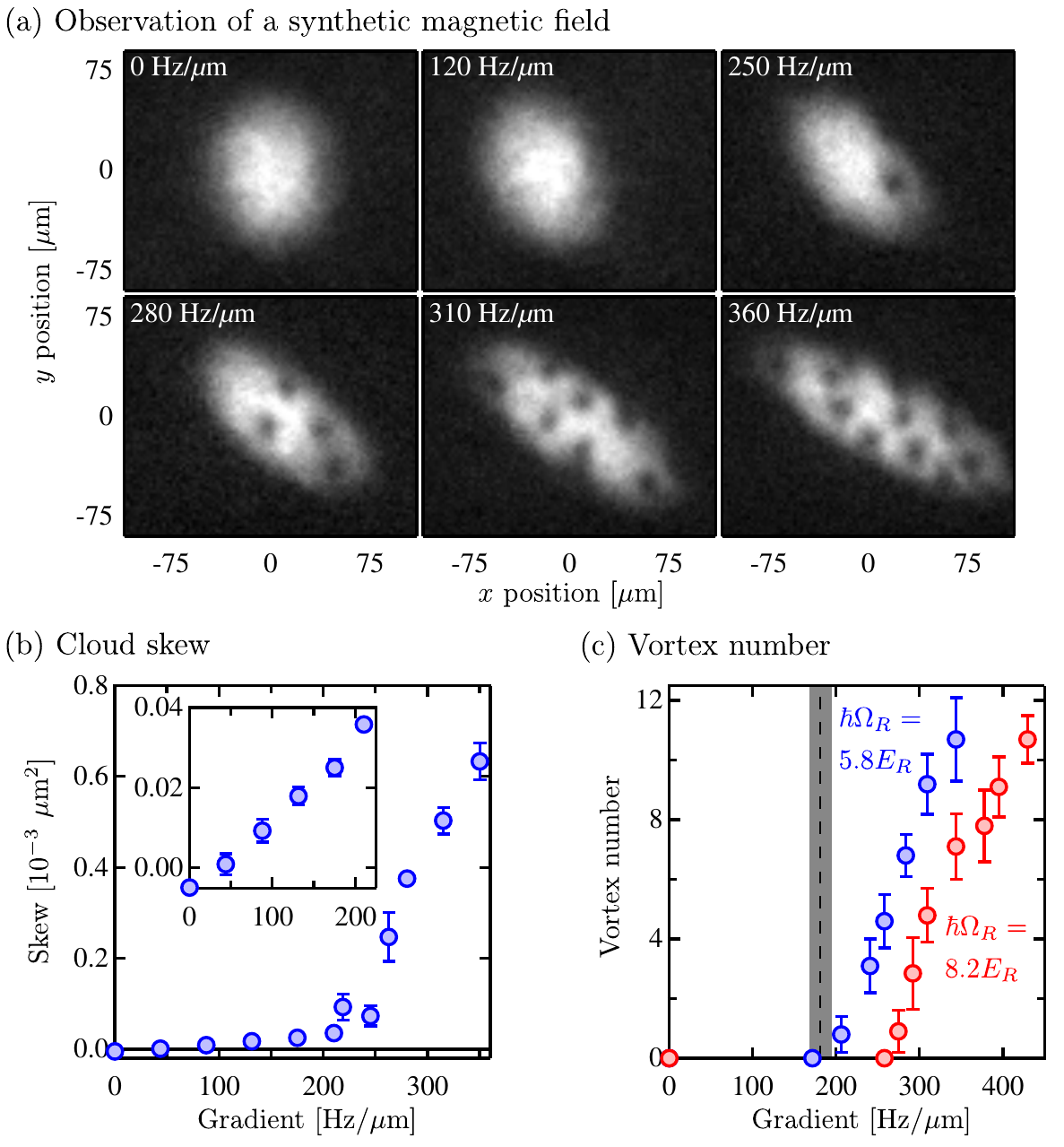}
\end{center}
\caption[Artificial magnetic field]{The artificial magnetic field.  (a)  A detuning gradient $\delta^\prime$ along $\ey$ from a magnetic field gradient, creates an artificial magnetic field.  This causes a shearing of the atomic cloud and allows the entry of vortices into the BEC.  (b)  With the gradient present, the vector potential depends on $y$ as  ${\mathcal A}_x\propto y$. When the Raman lasers are turned off, ${\mathcal A}_x\rightarrow0$.  This process introduces an electric field along $\ex$ that depends on $y$, and as a result the BEC undergoes a shearing motion during TOF.  (c) Above a critical gradient (grey band) it becomes energetically favourable for vortices to enter the BEC.  Data in (a) and (c) first appeared in Y.-J. Lin, {\it et al.} Nature. {\bf 462}, 628 (2009), Ref.~\cite{Lin2009b}.
}
\label{fig_MagneticField}
\end{figure}

With a synthetic  Abelian vector potential present, it is also natural to consider creating an artificial magnetic field ${\mathbfcal B}={\boldsymbol \nabla}\shorttimes{\mathbfcal A}$. The key ingredient for obtaining a non-zero effective magnetic field is the dressed state, given by a detuning which varies linearly along $\ey$,  with $\delta(y)=\delta' y$. This gives 
${\mathbfcal A}\approx -2\hbar\kr \delta/\Omega_R\ex$ when $\delta\ll\Omega_R$.

To create an artificial magnetic field such as in Ref. \cite{Lin2009b}, the geometry shown in Fig.~\ref{fig_layout} can be used, with the three $m_F$ states of $\Rb87$'s hyperfine ground state Zeeman split by a biasing magnetic field along $\ey$. The states are coupled with a pair of Raman lasers giving a momentum exchange along $\ex$.  Using a pair of quadrupole coils in an anti-Helmholtz geometry aligned along $\ez$, produces a magnetic field gradient of the form ${\bf B}_{\rm quad} \approx \beta^\prime(x\ex + y\ey -2z\ez)$ which adds to the bias field ${\bf B}_0 = B_0\ey$. The resulting Zeeman shift is proportional to $\left|{\bf B}_0 + {\bf B}_{\rm quad}\right|\approx B_0 + \beta^\prime y$ giving the desired detuning gradient along $\ey$, which in turn results in a synthetic magnetic field ${\mathbfcal B} = {\mathcal B}\ez$.  Figure~\ref{fig_MagneticField}a depicts images of atoms after TOF, in which the appearance of an artificial magnetic field is marked in two ways.

Firstly, the initially symmetric cloud acquires a shear as ${\mathcal B}$ increases.  With the gradient present, the vector potential depends on $y$ as ${\mathcal A}_x\propto y$. When the Raman lasers are turned off at the beginning of the time of flight, ${\mathcal A}_x\rightarrow0$.  This process introduces an electric field along $\ex$ proportional to $y$, and as a result the BEC undergoes a shearing motion during the expansion. 

Above a critical gradient vortices spontaneously enter into the non-rotating BEC.  In this regime, the BEC is described by a macroscopic wave function $\psi({\bf r})=|\psi({\bf r})|e^{i\phi({\bf r})}$, which obeys the Gross-Pitaevskii equation. The phase $\phi$ winds by
$2\pi$ around each vortex, with amplitude $|\psi|=0$ at the vortex center. The magnetic flux $\Phi_{{\mathcal B}}$ results in $N_v$ vortices and
for an infinite, zero temperature system, the vortices are arrayed in a lattice~\cite{Yarmchuk1979} with density ${\mathcal B}/h$. For finite
systems vortices are energetically less favourable, and their areal density is below this asymptotic value, decreasing to zero at a
critical field ${\mathcal B}_c$. 


\section{Non-Abelian gauge potentials and spin-orbit coupling}
\label{sect:soc}



\label{sect:soc}

Let us now turn to the case of non-Abelian geometric gauge potentials. As already mentioned
in Section \ref{sect:gaugefields}, they emerge when the centre
of mass motion of atoms takes place in a manifold of degenerate or
quasi-degenerate internal dressed states. The non-Abelian gauge
potentials provide a coupling between the centre of mass motion and
the internal (spin or quasi-spin) degrees of freedom, thus causing
an effective ``spin-orbit" coupling. In this Section, we shall
overview several schemes for creating these gauge potentials
in cold-atom setups. We refer to Sect. \ref{non-Abelian_Wilson} for a discussion on genuine non-Abelian gauge structures.

\subsection{The tripod scheme\label{sub:Tripod-scheme}}

\begin{figure}
\begin{centering}
\includegraphics[width=7cm]{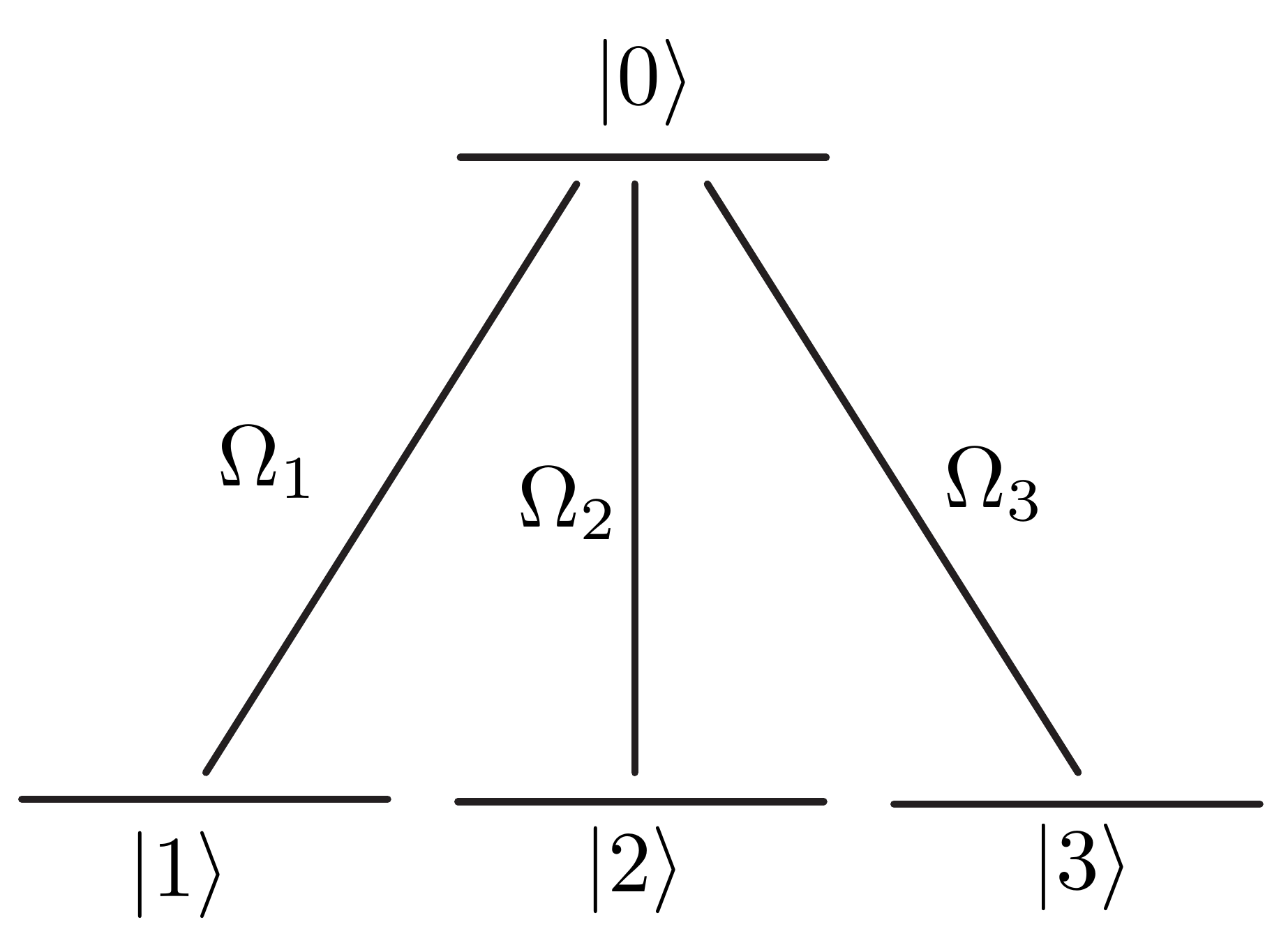}\caption{The tripod configuration of the atom-light coupling in which the atomic
ground states $|j\rangle$ (with $j=1,2,3$) are coupled to the excited
state $|0\rangle$ via the Rabi frequencies $\Omega_{j}$ of the laser
fields. }

\par\end{centering}

\label{tripod-figure} 
\end{figure}
Let us begin with a tripod setup (shown in Fig. \ref{tripod-figure})
for the atom-light coupling, comprising three lower atomic levels coupled with an excited level via the laser fields with the
Raman frequencies $\Omega_{j}$. Compared to the $\Lambda$ setup
analyzed in the previous Section, now there is an additional third
laser driving the transitions between an extra ground state $3$ and
the excited state $0$. The tripod scheme has two degenerate dark
states representing the superpositions of the three ground states
immune to the atom-light coupling. 

The tripod scheme was initially considered by Olshanii and co-authors
\cite{Olshanii91QO,Kulin1999} in the context of  velocity selective
coherent population trapping. In another development Klaas
Bergmann and co-workers theoretically \cite{Unanyan98OC,Unanyan99PRA,Unanyan2004}
and experimentally \cite{Theuer1990OE,Vewinger2003} explored the
adiabatic evolution of the dark-state atoms in the time-dependent laser
fields (the tripod-STIRAP), governed by the matrix valued geometric
phase \cite{Wilczek:1984}. Ruseckas et al considered
the adiabatic motion of the dark-state atoms in the spatially varying
laser fields \cite{Ruseckas2005} providing non-Abelian gauge potentials
for the centre of mass motion using the tripod scheme. 

Assuming exact resonance for the atom-light coupling, the Hamiltonian
of the tripod atom reads in the interaction representation 
\begin{equation}
\hat{M}=\frac{\hbar}{2}\Bigl(\Omega_{1}|0\rangle\langle1|+\Omega_{2}|0\rangle\langle2|+\Omega_{3}|0\rangle\langle3|\Bigr)+H.c.=\frac{\hbar}{2}\Omega|0\rangle\langle B|+H.c.\,,\label{eq:U-tripod}
\end{equation}
where $\Omega=\sqrt{|\Omega_{1}|^{2}+|\Omega_{2}|^{2}+|\Omega_{3}|^{2}}$
is the total Rabi frequency and $|B\rangle=\left(\Omega_{1}^{*}|1\rangle+\Omega_{2}^{*}|2\rangle+\Omega_{3}^{*}|3\rangle\right)/\Omega$
is the atomic bright state, representing a superposition of the atomic
ground states directly coupled to the excited state $|0\rangle$. 

The Hamiltonian $\hat{M}$ has two eigenstates $\left|\chi_{1}(\mathbf{r})\right\rangle \equiv|D_{1}\rangle$
and $\left|\chi_{2}(\mathbf{r})\right\rangle \equiv|D_{2}\rangle$
known as the dark (or uncoupled) states which are orthogonal to the
bright state $|B\rangle$, contain no excited state contribution and
are characterized by zero eigenenergies ($\hat{M}|D_{j}\rangle=0$). 
The bright state $|B\rangle\equiv|B\left(\mathbf{r}\right)\rangle$
is position dependent due to the position dependence of the amplitude
and the phase of the laser Rabi frequencies $\Omega_{j}\equiv\Omega_{j}\left(\mathbf{r}\right)$.
Thus the adiabatic elimination of the bright and excited states leads
 to the adiabatic centre of mass motion of the dark state atoms
affected by the vector potential $\hat{{\mathbfcal A}}^{\left(q\right)}$
(with $q=2$ since there are two dark states) to be labelled simply
by $\hat{{\mathbfcal A}}$. The explicit expressions for the vector
potential $\hat{{\mathbfcal A}}$ and the accompanying geometric scalar
potential are available in the previous original \cite{Ruseckas2005}
or review \cite{Juzeliunas2008-Chapter,Dalibard2011} articles. 

If two Rabi frequencies $\Omega_{1,2}$ represent co-propagating first
order Laguerre-Gaussian beams with opposite vorticity along the propagation
axis $z$ and the third Rabi frequency $\Omega_{3}$ represent the
first order Hermite-Gaussian beam propagating in $x$ direction, the
induced vector potential $\hat{{\mathbfcal A}}$ contains a contribution
due to a magnetic monopole \cite{Ruseckas2005,Pietila09PRL} times
a Pauli matrix. On the other hand, if all three beams co-propagate,
two of them carrying opposite vortices, the third beams being without
the vortex, a persistent current can be generated for ultracold atoms
\cite{Song2009}. 

By properly choosing the laser fields, the Cartesian components of
the vector potential $\hat{{\mathbfcal A}}$ do not commute, leading
to the non-Abelian gauge potentials (see also Sect. \ref{non-Abelian_Wilson}). We note that this can
happen even if the Rabi frequencies $\Omega_{j}$ of the tripod setup
represent (non-collinear) plane waves \cite{Jacob2007,Juzeliunas2008PRA,Juzeliunas2010,Juzeliunas2008PRL}
or the properly chosen standing waves \cite{Stanescu2007a,Stanescu2008,Vaishnav08PRL}.
 In that case, one can produce a uniform vector potential whose
Cartesian components are proportional to the Pauli matrices, thus generating
a spin-orbit coupling (SOC) of the Rashba-Dresselhaus type for
ultracold atoms. The rapidly progressing field of spin-orbit-coupled atomic gases will be further explored in the following.

\subsection{Spin-orbit coupling for ultracold atoms}

Spin-orbit coupling (SOC) is pervasive in material systems. In some
cases it leads to parasitic effects such as reduced spin coherence
times~\cite{Koralek2009}, while in other contexts, like topological
insulators, it is essential~\cite{Kane2005,Hasan2010}. Topological
insulators -- non-interacting fermionic systems -- represent a first
realization of time-reversal (TR) invariant systems with topological
order~\cite{Hasan2010}.  In analogy with the progression from the TR-violating
single-particle integer quantum Hall effect (IQHE) to the interaction
driven fractional quantum Hall effects (FQHEs), the next important
step is realizing strongly interacting cousins to the topological
insulators, of which topological superconductors are a first example~\cite{Schnyder2008,Sau2010}.
Ultracold atoms are an ideal platform to study strongly interacting
SOC systems, both with bosonic~\cite{Stanescu2008} and fermionic atoms~\cite{Zhu2006}.
Since ultracold atoms lack intrinsic SOC, numerous techniques (including
the tripod setup considered above) have been suggested for generating
SOC (generally equivalent to non-Abelian gauge potentials~\cite{Dalibard2011}),
with optical~\cite{Osterloh2005,Ruseckas2005,Zhu2006,Liu2009,Juzeliunas2010},
rf~\cite{Goldman2010a} or pulsed magnetic \cite{Anderson2013,Xu2013}
fields.  A first step towards this goal has been experimentally achieved with the recent implementation of an Abelian SOC (i.e. a SOC term generated along one spatial direction only) in several laboratories~\cite{Lin2011,Wang2012,Cheuk2012,Zhang2012PRL,Fu2013PRA,Fu2013,Zhang2013PRA,Qu2013,LeBlanc2013}, see Sec.~\ref{sub:Exp-SOC}.

In Sect.~\ref{sect:gaugefields}, we considered how non-trivial artificial
gauge fields can appear for atoms adiabatically moving in a restricted
set of  ``target'' states. Then in Sect.~\ref{sect:schemes}, we
studied an experimentally relevant case where the artificial gauge
field ${\mathbfcal A}$  was a vector of real numbers, i.e. a gauge potential related to an Abelian (U(1)) gauge structure.
Here we will focus on the most simple extension where the vector potential
is non-Abelian, but is spatially uniform, such as the one generated
using the tripod setup mentioned above. Because of the $-i\hat{\mathbfcal A}\shorttimes\hat{\mathbfcal A}/\hbar$
term appearing in the expression (\ref{eq:B}) for the non-Abelian magnetic field, a uniform
non-Abelian vector potential can have measurable effects \cite{Mead1992}.

SOC can be a simple example of non-Abelian vector potentials: suppose
$\hat{\mathbfcal A}$ is a spatially uniform vector potential with
non-commuting elements each described by $2\shorttimes2$ matrices
acting on a pseudospin-degree of freedom. We can expand the general
Hamiltonian 
\begin{eqnarray}
\hat{H} & =\frac{\hbar^{2}}{2m}\left[\mathbf{k}-\frac{\hat{\mathbfcal A}}{\hbar}\right]^{2}+V(\hat{{\bf r}})\\
 & =\frac{\hbar^{2}\mathbf{k}^{2}}{2m}-\frac{\hbar}{m}\left[\hat{\mathbfcal A}_{x}k_{x}+\hat{\mathbfcal A}_{y}k_{y}+\hat{\mathbfcal A}_{z}k_{z}\right]+\frac{1}{2m}\hat{\mathbfcal A}\cdot\hat{\mathbfcal A}+V({\bf r}),
\end{eqnarray}
 where each of the terms $\hat{A}_{j}k_{j}$ couple the spin to the
atom's linear motion.  In general, because $[\hat{\mathbfcal A}_i,\hat{\mathbfcal A}_{j}]\neq0$
for some $i,j$, this uniform vector potential cannot be eliminated
by a gauge transformation.

As we argued in Sect.~\ref{sect:gaugefields}, non-trivial gauge
fields require that the atomic motion be restricted to a target subspace
of the initial Hamiltonian. In the present case this implies that
the full Hamiltonian be spanned by a minimum of three internal states.
The tripod scheme -- involving four laser-coupled levels -- is the
most commonly used theoretical model in which non-Abelian gauge fields
appear. Here we will present a slightly simpler ``ring coupling''
model containing only three sequentially coupled levels~\cite{Campbell2011}
which displays the same physics (Interestingly, the familiar tripod
scheme~\cite{Ruseckas2005,Jacob2007,Stanescu2007,Juzeliunas2008PRA,Stanescu2008,Juzeliunas2010,Dalibard2011}
reduces to the three level ring model when the excited state is far
from resonance and thus can be adiabatically eliminated.)

We consider $3$ ground or metastable atomic ``spin'' states $\left\{ \ket{1},\ket{2},\ket{3}\right\} $
coupled together with complex valued matrix elements $\Omega_{j+1,j}=\frac{1}{2}\Omega\exp\left[i\left({\bf k}_{j}\!\cdot\!{\bf x}\right)\right]$,
linking each state to each other state. Here, $\Omega$ describes
the optical coupling strength, $\hbar{\bf k}_{j}$ is
the respective momentum acquired in the $j\rightarrow j+1$
atomic transition. Throughout this Section, spin indices are taken
${\rm mod}(3)$, implying periodic boundary conditions $\ket{4}=\ket{1}$
for spin states.

Including the motional degrees of freedom, the momentum representation
second-quantized Hamiltonian 
\begin{eqnarray}
\hat{\mathcal{H}}= & \int\frac{d^{2}{\bf k}}{(2\pi)^{2}}\sum_{j}\Bigg\{\left(\frac{\hbar^{2}\left|{\bf k}\right|^{2}}{2m}\right)\hat{\phi}_{j}^{\dagger}({\bf k})\hat{\phi}_{j}({\bf k})\nonumber \\
 & +\frac{\Omega}{2}\left[\hat{\phi}_{j+1}^{\dagger}({\bf k}+{\bf k}_{j})\hat{\phi}_{j}({\bf k})+{\rm h.c.}\right]\Bigg\}\label{eq:main}
\end{eqnarray}
 describes a system of 2D atoms in the momentum representation, where
all summations over $j$ range from $1$ to $3$. Here, $\{\phi_{j}^{\dagger}({\bf k})\}$
is the spinor field operator describing the creation of a particle
with momentum ${\hbar{\bf k}}$ in internal state $\ket{j}$. In what
follows, we require that $\sum{\bf k}_{i}=0$, so that no momentum
is transferred to an atom upon completing a closed-loop transition
$\ket{1}\!\rightarrow\!\ket{2}\!\rightarrow\!\ket{3}\!\rightarrow\!\ket{1}$.
In this case, the momenta-exchange can be represented in terms of
the differences ${\bf k}_{j}={\bf K}_{j+1}-{\bf K}_{j}$, and we require
${\bf K}_{j}$ to have zero average. The displacement vectors ${\bf K}_{j}=\sum_{l}l{\bf k}_{l+j-1}/3$
define these transformations explicitly.

In the spirit of Sects.~\ref{subsect:twolevel}
and~\ref{subsect:threelevel}, we substitute $\hat{\varphi}_{j}^{\dagger}({\bf q})=\hat{\phi}_{j}^{\dagger}({\bf q}+{\bf K}_{j})$
into the Hamiltonian in Eq.~(\ref{eq:main}) which separates into
an integral $\int\sum_{j,j^{\prime}}\hat{{\varphi}}_{j}^{\dagger}({\bf q}){\hat{H}}_{j,j^{\prime}}({\bf q})\hat{\varphi}_{j^{\prime}}({\bf q})d^{2}{\bf q}/(2\pi)^{2}$
over $3\!\times\!3$ blocks
\begin{eqnarray}
{H}_{j,j^{\prime}}({\bf q})= & \frac{\hbar^{2}\left|{\bf q}+{\bf K}_{j}\right|^{2}}{2m}\delta_{j,j^{\prime}}+\frac{\Omega}{2}\left(\delta_{j,j^{\prime}+1}+{\rm h.c.}\right)\label{EqnDisplaced}
\end{eqnarray}
where each block is labeled by the quasi-momentum ${\hbar{\bf q}}$. 

We can relate the coupling term in Eq.~(\ref{EqnDisplaced}) to the situation of a 1D periodic tight binding Hamiltonian
with a hopping matrix element $\Omega/2$, and where the three sites correspond to the three internal states of the atom. For $\Omega$ much larger than the kinetic energy the coupling term can be diagonalised using a basis conjugate to the spin-index
$j$ with field operators 
\begin{eqnarray}
\hat{\tilde\varphi}_{\ell}^{\dagger}({\bf q)} & =\frac{1}{3^{1/2}}\sum_{j=1}^{N}e^{i2\pi\ell j/N}\hat{{\varphi}}_{j}^{\dagger}({\bf q).}
\end{eqnarray}
The corresponding eigenenergies are given by $E_{\ell}=\Omega\cos(2\pi\ell/3)$, where $\ell\in\{0,1,2\}$ is analogous
to the usual crystal momentum.


The ground state is two-fold degenerate for $\Omega>0$ for states at $\ell=1$ and $\ell=2$. If the displacement vectors ${\bf K}_{j}$
are chosen such that ${\bf K}_{j}=-\kl\sin\left(2\pi j/3\right){\bf e}_{x}+\kl\cos\left(2\pi j/3\right){\bf e}_{y}$, the full Hamiltonian matrix becomes
\begin{eqnarray}
H_{\ell,\ell^{\prime}}({\bf q})= & \left({\bf q}^{2}+1+E_{\ell}\right)\delta_{\ell,\ell^{\prime}}+\left[\left(iq_{x}+q_{y}\right)\delta_{\ell-1,\ell^{\prime}}+{\rm h.c.}\right],\label{DressedFullHamiltonian}
\end{eqnarray}
where momentum is expressed in units of $\kl$ and energy in units of  recoil energy $\El=\hbar^{2}\kl^{2}/2m$. We will in the following consider the two nearly degenerate states with $\ell=1$ and $\ell=2$, yielding the pseudospins $\ket{\downarrow}$ and $\ket{\uparrow}$.

In the subspace spanned by the lowest energy pair of dressed states, we consequently get a zeroth order Hamiltonian of the Rashba form 
\begin{eqnarray}
\hat{H}^{(0)} & =\left|{\bf q}\right|^{2}\hat{1}+\left(\hat{\sigma}_{x}q_{y}-\hat{\sigma}_{y}q_{x}\right),\label{EqnRashba}
\end{eqnarray}
where $\hat{\sigma}_{x,y,z}$ are the standard Pauli matrices. For finite coupling $\Omega$ it is instructive to adiabatically eliminate the excited states order-by-order in perturbation theory, which gives additional effective terms $\hat{H}^{(n)}$ ~\cite{Campbell2011}.

\subsection{Experimental realisation of spin-orbit coupling} \label{sub:Exp-SOC}

The key to the experimental realization of spin-orbit coupling, is in the fact that the laser
geometry in Eq.~\eref{TwoByTwo} can also be used to obtain a coupling of the form,
\begin{eqnarray}
\hat{H} & =\frac{\hbar^{2}\hat{{\bf k}}^{2}}{2m}-\frac{\hbar^{2}\kr}{m}\hat{\sigma}_{y}\hat{k}_{x}+\frac{\hbar\delta}{2}\hat{\sigma}_{y}+\frac{\hbar\Omega_{R}}{2}\hat{\sigma}_{z}+U(\hat{\mathbf{r}})+\frac{\hbar^{2}\kr^{2}}{2m}\label{eq:onedimensionSOC}
\end{eqnarray}
where $f=1/2$ was chosen, together with a pseudo-spin
rotation $\hat{\sigma}_{z}\rightarrow\hat{\sigma}_{y}\rightarrow\hat{\sigma}_{x}\rightarrow\sigma_{z}$.
The Hamiltonian in Eq.~(\ref{eq:onedimensionSOC}) describes a spin-orbit coupled
system consisting of an equal sum of Rashba and Dresselhaus
terms. This corresponds to the Abelian gauge potential ${\mathbfcal A}=\hbar\kr\ex\sigma_y$ containing a single Cartesian component. The situation is non-trivial  due an additional Zeeman term $\frac{\hbar\Omega_{R}}{2}\hat{\sigma}_{z}$, and leads to a number of interesting single- and many-body effects recently studied both experimentally \cite{Lin2011,Wang2012,Cheuk2012,Zhang2012PRL,Fu2013PRA,Fu2013,Zhang2013PRA,Qu2013,LeBlanc2013} and theoretically\cite{Ho2012,Li2012PRL,Han:2012bis,Seo2012,Seo:2012bis,Li:2012ef,Martone:2012kl,Ozawa:2013ku,Li:2013ez}.

\begin{figure}[tb]
\includegraphics[width=12cm]{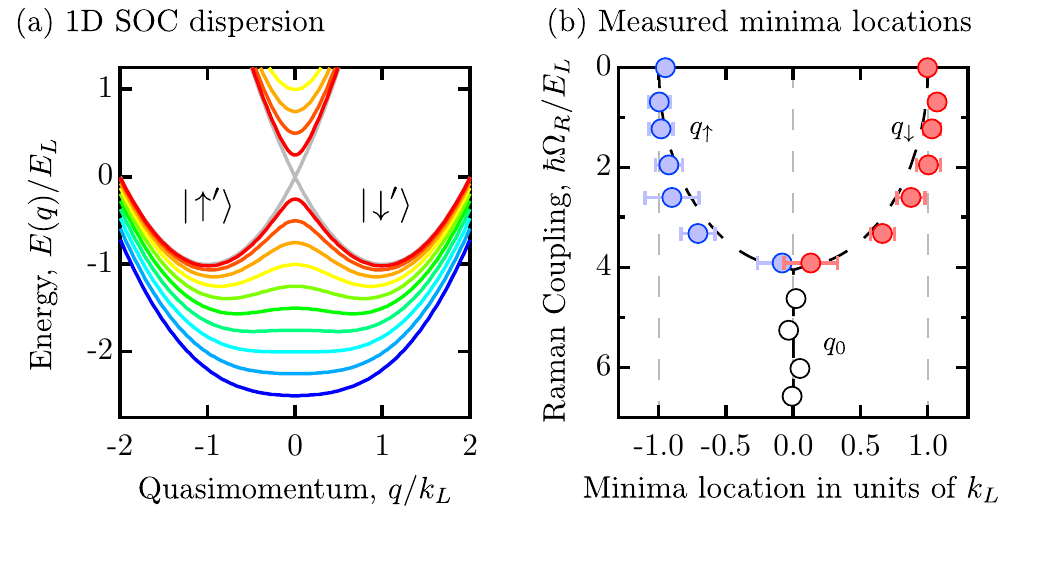}\caption[Spin orbit coupled BEC]{The spin-orbit coupled BEC. (a) The dispersion $E(q)$, where the transition between the double minimum and the single minima configurations are illustrated. (b)
The quasi momentum obtained by adiabatically loading a spin mixture
into a Raman dressed state with the desired $\Omega_{R}$. Data first
appeared in Y.-J. Lin, \textit{et al.} Nature. \textbf{471}, 83 (2011),
Ref.~\cite{Lin2011}. }
\label{fig_SOC} 
\end{figure}

In Fig.~\ref{fig_SOC}a it is shown how the spin-orbit coupling dispersion, as a function of the laser coupling strength, is dramatically different from that of a free particle. For small $\Omega_{R}$, the lowest dressed band consists of a double-well in momentum space \cite{Higbie2004},
with two distinct momenta where the group velocity is zero. The states near the two minima are the dressed spin states, $\ket{\uparrow'}$
and $\ket{\downarrow'}$. As $\Omega$ increases, the two dressed spin states merge. In this limit the picture corresponds to spinless bosons with a tunable dispersion relation\cite{Lin2009}.

In the experiment by Lin \textit{et al.} \cite{Lin2011} the effective two-level system was created in $\Rb87$'s $f=1$ manifold where the $\ket{m_{F}=-1}=\ket{\downarrow}$ and $\ket{m_{F}=0}=\ket{\uparrow}$ states were isolated. This could be achieved by using sufficiently large magnetic field ($g_{F}\mu_{B}B_{0}\approx4.81\MHz$) such that the quadratic Zeeman shift detuned the $\ket{m_{F}=+1}$ state appreciably from the Raman resonance. Figure~\ref{fig_SOC}a shows the quasimomentum measured by first preparing an equal mixture with $\ket{\downarrow}-\ket{\uparrow}$, and then slowly increasing $\Omega_{R}$. In this process, each dressed
spin state adiabatically follows its associated minima. The resulting momentum can be measured using time of flight, which will reveal the presence of a non-trivial spin-orbit coupling in the system. 

Up to now only a 1D spin-orbit coupling corresponding Abelian gauge potential has been generated \cite{Lin2011,Wang2012,Cheuk2012,Zhang2012PRL,Fu2013PRA,Fu2013,Zhang2013PRA,Qu2013,LeBlanc2013}
in spite of numerous proposal for creating the 2D \cite{Dudarev2004,Ruseckas2005,Stanescu2007a,Jacob2007,Juzeliunas2008PRA,Stanescu2008,Vaishnav08PRL,Juzeliunas2010,Campbell2011,Dalibard2011,Zhai2012JMPB,Xu2012,Galitski2013,Anderson2013,Xu2013,LiuLaw:2013,Sun2014} and 3D \cite{Anderson2012PRL,Anderson2013} Rashba type spin-orbit coupling (SOC) in cold
atomic gases. 
Implementation of the 2D or 3D Rashba SOC would allow for the study of rich ground
state physics proposed in systems of many-body fermions \cite{Liu2009,Vyasanakere:2011,Vyasanakere:2011bis,Jiang2011,Hu:2011,Liu:2012,Chen2012PRA,He2012,He2012a,Cui2013a,Zheng2013,LiuLaw:2013,Maldonado2013}
and bosons \cite{Stanescu2008,Sinha2011,Radic2011PRA,Wu11CPL,Hui2012,Gou2012,Kawakami2012,Ruokokosk2012,Xu2012PRA,Sedrakyan2012PRA,Song2012arXiv,Sun2014,Su2014},
of which many properties  have no solid-state physics analogue. This will be considered in the next Section.


\section{The effects of collisions in the presence of spin-orbit coupling}
\label{sect:interactions}



\label{sect:interactions}

Synthetic magnetic fields and spin-orbit couplings result from engineering the single particle Hamiltonian, however, ultracold gases can be strongly influenced by collisions.  Absent artificial gauge fields, these systems are a unique playground where the inter-particle interaction strength can be tuned virtually at will~\cite{Bloch2008a,Lewenstein2007}. In this Section, we describe how interacting BECs and degenerate Fermi gases behave in the presence of an external gauge potential,   focusing on the case where atoms are subjected to a synthetic SOC. The interplay between interactions and synthetic magnetic fields induced by rotation were described in the review by Cooper~\cite{Cooper2008}; for a more complete overview on interacting spin-orbit-coupled atomic gases, consider the reviews by H. Zhai~\cite{Zhai2012JMPB,Zhai2014-review} and X. Zhou et al \cite{Zhou2013}, as well as other articles recently published in the special issue on non-Abelian gauge fields edited by Gerbier et al. \cite{Gerbier:2013issue}.

\subsection{The weakly interacting Bose gas with SOC}

In a weakly interacting Bose gas where $a^3\rho\ll1$ ($a$ is the $s$-wave scattering length and $\rho$ is the atom density), only two-body scattering processes are relevant.  The interaction between atoms can be described by the contact potential
\be
V_{int}({\bf r}-{\bf r'})=\frac{4\pi\hbar^2 a}{m}\delta({\bf r}-{\bf r'})= g \, \delta({\bf r}-{\bf r'}).
\ee
In most cases, the dynamics is accurately described by the mean-field Gross-Pitaevskii equation~\cite{Bloch2008a}
\be
i\hbar\frac{\partial}{\partial t}\Psi (\bs r,t)=\Bigl (-\frac{\hbar^2}{2m}\nabla^2+V(r)+g|\Psi (\bs r,t)|^2 \Bigr) \Psi(\bs r,t),
\ee
a nonlinear \Schrodinger equation describing the condensate wavefunction $\Psi=\langle \hat \Psi \rangle$, with a potential $V(r)$ and a nonlinearity proportional to the density $\rho=|\Psi (\bs r,t)|^2$. Interactions -- captured by this nonlinearity -- change the dynamics dramatically compared to the non-interacting gas, as evidenced by the elementary excitation spectra\cite{stringari1996,ohberg1997,Fliesser1997} and the nucleation of vortices \cite{Madison2000}. 

In the context of artificial gauge potentials we ask: what effects are added for interacting gases with SOC? This scenario has been studied by a number of authors \cite{Stanescu2008,Wang2010,Sinha2011,Barnett:2012fm,Zhang2012,Ozawa2012a,Ozawa2012b,Zhou:2011,Cui:2013,He:2012,Liao:2013,Han:2012,Zhou:2013,Chen-2014},
showing that SOC dramatically alters the ground state properties. In such situations, the general second-quantized Hamiltonian is
\be
\hat H=\int d{\bf r} \, \hat \Psi^\dagger \Biggl [ \frac{1}{2m}({\bf p}-\hat{\mathbfcal A})^2+V(r)+\hat G(\hat \Psi, \hat \Psi^\dagger) \Biggr ] \hat \Psi = \hat H_0 + \hat H_{\text{int}}, \label{so-gp1}
\ee
where the SOC is described by the gauge potential $\hat{\mathbfcal A}$; $\hat \Psi$ is a multicomponent field operator; and $V(r)=m \omega^2 r^2/2$ is the confining potential with frequency $\omega$. The exact form of the interaction term $\Psi^\dagger\hat G(\hat \Psi, \hat \Psi^\dagger)\Psi$ depends on the details of the physical setup \cite{Bloch2008a,Wang2010,Sinha2011}. Typically, the field operator $\hat \Psi$ associated with a dressed state -- a superposition of the bare atomic levels -- and the two-body scattering lengths entering $\hat G(\hat \Psi, \hat \Psi^\dagger)$ depend on the different bare states.  In this dressed-state picture, this can yield spin-dependent collision terms, e.g.,  $\hat G_{\mu \nu}(\hat \Psi, \hat \Psi^\dagger) = g_{\mu \nu} \hat \Psi_{\mu} \hat \Psi^{\dagger}_{\nu}$ with $ g_{\mu \nu} \ne \text{const}$, independent of the gauge potential $\hat{\mathbfcal A}$, see Refs.~\onlinecite{Wang2010,Sinha2011}.  Furthermore (and beyond the scope of the discussion here) the atomic contact interaction can acquire momentum dependence -- a finite range -- from the momentum-dependence of the dressed states as expressed in the basis of bare atomic states~\cite{Williams2012}.

Here, we review the ground-state properties of a quasi-spin-1/2 system with isotropic 2D Rashba SOC, where $\hat \Psi = (\hat \Psi_1 , \hat \Psi_2)^{T}$, and where $\hat{\mathbfcal A}$ is simply expressed in terms of the Pauli matrices as $\hat{\mathbfcal A}= \hbar \kappa (\hat \sigma_x , \hat \sigma_y)$. 
\footnote{For the ground-state properties of interacting spin-1 and spin-2 SOC bosons, see Refs \cite{Gou2012,Ruokokosk2012,Xu2012PRA,Song2012arXiv,LanPRA2014}.%
} The synthetic gauge potentials giving rise to spin-orbit coupling are remarkably flexible in the sense that the precise shape of the gauge potential can often be controlled at will. The Rashba type coupling is one example, which is also often encountered in solid state scenarios. Such spin-orbit couplings can be achieved optically \cite{Ruseckas2005,Osterloh:2005,Juzeliunas2010}, but also using periodically-driven cold-atom systems, such as those based on pulsed-magnetic fields \cite{Anderson2013,Xu2013,Goldman:2014uz}, see also Section 8.

\begin{figure}[h!]
	\includegraphics[width=0.7\columnwidth]{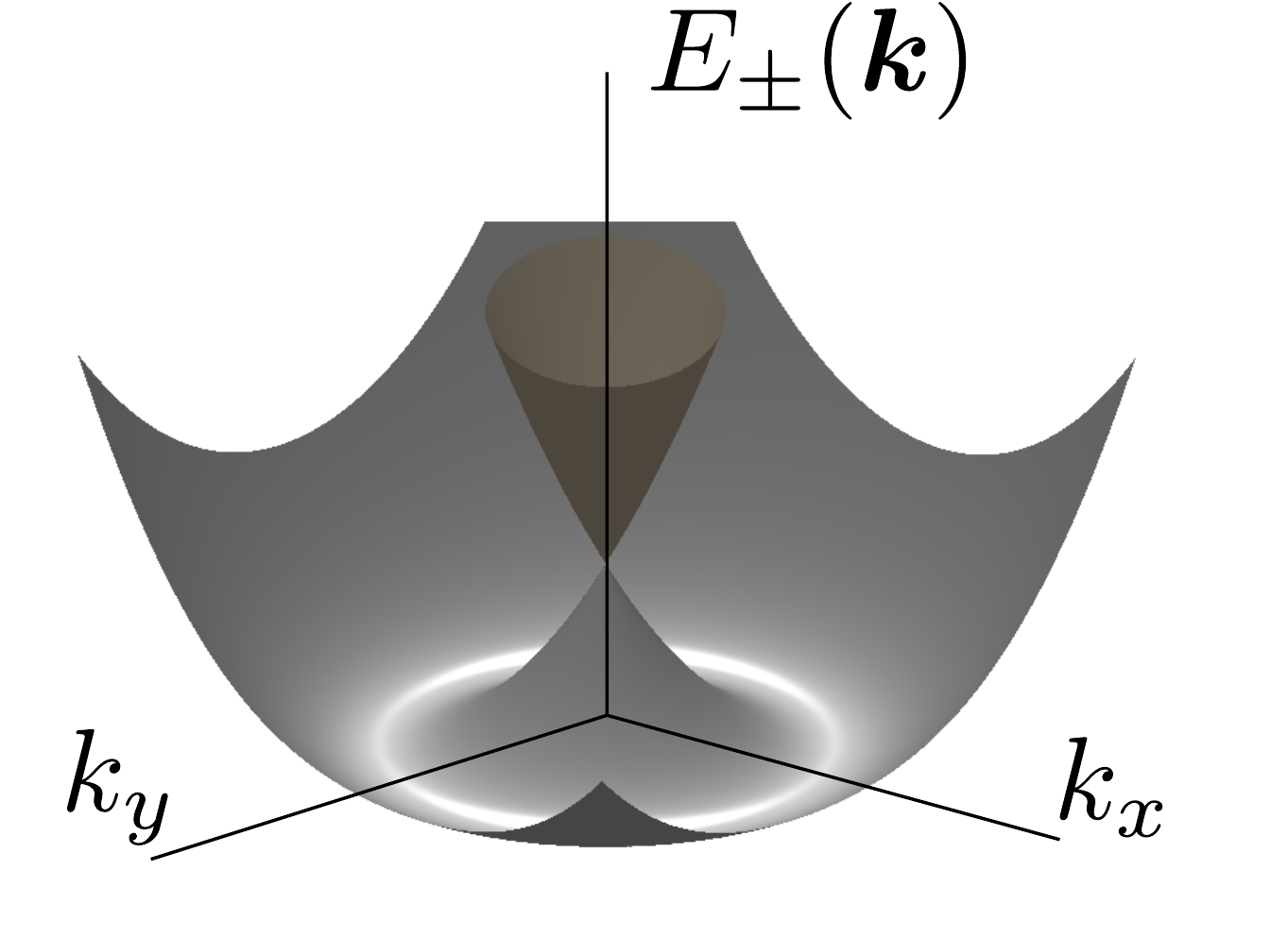}
	\caption{\label{figurebandSO} Dispersion for a homogeneous gas with Rashba SOC: $E_{\pm} (\bs k)$.  $\bs k$ denotes momentum in the $\ex$-$\ey$ plane. When $\hat{\mathbfcal A} = \hbar \kappa (\hat \sigma_x, \hat \sigma_y)$, the lowest band is minimal on the ring $\vert \bs k \vert=\kappa$.}
\end{figure}

In this configuration, the single-particle energy spectrum of the homogeneous ($V(r)=0$) system has the two energy branches $E_{\pm} (\bs k) \propto (\vert \bs k \vert  \pm \kappa)^2$ illustrated in Fig. \ref{figurebandSO}.  The main effect of the Rashba coupling $\hat{\mathbfcal A}$ is to replace the unique condensation point $k_{\text{min}} (\hat{\mathbfcal A}=0)=0$ by a continuum of minima located on the ring $k_{\text{min}}(\hat{\mathbfcal A}\ne0)=\kappa$, see Fig. \ref{figurebandSO}.  These degenerate single-particle ground states affect the many-body problem, where an interacting condensate's ground state will be chosen through a complex competition between the interactions and the  parameters of the non-interacting model (e.g. the trap frequency $\omega$ and the Rashba coupling strength $\kappa$).  Hence, in the presence of Rashba SOC, interactions play a major role even when they are weak  \cite{Zhai2012JMPB,Zhai2014-review,Zhou2013}.

\subsubsection{The homogeneous interacting system with Rashba SOC}
\label{homogSO}
The mean-field phase diagram of the homogenous, $V({\bf r})=0$, spin-orbit-coupled BEC is already altered by a simple interaction term of the form  \cite{Wang2010}
\be
\hat{H}_{\rm int} = \int d{\bf r} \, g \hat n_1 ^2 + g \hat n_2 ^2 + g_{12} \hat n_1 \hat n_2, 
\label{simpleterm}
\ee
where the two scattering parameters $g$ and $g_{12}$ account for spin-dependent collisions in a minimal manner, and where $\hat n_{\mu} = \hat \Psi^{\dagger}_{\mu} \hat \Psi_{\mu}$ denote the density operators. The mean-field phase diagram described below has been obtained by Wang \emph{et al.} \cite{Wang2010}, through a minimization of the Gross-Pitaevskii free energy for the two-component order parameter $\Psi (\bs r)=\langle \hat \Psi (\bs r) \rangle$, derived from the SOC Hamiltonian \eqref{so-gp1}-\eqref{simpleterm}. When $g_{12} < 2g$, the densities associated with the two spin components $\rho_{1,2}(\bs r)=\vert \Psi_{1,2} \vert^2$ are uniform in space, but the phase of the condensate $\Psi (\bs r)=\langle \hat \Psi (\bs r) \rangle$ varies periodically along one direction, spontaneously breaking rotational symmetry. This regime characterizes the ``plane wave phase,'' where the macroscopic wave function of the condensate takes the simple form \cite{Wang2010,Zhai2012JMPB,}
\be
\Psi (\bs r)= \sqrt{(\rho_1+\rho_2)/2} \, e^{i \kappa x} \, (1 , 1)^{T}\label{planewavecondensage}.
\ee
Without loss of generality, we have taken the plane-wave to be along $\ex$, reflecting the spontaneous symmetry breaking. When $g_{12} > 2g$, the condensate $\Psi (\bs r)$ is described by a superposition of two such plane waves, 
\be
\Psi (\bs r)= \frac{1}{2} \sqrt{\rho_1+\rho_2} \, \biggl [ e^{i \kappa x} \, \left(\begin{array}{c}1 \\1\end{array}\right) + e^{-i \kappa x} \,  \left(\begin{array}{c}1 \\-1\end{array}\right) \biggr ] =  \sqrt{\rho_1+\rho_2} \left(\begin{array}{c}\cos(\kappa x) \\i\sin(\kappa x)\end{array}\right), \label{stripecond}
\ee
yielding a periodic spatial modulation of the spin-density
$$\rho_{s}(\bs r)=\vert \Psi_{1} (\bs r) \vert^2 -\vert \Psi_{2} (\bs r)  \vert^2 = (\rho_1+\rho_2) \cos (2 \kappa x).$$
This phase is known as the ``stripe superfluid'' or ``standing wave phase''  \cite{Wang2010}. The mean-field phase diagram of the homogeneous system, obtained from the Gross-Pitaevskii equation derived from Eqs.~\eqref{so-gp1}-\eqref{simpleterm}
is robust against deformation of the SOC term $\hat{\mathbfcal A}$ (without loosing the cylindrical symmetry), and 
generally holds in the presence of additional fields, e.g., Zeeman terms \cite{Zhai2012JMPB}.

\subsubsection{The trapped interacting system with Rashba SOC}

The phase diagram is enriched by the harmonic confinement $V(r)=m \omega^2 r^2/2$ generally present in cold-atom experiments \cite{Bloch2008a}. Remarkably, novel phases emerge in trapped systems \cite{Sinha2011}, even under the strong  (and generally unphysical) assumption that all the interaction processes are described by a single  scattering parameter, i.e., when the interaction term reduces to the spin-independent form $\Psi^{\dagger} \hat G\Psi=g_{\text{eff}} (\rho_1(r)+\rho_2(r))^2$, where $g_{\text{eff}}$ is an effective scattering length describing all collisions \cite{Sinha2011}. This strong simplification aims to isolate the main effects introduced by the trap in spin-orbit coupled gases \footnote{The interplay between spin-dependent collisions and the effects of the trap could be investigated by introducing additional scattering parameters $g_{\mu \nu} \ne g_{\text{eff}}$ into the problem, in the spirit of Eq. \eqref{simpleterm}, which could lead to even more involved phase diagrams than the one presented here.}.  The external trap introduces an additional energy scale $\hbar\omega$ into the homogenous problem discussed above, hence it is natural to describe the resulting phase diagram in terms of the dimensionless parameters $\mathfrak{K}=\kappa l_0$ and $\mathfrak{g}=g_{\text{eff}} m / \hbar^2$, where $l_0=\sqrt{\hbar/m\omega}$. 

\begin{figure}[h!]
	\includegraphics[width=1\columnwidth]{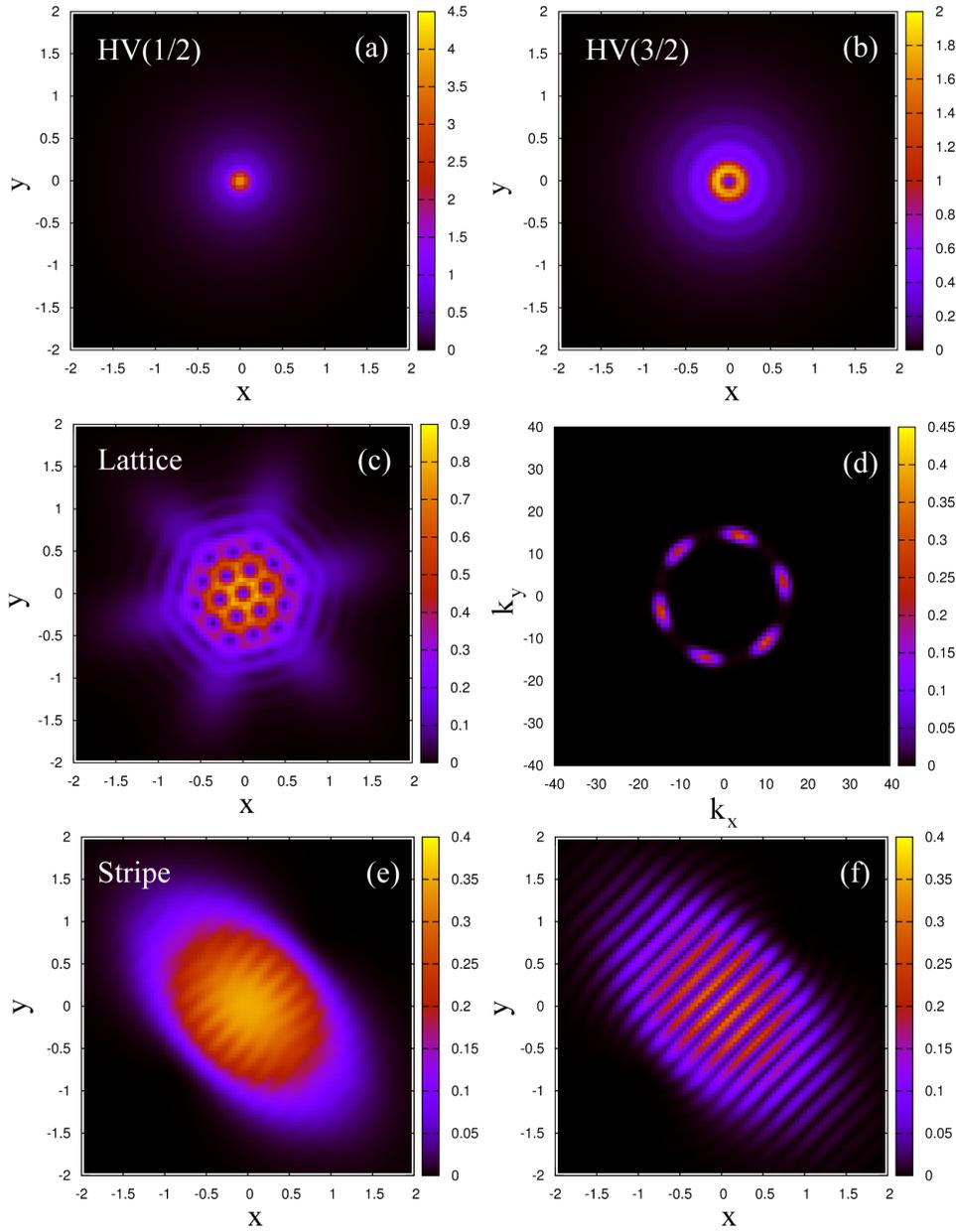}
	\caption{\label{luis1} Trapped spin-orbit coupled gas. (a)--(c),(e): The ground state density $\rho = \vert \Psi_1 (\bs r) \vert ^2 + \vert \Psi_2 (\bs r) \vert ^2$, for increasing interaction strengths $\mathfrak{g}=0.05, 0.1, 0.85, 2$. The dimensionless SOC strength is $\mathfrak{K}=15$. (d) The momentum distribution for $\mathfrak{g}=0.85$. (f) Density $\rho_1 = \vert \Psi_1 (\bs r) \vert ^2$ of a single component for $\mathfrak{g}=2$. Reproduced with permission from S. Sinha, R. Nath and L. Santos, Phys. Rev. Lett. {\bf 107}, 270401 (2011).}
\end{figure}

In the non-interacting regime where SOC  dominates ($\mathfrak{g}=0$, $\mathfrak{K} \gg 1$), the upper branch $E_{+} (\bs k)$ can be neglected,  and, going to the momentum representation, one obtains the single particle wavefunction $\Psi (\bs k)= \psi (\bs k) \, {\bf u}_{-}(\bs k)$,
where 
\be
{\bf u}_{-}(\bs k)=\frac{1}{\sqrt{2}}\left(\begin{array}{c} 1 \\ \frac{k_x+ik_y}{k}\end{array}\right)=\frac{1}{\sqrt{2}}\left(\begin{array}{c} 1 \\  e^{i \varphi}\end{array}\right)
\ee 
is an eigenstate of the homogeneous system in the lowest branch $E_{-} (\bs k)$; and $\varphi$ denotes the polar angle of $\bs k$. Adding the confining potential to the single-particle Hamiltonian, $H (\bs k) \propto (\hbar \bs k - \hat{\mathbfcal A})^2 - \nabla_k^2$, and solving the corresponding \Schrodinger equation using the ansatz $\psi (\bs k)= \sum_l k^{-1/2} f_l (\bs k) e^{i l \varphi}$, yields the eigenenergies \cite{Sinha2011}
\be
E_{nl}=\frac{(l+\frac{1}{2})^2}{2\mathfrak{K}^2}+n+\frac{1}{2}, \label{so-trap}
\ee
where $n$ labels the radial excitations around the minima of the mexican-hat potential $k \approx \kappa$ (see Fig. \ref{figurebandSO}) and $l \in \mathbb{Z}$. The degenerate ground states
\be
\Psi_{l=\{0,-1\}}(k)\propto e^{-(k-\kappa)^2 \l_0^2/2+i l \varphi}{\bf u}_- (\varphi)
\ee
have quantum numbers $n=0$ and $l= \{ 0, -1 \}$.  In real space, and using polar coordinates $(r,\theta)$, the two degenerate ground states become
\begin{align}
\Psi_0(r,\theta)&\propto \left(\begin{array}{c} J_0(\mathfrak{K} r) \\e^{i\theta}J_1(\mathfrak{K} r)\end{array}\right), & {\rm and} &&
\Psi_{-1}(r,\theta)&\propto \left(\begin{array}{c} e^{-i\theta}J_1(\mathfrak{K} r) \\  J_0(\mathfrak{K} r)  \end{array}\right)\label{d1}.
\end{align}
These half-vortex states \cite{Wu2011halfvortex,Stanescu2008,Sinha2011,Wang2010,Zhai2012JMPB,Zhai2014-review} appear in various contexts, including topological quantum computing \cite{Nayak:2008}, superfluid $^3$He \cite{Salomaa:1985} and triplet superconductors.  The degenerate states in Eq.~\eqref{d1} and any linear combination thereof,   yield the rotationally symmetric density distribution illustrated in Fig.~\ref{luis1} (a).  As pointed out in Ref. \onlinecite{Sinha2011}, this half-vortex regime (HV 1/2) survives for finite interaction strength $\mathfrak{g} < \mathfrak{g}_1$, where $\mathfrak{g}_1 \sim \mathfrak{K}^{-2}$ is related to the energy difference between the ground ($\vert l+1/2 \vert =1/2$) and higher energy ($\vert l+1/2 \vert =3/2$) states.  Indeed, for large SOC strength $\mathfrak{K}^{2} \gg 1$, the energy difference between successive angular excitations [Eq. \eref{so-trap}] is small compared to radial excitations.  Thus, the population of higher angular momentum states ($l\ge 1$) should be favoured at greater, but still reasonably small, interaction strength $\mathfrak{g} \ge \mathfrak{g}_1$\cite{Sinha2011}. A condensate formed with higher angular momentum states satisfying $\vert l+1/2 \vert =3/2$, called the (HV 3/2)--phase, occurs in a certain range of the interaction strength $\mathfrak{g}_1 <\mathfrak{g} < \mathfrak{g}_2$, and has the ring-shaped density distribution depicted in Fig. \ref{luis1} (b).

For larger interaction strength, $\mathfrak{g} > \mathfrak{g}_2$, the gas enters a ``lattice phase'',   where its density is modulated in the form of a hexagonal lattice, see Fig. \ref{luis1} (c).  Interestingly, rotational symmetry is broken and the lattice period is independent of the interaction strength $\mathfrak{g}$.  In fact, the distance between the lattice minima is solely dictated by the SOC strength $\mathfrak{K}$. Consequently, the resulting momentum distribution has a robust ring-like structure with six peaks, see Fig. \ref{luis1} (d). 

\begin{figure}[h!]
	\includegraphics[width=1\columnwidth]{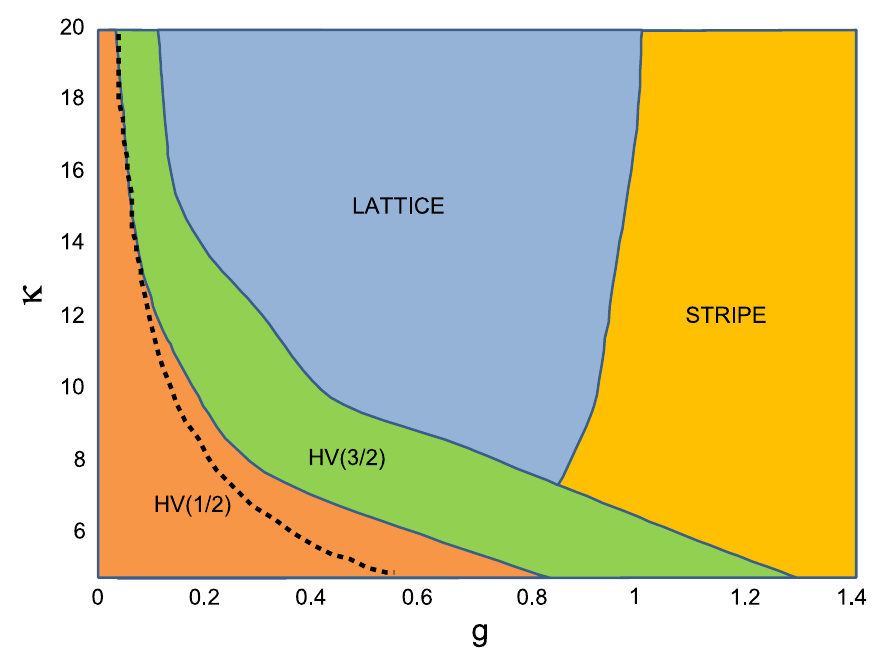}
	\caption{\label{luis3} Ground state phase diagram of the harmonically trapped spin-orbit coupled Bose gas. Reproduced with permission from S. Sinha, R. Nath and L. Santos, Phys. Rev. Lett. {\bf 107}, 270401 (2011).}
\end{figure}

The three phases discussed above all have in common that their densities satisfy (continuous or discrete) rotational symmetry. This is no longer the case for large interaction strength ($\mathfrak{g} \gg \mathfrak{g}_2$).  In this ``highly"-interacting regime, the spin-orbit coupled cloud enters a ``striped phase,'' where the density is sinusoidally modulated in a specific direction, see  Figs. \ref{luis1} (e)-(f).  This anisotropic state shares similarities with the ``stripe phase'' in the homogeneous system with spin-dependent interactions $g_{12} > g$ (see Eq. \eqref{stripecond} and Sect.~\ref{homogSO}). Mapping out the full phase diagram for the trapped spin-orbit coupled gas, by scanning the dimensionless parameters $\mathfrak{g}$ and $\mathfrak{K}$, reveals a surprisingly intricate picture, see Fig. \ref{luis3}. For readers further interested in the effects of finite temperature, anisotropic Rashba coupling $\hat{\mathbfcal A} = \hbar (\kappa_x \hat \sigma_x, \kappa_y \hat \sigma_y)$ and beyond-mean-field descriptions, we refer to Ref. \cite{Zhai2012JMPB,Zhai2014-review}, and references therein. Finally, the interplay between SOC and dipolar interactions, leading to diverse spin and crystalline structures, have been explored in Refs. \cite{Gopalakrishnan:2013,Wilson:2013}.

\subsection{Interacting Fermi gases with SOC}

Degenerate atomic Fermi gases are a remarkably rich platform for studying strongly-correlated phases common in condensed-matter systems \cite{Bloch2008a,Chin2008}. Today, the atomic systems are widely investigated, both theoretically and experimentally, to deepen our understanding of (high-$T_c$) superconductivity and topological phases (see also Sect.~\ref{sect:simulation}). Degenerate Fermi gases offer unique advantages compared to their solid-state brethren.  The interactions between fermions can be tuned at will through Feshbach resonances \cite{Bloch2008a,Chin2008}, leading to rich phase diagrams that describe the evolution from strongly repulsive to strongly attractive Fermi systems. For two-component (spin-1/2) fermions interacting through $s$-wave collisions, the system evolves from a BCS phase of Cooper pairs (a fermonic superfluid) when the scattering length is small and negative $a_s<0$, to a BEC of tightly-bound molecules (a bosonic superfluid) when $a_s >0$. The crossover between the BCS and the BEC phases \cite{Greiner2003a,Jochim2003,Zwierlein2003} takes place as $1/a_s$ travels through zero (i.e. the unitary limit $a_s=\infty$), the threshold for the formation of bound molecular states in the two body problem \cite{Bloch2008a,Chin2008}. Here we discuss how these phase transitions are modified by the inclusion of synthetic SOC.  This problem was already considered in solid-state physics, where it was shown that Rashba SOC significantly alters the physical properties of superconductors displaying strong electric fields near their surface \cite{Gorkov:2001}. For non-zero SOC, the Cooper-pair wavefunction is a mixture of singlet and triplet components, which results in a non-trivial and anisotropic spin susceptibility tensor that reflects the response of the system to an external magnetic field. \\

Vyasanakere and Shenoy \cite{Vyasanakere:2011} first demonstrated that the inclusion of a SOC term drastically modifies the properties of the interacting Fermi gas, especially on the BCS side of the resonance (where the scattering length is negative,
$a_s <0$).  In the presence of a non-Abelian gauge potential, two-body bound-states are present \emph{even in the BCS side} $a_s <0$ of the resonance. In other words, the SOC shifts the threshold for the formation of bound-states towards finite negative values $a_s <0$. Moreover, for certain configurations of the gauge potential (e.g. sufficiently large SOC strength), bound states are formed for \emph{any} value of the scattering length. These SOC-induced bound states, referred to as ``rashbons", are described by a two-body wavefunction featuring singlet and triplet components, similarly to the Cooper-pair wave function in the presence of SOC \cite{Gorkov:2001,Yu:2011}. The rashbons lead to superfluids with nematic spin structures at low temperature, similar to superfluid $^3$He \cite{Vyasanakere:2011}. Since bound states are allowed within the BCS regime and the binding energy $E_b$ strongly depends on the gauge potential's strength, a subsequent work by Vyasanakere, Zhang, and Shenoy \cite{Vyasanakere:2011bis} showed that a BCS-BEC crossover could be induced by varying the Rashba spin-orbit strength while keeping the scattering length $a_s<0$ \emph{fixed} within the BCS regime. This Rashba-induced transition, from the BCS state to the molecular BEC formed by rashbons, is accompanied by a topological modification of the Fermi surface \cite{Vyasanakere:2011bis}. The general structure of the two-body wavefunction, with singlet and triplet components, is maintained during the transition: as the Rashba coupling is increased, the Cooper-pair wave function converges towards the rashbon wave function, suggesting that the crossover nature of the transition is preserved \cite{Vyasanakere:2011bis,Yu:2011}. The crossover has been further established, by confirming that the ground state is indeed protected by a gap throughout the transition \cite{Yu:2011}. A complete description of the Rashba-induced BCS-BEC crossover, in terms of various gauge potential configurations, is presented in Ref.~\onlinecite{Vyasanakere:2011bis} (see also Ref. \cite{Shenoy2013}). 

Surprisingly, at zero temperature, the condensed $\rho_c/\rho$ and superfluid $\rho_s/\rho$ fractions behave oppositly as the Rashba coupling strength $\kappa$ is increased, see Ref. \onlinecite{Zhou2012,He2012,He2012a}.  The condensate fraction increases monotonically  with $\kappa$, eventually reaching $\rho_c/\rho \rightarrow 1$. This enhanced condensate formation, even present for $a_s <0$ (BCS side), is compatible with the formation of the rashbons discussed above.  In contrast, the superfluid fraction $\rho_s/\rho$ is generally \emph{reduced} by Rashba SOC. Depending on the value of the scattering length $a_s$, the fraction $(\rho_s/\rho) (\kappa)$ decreases monotonically or non-monotonically with increasing Rashba SOC.  Superfluidity is never totally suppressed by Rashba SOC \cite{Zhou2012}, furthermore, on the BCS side of resonance, the superfluid transition temperature $T_c$ as well as the pairing gap $\Delta$ increase as a function of the Rashba coupling strength \cite{Yu:2011}. These two quantities respectively converge towards the significantly higher molecular-BEC superfluid temperature $T_c (\kappa) \rightarrow T_c^{\text{BEC}} \gg T_c^{\text{BCS}} (\kappa =0)$ and the two-body binding energy $\Delta (\kappa) \rightarrow E_{b} \gg \Delta (\kappa=0)$. L. He and X.-G. Huang \cite{He2012} investigated the Rashba-induced BCS-BEC crossover in  two dimensions, where the Berezinskii-Kosterlitz-Thouless transition temperature is also enhanced by the Rashba coupling. Similar properties have been studied for Fermi gases trapped in optical lattices, where the interplay between interactions and the pseudo-relativistic (Dirac) spectrum generated by the Rashba coupling leads to interesting behavior\cite{Sun:2013}. Taken together, these results highlight the major deviations from standard superconductivity, which is offered by the large and tunable Rashba coupling possible in atomic systems.  Methods for probing the unusual properties of atomic BCS superfluids in the presence of Rashba coupling and rashbon condensates (i.e. molecular BEC induced by Rashba coupling in the BCS limit $a_s<0$) have been proposed in Ref. \onlinecite{Hu:2011}, based on Bragg spectroscopy and density profiles.  Other effects related to the presence of Rashba (or equal Rashba-Dresselhaus) SOC during the BCS-BEC crossover could also be measured through the isothermal compressibility and the spin susceptibility \cite{Han:2012bis}. \\

In most studies, the scattering length $a_s$ was chosen to be independent of the synthetic SOC, and thus, it is treated as a constant along the Rashba-induced BCS-BEC crossover. Refs.~\cite{Gopalakrishnan:2011,Ozawa2011,Maldonado2013a} describe how interactions are modified in the presence of strong Rashba coupling, offering a relation between physical scattering lengths and mean-field interactions. \\

Another important topic concerns the realization of topologically-ordered superfluids using Fermi gases subjected to Rashba SOC \cite{Sato:2009,Sau:2011,Liu:2012,Zhang:2008,Gong2011,Iskin2011,Seo2012,Seo:2012bis}.  This line of research is motivated by the possibility to create and control non-Abelian excitations such as Majorana fermions in a versatile setup, with possible application to topological computation (see Ref. \onlinecite{Nayak:2008} and Sect.~\ref{sect:simulation}). To achieve this goal in a Fermi gas, the following basic ingredients are required: (a) a Rashba SOC $\hat{\mathbfcal A} \sim \kappa \, \hat \sigma_{\mu} \bs e_{\mu}$ (or in 1D, any form of SOC), (b) a Zeeman term $\hat H_Z= h_Z \hat \sigma_z$ and (c) $s$-wave interactions with scattering length $a_s$ \cite{Sau2010,Sau:2011}. Based on this minimal and realistic system, several authors have investigated the rich interplay between the Rashba-induced BCS-BEC crossover, the population imbalance produced by the Zeeman term, and the emergence of topological order, which could be observed by varying the control parameters $(h_z,\kappa,a_s)$ \cite{Seo2012,Gong2011,Iskin2011,Liao:2012}. These simple systems present a new path for exploring the FFLO finite-momentum paired states \cite{Zheng2012a,Zheng2012b,Dong2012,Shenoy2012,Liu2013}. Finally, we note that finite-momentum ground-states can be generated in Rashba-coupled Fermi gases also in the \emph{absence} of  a Zeeman splitting term, provided that the collisional interaction is strong enough \cite{Maldonado2013}.


\section{Gauge potentials in optical lattices: Engineering the Peierls phases}
\label{sect:lattices}





In the previous Sections, we have considered the creation of artificial gauge fields for ultracold atoms moving in continuous space (i.e. typically, atoms in a harmonic trap). Here, we discuss the emergence of lattice gauge structures (e.g. artificial magnetic fluxes), which can be realized by taking advantage of the discrete motion of atoms in optical lattices. We begin with a brief overview of lattice gauge structures and lattice models, and we then describe their experimental implementation with optical lattices.

\subsection{Introduction: Optical lattices}

\label{sect:lattices}


In theoretical physics, lattice models are generally developed for two different reasons. The first motivation for discretizing configuration space is to apply analytical or numerical tools that are unavailable for continuum systems.  The results obtained from these lattice descriptions are relevant for large systems with vanishingly small lattice periods. A prime example of this approach are lattice gauge theories, which deepen our understanding of Quantum Electrodynamics (QED) and Quantum Chromodynamics (QCD)~\cite{Zinn,Wilson1974,Gupta,Kogut1983}. 

In contrast, many systems have crystalline order and are therefore quite naturally described by lattice models.  This situation frequently arises in condensed-matter systems, where electrons move about in the periodic potential formed by an orderly array of ions~\cite{Ashcroft} which together comprise a crystalline material.  Here, the lattice is physical: the lattice spacing is finite and the hopping parameters are determined by the material under scrutiny.  When electrons at the Fermi-energy are tightly bound to the ionic potential, the system is well described by the Fermi-Hubbard Hamiltonian
\begin{equation}
\hat{H}= -J \sum_{\langle j,k \rangle, \sigma} \bigl ( \hat{c}_{j,\sigma}^{\dagger} \hat{c}_{k, \sigma} + {\rm h.c.} \bigr )+\frac{U}{2} \sum_j   \hat{c}_{j,\uparrow}^{\dagger}  \hat{c}_{j,\downarrow}^{\dagger}  \hat{c}_{j,\downarrow}  \hat{c}_{j,\uparrow},\label{hubbard}
\end{equation}
a single-band tight-binding model, in which $\langle j,k \rangle$ indicates that the summation is over neighboring lattice sites.   Here, $J$ and $U$ describe the tunneling amplitude and on-site interactions, respectively; and $\hat{c}_{j,\sigma}^\dagger$ describes the creation of a fermion at lattice site $j$ with spin $\sigma=(\uparrow, \downarrow)$. The Hubbard model, and its generalizations, have shed light on numerous physical phenomena such as metal-insulator transitions~\cite{Mott1968}, high-$T_c$ superconductivity~\cite{Lee2006}, transport properties in graphene~\cite{CastroNeto2009}, and recently topological insulating phases~\cite{Hasan2010,Qi2011,Qi2008}.  The emulation of lattice models, using quantum analog simulators~\cite{Feynman1982}, makes possible the exploration of effects common in condensed-matter and high-energy systems. 

Ultracold atoms trapped in optical lattices can nearly perfectly realize the Hubbard model~\cite{Jaksch1998,Jaksch2005} in the laboratory~\cite{Greiner2002,Jordens2008}.  Spinless bosonic or fermionic atoms are described by the general Hamiltonian
\begin{eqnarray}
H=&\int d\bs{x} \, \hat \psi^{\dagger} (\bs r) \biggl [ \frac{\bs p ^2}{2 m} + V_{{\rm opt}} (\bs r) + V_{{\rm conf}} (\bs r) \biggr ] \hat \psi (\bs r) \nonumber \\
&+ g  \int d\bs{x} \ \hat \psi^{\dagger}(\bs r) \hat \psi^{\dagger}(\bs r)  \hat \psi (\bs r) \hat \psi (\bs r), \label{hamatoms}
\end{eqnarray}
where $\psi^\dagger(\bs r)$ describes the creation of an atom at position ${\bf x}$; $V_{{\rm opt}} (\bs r)$ is a periodic optical potential; $V_{{\rm conf}} (\bs r)$ is an external confining potential; and $g$ characterizes the interatomic interaction.  As a consequence of $V_{{\rm opt}} (\bs r)$'s lattice structure, it is convenient to express the field operator $\hat \psi^\dagger (\bs r)$ in terms a sum over Wannier orbitals $w_{\lambda} (\bs r - \bs r_j)$ at site $\bs r_j$ and band $\lambda$, which together constitute a set of orthogonal functions~\cite{Wannier:1937,Ashcroft,Kohn:1959,Jaksch1998,Jaksch2005}.  For a sufficiently deep lattice potential, this expansion can be restricted to the lowest band, namely $\hat \psi^\dagger (\bs r)\approx \sum_j \hat c^\dagger_j w^*(\bs r- \bs r_j)$, in which case the general Hamiltonian \eref{hamatoms} reduces to the tight-binding Hamiltonian 
\begin{equation}
\hat{H}=  \sum_j \epsilon_j  \hat{c}_{j}^{\dagger}  \hat{c}_{j}-J \sum_{\langle j,k \rangle} \bigl ( \hat{c}_{j}^{\dagger} \hat{c}_{k} + {\rm h.c.} \bigr )+\frac{U}{2} \sum_j   \hat{c}_{j}^{\dagger}  \hat{c}_{j}^{\dagger}  \hat{c}_{j}  \hat{c}_{j} \,,\label{hubbard2}
\end{equation}
a spinless bosonic Hubbard Hamiltonian~\eref{hubbard}, where $U \propto g$ is the on-site interaction strength and $\epsilon_j \approx V_{{\rm conf}} (\bs r_j)$ is the energy offset  \cite{Jaksch1998}. The atomic Hubbard Hamiltonian \eref{hubbard2} features the tunneling amplitudes $J_{jk}\!=\!- \int d\bs{x} \, w^{*}(\bs r_j) \left[\bs p ^2/2m + V_{{\rm opt}} (\bs r)\right] w(\bs r_k)$ between the nearest-neighboring sites $J_{jk}\equiv J$, which are typically much larger than the tunneling matrix elements between more distant sites \cite{Ashcroft}.  Spinful Hubbard models can be implemented using fermionic or bosonic atoms with two or more internal states~\cite{Altman:2003}. Experimental realization of Hubbard type systems with ultracold atoms in optical lattices has produced  important insight, starting with the experimental observation of quantum phase transitions: superfluid to Mott insulator for bosons \cite{Greiner2002} and Fermi-liquid to Mott insulator for fermions \cite{Jordens2008}.  In addition, different lattice topologies give rise to  pseudo-relativistic Dirac fermions~\cite{Tarruell2011}, and systems far from equilibrium can be realized~\cite{Schneider2012}.

Today, significant experimental effort is being devoted to realizing strong synthetic magnetic fields and to creating new kinds of spin-orbit couplings in optical lattices \cite{Aidelsburger:2011,Aidelsburger:2013,Ketterle:2013}.  Such systems are well suited for studying quantum Hall states, topological insulators and superconductors (see Section \ref{sect:simulation}). Optical lattices penetrated by staggered magnetic fluxes have also been realized with a view to studying frustrated magnetism \cite{Eckardt2010,Struck2011,Struck:2013}.  In this Section, we focus on different experimental methods for creating synthetic gauge fields in optical lattices.  First, we recall how general gauge structures are defined in the lattice framework, mainly discussing the case of a uniform magnetic field (\emph{i.e.}, the Hofstadter model \cite{Hofstadter:1976}).  We then comment on generalizations of this concept to other families of lattice gauge structures, such as non-Abelian gauge fields. Finally, we present methods for  implementing the Hofstadter model and its generalization in the lab.

\subsection{Gauge structures on the lattice}

Here, we review the physics of quantum particles constrained to evolve on a lattice and subjected to classical (static) gauge fields: firstly, focusing on Abelian gauge potentials, i.e. magnetic fields, in two-dimensional lattices, and then moving to non-Abelian gauge fields.

\subsubsection{The Abelian case}

\begin{figure}[tb]
\begin{center}
\includegraphics[width=5in]{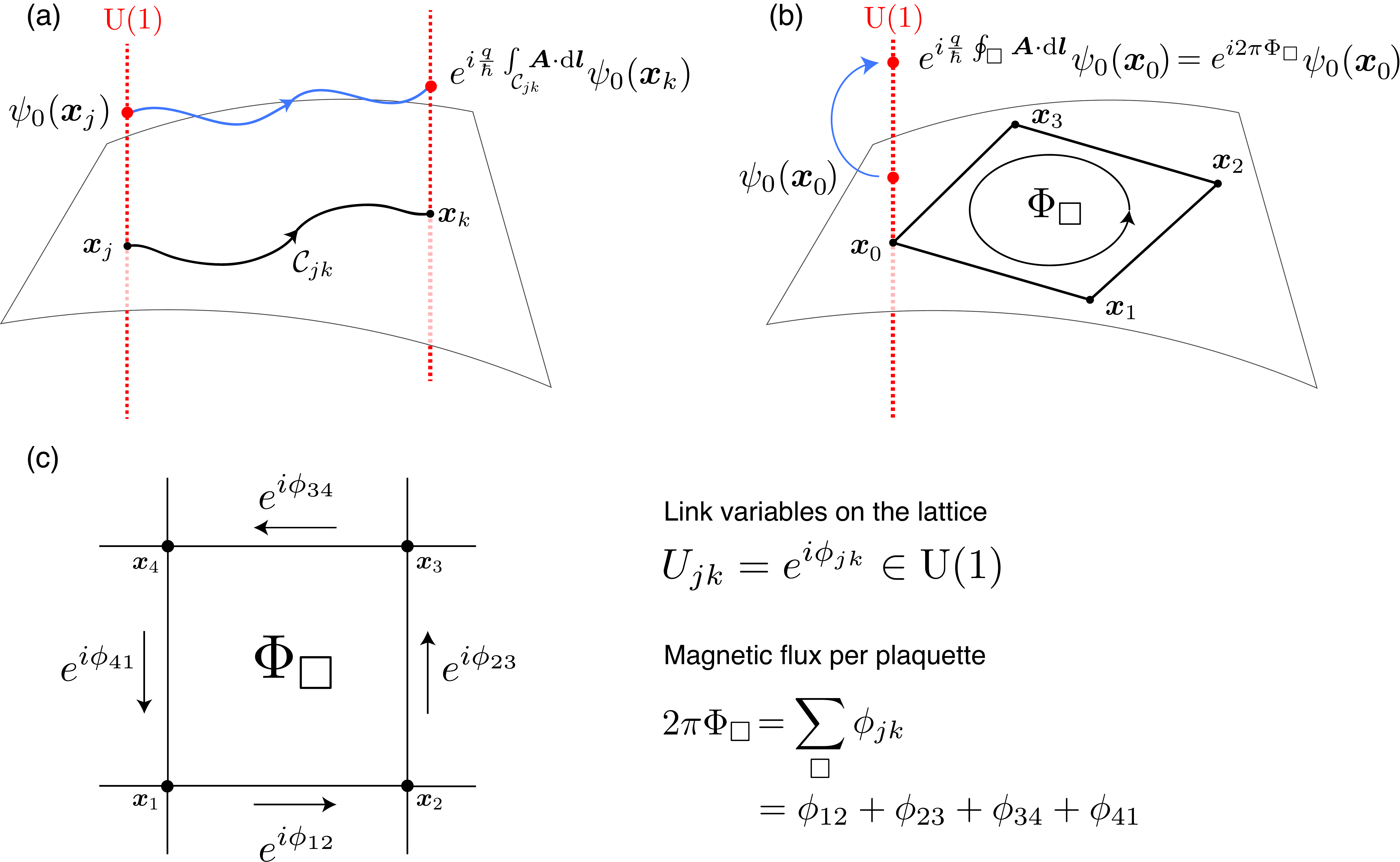}
\end{center}
\caption[Particle in a magnetic field]{Quantum particle in a magnetic field $\mathbfcal{B}=\bs{\nabla} \times \mathbfcal{A}$.  (a) When a particle follows a path $\mathcal{C}_{jk}$, from a reference point $\bs{x}_j$ to a point $\bs{x}_k$, it acquires a ``magnetic phase factor" $U_{jk}=\exp (i \phi_{jk}) \in \text{U (1)}$ determined by the gauge potential $\mathbfcal{A}$. (b) When a particle performs a loop $\square$, it realizes the Aharonov-Bohm effect. The wave function acquires a Berry phase proportional to the number of magnetic flux quanta $\Phi_{\square}$ penetrating the region enclosed by the loop. The loop $\square$ defines a U(1) transformation, called holonomy \cite{nakahara}. (c) Gauge structures on the lattice: the links $j-k$ are all associated with link variables $U_{jk}=\exp (i \phi_{jk}) \in \text{U (1)}$, which are directly related to the magnetic flux $\Phi_{\square}$ penetrating the plaquettes, see Eq. \eqref{fluxdef}. 
}
\label{fig_gaugelattice}
\end{figure}

Consider a particle subject to a gauge potential $\mathbfcal{A}\equiv q\bs{A}$ (where the coupling constant $q$ would be the electric charge in conventional electromagnetism) confined to the $\ex-\ey$ plane $\mathbb{R}^2$, described by the single-particle Hamiltonian
\be
H=\frac{1}{2m} \left[ \bs p - \mathbfcal{A} (\bs r) \right] ^2,\label{gaugecont}
\ee
When the particle follows a path from a reference point $j$ to a point $k$, whose coordinates are denoted $\bs r_j$ and $\bs r_k$, respectively, it acquires a ``magnetic phase factor'' \cite{Berry1980,Dirac1931,Aharonov1959,Berry:1984}
\be
\psi (\bs r_k)= \exp\left(\frac{i}{\hbar} \int_j^k \mathbfcal{A} \cdot {\rm d} \bs l\right) \psi_0 (\bs r_k)=U_{jk} \psi_0 (\bs r_k),\label{linkvardef}
\ee
where $\psi_0 (\bs r_k)$ denotes the wavefunction in the absence of the gauge potential, as in Fig. \ref{fig_gaugelattice} (a). Here, we introduced the \emph{link variable}  \cite{Zinn}
\be
U_{jk}=\exp\left(\frac{i}{\hbar} \int_j^k \mathbfcal{A} \cdot {\rm d} \bs l\right) = e^{i \phi_{jk}} = \bigl ( U_{kj}\bigr)^{*} \in {\rm U}(1),\label{linkvardef2}
\ee 
embodying the effect of the gauge potential on the fictitious link connecting the points $j$ and $k$. From the perspective of fibre-bundle theory, the link variables define the parallel transport on a principle fibre bundle $P ({\rm U(1)}, \mathbb{R}^2)$, where the connection $\mathbb{A}=i \mathcal{A}_{\mu} dx^{\mu}$ is determined by the gauge potential $\mathbfcal{A}$ (cf. Ref. \onlinecite{nakahara}). The link variables $U_{jk}=\exp\left(i \phi_{jk}\right)$ are generally called ``Peierls phases'' in condensed-matter physics (see Sect.~\ref{sectionmodels}), where they naturally appear in the description of solids subjected to magnetic fields \cite{Luttinger:1951,Hofstadter:1976}.  Under gauge transformations, the link variables \eref{linkvardef2} change according to
\begin{eqnarray}
&\mathbfcal{A} \rightarrow \mathbfcal{A}'= \mathbfcal{A} + \bs \nabla \chi \\
&U_{jk} \rightarrow U_{jk}'=U_{jk}\exp\left[i \frac{\chi (\bs r_k) - \chi (\bs r_j)}{\hbar} \right]. \label{transf}
\end{eqnarray}

To develop lattice models, we introduce the notion of a \emph{plaquette}, a closed region in space delimited by a set of points $\{ \bs r_1, \bs r_2 , \bs r_3 , \dots , \bs r_L \}$,  connected by links. When the particle performs a loop $\square$ around such a plaquette, it acquires an Aharonov-Bohm phase \cite{Berry:1984,Aharonov1959}
\be
\psi(\bs r_1) \overset{\square}{\longrightarrow}  \exp\left(\frac{i}{\hbar} \oint_{\square} \mathbfcal{A} \cdot {\rm d} \bs l\right) \psi(\bs r_1)=  \exp\left(2 \pi i\Phi_{\square}\right) \psi(\bs r_1),
\ee
where $\Phi_{\square}=h^{-1} \int \mathbfcal{B} \cdot {\rm d} \bs S$ is the number of magnetic flux quanta $\Phi_0=h$ penetrating the plaquette $\square$; and $\mathbfcal{B} = \bs \nabla \times \mathbfcal{A}$ is the magnetic field associated with the gauge potential $\bs A$, see Fig. \ref{fig_gaugelattice}bc.  This is the well-known Aharonov-Bohm effect \cite{Aharonov1959,Gerry1979}, which can also be expressed in terms of the link variables $U_{jk}=\exp\left(i \phi_{jk}\right)$ defined in Eq. \eref{linkvardef2} as
\be
e^{i 2 \pi \Phi_{\square}}=U_{12} \, U_{23} \, U_{34} \dots U_{L-1 \, L} \, U_{L1}=\prod_{\square}U_{jk}=\exp\left(i \sum_{\square} \phi_{jk}\right).\label{fluxdef}
\ee
Using Eq.~\eref{transf}, we see that the magnetic flux obtained through the ``loop" product \eref{fluxdef} is a gauge-invariant quantity associated with the link variables 
\be 
e^{i 2 \pi \Phi'_{\square}}=\prod_{\square}U_{jk}'= 
\underbrace{\exp\left\{\frac{i}{\hbar} \sum_{\square} \left[ \chi (\bs r_k) - \chi (\bs r_j)\right]\right\}}_{=1} \, \prod_{\square}U_{jk} =e^{i 2 \pi \Phi_{\square}}.
\label{invariance}
\ee
Here, $\tilde \Phi= \Phi + N$ is physically equivalent to $\Phi$ for $N \in \mathbb{Z}$, since $\exp\left(i 2 \pi \tilde \Phi\right)=\exp\left(i 2 \pi \Phi\right)$. \\

These general considerations suggest that the lattice description of a quantum system subject to a gauge potential $\mathbfcal{A}$ should include 
\begin{enumerate}[(a)]
\item a set of lattice sites $\{ \bs r_j \}$, 

\item a set of links $\{ j - k \}$ connecting the sites and defining plaquettes $\square$, 
\item a set of link variables $\{ U_{jk} \}$ associated with the links $\{ j - k \}$, see Fig. \ref{fig_gaugelattice} (c). 
\end{enumerate}
Here, the information in $\mathbfcal{A}(\bs r)$ is fully contained in the link variables $\{ U_{jk} \}$. The physical gauge-invariant quantity is the magnetic flux penetrating each plaquette $\Phi_{\square}$, which can be derived from the link variables [see Eq. \eref{fluxdef}, and Fig. \ref{fig_gaugelattice}c].  The gauge-invariance relation \eref{invariance} is interesting from a quantum-simulation point of view, as it defines the simplest set of link variables $\{ U_{jk} \}$ for any magnetic flux configuration. The generalization of these concepts to higher dimensions is straightforward.

$B \sim50\ {\rm T}$ is the largest magnetic field that can routinely be applied to materials; because the typical lattice spacing is $a \approx 10^{-10}\ {\rm m}$, the flux per elementary plaquette $\Phi_{\square} \sim 10^{-4}$ is tiny.  At higher fields, with $\Phi \sim 0.1 - 1$ (requiring a fantastically large magnetic field $B \sim 10^{4}\ {\rm T}$), the interplay between magnetic and lattice structures is captured by the Hofstadter model discussed in Sect.~\ref{hofsection}.  In contrast, synthetic magnetic fields for cold atoms trapped in optical lattices (Sect.~\ref{experimentallattice}) can be engineered in the high-flux regime $\Phi \sim 0.1 - 1$.

\subsubsection{The non-Abelian case}

The proceeding discussion focused on Abelian gauge potentials and magnetic fields, where the link variables $U_{jk}\in {\rm U(1)}$ are phase factors that do not couple to internal degrees of freedom.  These concepts can be generalized to a wide family of multi-component systems with electronic spin, color, flavor or atomic spin degrees of freedom.  In the following, we will generically refer to these as ``spin'' degrees of freedom.  In this context, gauge potentials and link variables may act differently on the different spin states, and they are therefore  matrix-valued objects (see Refs.~\onlinecite{Mead1992,Zinn}).  The link variables belong to a Lie group, e.g. $\hat U_{jk} \in {\rm U(N)}$ or $\hat U_{jk} \in {\rm SU(N)}$, while the non-Abelian gauge potential is an element of the corresponding Lie algebra, e.g. $\hat{\mathcal{A}}_{\mu} \in \mathfrak{u}(N)$ or $\hat{\mathcal{A}}_{\mu} \in \mathfrak{su}(N)$. A formal generalization of Eq. \eref{linkvardef2} defines the link variable 
\be
\hat U_{jk}=\mathcal{P} \biggl \{ \exp\left[\frac{i}{\hbar} \int_j^k \hat{\mathbfcal{A}}(\bs r) \cdot {\rm d} \bs l\right] \biggr \},\label{linkvardef3}
\ee 
in terms of the non-Abelian gauge potential $\hat{\mathbfcal{A}}(\bs r)$ (the coupling constant is again subsumed into the definition of $\hat{\mathbfcal{A}}$).  The path-ordered integral $\mathcal{P}(\cdot)$ is required because the matrices $\hat{\mathcal A}_{x,y}(\bs r)$ at different points of the path do not necessarily commute \cite{Zinn,Gupta,Kogut1983}.  The link variable $\hat U_{jk}$ and the non-Abelian gauge structures are clearly connected in the continuum limit, where the sites $\bs r_{j,k}$ are sufficiently close to each other, i.e. $\bs r_j = \bs r$ and $\bs r_k=\bs r + \epsilon {\bf e}_{\mu}$, where $\epsilon \ll 1$, ${\bf e}_{\mu}$ is the unit vector along the $\mu$ direction and $\mu=x,y$.  In this case, the link variables are \cite{Wilson1974,Gupta,Zinn,Kogut1983}
\begin{eqnarray}
&\hat U_{\bs r , \bs r + \epsilon {\bf e}_{\mu}}= \hat U_{\mu} (\bs r)=\exp\left[i \frac{\epsilon}{\hbar} \hat{\mathcal A}_{\mu} (\bs r ) \right], \label{defna}  \\
&\hat U_{\bs r + \epsilon {\bf e}_{\mu},  \bs r }= \hat U_{-\mu } (\bs r +  {\bf e}_{\mu})=\exp\left[-i \frac{\epsilon}{\hbar} \hat{\mathcal A}_{\mu} (\bs r ) \right], \nonumber
\end{eqnarray}
which lead to the ``loop" product around a unit square plaquette,
\begin{eqnarray}
&\hat U_{x}(\bs r) \hat U_{y}(\bs r + \epsilon {\bf e}_{x}) \hat U_{-x}(\bs r + \epsilon ({\bf e}_{x}+{\bf e}_{y}) ) \hat U_{- y}(\bs r + {\bf e}_{y}) \nonumber \\
&=\prod_{\square} \hat U_{jk}=\exp\left[{\frac{i \epsilon ^2}{\hbar} \hat{\mathcal{F}}_{x y}} (\bs r)\right].\label{naloop}
\end{eqnarray}
Equations \eref{defna}-\eref{naloop} define the Yang-Mills field strength \cite{Yang1954} (analogous to the antisymmetric tensor $\hat{{\mathcal{F}}}_{kl}$ defined by Eq.\eref{eq:v-k--v-l-commutators} for the geometric gauge potentials) 
\be
\hat{\mathcal{F}}_{\mu \nu} (\bs r)= \partial_{\mu} \hat{\mathcal{A}}_{\nu} (\bs r) - \partial_{\nu} \hat{\mathcal{A}}_{\mu} (\bs r) + \frac{i}{\hbar} [\hat{\mathcal{A}}_{\mu} (\bs r), \hat{\mathcal{A}}_{\nu} (\bs r)],\label{fsdef}
\ee
to leading order in $\epsilon$ (cf. Refs.  \onlinecite{Zinn} and \onlinecite{Kogut1983} for derivations).  In general, the commutator, $[\hat{\mathcal{A}}_x,\hat{\mathcal{A}}_y]\ne0$, which is the hallmark of non-Abelian gauge theories \cite{Zinn}. The loop product \eref{naloop} thus generalizes Eq. \eref{fluxdef} to the non-Abelian case. This result connects the link variables, defined in the lattice context, and the field strength which plays a fundamental role in continuum theories.

As for the Abelian case, a lattice system with a non-Abelian gauge structure can be entirely described by (a) the lattice topology: the sites coordinates, and the links, delimiting the unit plaquettes; and (b) the link variables $\{ \hat U_{jk} \}$ associated with the links. Under local gauge transformations, the link variables $\hat U_{jk} \in U(N)$ transform according to \cite{Zinn,Kogut1983}
\be
\hat U_{jk} \longrightarrow \hat U_{jk}'= \hat T_j \hat U_{jk} \hat T_k^{\dagger},
\ee
where $\hat T_k \in U(N)$ is a local transformation (a rotation in spin space at the lattice site $\bs r_k$).  Therefore, the loop product around a unit plaquette, delimited by the lattice sites $\{ \bs r_1, \dots , \bs r_L \}$,
\be
\hat U_{\square}= \prod_{\square} \hat U_{jk}=\hat U_{12} \, \hat U_{23} \, \hat U_{34} \dots \hat U_{L-1 \, L} \, \hat U_{L1},\label{loop_operator_definition}
\ee 
is not in general gauge invariant \cite{Zinn, Mead1992}, since
\be
\hat U_{\square}\longrightarrow \hat U_{\square}'= \hat T_1 \hat U_{\square}\hat T_1^{\dagger}.\label{loop_not_invariant}
\ee
This is related to the fact that the field strength $\hat{\mathcal{F}}_{\mu \nu}$, or curvature \cite{nakahara}, of the continuum theory is not gauge-invariant.  We therefore introduce the Wilson loop \cite{Wilson1974,Goldman:2009,Zinn}, a gauge-invariant quantity associated to each plaquette $\square$,
\be
W(\square)={\rm Tr} \bigl (\hat U_{\square} \bigr ),\label{wilsonloop}
\ee
where ${\rm Tr}(\cdot)$ indicates the trace of the $N \times N$ matrix. The gauge invariant Wilson loop \eref{wilsonloop} generalizes the notion of flux per plaquette $\Phi_{\square}$ to  non-Abelian gauge fields.

\subsubsection{Genuine non-Abelian structures and the Wilson loop}\label{non-Abelian_Wilson}



Genuine non-Abelian properties result from the non-commutativity of the gauge structure. When a particle performs two successive loops $\gamma_{1}$ and $\gamma_{2}$, that both start and end at the same point $\bs r_0$, it will acquire a geometric phase factor $U \in U(N)$. If, for every pair of loops $\gamma_{1,2}$, this geometric phase factor does not depend on the order of the operations $U=U_{12} = U_{21}$ (where $U_{12}$ [resp. $U_{21}$] corresponds to the situation where the loop $\gamma_{1}$ [resp. $\gamma_{2}$] has been performed first), then the system is \emph{Abelian} in nature. The observable effects related to these phase factors are those of a \emph{commutative} gauge theory. On the other hand, if the phase factors differ $U_{12} \ne U_{21}$ for some loops $\gamma_{1}$ and $\gamma_{2}$, then the gauge theory is genuinely \emph{non-Abelian}. The manifestations of the underlying non-Abelian gauge structure do not have any Abelian counterpart  \cite{Zinn,Wilson1974,Gupta,Kogut1983,Mead1992,Zhang:2008na}. 

In continuum gauge theories, the non-Abelian property $U_{12} \ne U_{21}$ can be traced back to the non-commutativity of the field strength,
 $[\hat{\mathcal{F}}_{\mu \nu} (\bs r), \hat{\mathcal{F}}_{\mu^{\prime} \nu^{\prime}} (\bs r^{\prime})] \ne 0$, which naturally stems from the non-commutativity of the gauge potential's components $[\hat{\mathcal{A}}_{\mu}, \hat{\mathcal{A}}_{\nu}] \ne 0$, see Ref. \cite{Mead1992}. 

As already pointed out in Section \ref{abvsnonab},  the criterion $[\hat{\mathcal{A}}_{\mu}, \hat{\mathcal{A}}_{\nu}] \ne 0$ is generally used in the literature to specify ``non-Abelian gauge potentials", but it is not sufficient to attest that the system hosts genuine non-Abelian properties, which are captured by the field strength $\hat{\mathcal{F}}_{\mu \nu}$ or the loop operators $U$, as described above \cite{Mead1992,Zhang:2008na}. To avoid any ambiguity, we stress that the term \emph{non-Abelian gauge fields} should be entirely based on the non-commutativity of the field strength or loop operators [see also Eq.\eqref{non-comm_property} below], as described in this Section.

An analogous criterium can be introduced in the lattice framework, in terms of the loop operator $\hat U_{\square}$ defined in Eq. \eqref{loop_operator_definition}. The lattice gauge structure associated with the link variables $\{ \hat U_{jk} \}$ is said to be \emph{genuinely non-Abelian} if there exists a pair of loops $\hat U_{\square_1}$ and $\hat U_{\square_2}$, both starting and ending at some site $j=1$, such that
\be
\hat U_{\square_1} \hat U_{\square_2} \ne \hat U_{\square_2} \hat U_{\square_1}.\label{non-comm_property}
\ee
Note that the non-Abelian criterium, based on the non-commutativity property \eqref{non-comm_property}, is gauge invariant, as can be verified using Eq. \eqref{loop_not_invariant}. Therefore, the link variables $\{ \hat U_{jk} \}$ which lead to non-commutating loop operators, provide genuine non-Abelian effects \cite{Mead1992,Zhang:2008na,Osterloh:2005} on the lattice. \\

Besides, one can introduce a simple criterium to detect lattice configurations in which all the loop matrices $\hat U_{\square}$ can be simultaneously  gauge-transformed into a simple phase factor
\be
\hat U_{\square}=\exp (i 2 \pi \Phi) \hat{1}_{N \times N} , \quad \forall \, \text{ loops } \, \square. \label{trivial_lattice_gauge}
\ee 
If such a reduction \eqref{trivial_lattice_gauge} was possible over the whole lattice, then the multi-component system would behave as a collection of uncoupled Abelian subsystems subjected to the same flux $\Phi$ \cite{Goldman:2009}. On the square lattice, denoting $\hat U_{x,y}$ the link variables along $\ex$ and $\ey$, the condition $[\hat U_x, \hat U_y]\ne 0$ would constitute a natural, but incomplete criterium to detect non-Abelian structures.  For example, the link variables $\hat U_{x,y}=\exp (i \pi \hat \sigma_{x,y} /2)$ have $[\hat U_x, \hat U_y]= - 2i  \hat\sigma_z\ne 0$, while the loop matrix around a unit cell is $\hat U_{\square}=\exp (i \pi )  \hat{1}_{2 \times 2}$:  This configuration therefore corresponds to a two-component lattice subjected to a uniform (Abelian) flux per plaquette $\Phi=1/2$ \footnote{Note that the mapping from N uncoupled lattices penetrated by a same magnetic flux $\Phi$ to a SU(N) tight-binding model with constant hopping operators $\hat U_{x,y} \in \text{SU(N)}$ has been explored in Ref.  \cite{Barnett2012PRL}.}. As shown in Ref. \cite{Goldman:2009}, the Wilson loop \eref{wilsonloop} provides an unambiguous criterium  to determine whether the set $\{ \hat U_{jk} \}$ leads to uncoupled Abelian subsystems captured by Eq. \eqref{trivial_lattice_gauge}. Indeed, the loop operator $\hat U_{\square} \in U(N)$ reduces to a simple phase factor $\hat U_{\square}=\exp (i 2 \pi \Phi) \hat{1}_{N \times N}$ if and only if the Wilson loop satisfies $\vert W(\square) \vert=N$. 

The criterium based on non-trivial Wilson loops $\vert W(\square) \vert \ne N$ is a necessary condition for attesting that the link variables $\{ \hat U_{jk} \}$ produce a non-Abelian gauge field [Eq. \eqref{non-comm_property}]. However, we stress that this condition is not \emph{sufficient}. For instance, a spin-1/2 lattice satisfying $\hat U_{\square}=\exp (i 2 \pi \Phi \hat \sigma_z)$ in all its plaquettes is clearly Abelian in the sense of Eq. \eqref{non-comm_property}, while it is generally associated with a non-trivial Wilson loop $\vert W(\square) \vert \ne 2$, since the different spin components feel an opposite magnetic flux $\Phi \hat \sigma_z$, i.e. the trivialization \eqref{trivial_lattice_gauge} is not satisfied in this case. We note that in the context of the non-Abelian Aharonov-Bohm effect \cite{Horvathy:1986}, such a gauge field, although \emph{Abelian}, is loosely referred to as a ``non-Abelian flux". Again, to avoid any ambiguity, we stress that  \emph{non-Abelian gauge fields} defined on a lattice are those that satisfy the strict criterium based on non-commutating loop operations [Eq.\eqref{non-comm_property}].

\subsection{The lattice Hamiltonians: a few models}
\label{sectionmodels}

The non-relativistic Abelian and non-Abelian lattice gauge structures discussed above derive from applied gauge potentials (or equivalently, link variables); these are classical and non-dynamical and are captured by the non-relativistic tight-binding Hamiltonian 
\begin{equation}
\hat{H}= -J \sum_{j,k} \sum_{\sigma, \sigma '} \hat{c}_{j,\sigma}^{\dagger} \bigl ( U_{jk} \bigr )_{\sigma \sigma'} \hat{c}_{k, \sigma '} + {\rm h.c.}.\label{hubbardgauge}
\end{equation}
In this general expression, the operator $ \hat{c}_{j,\sigma}^{\dagger}$ describes the creation of a particle at the lattice site $\bs r_j$, in the spin state $\sigma=1, \dots, N$, and the link variables $U_{jk} \in U(N)$ contribute to the hopping between connected sites $(j,k)$. The tunneling's strength is characterized by the hopping rate matrix element $J$. For simplicity, we take $J$ to be uniform; and since we are focusing on non-interacting particles, we omitted the interaction term in Eq. \eref{hubbardgauge}. In the next paragraphs, we review a few relevant lattice models corresponding to specific configurations of the lattice and link variables.

\subsubsection{A uniform magnetic flux through the lattice: The Hofstadter model}
 \label{hofsection}

\begin{figure}[tb]
\begin{center}
\includegraphics[width=6.in]{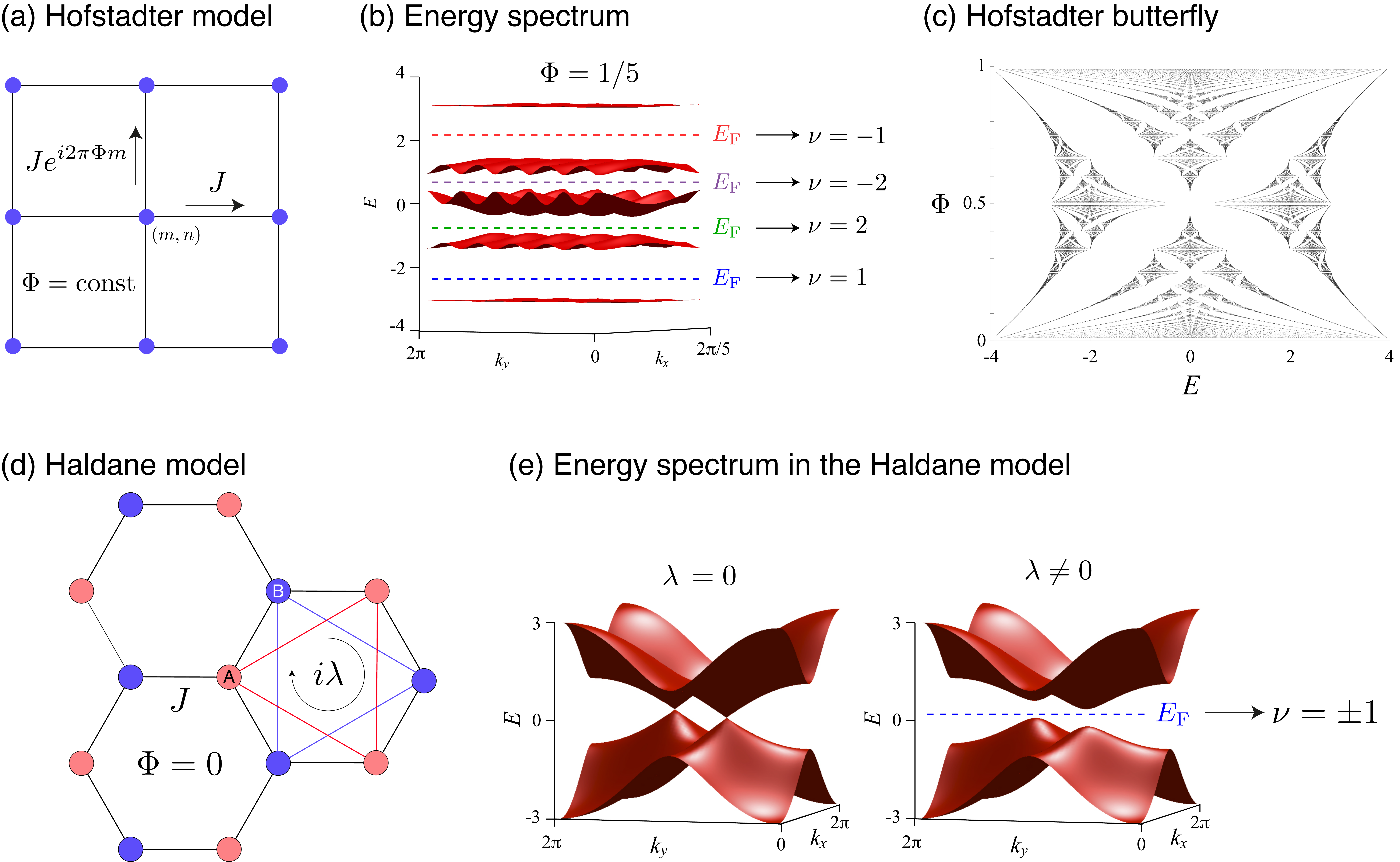}
\end{center}
\caption[Hofstadter and Haldane models]{(a) The Hofstadter model on the square lattice. (b) Bulk energy spectrum of the Hofstadter model for $\Phi=1/5$.  The quantized Hall conductivity $\sigma_H= (e^2/h) \, \nu$ is shown for Fermi energies located within the bulk gaps. (c) The Hofstadter butterfly and its energy spectrum as a function of the magnetic flux $\Phi$. (d) The Haldane model on the honeycomb lattice. (e) Energy spectrum of the Haldane model. In the absence of NNN hopping ($\lambda=0$), the spectrum has two Dirac cones. In the presence of  complex NNN hopping ($\lambda \ne 0$), a bulk gap opens, leading to an anomalous quantum Hall phase with the Chern number $\nu = \pm 1$. Note that the lowest bands of the Haldane and Hofstadter models presented in (b) and (e) are topologically equivalent. However, in contrast with the Haldane model, the lowest band of the Hofstadter spectrum shown in (b) exhibits an almost completely flat dispersion; these topological-flat-band configurations are good candidates for realizing fractional Chern insulators \cite{Parameswaran}.
}
\label{fig_hofstadter}
\end{figure}

A square lattice with a uniform magnetic field ${\mathbfcal{B}}\!=\! \bs \nabla \times \bs {\mathbfcal{A}}\!=\!B {\bf e}_z$ is an iconic model -- called the Hofstadter model \cite{Hofstadter:1976} -- of a two-dimensional electron gas (2DEG) in a strong magnetic field, and is the most simple case of a non-trivial gauge field in a lattice.  Here, particles undergo the traditional (i.e. Abelian) Aharonov-Bohm effect when they circulate around the unit square plaquettes, $\psi (\bs r) \rightarrow \psi (\bs r) \exp (i 2 \pi \Phi)$.  This model is described by Eq. \eref{hubbardgauge}, with the specific form 
\begin{equation}
\small{\hat{H}\!=\! -J \sum_{m,n} e^{i \phi_{x} (m,n)} \hat{c}_{m+1,n}^{\dagger}   \hat{c}_{m,n} + e^{i \phi_{y}(m,n)} \hat{c}_{m,n+1}^{\dagger}   \hat{c}_{m,n} + {\rm h.c.}}, \label{hofstadter}
\end{equation}
where the lattice sites are located at $(x,y)=(m a, n a)$, $m,n \in \mathbb{Z}$; and $a$ is the lattice period. Here, the link variables $U_{jk}=\exp \bigl [i \phi_{x,y} (m,n) \bigr] \in {\rm U(1)}$,  commonly known as ``Peierls phases", satisfy \eref{fluxdef},
\begin{eqnarray}
\prod_{\square} U_{jk}&= e^{i \left[(\phi_{x} (m,n) + \phi_{y} (m+1,n) - \phi_{x} (m+1,n+1) - \phi_{y} (m,n+1)  \right]}, \nonumber \\
&=e^{i 2 \pi \Phi}.
\end{eqnarray}
In a tight-binding treatment \cite{Ashcroft,Luttinger:1951,Hofstadter:1976}, Eq. \eref{hofstadter} can be obtained through the \emph{Peierls substitution} $E_0(\hbar \bs k \rightarrow \hat{\bs p} -{\mathbfcal {A}})$, where $E_0 (\bs k)$ is the tight-binding model's field-free dispersion relation \footnote{In the case of a square lattice, the field-free dispersion relation is $E_0 (\bs k)= - 2 J \left ( \cos (k_x a) + \cos (k_y a) \right )$, where $a$ is the lattice spacing. Performing the Peierls substitution, $\hbar \bs{k} \rightarrow \hat{\bs{p}} - \bs{A}$ yields the effective single-particle Hamiltonian $\hat h= -J \left [ e^{i \phi_x (\bs r)} \hat{T}_a^x +  e^{i \phi_y (\bs r)} \hat{T}_a^y + \text{h.c.} \right ]$, where $\phi_{x,y} (\bs r) = (a/\hbar) A_{x,y} (\bs r)$ and where $\hat{T}_a^{x,y}$ are translation operators on the lattice, $\hat{T}_a^{\mu} \psi (\bs r)= \psi (\bs r - a \bs{1}_{\mu})$  \cite{Luttinger:1951,Hofstadter:1976}.}. The Hofstadter Hamiltonian \eref{hofstadter} can alternately be obtained from the continuum Hamiltonian \eref{gaugecont}, by discretizing the spatial coordinates and the derivative operators \cite{Gagel:1995}.  This continuum approach is only rigorously valid in the limit $a \rightarrow 0$, namely for small flux $\Phi=B a^2/\Phi_0$, where the energy structure only slightly deviates from the Landau levels. Away from this continuum limit, $\Phi \sim 1$, the tight-binding Hamiltonian \eref{hofstadter} displays new features originating from the underlying lattice structure. The hopping phases in the Landau gauge are
\begin{equation}
\phi_x=0 {\rm  , } \, \phi_y (m)= 2 \pi \Phi m, \text{ see Fig. \ref{fig_hofstadter} (a);} \label{landaugauge}
\end{equation}
for a rational flux $\Phi=p/q$, with $p,q \in \mathbb{Z}$, the Hamiltonian commutes with the magnetic translation operators $T_x^q \psi (m,n)= \psi (m+q,n)$ and $T_y \psi (m,n)= \psi (m,n+1)$. In this gauge, the system is described by $q \times 1$ magnetic unit cells (gauge dependent) and its energy spectrum splits into $q$ (gauge independent) subbands, see Fig. \ref{fig_hofstadter}b. The spectrum has a fractal (self-similar) pattern set of eigen-energies, called the Hofstadter butterfly \cite{Hofstadter:1976}, see Fig. \ref{fig_hofstadter}c, and it has quantum Hall phases with topological order, to be discussed in Sect.~\ref{quantumsimulation} [see also Fig. \ref{fig_hofstadter} (b)].  At the special ``$\pi$-flux" case $\Phi=1/2$, the Hofstadter lattice reduces to a two-band model displaying conical intersections at zero energy \cite{Kohmoto:1989,Hatsugai:2006,Lim2010}, similar to graphene's pseudo-relativistic spectrum \cite{Wallace1947,CastroNeto2009}. In this singular case, the system satisfies time-reversal symmetry $H (\Phi=1/2) \equiv H (- \Phi)$. The Hofstadter model can easily be extended to other two-dimensional lattices \cite{Kimura:2002,Aoki:1996,Vidal:1998,Goldman:2011,Bercioux:2011}, and in particular to the honeycomb lattice \cite{Rammal:1985,Hatsugai:2006}, making it an efficient tool to investigate the electronic properties of graphene or other exotic materials in strong magnetic fields \cite{Goerbig:2011}. The effects of interactions in the Hofstadter model have been analyzed in Refs. \cite{Palmer:2008,Powell:2010ck,Powell:2011ek,Harper:2014vi}. The Hofstadter model has been realized with cold atoms in modulated optical lattices \cite{Aidelsburger:2013,Ketterle:2013,Aidelsburger:2014}.

\subsubsection{Local magnetic flux: The Haldane model}
\label{haldanesection}
In the early 1980's, the quantum Hall effect was discovered in two-dimensional electronic systems at modest magnetic fields \cite{vonKlitzing:1986}, and until a seminal work by Haldane \cite{Haldane:1988}, it was believed that the presence of a uniform magnetic field was necessary to produce this effect. In Ref. \onlinecite{Haldane:1988}, Haldane introduced a lattice model demonstrating that the essential ingredient was not the magnetic field, but rather the associated breaking of time-reversal-symmetry (TRS).  In Haldane's model, while TRS is broken \emph{locally}, the magnetic flux per unit cell is zero.  Figure \ref{fig_hofstadter}d shows this model's honeycomb lattice topology, featuring both nearest-neighbor (NN) and next-nearest-neighbor (NNN) hopping.  While the NN hopping is trivial (positive and real valued), the NNN hopping terms are accompanied by non-trivial Peierls phases, indicating the presence of a TRS-breaking gauge potential. The corresponding Hamiltonian \eref{hubbardgauge} is
\begin{eqnarray}
  &\hat H
  = - J \sum_{\langle k,l \rangle} \hat c_{k}^{\dagger} \hat c_{l}+ i   \lambda \sum_{\langle\!\langle k,l \rangle\!\rangle}
      \nu_{k,l} \hat c_{k}^{\dagger} \hat c_{l}, \label{haldane}
    \end{eqnarray}
where $\langle k,l \rangle$ and $\langle\!\langle k,l \rangle\!\rangle$ denotes the NN and NNN sites respectively of a honeycomb lattice, and $\nu_{k,l}=\pm 1$ depending on the orientation of the path connecting the NNN sites $\langle\!\langle k,l \rangle\!\rangle$.   The Peierls phases $U_{kl}=\exp (i \pi \nu_{k,l} /2)$ are complex for NNN sites, leading to non-zero \emph{local} magnetic flux $\Phi_{\alpha}$ within triangular ``subplaquettes" $\alpha$, even though the total flux $\Phi= \sum_{\alpha} \Phi_{\alpha}=0$ inside each honeycomb unit cell. For $\lambda \ne 0$, an energy gap opens in the spectrum, see Fig. \ref{fig_hofstadter}e, leading to non-trivial topological orders and quantum ``anomalous" Hall phases (see also Sect.~\ref{quantumsimulation}). The Haldane model played a surprising role in the prediction of the quantum spin Hall effect, an early precursor to TR preserving topological insulators (see below). The Haldane model has been recently implemented with cold atoms using time-modulated optical lattices \cite{Jotzu}.

\subsubsection{Spin-orbit coupling on the lattice: The Kane-Mele and the spin-dependent Hofstadter models}
\label{spinorbitlattice}

The Kane-Mele model was introduced as a tight binding framework to explore the effects of spin-orbit couplings (SOCs) in graphene \cite{Kane:2005,Kane:2005bis}. Based on this model, Kane and Mele predicted a quantum spin Hall effect, leading to the current interest in TR-invariant topological insulatrs (cf. also Refs. \onlinecite{Bernevig:2006,Hasan2010,Qi2011} and \onlinecite{Qi2008}). Interestingly, the intrinsic spin-orbit coupling term can be modeled by a NNN Haldane hopping term, acting oppositely on the two spin components $\sigma=\uparrow, \downarrow$. In the presence of SOC, each spin component feels a local ``magnetic" flux, yet TRS is preserved because the fluxes are opposite for  $\uparrow$ and $\downarrow$.  The corresponding link variables can be represented as $2\times2$ matrices, with $\hat U_{kl}^{{\rm SO}}=i\nu_{k,l} \hat \sigma_z$. In solids, the Rashba SOC produced by an external electric field \cite{Bychkov1984,Winkler03Review}, is modeled by a NN hopping term that mixes the spin components non-uniformly. It involves SU(2) link variables such as $\hat U_{kl}^{{\rm R}} = (\hat{\bs \sigma} \times \hat{\bs d}_{kl})_z \propto \hat \sigma_{x,y}$, where $\hat{\bs d}_{kl}$ denotes the unit vector along the link connecting the NN sites $k,l$. The total Kane-Mele Hamiltonian is
\begin{eqnarray}
 \hat H
  =& - J \sum_{\langle k,l \rangle} \hat c_{k}^{\dagger} \hat c_{l}+ i   \lambda_{{\rm SO}} \sum_{\langle\!\langle k,l \rangle\!\rangle}
      \nu_{k,l} \hat c_{k}^{\dagger} \,  \hat \sigma_z \, \hat c_{l}  \nonumber \\
      &+  i   \lambda_{{\rm R}} \sum_{\langle k,l \rangle}
    \hat c_{k}^{\dagger} \,  (\hat{\bs \sigma} \times \hat{\bs d}_{kl})_z \, \hat c_{l}, \label{kanemele}
    \end{eqnarray}
where the three terms correspond to: direct hopping between NN sites, intrinsic spin-orbit coupling, and  Rashba spin-orbit coupling \cite{Kane:2005,Kane:2005bis}. The Kane-Mele model for graphene can therefore be viewed as the direct SU(2) analogue of the Haldane model \eref{haldane} with an extra Rashba term characterised by the strength $ \lambda_{{\rm R}} $ featured in Eq.\eref{kanemele}.  Just as the Haldane model produces quantum Hall states, the Kane-Mele model produces quantum \emph{spin} Hall states, which can be viewed as two superimposed, and opposite, spin-filtered QH phases. A three-dimensional generalization of the Kane-Mele model \eref{kanemele}, where the link variables $\hat U_{kl}^{{\rm SO}}$ are defined along the links of a diamond lattice \cite{Fu:2007} reveals the exciting physics of 3D topological insulators \cite{Hasan2010,Qi2011,Qi2008}.

In the same spirit, Goldman et al. \cite{Goldman:2010} introduced a spinful SU(2) analogue of the Hofstadter model on a square lattice.  In this model, particles with spin $\sigma=\uparrow, \downarrow$ experience a uniform magnetic flux per plaquette $\Phi_{\uparrow}=-\Phi_{\downarrow}$ opposite in sign for the two spin components. The corresponding Hamiltonian is
\begin{eqnarray}
\small{\mathcal{H}\!=\!-J\sum_{m,n} \hat c_{m+1,n}^{\dagger} e^{i 2\pi \gamma \hat \sigma_x} \hat c_{m,n} \!+\!\hat c_{m,n+1}^{\dagger} e^{i  2\pi m \Phi \hat \sigma_z} \hat c_{m,n}+{\rm h.c.}},
\label {goldham}
\end{eqnarray}
where the SU(2) link variables $\hat U_{x},\hat U_y(m)$ act on the two-component field operator $\hat c_{m,n}$, defined at lattice site $(x,y)=(m a, n a)$.
For $\gamma=0$, this models corresponds to two decoupled copies of the spinless Hofstadter model \eref{hofstadter}-\eref{landaugauge}. The effect of the link variable $\hat U_y (m) \propto \hat \sigma_z$ is therefore analogous to the intrinsic spin-orbit coupling in Eq. \eref{kanemele}. For $\gamma \ne 0$, the two spin components are mixed as they tunnel from one site to its nearest neighbor: the link variable $\hat U_x \propto \sin (2 \pi \gamma)  \hat \sigma_x$ plays a role similar to the Rashba coupling in Eq. \eref{kanemele}. This model therefore captures the essential effects of the Kane-Mele TRS-invariant model in a multi-band framework, but offers the practical advantage of only involving NN hopping on a square lattice. The optical-lattice implementation of this model has been reported by  Aidelsburger et al. \cite{Aidelsburger:2013} and investigated by Kennedy et al. \cite{Kennedy:2013}. The effects of interactions in the SU(2) Hofstadter model have been analyzed by Cocks, Orth and coworkers in Refs. \cite{Cocks:2012,Orth:2013}.

\subsubsection{The square lattice subjected to a non-Abelian gauge potential}

The interplay between Abelian and non-Abelian gauge fields can be studied in a simple two-component model defined on a square lattice~\cite{Goldman:2009,Goldman:2009prl}, where the link variables along $\ex$ and $\ey$ are
\begin{align}
\hat U_x&= e^{i \alpha \hat \sigma_y},& {\rm  and} && \hat U_y (m)&= e^{i \beta \hat \sigma_x} e^{i 2 \pi \Phi m}.
\end{align}
The Abelian part of the gauge structure given by the U(1) phase $\phi_y(m)=\exp (i 2 \pi \Phi m)$ in $\hat U_y$, corresponds to a uniform magnetic flux per plaquette $\Phi$ (cf. Sect.~\ref{hofsection}). The SU(2) part is controlled by the parameters $\alpha$ and $\beta$.  In two limiting cases this system is purely Abelian, i.e. $\vert W (\square) \vert=2$: when $\alpha={\rm integer} \times \pi$ or $\beta={\rm integer} \times \pi$; and when $\alpha={\rm integer} \times \pi/2$ and $\beta={\rm integer} \times \pi/2$.  In both cases, the system reduces to two uncoupled (Abelian) models. For arbitrary values of $\alpha, \beta$ the system features non-Abelian fluxes [see Section \ref{non-Abelian_Wilson}]: the Wilson loop is uniform and non-trivial $\vert W (\square) \vert \ne 2$. This model was investigated in the contexts of quantum Hall physics \cite{Goldman:2009prl,Zamora:2011,Palmer:2011,Mei:2011,Grass:2011,Komineas:2012}, transition to the Mott-insulating phase \cite{Gra:2011} and non-Abelian anyonic excitations \cite{Burrello:2010}. Alternative non-Abelian lattice models have been proposed and studied \cite{Osterloh:2005,Vozmediano:2010,Goldman:2012,Satija:2008,Goldman:2007,Goldman:2007bis,Beugeling:2012,Maraner:2009}, showing the rich properties stemming from non-Abelian structures in non-relativistic quantum systems.

\subsection{Experimental realizations using optical lattices}
\label{experimentallattice}
 
\subsubsection{Laser-assisted-tunneling using different internal states}\label{laser-assisted-section}

Let us now consider the schemes enabling to create lattice gauge structures  for ultracold atoms moving in optical lattices. In this context, the main ingredient  for generating Peierls phases along the links of the lattice is the so-called \emph{laser-assisted-tunneling method}, initially introduced by Ruostekoski-Dunne-Javanainen \cite{Ruostekoski:2002,Ruostekoski:2008ig}, and Jaksch-Zoller \cite{Jaksch:2003}. This method is based on the possibility to couple atoms living on neighboring  sites of an optical lattice, hence controlling their tunneling over the lattice, as illustrated in Fig. \ref{fig_assisted_shaking} (a). Peierls phases are then engineered by controlling the phase of the coupling, which allows to simulate lattice (tight-binding) Hamiltonians with non-trivial link variables $U_{jk}$, for example leading to the Hofstadter Hamiltonian \eref{hofstadter} in 2D. This method was  further developed by several authors, such as Mueller \cite{Mueller2004}, Gerbier-Dalibard \cite{Gerbier:2010}, Anisimovas et al. \cite{Anisimovas:2014ga}, Goldman et al. \cite{Goldman:2010,Goldman:2013Haldane} and Mazza et al. \cite{Bermudez:2010,Mazza:2012}. This method, which can be extended to non-Abelian structures \cite{Osterloh:2005,Mazza:2012,Goldman:2013Haldane} and more exotic lattice geometries \cite{Gorecka:2011}, can be summarized as follows: 

\begin{enumerate}[(a)]
\item Gauge fixing: what link variables $U_{jk}$ should be generated?
\item Prevent spontaneous hopping along the links for which $U_{jk} \ne 1$.
\item Atom-light coupling: induce the hopping externally and engineer the desired link variables $U_{jk}$ by tuning the coupling lasers, see Fig. \ref{fig_assisted_shaking} (a).
\end{enumerate}

Any physical scheme accomplishing the crucial steps (b)-(c), relies on the specific properties of the atoms used in the experiment.  For instance, Gerbier and Dalibard proposed an elegant method exploiting the unique properties of alkaline-earth or Ytterbium atoms \cite{Gerbier:2010} [see also Anisimovas et al. \cite{Anisimovas:2014ga}]. In their  proposal, the induced-hopping (c) involves a coupling between the ground state and a long-lived metastable excited electronic state. For these atomic species, metastable states have remarkably long lifetimes, and thus, the single-photon transitions to the excited state have greatly reduced spontaneous emission rates (as compared, e.g.,  to schemes based on two-photon transitions). For the more common alkali atoms, e.g. Li, K, Rb, alternative schemes based on Raman couplings, i.e. two-photon transitions, are required \cite{Jaksch:2003,Aidelsburger:2011,Aidelsburger:2013,Aidelsburger:2013b,Ketterle:2013}. 

\subsubsection{Laser-assisted-tunneling and shaking methods}

\begin{figure}[tb]
\begin{center}
\includegraphics[width=4in]{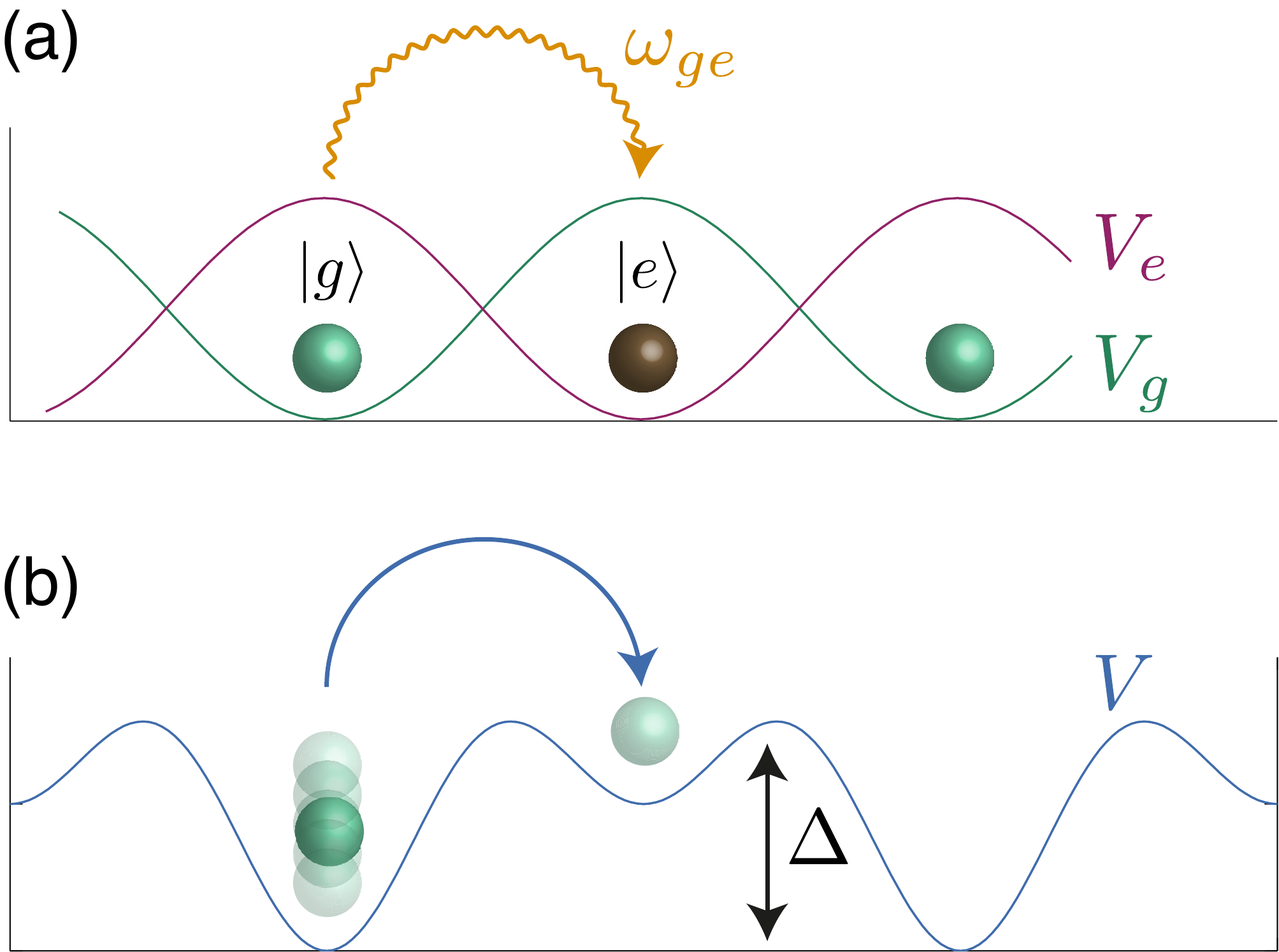}
\end{center}
\caption[Assisted tunneling and shaking]{ (a) Laser-assisted tunneling. Two internal states of an atom, denoted $\vert g \rangle$ and $\vert e \rangle$, are trapped in state-dependent optical lattices $V_{g,e}$. A resonant laser with frequency $\omega_{ge}$ couples the two states and induces effective tunneling matrix elements  \cite{Jaksch:2003,Gerbier:2010}. The configuration where $V_g=-V_e$ can be simply implemented for Yb atoms, by using lasers at the ``anti-magic" wavelength \cite{Gerbier:2010,Goldman:2013Haldane,Anisimovas:2014ga}. (b) Shaking an optical superlattice. Atoms are trapped in a superlattice potential $V$, displaying an energy offset $\Delta$ between neighboring lattice sites and chosen so as to inhibit the hopping [i.e. $\Delta \gg J$, where $J$ is the natural hopping amplitude]. A resonant modulation of the lattice potential, with frequency $\omega \approx \Delta/\hbar$, re-activates the hopping in a controlled manner \cite{Kolovsky:2011,Creffield:2013gp,Creffield:2014vw,Bermudez:2011prl,Bermudez:2012njp,Baur:2014ux}. The lattice modulation can be created by two (far-detuned) running-wave beams with frequencies $\omega_{1,2}$, such that $\omega_2 - \omega_1 = \Delta / \hbar$. The effective tunneling matrix element can be decorated with  non-zero Peierls phases, when the modulation further transfers momentum $\delta \bs k$ to the atoms, which can be realized using ``Raman" lasers with wave vectors $\bs k_{1,2}$, such that $\delta \bs k=\bs k_{2}- \bs k_{1}$ (see Section \ref{mainingredients}). An extension of this scheme was recently implemented in the Munich \cite{Aidelsburger:2011,Aidelsburger:2013,Aidelsburger:2013b} and MIT \cite{Ketterle:2013} experiments. See also Fig. \ref{fig_assistedtunneling}.
}
\label{fig_assisted_shaking}
\end{figure}

We point out that the concept of laser-induced tunneling, which is based on the coupling of atoms living on neighboring lattice sites [Fig. \ref{fig_assisted_shaking} (a)], is intimately related to methods exploiting time-periodic modulations (``shaking") of the optical lattice \cite{Eckardt:2005,Eckardt2010,Hemmerich2010,Kolovsky:2011,Struck2012,Arimondo2012}. This analogy is illustrated in Fig. \ref{fig_assisted_shaking} (b).  Here, atoms are trapped in an optical superlattice potential, where neighboring sites are shifted in energy by the offset $\Delta$ so as to inhibit the natural hopping. A resonant modulation of the lattice  then allows to re-establish the hopping in a controllable manner. In this framework, and in contrast with the standard laser-assisted method described in Sect. \ref{laser-assisted-section}, the atoms can be prepared in a \emph{single} internal state. Designing a modulation that also transfers momentum to the atoms, e.g. using two ``Raman" laser beams, also allows to generate Peierls phases in the effective hopping matrix elements (see Section \ref{mainingredients} and works by Kolovsky \cite{Kolovsky:2011}, Creffield-Sols \cite{Creffield:2013gp,Creffield:2014vw}, Bermudez et al. \cite{Bermudez:2011prl,Bermudez:2012njp}, Lim et al. \cite{Lim2008}, Baur-Schleier-Smith-Cooper \cite{Baur:2014ux} and Goldman et al. \cite{GoldmanDalibard:2015}.). This ``shaking" scheme was successfully implemented in the Munich \cite{Aidelsburger:2011,Aidelsburger:2013,Aidelsburger:2013b,Aidelsburger:2014} and MIT \cite{Ketterle:2013} experiments to imprint space-dependent Peierls phase in a 2D optical lattice, with a view to creating the Hofstadter model \eqref{hofstadter} with cold atoms. We point out that off-resonant potential modulations can also be considered to engineer Peierls phases in optical lattices, as recently demonstrated in Hamburg \cite{Struck2011,Struck:2013} [see Section \ref{sec:other schemes} for a more detailed discussion on periodically-driven cold-atom systems]. \\

Below, we present the main ingredients for engineering Abelian gauge structures in optical lattices, based on the experiment performed in Munich in 2011  \cite{Aidelsburger:2011,Aidelsburger:2013b}. The generalization to non-Abelian (matrix-valued) link variables $\hat U_{jk} \in {\rm U}(N)$ is briefly discussed in Sect.~\ref{matrixlinkvariables}.  Other interesting schemes are further discussed in Section \ref{sec:other schemes}. 

\subsubsection{Laser-assisted tunneling methods: the main ingredients}
\label{mainingredients}

\paragraph{Choosing the gauge} The first step consists in selecting the gauge, thus dictating the explicit form of the link variables $U_{jk}$.  For the Hofstadter model, we typically choose the Landau gauge \eref{landaugauge}, in which the link variables $U_{jk}=\exp \bigl [i \phi_{x,y} (m,n) \bigr]$ are non-trivial along $\ey$ only (i.e. $\phi_{x}=0$ and thus the hopping along $\ex$ can be the lattice's native tunneling). It is convenient to create non-trivial link variables $U_{jk}$ along the links connecting the NN sites of the lattice~\cite{Jaksch:2003,Gerbier:2010,Alba:2011,Goldman:2013}, so as to maximize the overlap between the Wannier functions (see Eq. \eqref{effectivehopping} below). \\

\begin{figure}[tb]
\begin{center}
\includegraphics[width=6in]{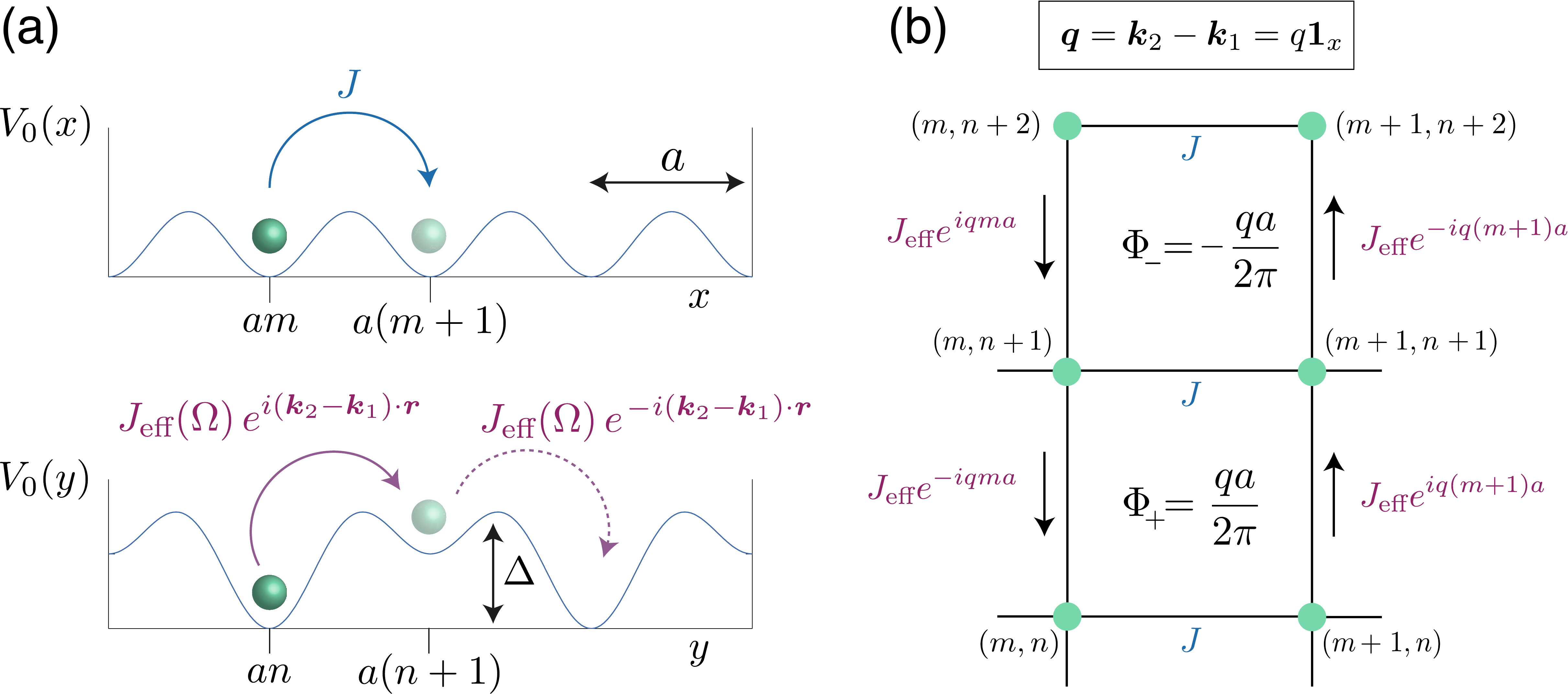}
\end{center}
\caption[Assisted tunneling method]{The laser-assisted tunneling method for creating Peierls phases in optical lattices. (a) The natural tunneling with amplitude $J$ is permitted along $\ex$, and is inhibited along $\ey$ by a superlattice potential (with offset $\Delta$). A pair of lasers, with wave vectors $\bs k_{1,2}$ and frequencies $\omega_{1,2}$, induce hopping along $\ey$ when $\omega_1 - \omega_2 \approx \Delta /\hbar$. The hopping matrix elements $J_{\text{eff}} \exp [i \phi (\bs r)]$ along $\ey$ realize the Peierls substitution. (b) When $\bs k_2 - \bs k_1 = q \bs{1}_x$, this generates synthetic magnetic fluxes $\Phi = \pm q a/ 2 \pi$ in the plaquettes (in units of the flux quantum).  In this configuration, the flux per plaquette $\Phi$ is staggered -- alternating sign along $\ey$.  A ``flux rectification" leading to a uniform magnetic flux over the lattice necessitates a slightly alternative setup (see the experiments in Refs. \cite{Aidelsburger:2013b,Ketterle:2013} and proposals in Refs. \cite{Jaksch:2003,Gerbier:2010,Dalibard2011}). 
}
\label{fig_assistedtunneling}
\end{figure}

\paragraph{Preventing natural hopping} The native hopping along the links for which $U_{jk} \ne 1$ must be eliminated.  For example, in the Landau gauge \eref{landaugauge}, this would be the natural hopping along $\ey$ of a square lattice.  This simple, but important, task allows the subsequent external induction and control of hopping.  In the Munich experiment of 2011 \cite{Aidelsburger:2011}, spontaneous hopping is prohibited using a superlattice along $\ey$, namely, by applying an energy offset $\Delta$ much larger than the natural hopping amplitude $J$ between alternating sites, see Fig. \ref{fig_assistedtunneling}a.  

Schemes involving state-dependent lattices are also envisaged \cite{Jaksch:2003,Gerbier:2010}, where atoms located in neighboring sites experience distinct optical potentials, see Fig. \ref{fig_assisted_shaking} (a).  State-dependent lattices are convenient for Ytterbium atoms \cite{Gerbier:2010}, and for bosonic alkali atoms.  However, for fermionic alkali atoms, they generally lead to large spontaneous rates, and thus alternative schemes are required \cite{Goldman:2010}. 

\paragraph{Induced-tunneling} The tunneling between neighboring sites $j \rightarrow k$ must be induced and controlled externally, leading to an effective hopping matrix element $J^{{\rm eff}}_{j\rightarrow k}$. In this Section, we consider the modulated-superlattice configuration used in the Munich experiment \cite{Aidelsburger:2011,Aidelsburger:2013}, which is similar to the schemes implemented in the subsequent Munich and MIT experiments \cite{Aidelsburger:2013b,Ketterle:2013, Aidelsburger:2014}. We refer the reader to Refs. \cite{Jaksch:2003,Gerbier:2010,Dalibard2011, Goldman:2013Haldane,Anisimovas:2014ga} for a description of the laser-induced tunneling methods using different internal states together with state-dependent lattices and inter-species coupling. In the superlattice illustrated in Fig. \ref{fig_assistedtunneling} (a), atoms in neighboring sites occupy Wannier functions $w_l (\bs r - \bs{r}_j)$ and $w_h (\bs r - \bs{r}_k)$, respectively (here the indices $l$ and $h$ refer to the low energy and high energy sites in the superlattice). For a given energy offset $\Delta$, tunneling can be induced by an external time-dependent perturbation 
\be
V_{{\rm coupl}}(\bs r,t)= \hbar \Omega \cos (\bs q \cdot \bs r - \omega_L t),\label{couplingraman}
\ee 
from a pair of lasers with wave vectors $\bs k_{1,2}$ and frequencies $\omega_{1,2}$.  The $\hbar \omega_L=\hbar (\omega_1 - \omega_2) \approx \Delta$ energy difference allows resonant coupling between the staggered lattice sites, see Fig. \ref{fig_assistedtunneling} (a). Here, $\Omega$ denotes the Rabi frequency characterizing the strength of the atom-light coupling. The key ingredient of the scheme is that the momentum transfer $\hbar \bs q =\hbar (\bs k_2 - \bs k_1)$ can be adjusted by tuning the angle between the laser beams. The resulting hopping amplitude, from a \emph{low} energy site $\bs r_j$ to a \emph{high} energy site $\bs r_k$, is then given by the integral \cite{Aidelsburger:2011} (see also \cite{Jaksch:2003,Gerbier:2010})
\begin{align}
J^{{\rm eff}}_{j\rightarrow k}&= \frac{\hbar \Omega}{2} \int  w^{*}_h (\bs r - \bs{r}_k) w_l (\bs r - \bs{r}_j) e^{i \bs q \cdot \bs r}     {\rm d}\bs r , \notag \\
& = J^{{\rm eff}}_{0} e^{i \bs q \cdot \bs r_j} = J^{{\rm eff}}_{0} \, U_{jk}, \label{effectivehopping}
\end{align}
where $\bs \delta_{kl}= \bs r_k - \bs r_j=\bs a$ is the vector connecting neighboring sites $j$ and $k$. The effective hopping is therefore characterized by the amplitude $J^{{\rm eff}}_{0}$, along with a complex Peierls phase factor determined by the momentum transfer $\hbar \bs q$, see Fig. \ref{fig_assistedtunneling} (a).  In other words, the time-dependent perturbation \eref{couplingraman} generates a link variable given by $U_{jk}=\exp \bigl [i \phi (\bs r_j) \bigr]$, where $\phi (\bs r_j)= (\bs k_2 - \bs k_1) \cdot \bs r_j$.  The effective hopping in Eq. \eref{effectivehopping} can also be written in the ``symmetric" notation
\be
J^{{\rm eff}}_{j\rightarrow k} (\bs r_j,\bs r_k)= \frac{\hbar \Omega}{2} e^{i\bs q \cdot (\bs r_j + \bs r_k)/2}  \int  w^{*}_h (\bs r - \bs \delta_{kl}/2) w_l (\bs r + \bs \delta_{kl}/2) e^{i \bs q \cdot \bs r} {\rm d}\bs r  =  \tilde{J}^{{\rm eff}}_{0} e^{i\bs q \cdot (\bs r_j + \bs r_k)/2},
\ee
which is appropriate for certain geometries, especially for the non-square lattices \cite{Alba:2011,Juzeliunas2013Physics,Goldman:2013,Goldman:2013Haldane}. 
Importantly, the effective hopping from a \emph{high} energy site $\bs r_k$ to a \emph{low} energy site $\bs r_j$
\begin{eqnarray}
J^{{\rm eff}}_{k\rightarrow j} (\bs r_k) &= \bigl (J^{{\rm eff}}_{j\rightarrow k} (\bs r_j) \bigr ) ^* \nonumber \\
&= J^{{\rm eff}}_{0}  e^{- i \bs q \cdot \bs r_j}= \tilde{J}^{{\rm eff}}_{0} e^{-i\bs q \cdot (\bs r_j + \bs r_k)/2},\label{oppositehopping}
\end{eqnarray}
results from a momentum transfer $-\hbar \bs q$ reversed in sign as compared to the {\it low}-to-{\it high} case\footnote{Here, we assumed that the overlap integrals $J^{{\rm eff}}_{0} $ and $\tilde{J}^{{\rm eff}}_{0}$ are real \cite{Jaksch:2003,Gerbier:2010,Aidelsburger:2011,Goldman:2013Haldane}.}. Finally, we point out that the effective tunneling amplitude can be expressed in terms of a Bessel function of the first kind $J^{{\rm eff}}_{0} = J \mathcal{J} (\kappa)$, with $\kappa \propto\Omega/\Delta$, as expected for shaken lattice systems (see Refs. \cite{Kolovsky:2011,Aidelsburger:2013,Aidelsburger:2013b,Ketterle:2013},  and also Section \ref{sec:shaking lattice} where off-resonnant driving is discussed in a related context).

\subsubsection{Flux configurations on the square lattice}

\paragraph{Staggered flux configuration}

Equations~\eref{effectivehopping}-\eref{oppositehopping} and Fig. \ref{fig_assistedtunneling} show that the induced tunneling is necessarily accompanied with alternating Peierls phases: $\phi_y (m,n)= - \phi_y (m,n+1)=\phi_y (m,n+2)=\dots$. Generating such Peierls phases on a square lattice, 
\begin{equation}
\phi_x=0 {\rm  , } \, \phi_y (m,n)= (-1)^{n} \, 2 \pi \Phi m,\label{landaugauge2}
\end{equation}
leads to a staggered flux configuration, where successive plaquettes  along $\ey$ are penetrated by fluxes $\pm \Phi$. Here, the synthetic magnetic flux $\Phi$ is governed by the Raman coupling lasers $\Phi \sim a \vert \bs q \vert=a \vert \bs k_2-\bs k_1 \vert$, and it can thus be set in the high-flux regime $\Phi \sim 0.1 - 1$, equivalent to the effects of a gigantic magnetic field $B \sim 10^4\ {\rm T}$ for electrons in a crystalline lattice. However, this staggered configuration does not break time-reversal symmetry, and is only equivalent to the Hofstadter model \eref{landaugauge} in the ``$\pi$-flux" limit $\Phi \rightarrow 1/2$. The experimental realization of the staggered flux model in 2011 \cite{Aidelsburger:2011}  constituted a first important step towards the realization of the Hofstadter model (i.e. uniform magnetic flux over the whole lattice). Staggered magnetic fluxes have also been realized in off-resonant shaken triangular optical lattices in 2013  \cite{Struck:2013}. 

\paragraph{The uniform flux configuration}

This alternating field must somehow be rectified to produce a uniform field.  This rectification requires individually addressing successive hoppings $J^{{\rm eff}}_{j\rightarrow k} (\bs r_j)$ and $J^{{\rm eff}}_{k\rightarrow j'} (\bs r_k)$, for example using more elaborate potential landscapes and additional laser frequencies \cite{Gerbier:2010,Mazza:2012,Goldman:2013Haldane,Aidelsburger:2013,Ketterle:2013}. In such landscapes, the energy offsets $\Delta_1$ and $\Delta_2$ between successive NN sites are different and the induced hoppings are produced independently.  This ``flux rectification", which is necessary to break time-reversal symmetry and produce quantum Hall states on the square lattice, has been realized in 2013 by the groups of I. Bloch \cite{Aidelsburger:2013} and W. Ketterle \cite{Ketterle:2013}, using a potential gradient to individually address adjacent tunneling processes  (as originally suggested in the Jaksch-Zoller proposal \cite{Jaksch:2003}). The uniform-flux configuration has also been realized with an all-optical setup, which has been used to extract the Chern number of Hofstadter bands \cite{Aidelsburger:2014}.

\subsubsection{Flux configurations on the honeycomb lattice}
\label{albasection}

Additional coupling lasers (see above) are unneeded for special lattice geometries, such as the honeycomb lattice. As was shown in Refs. \cite{Alba:2011}, \cite{Goldman:2013} and \cite{Anisimovas:2014ga}, it is possible to generate local Haldane-like fluxes using two intertwined triangular sublattices, denoted $A$ and $B$, coupled together by a single laser coupling $A \leftrightarrow B$. In this setup, the natural hopping between the NNN sites of the honeycomb lattice (i.e. the NN sites of the triangular sublattices $A$ and $B$) remains, while the laser-induced hopping with Peierls phases acts between NN sites. For arbitrary values of the coupling lasers wave vectors $\bs k_1 - \bs k_2 = \bs q$,  this configuration has non-trivial fluxes inside the subplaquettes of the unit hexagonal cells, and therefore reproduces the Haldane model's quantum Hall phases. A complete description of the phase diagrams and flux configurations stemming from this cold-atom setup can be found in Ref.~\onlinecite{Goldman:2013}. This scheme could also be easily implemented with Yb atoms, in which case it is convenient to set the lattice potential at a so-called ``anti-magic" wave-length \cite{Gerbier:2010}. In this case, the two internal states are automatically trapped in two intertwined triangular and honeycomb lattices, and the coupling can be directly addressed with a single-photon transition, see Anisimovas et al. \cite{Anisimovas:2014ga}. The Haldane model has been realized experimentally using time-modulated honeycomb optical lattices \cite{Jotzu}, as already mentioned in Section \ref{haldanesection}.\\

This honeycomb system could be also extended to the spinful Kane-Mele model for $Z_2$ topological insulators \cite{Kane:2005}. In this case, each sublattice must trap atoms in two internal atomic states (see Sect.~\ref{matrixlinkvariables}) coupled independently by lasers such that the tunneling operators ($2 \times 2$ matrices acting between NN sites $j$ and $k$) are $\hat U (j, k)= \exp (i \hat\sigma_z \delta \bs p \cdot (\bs r_{j} + \bs r_{k})/2)$. In this spinful honeycomb lattice configuration, non-trivial $Z_2$ topological phases, featuring helical edge states exist \cite{Goldman:2013}. This system can include time-reversal breaking perturbations, testing the robustness of the $Z_2$ topological phases and exploring phase transitions between helical and chiral edge textures \cite{Goldman:2012,Beugeling:2012}.

\subsubsection{Matrix link variables: non-Abelian gauge potentials}
\label{matrixlinkvariables}

The general atom-coupling method presented in Sect.~\ref{mainingredients} for creating Peierls phases $U_{jk}=\exp \bigl (i \phi (\bs r) \bigr)$ can be extended to produce matrix-valued link variables [e.g. $\hat U_{jk} \in {\rm U} (N)$].  For this, each lattice site must host $N$ nearly-degenerate spin states with index $\tau=1, \dots , N$. Using external (real) magnetic fields to lift any degeneracies, each state $\tau$ can then be individually trapped and coupled to other sublevel states $\tau \ne \tau'$. In particular, the effective tunneling of an atom in a specific internal state $\tau$, from site $j$ to $k$, could then be induced and controlled individually by external couplings \cite{Osterloh:2005,Goldman:2009,Goldman:2009prl,Goldman:2013Haldane}. In principle, the coupling can be chosen to flip the atomic spin $\tau \rightarrow \tau' \ne \tau$ during the hopping process, resulting in a non-diagonal hopping matrix $\hat J^{\tau \tau'}_{j \rightarrow k} \not\propto \hat{1}_{N \times N}$. Considering the simplest case $N=2$, this scheme can produce synthetic spin-orbit coupling terms with $\hat U_{jk} \sim \hat{\sigma}_{\mu}$. Such non-Abelian gauge potentials, based on spin-dependent hopping, lead to the spin-1/2 models presented in Sect.~\ref{spinorbitlattice}, making possible topological insulators in 2D and 3D optical lattices \cite{Goldman:2010, Bermudez:2010,Mazza:2012,Kennedy:2013,Aidelsburger:2013}. Non-trivial hopping operators along both spatial directions, using laser-induced tunneling techniques, generally require checkerboard lattices  \cite{Goldman:2013Haldane}.

\subsection{Other relevant schemes \label{sec:other schemes}}

While our list of ingredients for inducing non-trivial link operators $U_{jk}$ in optical lattices was based on laser-induced tunneling methods, alternative methods have been proposed and some of them have been experimentally realized.  This Section reviews these schemes.

\subsubsection{Shaking the lattice} \label{sec:shaking lattice}

A quite different strategy consists in using off-resonant
periodically-driven optical lattices \cite{Arimondo2012,Windpassinger2013RPP}.
One can generate artificial magnetic flux in such lattices by combining
lattices and time-dependent quadrupolar potentials \cite{Goldman:2014uz,Sorensen2005},
by modulating the lattice depth (i.e. the tunneling amplitude) in
a directional manner \cite{Kitagawa2010} or by shaking optical lattices
\cite{Eckardt:2005,Eckardt2010,Hemmerich2010,Kolovsky:2011,Struck2012,Arimondo2012}.
Similarly, spin-orbit couplings could be generated by subjecting an
optical lattice to time-dependent magnetic fields \cite{Goldman:2014uz,Xu2013,Anderson2013}.
A general formalism describing periodically-driven quantum systems
and effective gauge structures can be found in Refs. \cite{Kitagawa2010,Goldman:2014uz}.

Motivated by the recent experiments in Hamburg \cite{Struck2011,Struck2012,Struck:2013},
let us consider a scheme based on shaken optical lattices, as already 
outlined in Sect. \ref{sub:Shaking}. The method relies on off-resonant
modulations of the lattice potential $V_{\text{OL}}(\mathbf{r}^{\prime})$,
where $\mathbf{r}^{\prime}\equiv\mathbf{r}^{\prime}(t)=\bs{r}-\bs{r}_{0}(t)$ is the position vector in the
oscillating frame of reference [see Eq.~\eqref{eq:shifted_frame}]. The displacement vector $\bs{r}_{0}(t)=\bs{r}_{0}(t+T)$
is time-periodic, with period $T=2\pi/\omega$, and is given by Eq.~(\ref{eq:linear-circular})
for linear or circular harmonic shaking. As presented in Sect. \ref{sub:Shaking}, the transformation to
a non-inertial frame of reference modifies the Hamiltonian by
 adding a spatially homogeneous vector potential ${\mathbfcal A}(t)$ 
to the momentum, see Eq.~(\ref{eq:H-shaken-frame-1}):
\begin{equation}
H^{\prime}=\frac{\left({\bf p}-{\mathbfcal A} (t)\right)^{2}}{2m}+V_{\text{OL}}\,\left(\mathbf{r}\right)\,,\quad\mathrm{with}\quad{\mathbfcal A}=m\dot{\mathbf{r}}_{0}\left(t\right).\label{eq:H'-lattice}
\end{equation}
The full Hamiltonian $H^{\prime}$ shares the same spatial periodicity
as the lattice potential $V_{\text{OL}}\,\left(\mathbf{r}\right)$, 
hence, the quasimomenta are still good quantum numbers for the instantaneous
eigenstates and eigen-energies of the Hamiltonian $H^{\prime}$. Since
the Hamiltonian is also time-periodic, it is appropriate to deal with
the quasi-energies and the corresponding eigenstates of the  
effective (Floquet) Hamiltonian $\hat{H}_{\text{F}}$ [defined below in Eq.~(\ref{eq:U(T)})] when analyzing the topological
properties and effective gauge structures induced by the driving \cite{Kitagawa2010,Galitski2011NP,Goldman:2014uz}. 

Let us consider a situation where the potential $V_{\text{OL}}\,\left(\mathbf{r}\right)$
traps atoms in a deep 1D optical lattice, directed along $x$, which is 
subjected to a harmonic modulation $x_{0}\left(t\right)=\kappa\sin\left(\omega t\right)$.
Using the tight-binding approximation, the second-quantized lattice Hamiltonian reads 
\begin{equation}
\hat{H}(t)=-J\left(\hat{T}e^{i{\mathcal{A}}\left(t\right)  a/\hbar}+\hat{T}^{\dagger}e^{-i{\mathcal{A}}\left(t\right)  a/\hbar}\right),\quad\hat{T}=\sum_{j}\hat{a}_{j+1}^{\dagger}\hat{a}_{j},\label{ham_tb_1D}
\end{equation}
with ${\mathcal{A}}\left(t\right)=m\dot{x}_{0}\left(t\right)=m\kappa\omega\cos\left(\omega t\right)$,
where the operator $\hat{a}_{j}^{\dagger}$ creates a particle at
lattice site $x=ja$, $a$ is the lattice spacing and $J$ is the
tunneling matrix element between the neighboring sites. The time-dependent phase factors
$\exp\left(\pm i{\mathcal{A}}\left(t\right)  a/\hbar\right)$ accompanying the hopping terms
appear by simple application of the Peierls substitution \cite{Hofstadter:1976,Luttinger1951}. It is to be emphasized that these ``time-dependent Peierls phases" oscillate fast in time, and in particular, they should not be mistaken with the standard Peierls phases associated with (synthetic) magnetic flux discussed in Section \ref{hofsection}. 

The effective (Floquet) Hamiltonian $\hat{H}_{\text{F}}$ ruling the time-averaged
dynamics can be defined through the evolution operator over a cycle \cite{Kitagawa2010,Galitski2011NP,Goldman:2014uz}
\begin{equation}
\hat{U}(T)=\mathcal{T}\exp\left(-i\int_{0}^{T}\hat{H}(\tau)\text{d}\tau\right)=\exp\left(-iT\hat{H}_{\text{F}}\right),\label{eq:U(T)}
\end{equation}
where $\mathcal{T}$ denotes time-ordering. In the absence of additional confining trap (e.g. a harmonic trap $V_{\text{conf}}\sim x^2$), and imposing the periodic boundary conditions, one verifies that the 1D tight-binding Hamiltonian in Eq.~(\ref{ham_tb_1D}) commutes with itself at different times. In this case, the time-ordering can be omitted in Eq.~(\ref{eq:U(T)}), and the effective Hamiltonian $\hat{H}_{\text{F}}$ is then simply obtained through the time-average 
\begin{align}
\hat{H}_{\text{F}}&=\left(1/T\right)\int_{0}^{T}\hat{H}(\tau)\text{d}\tau , \label{first_order_average}\\
&=-J\mathcal{J}_{0}(\xi_{0})\left[\sum_{j}\hat{a}_{j+1}^{\dagger}\hat{a}_{j}+a_{j-1}^{\dagger}\hat{a}_{j}\right],\quad\xi_{0}=m\kappa a\omega/\hbar\,,
\label{ham_tb_1D_F}
\end{align}
where $\mathcal{J}_{0}(\xi_{0})$ denotes the Bessel function of the
first kind \cite{Eckardt:2005,Kolovsky:2011,Goldman:2014uz}. 

Thus, modulating an optical lattice enables one to control the sign and
amplitude of the tunneling between the neighboring sites: $J_{j\rightarrow k}^{{\rm eff}}=-J\mathcal{J}_{0}(\xi_{0})$.
This can have non-trivial consequences for triangular optical lattices,
where a change of sign $\mathcal{J}_{0}(\xi_{0})<0$ leads to staggered
synthetic magnetic fluxes and frustrated magnetism \cite{Struck2011}. The effects of additional potentials, which would require time-ordering in Eq.~(\ref{eq:U(T)}), can be evaluated by considering a perturbative treatment in $(1/\omega)$ \cite{Maricq:1982wf,Rahav:2003it,Goldman:2014uz,Eckardt-2014-unpubl}. In particular, we note that the cycle-averaged effective Hamiltonian in Eq. \eqref{first_order_average} corresponds to the lowest-order term of the Magnus expansion \cite{Maricq:1982wf}, which is generally relevant for sufficiently short periods $T$.

To induce complex effective Peierls phase factors, $J_{j\rightarrow k}^{{\rm eff}}\rightarrow J_{j\rightarrow k}^{{\rm eff}}e^{i\theta_{jk}}$,
non-sinusoidal driving is necessarily required, so that certain temporal
symmetries should be broken by the forcing \cite{Struck2012}. The
specific protocol implemented by Struck et al. \cite{Struck2012},
consists in a sinusoidal forcing over a period $T_{1}$ interrupted by a
short periods of rest $T_{2}$, so the lattice perturbation has a total
period of $\tau=T_{1}+T_{2}$ and a zero mean value. Under such conditions,
the effective tunneling operator is \cite{Struck2012} 
\begin{equation}
\frac{J_{j\rightarrow k}^{{\rm eff}}(\xi_{0},\omega)}{J}=\frac{T_{2}}{\tau}e^{iK(T_{1}/\tau)}+\mathcal{J}_{0}(\xi_{0})\frac{T_{1}}{\tau}e^{-iK(T_{2}/\tau)},
\end{equation}
leading to a constant but complex valued effective Peierls phase.
This scheme was realized experimentally in Hamburg, for bosons in
a 1D optical lattice, where the phase $\theta$ affected the lowest
Bloch band dispersion through $E(k)=-2\vert J_{j\rightarrow k}^{{\rm eff}}(\xi_{0},\omega)\vert\cos(ka-\theta)$.
In this setup, the effective Peierls phase $\theta$ can therefore
be evaluated by measuring the quasimomentum distribution of the BEC,
whose superfluid ground state is reached at a finite value dictated
by $\theta$ \cite{Struck2012}.

The extension of this method to generate synthetic (staggered)
magnetic fluxes in 2D optical lattices has been implemented for a
triangular optical lattice \cite{Struck2011,Struck:2013}. Similar
schemes can be envisaged to produce synthetic spin-orbit couplings, uniform magnetic fields and topological insulating states with shaken optical lattices \cite{Hauke:2012}. Recently, and in
direct analogy with solid-state-physics Floquet topological states
\cite{Oka2009,Kitagawa2010,Cayssol:2013gk,Grushin:2014gt,Eckardt-2014-unpubl}, it has
been suggested that (circularly) shaken hexagonal optical lattices could reproduce
the physics of graphene subjected to circularly polarized light {[}see
Sect. \ref{sub:Shaking} and Refs. \cite{Baur:2014ux,Zhai2014-Floquet}{]},
with a view to realizing the Haldane model \cite{Haldane:1988} and
topological bands with cold atoms. 

In general, the great versatility of periodically-driven systems can be exploited to generate a wide family of gauge structures in cold-matter systems. Let us illustrate this fact by considering a general static Hamiltonian $\hat H_0$ subjected to a single-harmonic modulation $\hat V(t)= \hat A \cos (\omega t) + \hat B \sin (\omega t) $, where $\hat A$ and $\hat B$ are arbitrary operators. Following Ref. \cite{Goldman:2014uz}, the effective Hamiltonian reads
\begin{align}
&\hat H_{\text{eff}}\!=\! \hat H_0 \!+\! \frac{i}{2   \omega} [\hat A, \hat B] \!+\! \frac{1}{4 \omega ^2}\! \left (  [[ \hat A, \hat H_0], \hat A] \!+\!  [[ \hat B, \hat H_0], \hat B] \right ) \!+\! \mathcal{O} (1/ \omega ^3). 
\label{Floquet_perturb}
\end{align}
Hence, convenient choices for the static Hamiltonian $\hat H_0$ and ``pulsed" operators $\hat A$ and $\hat B$ can generate a large variety of gauge structures, such as spin-orbit coupling [e.g. $[\hat A, \hat B] \sim p_x \sigma_x + p_y \sigma_y$] and orbital magnetism [e.g. $[\hat A, \hat B] \sim x p_y - y p_x$].\\

The main advantage of the shaking method is that it can be applied to
different lattice geometries and various atomic species.
In this sense, it is particularly interesting when considering the
physics of fermionic species, where state-dependent lattices and Raman
coupling (cf. Sect.~\ref{chip}) are generally associated with higher
spontaneous emission rates. However, periodically-driven systems might
also suffer from heating issues, which could naturally emanate from
the external forcing. For instance, inter-particle collisions due
to the micro-motion -- which is inherent to the fast modulations captured by the time-dependent Hamiltonian, e.g. Eq.(\ref{ham_tb_1D}) --  
or transitions to higher energy bands could potentially
lead to uncontrollable heating processes. Moreover, the manner by
which a driven quantum system absorbs energy strongly depends on its
characteristics (e.g. energy spectrum), which indicates that this
question should generally be addressed on a case-by-case basis. The
thermodynamics of periodically-driven systems is an important subject,
which has been recently studied in Refs. \cite{DAlessio:2013fv,Lazarides:2013uh,Vorberg:2013df,Langemeyer:2014ww}.

\subsubsection{Using RF fields \label{sec:RF fields}}

Recently, Jimenez-Garcia et al. proposed and implemented an original setup producing periodic potential for cold atoms~\cite{Jimenez-Garcia2012}, with naturally complex tunneling operators $J^{{\rm eff}}_{j\rightarrow k} e^{i \theta}$. In other words, this scheme produces the lattice and the Peierls phases $\theta$ \emph{simultaneously}. In this method, $^{87}$Rb atoms in the $f=1$ ground level are concurrently subjected to a radio-frequency (RF) magnetic field and to two counter-propagating Raman laser beams with wave number $k_{{\rm R}}$. The Raman and RF fields resonantly coupled the three spin states $\vert m_F=0, \pm 1\rangle$, yielding the coupling Hamiltonian, see Secs.~\ref{sec:TwoRaman-general} and \ref{sub:A-pair-of}, 
\be
\hat H_{{\rm RF + R}} (\bs r) = \bs{\Omega} (\bs r) \cdot \hat{\bs{F}} + {\rm constant}, \label{rfandraman}
\ee
where $\hat{\bs{F}}$ is the angular momentum operator and 
$$
\bs{\Omega} (\bs r)\equiv\bs{\Omega} (x)= \frac{1}{\sqrt{2}}(\Omega_{{\rm RF}} + \Omega_{{\rm R}} \cos ( 2 k_{{\rm R}} x) , - \Omega_{{\rm R}} \sin ( 2 k_{{\rm R}} x) , \sqrt{2} \delta).
$$
Here $\Omega_{{\rm RF}}$ and $\Omega_{{\rm R}}$ are the coupling strength, or Rabi frequencies, associated with the RF and Raman fields, and $\delta$ is the detuning from Raman resonance \cite{Jimenez-Garcia2012}. The two non-trivial eigenvalues obtained of \eref{rfandraman} are
\be
\Omega (x) = \pm \sqrt{\Omega_{{\rm RF}} \Omega_{{\rm R}} \cos ( 2 k_{{\rm R}} x) + {\rm constant}}, \label{rflattice}
\ee
producing a periodic potential along $\ex$ when $\Omega_{{\rm RF}}, \Omega_{{\rm R}}  \ne 0$. Importantly, Eq. \eref{rfandraman} describes a spin-1 particle subjected to an effective space-dependent magnetic field $\bs B_{{\rm eff}} (\bs r)= \hbar  \bs{\Omega} (\bs r)/g_F \mu_B$, where $g_F$ and $\mu_B$ are the Land\'e factor and Bohr magneton, respectively. Now, according to Berry \cite{Berry:1984}, an eigenstate of Hamiltonian \eref{rfandraman} acquires a geometric (Berry) phase $\theta$ when a loop $\gamma$ is performed in the parameter space spanned by the ``magnetic" field $\bs B_{{\rm eff}} (\bs r)$. Here, $\bs B_{{\rm eff}} (\bs r)$ is space-dependent, and therefore, the spin will precess and make a loop each time an atom hops from one lattice site to its neighboring sites (the lattice structure being dictated by Eq. \eref{rflattice}). The hopping process is therefore naturally accompanied with a non-trivial ``Peierls" phase $\theta$, proportional to the solid angle subtended $B_{{\rm eff}} (\bs r)$. This strategy relies on the fine tuning of both the Raman and RF fields, and simultaneously produces a periodic structure intrinsically featuring Peierls phases, as demonstrated experimentally at NIST \cite{Jimenez-Garcia2012}.

Extensions of this setup could realize non-trivial magnetic flux configurations in 2D lattices, through space-dependent Peierls phases.  Again, this requires more complicated but realistic arrangements, for example, combining a state-independent optical lattice along one direction with a state-dependent ``vector" lattice along the other \cite{Jimenez-Garcia2012}. Alternatively, one could make use of the optical flux lattices \cite{Cooper2011a,Cooper2011,Juz-Spielm2012NJP,CooperDalibard:2012,CooperMoessner:2012} considered in Sec.~\ref{sub:Flux-latt}. \\

\subsubsection{On a chip}
\label{chip}

Most proposals based on laser-coupling methods for the alkali atoms, involve two-photon Raman transitions coupling different internal atomic states. In this context, a crucial experimental issue concerns the minimization of undesired heating stemming from spontaneous emission, which cannot be reduced even using large detuning of the Raman lasers, as pointed out at the end of the Sec. \ref{sec:TwoRaman-general}.  
 This drawback is particularly severe for alkali fermions, e.g. $^{40}$K and $^{6}$ Li, where the  possible detunings are so small that they necessarily imply large spontaneous emission rates.
Therefore alternative methods are required to investigate the physics of fermionic atoms subjected to synthetic gauge potentials. Several solutions have already been invoked above, such as exploiting the coupling to long-lived electronically excited states (e.g. using Ytterbium) or engineering the gauge potential through lattice-shaking methods. 

Another solution consists in trapping and coupling different hyperfine levels using the magnetic (Zeeman) potentials produced by an array of current-carrying wires, therefore avoiding the use of optical state-dependent lattices and Raman coupling \cite{Goldman:2010}. Here we describe how such a setup could be implemented to reproduce the Abelian gauge structure \eref{landaugauge} of the Hofstadter model (a generalization of this atom-chip method has been described in Ref. \onlinecite{Goldman:2010} to produce the SU(2) Hamiltonian \eref{goldham}). First, consider two atomic states $\ket{g} = \ket{f=1/2,m_F=1/2}$, $\ket{e} = \ket{3/2,1/2}$ of $^{6}$Li, trapped in a primary (state-independent) optical lattice potential $V_1(x)=V_x\sin^2(k x)$ along $\ex$. Importantly, the corresponding pair of laser beams are incident on an atom chip's reflective surface, trapping the atoms about 5 $\mu$m above the device. The atom chip is consists of an array of conducting wires, with alternating currents $\pm I$ \cite{Folman:2008}, which traps the atoms in a deep 1D Zeeman lattice along $\ey$. Since the chosen states $\vert e,g \rangle$ have opposite magnetic moments, the Zeeman shifts produce a state-dependent lattice along this direction. Then, additional \emph{moving} Zeeman lattices \cite{Folman:2008}, with space-dependent currents $I_m = I \sin (2 \pi \Phi m - \omega t)$, act as a time-dependent (RF) perturbation directly coupling the internal states (in direct analogy with \eref{couplingraman}), and thereby producing the desired space-dependent Peierls $\phi_y (m)= 2 \pi \Phi m$. Such an elegant and versatile method, suitable both for bosonic and fermionic species, is currently in development at NIST, Gaithersburg.

\subsubsection{Immersion into a rotating BEC}

Jaksch and Klein proposed immersing the optical lattice and the atoms of interest ($A$-atoms) into a BEC formed by $B$-atoms \cite{Klein:2009}. Then, the interaction between the $A$ and $B$ atoms result in phonon excitations, which in turn induce effective interactions between $A$-atoms located at NN sites \cite{Klein:2009}. When $B$-BEC is made to rotate, the resulting effective interactions imprint space-dependent Peierls phases to the tunneling matrix elements of the $A$-atoms. Such a system could potentially lead to the realization of a synthetic magnetic field in the dynamics of the $A$-system.

\subsubsection{Quasi-2D gauge structures using 1D optical lattices}
\label{quasi2D}

Finally, interesting properties of 2D systems can be captured by 1D systems, through the concept of dimensional reduction \cite{Qi2008}. This fact, which is particularly relevant for topological systems, offers an interesting route for the exploration of topological order using 1D optical lattices \cite{Lang:2012,Mei:2012}. Also, this strategy suggests an alternative way to reproduce the effects of magnetic fields in a simple optical-lattice environment.

Let us illustrate this concept for 2D topological systems, namely, systems exhibiting robust edge states at their 1D boundaries with energies $E_{\text{edge}}$ located within bulk gaps \cite{Qi2011}. First, suppose that the system is defined on a 2D square lattice and described by a general Hamiltonian
\be
\hat H = \sum_{j,k} \hat c^{\dagger}_{j} \,    h_{j k} \,  \hat c_{k}, \label{eq:full2D}
\ee
where $c^{\dagger}_{j}$ creates a particle at lattice site $\bs r_{j}=(m,n)$ and $m,n=1, \dots , L$. In this open 2D geometry, topological edge states are localized along the single 1D boundary delimiting the large square $L \times L$ \cite{Qi2008}. Now, consider the cylindrical geometry obtained by identifying the opposite edges at $n=1$ and $n=L$, namely, by imposing periodic boundary conditions along the $y$ direction only. In this geometry, and considering Bloch's theorem, the system is well described by the spatial coordinate $m=1, \dots, L$ and the quasi-momentum $k_y= (2 \pi/L) l$, where $l=1, \dots, L$. In particular, the system Hamiltonian \eqref{eq:full2D} can be decomposed as a sum 
\be
\hat H = \sum_{m=1}^L \sum_{k_y} \hat H (m; k_y) = \sum_{k_y} \hat H_{\text{1D}} (k_y),\label{eq:full1D}
\ee 
which indicates that the 2D system can be partitioned into independent 1D chains \cite{Qi2008}. The edge states are now located at the two opposite edges of the cylinder, defined at $m=1$ and $m=L$, and their dispersion relations are expressed as $E_{\text{edge}}=E_{\text{edge}}(k_y)$. The general dimensional-reduction strategy can be formulated as follows: the energy spectrum and edge-state structures emanating from the 2D Hamiltonian \eqref{eq:full2D}-\eqref{eq:full1D} could be captured by a family of 1D models with Hamiltonians $\hat H_{\text{1D}} (\theta)$, where the controllable parameter $\theta \in [0 ,  2 \pi]$ should be identified with the quasi-momentum $k_y$. In other words, engineering a 1D system with tunable Hamiltonian $\hat H_{\text{1D}} (\theta)$ would provide useful informations related to the full 2D system of interest \eqref{eq:full2D}.

To be specific, let us consider the Hofstadter model introduced in Sect.~\ref{hofsection} and described by the Hamiltonian \eqref{hofstadter} - \eqref{landaugauge}. Using the Landau gauge, the single-particle wave function can be written as $\psi (m,n)= \exp (i k_y n) u (m)$, where the function $u (m)$ satisfies a 1D Schr\"odinger equation and where $k_y$ is the quasi-momentum along $\ey$. The Schr\"odinger equation associated with this ``Hofstadter cylinder" takes the form of the Harper-Aubry-Andr\'e equation \cite{Hofstadter:1976,Aubry:1980,Hatsugai:1993}
\begin{equation}
E \psi (m)= - J \bigl [ \psi (m+1) + \psi (m-1) + 2 \cos (2 \pi \Phi m - k_y) \psi (m)  \bigr ],\label{aubry}
\end{equation}
where $m=1, \dots, L$. Solving this equation yields the projected bulk bands of the Hofstadter model $E (k_x,k_y) \rightarrow E(k_y)$, which display $q-1$ bulk gaps for $\Phi=p/q \in \mathbb{Q}$. Moreover, since the cylinder is a partially opened geometry (with edges), new states with energies $E_{\text{edge}}(k_y)$ are located within the bulk gaps. These states are topological edge states, which play an important role in the quantum Hall effect \cite{Hatsugai:1993}. A typical spectrum, showing the bulk and edge states dispersions, is shown in Fig. \ref{fig_edgestates}. Originally, the Aubry-Andr\'e model \eqref{aubry} was studied as a simple model for Anderson localization \cite{Aubry:1980}; in this 1D model, $k_y$ is an adjustable parameter (not related to quasi-momentum), and it is therefore treated on the same level as the parameter $\Phi$. Recently, this model has been realized with cold atoms \cite{Roati:2008}, in a quasi-periodic 1D lattice created by interfering two optical lattices with incommensurate wave numbers $k_{1,2}$. In this context, the parameters in Eq. \eqref{aubry}, $\Phi=k_2/k_1$ and $k_y$, are tuned by the lasers creating the two lattices \cite{Roati:2008}. This reproduces the dimensional reduction of the Hofstadter optical lattice, where the synthetic magnetic flux $\Phi$ can be easily adjusted and where the ``quasi-momentum" $k_y$ is fixed by the laser phases. Considering fermionic atoms in such a 1D setup, and setting the Fermi energy inside a bulk gap, it is possible to populate topological edge states for a certain range of the tunable phase $k_y$ \cite{Lang:2012,Kraus:2012prl}. 

Extending this scheme to two-component 1D optical lattices, and considering well-designed state-dependent bichromatic optical lattices \cite{Mei:2012}, Mei et al. obtained the dimensional reduction of the $Z_2$ topological insulator in Eq. \ref{goldham}. By sweeping the tunable phase $k_y$, one is then able to transfer topological (helical \cite{Kane:2005}) edge states with opposite spin, from one edge to the other \cite{Mei:2012}. The interacting bosonic version of the 1D bichromatic optical lattice described above has been explored by Deng and Santos \cite{Deng:2013}, where the topological phase diagram has been obtained in terms of the interaction strength, the atomic filling factor and the strength of the auxiliary lattice. 

The dimensional reduction strategy therefore allows to access gauge structures and topologically ordered phases in a rather simple manner. These methods could be extended to access unobserved phenomena, such as the 4D quantum Hall effect \cite{Kraus:2013}.\\

 Finally, 2D atomic lattice systems could be realized through the concept of \emph{synthetic dimensions} \cite{Boada:2012}, where the dimensionality of the physical optical lattice $D_{\text{phys}}$ is augmented by a synthetic dimension spanned by the internal states of the atoms. For instance, a 2D (semi-)synthetic lattice could be obtained from a $D_{\text{phys}}=1$ optical lattice with $L$ lattice sites (with spatial coordinates $x=ma$, where $a$ is the physical lattice spacing and $m=1, 2 , \dots , L $), and filled with a $N$-component atomic gas: in such a configuration, the 2D synthetic lattice is characterized by $L \times N$ lattice sites located at the coordinates $\bs r_{\text{site}}= (m,n)$, where $m=1, \dots , L$ and $n=1, \dots , N$. In synthetic lattices, the hopping is natural along the physical direction $x$, while it is assisted along the synthetic (spin) direction through atom-light coupling (e.g. Raman transitions). The realization of synthetic magnetic fluxes in synthetic lattices has been described by Celi et al. in Ref. \cite{Celi:2013}, where it was shown that such setups could be exploited to observe the Hofstadter butterfly spectrum and the evolution of chiral (topological) edge states, using atoms with $N>2$ internal states. The propagation of chiral edge states in ladder systems ($N=2$) penetrated by synthetic magnetic fields has been studied by H{\"u}gel and Paredes \cite{Hugel:2013}. An experimental implementation of of such a ladder scheme (without involvement of the synthetic dimension) has been recently reported by Atala et al. \cite{Atala:2014uc}.


\section{Probing the effects of synthetic gauge potentials: a quantum simulation perspective}
\label{sect:simulation}


\label{sect:simulation}

\label{quantumsimulation} 
The addition of synthetic gauge potentials can modify the properties of the atomic system significantly. By appropriately tuning this gauge field, single- and many-body configurations can be obtained, leading to a plethora of interesting quantum phases.  Numerous cold-atom systems can realize quantum Hall physics, where Landau-like levels emerge from a synthetic magnetic field, e.g. through rotation \cite{Cooper2008} or laser-induced methods \cite{Dalibard2011}. Similarly, optical-lattice setups subjected to synthetic magnetic fluxes lead to band structures characterized by non-zero Chern numbers or Wilson loops, offering an alternative route to reach quantum Hall (Chern insulating) phases. Cold-atom simulators of quantum Hall states are motivated by the fact that these controllable systems could offer an instructive insight on the physics of QH liquids, such as those featuring non-Abelian excitations \cite{Cooper2008,Burrello:2010,JuliaDiaz:2012njp,CooperDalibard:2012}. We note that QH photonics systems offer an alternative route towards this goal \cite{Carusotto:2013gh,Hafezi:2011dt,Rechtsman:2013fe,Ozawa:2013tt,Hafezi:2013cm,Umucallar:2012bo,Umucallar:2013df,Haldane:2008cc}.

The full set of non-interacting topological phases of matter have now been classified \cite{Hasan2010,Qi2011}. This ``Periodic Table" includes the $Z_2$ topological insulators, which have been experimentally observed in 2D and 3D materials \cite{Konig2007,Hsieh2008,Chen:2009}, but also includes topological insulating and superconducting phases that are not known to exist in materials \cite{Hasan2010,Qi2011}.  Cold-atom setups subjected to gauge potentials might be tailored to access these phases, including those exhibiting Majorana zero modes \cite{Sau:2011,Liu:2012} (see also the proposals to realize topological Kitaev-like chains using atomic quantum wires \cite{Jiang:2011,Kraus:2012,KrausDalmonte:2013,Nascimbene:2012}). 

Gauge structures have been experimentally realized in several laboratories (see previous Sections), and a first observation of the Zak (geometric) phase \cite{Xiao2010} has been reported for a 1D optical lattice reproducing the Rice-Mele model \cite{Atala:2012}. However, going beyond this first step , the detection of quantized topological invariants in atomic setups remains an important issue. 

In this Section, we review methods for identifying topological matter using probes that are available in existing cold-atom laboratories.  While we focus on detecting topological order, synthetic gauge potentials can also lead to other interesting effects, such as pseudo-relativistic band structures [see the review \cite{Zhang2012FP} and the laboratory measurement of Zitterbewegung~\cite{LeBlanc2013}] and vortex physics (see the review \cite{Cooper2008} and also Refs. \onlinecite{Bhat:2006vortex,Goldman:2007vortex,Lim2008,Goldbaum:2009,Zhai:2010}). Finally, the quantum simulation of quantum field theories, such as encountered in high-energy physics, is discussed in Sect. \ref{sect:dynamical}.
 
\subsection{Probing quantum Hall physics in synthetic magnetic fields: From atomic Landau levels to strongly-correlated states} 
 
In 2D electronics systems, the Lorentz force from a perpendicular magnetic field deflects electrons perpendicular to a driving electric field, resulting in a non-zero transverse Hall conductivity \cite{Ashcroft}. Thus the associated transverse current is a natural signature of synthetic magnetic fields in cold-atom systems.  However, transport measurements are relatively complicated to perform in cold-atom experiments, as it requires the challenge of first connecting reservoirs to the system and then detect currents \cite{Brantut2012}. Recently, an analogue of a transport experiment in a flattened BEC subjected to a synthetic magnetic field was performed in the AC limit -- analogous to capacitively contacting an electronic system.  In this scheme \cite{LeBlanc2012}, currents are induced by modulating the external confining potential, allowing the full reconstruction of the resistivity tensor (including the Hall response).

The quantum mechanical problem of a particle in a magnetic field has the well-known Landau levels as its eigen-energies, which lie at the root of the quantum Hall effect \cite{vonKlitzing:1986}.  For non-interacting fermions in a synthetic magnetic field, the ``LLL-regime" (where only the lowest Landau level (LLL) is occupied) could be detected in time-of-flight experiments, by simply observing that the LLL states expand much faster than the states in higher Landau levels \cite{Ohberg:2005}. In the weakly-interacting regime of cold bosonic gases in a synthetic magnetic field, the mean field ground state is characterized by a macroscopic occupation of the LLL, giving rise to a triangular vortex lattice in the particle density (see Sect.~\ref{ian:vortex} and \cite{Cooper2008}). As the number of vortices increases, the atomic analogue of the QH filling factor decreases and the system enters the strongly-correlated regime where a family of FQH-like liquids have been predicted \cite{Cooper2008,Hafezi2007,Palmer:2008,JuliaDiaz:2012njp,JuliaDiaz:2011pra,CooperDalibard:2012}.  A thoughtful description of the atomic Landau problem is given in Ref.~\cite{Cooper2008}, where observable signatures are discussed both for the weakly and strongly interacting regimes. The stabilization of FQH atomic states in realistic conditions, e.g. by significantly increasing the incompressibility gap with respect to typical experimental temperatures, still constitutes a fundamental issue \cite{Hafezi2007,Cooper2008,JuliaDiaz:2013,Roncaglia:2011,CooperDalibard:2012}. Once realized in laboratories, the FQH liquids could be distinguished in experiments owing to their incompressible nature, manifested as plateaus in the atomic density distribution \cite{Cooper:2005}. Additional signatures of these strongly-correlated states are present in transport measurements with fractional transverse (Hall-like) conductivities \cite{Bhat:2007,Palmer:2008}; through density--density correlation functions \cite{Read:2003}; or by the response of the atomic cloud to quasihole excitations induced by an external laser beam \cite{Gra:2012}. Finally, the topological order associated with these FQH states could be evaluated by directly detecting the topological edge states (see Ref. \onlinecite{Wen:1995} and Sect.~\ref{detectedge} below).

\subsection{Identifying topological order} 

Diverse theoretical proposals to simulate topological phases with cold atoms exist. These proposals describe techniques for creating: Chern insulators \cite{Jaksch:2003,Goldman:2007,Shao:2008,Stanescu:2009,Li:2009,Liu:2010,Stanescu2010,Alba:2011,Cooper2011a,Cooper2011,CooperDalibard:2012,CooperMoessner:2012,Goldman:2013}; $Z_2$ topological insulators in 2D \cite{Goldman:2010,Beri:2011,Mazza:2012,Mei:2012} and 3D lattices \cite{Bermudez:2010}; (interaction-induced) topological Mott insulators \cite{Dauphin:2012,Sun:2012}; $Z$ (class AIII) topological insulators in 1D \cite{XJLiu:2012}; and topological superconductors \cite{Kraus:2012,KrausDalmonte:2013,Nascimbene:2012,Liu:2012,Liu:2013}. Recently, the 1D Rice-Mele model was realized in a 1D optical lattice \cite{Atala:2012}, where its non-trivial topological phase has been identified through the experimental determination of the Zak phase. 

In general, detecting topological properties is subtle.  Topological phases are all specified by two ``holographic" characteristics \cite{Hasan2010,Qi2011}: (1) in a topologically-ordered bulk, there exists a non-zero topological invariant associated with the bulk states \cite{Kohmoto:1985}, and (2) topological edge states are spatially localized on the periphery of the system with energies within the bulk gaps \cite{Hatsugai:1993}. Both properties are robust and simultaneously survive as long as the bulk gaps remain open and temperatures are small compared to the gaps. In the following, we consider non-interacting fermionic systems with the Fermi energy tuned inside such a topological gap. 

Before reviewing proposals for measuring topological properties in atomic systems, let us briefly summarize the state-of-the-art in solid-state systems.  Topological invariants have been revealed through transport measurements, e.g., by measuring the Hall conductivity in the integer QH regime \cite{Thouless1982,Kohmoto:1985}, or in the quantum spin Hall (QSH) regime of topological insulators \cite{Hasan2010,Qi2011}. Besides, edge states have been identified through spectroscopy \cite{Hasan2010,Qi2011} and interference methods \cite{Ji:2003,Karmakar:2011}. Very recently, QSH helical edge states have been directly imaged using a cryogenic microwave impedance microscope \cite{Kundhikanjana:2012}. 

Performing transport measurements in atomic systems is not as straightforward as in condensed-matter setups. However, cold-atoms setups offer complimentary techniques that can be exploited to measure topological invariants and identify topological edge states. In the following sections, we sketch several methods allowing to identify the Chern number $\nu$, which classifies Chern (QH) insulating phases  \cite{Thouless1982,Kohmoto:1985}. Generalizations of these methods to detect other topological classes are also discussed.

\subsubsection{Atomic Chern insulators: measuring the Chern number and topological edge states}

\paragraph{Density plateaus in the Hofstadter optical lattice, the Streda formula and signatures in time-of-flight experiments}

The Hofstadter butterfly is the spectrum of a particle on a 2D lattice subjected to a uniform magnetic field \cite{Hofstadter:1976}. When the magnetic flux per plaquette is rational, $\Phi=p/q \in \mathbb{Q}$, this spectrum splits into $q$ sub-bands $E_{\lambda} (\bs k)$, where $\lambda=1, \dots , q$ and where $\bs k=(k_x,k_y)$ is the quasi-momentum, see Fig. \ref{fig_hofstadter}b. Each bulk band $E_{\lambda} (\bs k)$ is associated with a Chern number $N_{\lambda}$, a topological index which remains constant under external perturbations as long as the bulk gaps do not close \cite{Kohmoto:1985}. The Chern number $\nu_{\lambda}$ is an integer given by the integral
\begin{eqnarray}
\nu_{\lambda} &= \frac{1}{2 \pi} \int_{BZ} \mathcal{F}_{\lambda} (\bs k) d \bs k \nonumber \\
&= \frac{i}{2 \pi} \int_{\mathbb{T}^2} \langle \partial_{k_x} u_{\lambda} (\bs k) \vert \partial_{k_y} u_{\lambda} (\bs k) \rangle - (\partial_{k_x} \leftrightarrow \partial_{k_y}) d^2 \bs{k},\label{chernnumber}
\end{eqnarray}
over the first Brillouin zone (BZ) of the Berry's curvature $\mathcal{F}_{\lambda}$ associated with the band $E_{\lambda} (\bs k)$.  Here, $\vert u_{\lambda} (\bs k) \rangle$ is the single-particle state in the $E_{\lambda}$ band with crystal momentum ${\bf k}$. This topological order strikingly manifest itself when the Fermi energy $E_F$ resides in a bulk energy gap. The transverse (Hall) conductivity is then quantized as
\begin{equation}
\sigma_H= \sigma_0 \,  \nu= \sigma_0 \sum_{\lambda < E_F} \nu_{\lambda},\label{Halleq}
\end{equation}
where the sum includes the contribution of all occupied bulk bands, and $\sigma_0$ is the conductivity quantum, see Fig. \ref{fig_hofstadter}b. Interestingly, this quantized quantity is related to the particle density $n (\bs{x})$, making its detection with cold atoms particularly practical \cite{Umucallar:2008}. The connection to density is based on the Streda formula \cite{Streda:1982,Kohmoto:1989}, which re-expresses the quantized Hall conductivity as the derivative $\sigma_H \propto \partial N / \partial B$, where $N$ is the number of states lying below the Fermi energy and $B$ is the  magnetic field. For optical lattice experiments with a smooth confining potential $V_{c} (r)$, the spatial density profile $n (r)$ in the local-density approximation is \cite{Gerbier:2010}
\begin{equation}
n(r)= \int dE \, D (E) \, \Theta [ E_F - V_{c} (r) - E],\label{LDA:chern}
\end{equation} 
where $D(E)$ is the homogeneous-system density of states. Thus, the density $n (r)$ counts the number of states below the ``local chemical potential" $\mu (r)= E_F - V_{c} (r)$.  In the presence of a uniform synthetic magnetic flux $\Phi \approx p/q$, the bulk energy spectrum $E (\bs k)$ associated with the homogeneous system is split into $q$ bulk bands. According to Eq. \eqref{LDA:chern}, this splitting will produce  $q-1$ plateaus in the density profile $n (r)$ \cite{Gerbier:2010}. One can thus associate each density plateau with one of the $q-1$ bulk gaps characterizing the homogeneous-system spectrum $E (\bs k)$. Then, by comparing the density plateaus $n_{1,2}$ obtained from two different configurations of the magnetic flux $\Phi_{1,2}$ but corresponding to the opening of the same bulk gap in the bulk spectrum $E (\bs k)$, one obtains the analogue of the Streda formula for the Hofstadter optical lattice \cite{Umucallar:2008}. Identifying the plateaus $n_{1,2}$ corresponding  to the same $r$th bulk gap, one obtains the integer
\begin{eqnarray}
\nu =\frac{\Delta n}{\Delta \Phi} = \frac{n_2 - n_1}{\Phi_2 - \Phi_1} = \sum_{\lambda=1}^{r} \nu_{\lambda},
\end{eqnarray}
analogous to the quantized Hall conductivity of an electronic system with the Fermi energy set within the $r$th gap \cite{Umucallar:2008}. This method allows access to the sum of Chern numbers $\sum_{\lambda=1}^{r} \nu_{\lambda}$, by comparing two measurements of atom density at different values of synthetic magnetic flux, and offers a simple method to identify topological order \cite{Umucallar:2008} and phase transitions  \cite{Bermudez:2010NJP}.\\

Moreover, it was shown in Ref. \onlinecite{Zhao:2011} that the topological quantity $\nu$ could also be revealed in the momentum density $\rho (\bs k)$ of the same Hofstadter optical-lattice. Indeed, under specific conditions, i.e. in the limit of large hopping anisotropy or for small synthetic flux per plaquette $\Phi$, the images obtained from time-of-flight experiments should display oscillations whose periodicity can be related to the value of $\nu$ \cite{Zhao:2011}. 

\paragraph{Measuring the winding number of a Haldane-Chern insulator}

The Haldane model \cite{Haldane:1988} and its cold-atom generalizations \cite{Shao:2008,Stanescu:2009,Li:2009,Stanescu2010,Liu:2010,Alba:2011,Juzeliunas2013Physics,Goldman:2013} are all described by two-band Hamiltonians of the form
\begin{equation}
H (\bs k) = \epsilon (\bs k) \hat{1} + \bs d (\bs k) \cdot \bs \hat\sigma,
\end{equation}
where $\hat{\bs{\sigma}}=\hat{\sigma}_{x,y,z}$ is the vector of Pauli matrices. The Berry's curvature $\mathcal{F}$ associated with the lowest energy band $E_{-} (\bs k)$, and the related Chern number $\nu$ in Eq. \ref{chernnumber}, can be expressed in terms of the normalized vector $\bs n (\bs k) = \bs d (\bs k) / \vert \bs d (\bs k) \vert$ \cite{Qi2008}, 
\begin{eqnarray}
&\mathcal{F} (\bs k) = \frac{1}{2} \bs n \cdot (\partial_{k_x} \bs n \times \partial_{k_y} \bs n)  \\
&\nu = \frac{1}{4 \pi} \int_{\mathbb{T}^2} \bs n \cdot (\partial_{k_x} \bs n \times \partial_{k_y} \bs n) d \bs k.
\end{eqnarray}
From this, we notice that the Chern number $\nu$ measures the number of times the vector field $\bs n (\bs k)$ covers the unit sphere as $\bs k$ is varied over the entire Brillouin zone $\mathbb{T}^2$. Therefore, for a system prepared in a phase $\nu \ne 0$, an experimental measurement of $\bs n (\bs k)$ would depict a Skyrmion pattern on a ``pixelated" Brillouin zone, leading to an approximate measure of the Chern number \cite{Alba:2011,Juzeliunas2013Physics,Goldman:2013}. For the specific Haldane-like model introduced by Alba et al. and discussed in Sect.~\ref{albasection}, the vector field $\bs n (\bs k)$ can be reconstructed from spin-resolved momentum densities $\rho_{A,B} (\bs k)$ associated with the two atomic species present in the lattice \cite{Alba:2011,Juzeliunas2013Physics,Goldman:2013}. This model therefore offers a simple platform to measure the Skyrmion patterns and topological index $\nu$ from spin-resolved time-of-flight images.

\paragraph{Semiclassical dynamics, the Berry's curvature and the Chern number}

The equations of motion of a wave packet evolving on a lattice, centered at position $\bs r$ with crystal-momentum $\bs k$, and driven by an external force $\bs F$, are given by \cite{Xiao2010}
\begin{eqnarray}
\dot{\bs r}= \frac{1}{\hbar} \frac{\partial E (\bs k)}{\partial \bs k} - \left[\dot{\bs k} \times  \bs{1}_z \right] \mathcal{F}  (\bs k)  \nonumber \\
\hbar \dot{\bs k}= \bs F , \label{blochoscil}
\end{eqnarray}
where $\mathcal{F} (\bs k)$ is the Berry's curvature introduced in Eq. \eqref{chernnumber}, and where $ E (\bs k)$ is the band structure characterizing the lattice system. In Ref. \cite{Price:2012}, Price and Cooper showed that cold atoms undergoing Bloch oscillations \cite{BenDahan:1996} can follow trajectories \eqref{blochoscil} where the band structure's contribution to the velocity $\propto \partial E (\bs k)/\partial \bs k$ vanishes. Following this protocol, a measure of the mean velocity for many trajectories gives the Berry's curvature $\mathcal{F} (\bs k)$ over a ``pixelated" Brillouin zone \cite{Price:2012}. By properly adjusting the path undergone by the wave packet, the Chern number  \eqref{chernnumber} can be evaluated. In principle, this method could be applied to any lattice system. \\

A similar scheme was recently proposed by Abanin et al. \cite{Abanin:2012}, where Bloch oscillations are combined with interferometry techniques to determine the Berry's curvature and the Chern number of 2D optical lattice systems. This method is based on the measure of the Zak phase \cite{Xiao2010}, which has already been experimentally implemented in a 1D system \cite{Atala:2012}. In Ref. \cite{Liu:2013order}, Liu et al. introduced a method to measure the Chern number based on spin-resolved Bloch oscillations, observing that the topological index can be obtained by measuring the spin-polarization of the atomic gas at specific (highly symmetric) points within the Brillouin zone.\\

Finally, it was shown by Dauphin and Goldman that the Chern number (or equivalently the Hall conductivity) could be measured by imaging the center-of-mass displacement of a Fermi gas subjected to a constant force $\bs F= F \bs 1_y$. Setting the Fermi energy within a topological bulk gap, the contribution from the group velocity naturally vanishes and the displacement along the transverse direction $x$ is then directly proportional to the force multiplied by the Chern number \cite{Dauphin:2013}. This simple and direct scheme is robust against perturbations and it could be implemented in any cold-atom setup hosting Chern insulating phases to detect non-trivial topological order. We stress that this method requires high-resolution microscopes to measure the (integral) Chern number with high precision, i.e. $\nu_{\text{measured}} = \nu \pm 0.01$, the mean displacement being of the order of a few tens of lattice sites after reasonable experimental times \cite{Dauphin:2013}.

\paragraph{The hybrid-time-of-flight measurement}

Wang et al. \cite{Wang:2013} showed that the Chern number could be read out from hybrid time-of-flight (TOF) images, by detecting atom density after suddenly releasing the external confining potential $V_{trap} (x,y)$ along one direction only, for example along $\ey$. By combining \emph{in situ} imaging along $\ex$ and TOF imaging along the release direction $\ey$, such an experiment would give access to the hybrid particle density $\rho (x, k_y)$, where $k_y$ is the crystal-momentum along $\ey$.  The relation between the hybrid density and the Chern number are related through dimensional reduction (Sect.~\ref{quasi2D}). In this picture, the 2D Hofstadter model is viewed as a 1D Harper-Aubry-Andr\'e lattice directed along $\ex$ and described by a Hamiltonian $\hat{H} (\Phi,k_y)$, where $\Phi$ (the flux) and $k_y$ (crystal-momentum) are both interpreted as parameters (see Eq. \ref{aubry}). The proposal of Ref. \onlinecite{Wang:2013} exploits a relation between the electric polarization of electronic systems to their quantized Chern number \cite{Xiao2010,KingSmith:1993}. The Chern number measures the charge transported from one boundary of the 1D lattice to the other under the variation of the parameter $k_y=0 \rightarrow 2 \pi$, i.e. after a full cycle in the Brillouin zone. In the cold-atom framework, the measured hybrid density $\rho (x, k_y)$ permits a numerical reconstruction of this transport property, thereby providing an efficient way to directly evaluate the Chern number of topologically-ordered optical lattices \cite{Wang:2013}.

\paragraph{Detecting topological edge states in atomic Chern insulators}\label{detectedge}

\begin{figure}[tb]
\begin{center}
\includegraphics[width=5.5in]{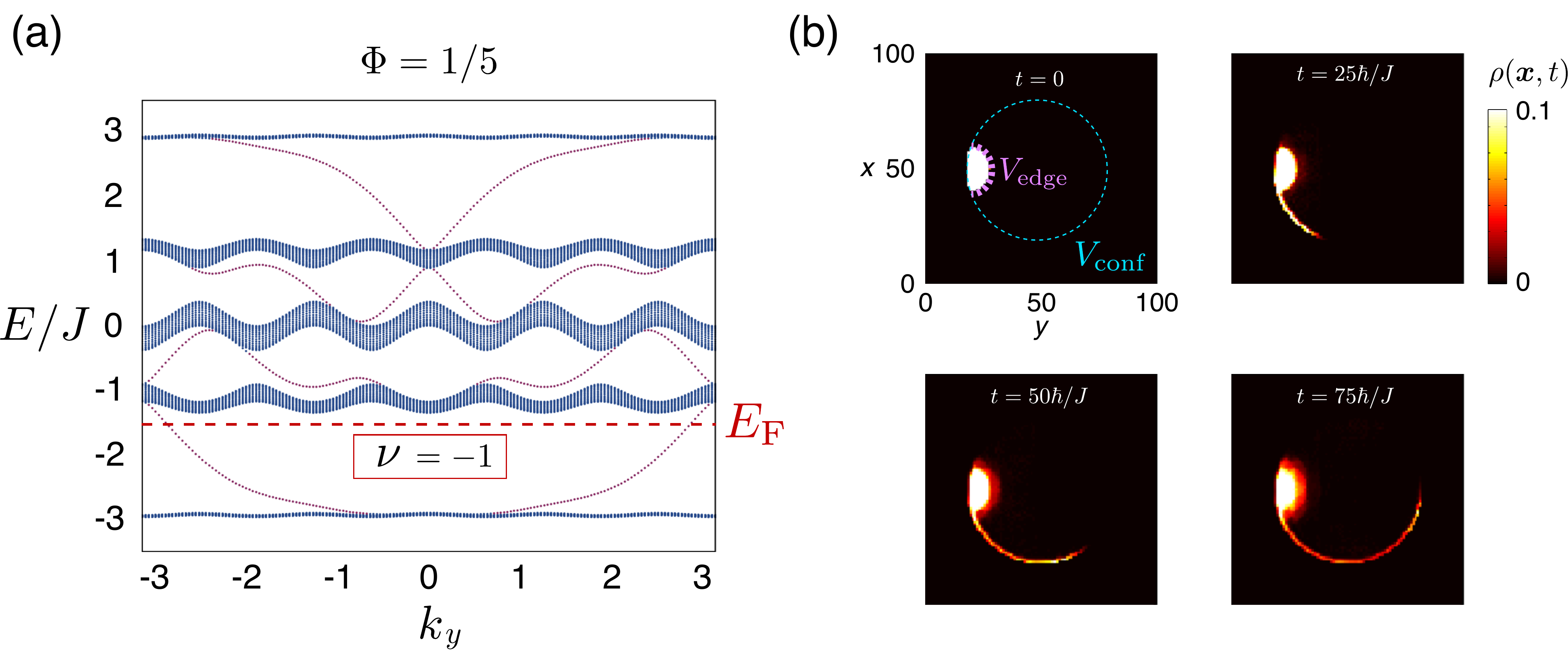}
\end{center}
\caption[]{(a) The Hofstadter model's energy spectrum with open boundary conditions along $\ex$ (from $x=0$ to $x=L$), and with periodic boundary conditions along $\ey$: the spectrum $E(k_y)$ is shown as a function of the quasi-momentum $k_y$ for $\Phi=1/5$. The projected bulk bands $E(k_x,k_y) \rightarrow E(k_y)$, shown in Fig. \ref{fig_hofstadter}b are plotted in blue. The red dispersion branches within the bulk gaps correspond to propagating edge modes near $x=0$ and $x=L$. When the Fermi energy is within the first bulk gap, edge modes up to the Fermi energy are populated, one each on the system's top and bottom. The number of edge-state modes within each gap is in agreement with the Chern numbers presented in Fig. \ref{fig_hofstadter} b. (b) An atomic Fermi gas with $E_F$ as in (a), was initially confined by a potential wall $V_{\text{edge}}$ in a small region located in the vicinity of the circular edge. After releasing the wall $V_{\text{edge}}$, the edge states propagate along the circular edge created by the potential $V_{\text{conf}}(r)$, making them directly observable in in-situ images of the spatial density $\rho (\bs x,t)$ \cite{GoldmanDalibard:2012}. 
}
\label{fig_edgestates}
\end{figure}

Topological edge states are populated when the Fermi energy is located within a topological bulk gap \cite{Hasan2010,Qi2011}. In Chern insulators, as described by the Hofstadter \cite{Hofstadter:1976} or Haldane \cite{Haldane:1988} models, for example, this happens when the sum of Chern numbers associated with the bulk bands lying below the gap is non-zero
\begin{equation}
\nu = \sum_{\lambda < E_F} \nu_{\lambda} \ne 0.
\end{equation}
According to the bulk-edge correspondence, the topological index $\nu$  corresponds to the number of edge-modes present within the bulk gap (these modes are responsible for the quantized Hall conductivity in Eq. \eqref{Halleq}, see Ref. \onlinecite{Hatsugai:1993}). The bulk-edge correspondence is illustrated in Fig. \ref{fig_edgestates}a. In general, the number of occupied edge modes within a bulk gap $N_{edge}$ and below $E_F$ contains a very small fraction of the total number of particles in the system $N_{tot}$.  In a circular atomic Chern insulator, produced by an external confining potential $V_c(r)$, the ratio of edge to total states is $N_{edge}/N_{tot} \sim a/R_F$, where $R_F$ is the Fermi radius of the system and $a$ is the lattice spacing \cite{Goldman:2012epj}. For typical systems with $R_F \sim 100 a$ and $N \sim 10^{4}$ particles, only a few tens of atoms will occupy edge modes. This simple observation indicates that the direct detection of topological edge states is a subtle and challenging task.

Identifying bulk topological order using the system's edges requires measuring characteristics of topological edge states distinct from those of the many bulk states. In Chern insulators, all the edge states present within a bulk gap propagate in the same direction -- they are chiral -- and their dispersion relation is approximately linear. Therefore, with the Fermi energy in a bulk gap, the dispersion relations of low-energy excitations give a clear signature for the presence (or absence) of topological edge states \cite{Scarola:2007,Liu:2010,Stanescu2010,Goldman:2012prl}. In principle, Bragg spectroscopy \cite{Stenger:1999} -- a technique based on momentum-sensitive light scattering which, for example, found application in measuring BEC's collective modes -- could offer such a probe \cite{Liu:2010,Stanescu2010}. Unfortunately, the number of particles $\sim N_{edge}$ excited by a Bragg probe focused on the cloud's edge (e.g. using high-order Laguerre-Gaussian beams \cite{Goldman:2012prl}) would be extremely small compared to the total number of particles $N_{tot}$, making the Bragg signal undetectable \cite{Goldman:2012prl,Goldman:2012epj}.  To overcome this drawback, a ``shelving method" was proposed in Refs. \cite{Goldman:2012prl,Goldman:2012epj}. In this scheme, the Bragg probe transfers energy and angular momentum to atoms located in the vicinity of the Fermi radius and simultaneously changes their internal states. This completely removes the edge states from the cloud -- allowing imaging on a dark background unpolluted by the untransferred atoms -- enabling the detecting of the edge-mode's dispersion relation from the Bragg signal \cite{Goldman:2012prl,Goldman:2012epj}.

Another method forces the edge states to propagate in a region that is unoccupied by the bulk states, allowing for a direct imaging of these topological states. Such a method was proposed in Ref. \cite{GoldmanDalibard:2012}, where the atomic cloud is initially shaped by large repulsive walls and prepared in a Chern insulating phase. After suddenly removing the walls, the chiral edge states propagate in a chiral manner along the circular edge of the cloud, while the bulk states tend to fill the initially vacant regions. This method is particularly efficient starting from a topological flat band, in which case the bulk states remain immobile for long times, allowing for the clear imaging of the propagating edge states, see Fig. \ref{fig_edgestates}b.  Schemes for isolating the edge state signal in the case of dispersive bulk states were also proposed in \cite{GoldmanDalibard:2012}, based on their chiral nature. Other methods to launch the edge state currents, by quenching the parameters of the microscopic Hamiltonian, were proposed in Ref. \cite{Killi:2012}.

Topological edge states generally survive in smooth confining potentials \cite{Stanescu:2009,Stanescu2010,Goldman:2012prl,Goldman:2012epj,GoldmanDalibard:2012,Buchhold:2012}. However, their angular velocity dramatically decreases and their localization length increases as the potential is smoothened. The use of sharp boundaries is therefore preferable to detect this edge-states physics in experiments \cite{Goldman:2012prl,Goldman:2012epj,GoldmanDalibard:2012}. The detection of atomic topological edge states has been further studied in Refs. \cite{Javanainen:2003dy,Ruostekoski:2008ig,Barnett:2013ce,Celi:2013,Hugel:2013,Reichl:2014vj}. 

Finally, we stress that extending the schemes to probe topological invariants and edge states in static systems to the case of periodically-driven systems, should be handled with care. Indeed, any  potential $\hat V_{\text{probe}}$ associated with a probing protocol, e.g. a static force to measure the Chern number, or a walking potential to probe edge-state dispersion relations, will potentially alter the effective Hamiltonian, and hence, the quasi-energy band structure \cite{Goldman:2014uz}. In other words, measuring the topological order associated with an effective Hamiltonian may destroy it.

\subsubsection{Simulating $Z_2$ topological insulators and axion electrodynamics}

Adding synthetic spin-orbit couplings to 2D and 3D optical lattices opens the possibility to simulate and detect the unusual properties of $Z_2$ topological insulators \cite{Goldman:2010,Bermudez:2010,Beri:2011,Mazza:2012}. The methods for detecting the chiral edge states of Chern atomic insulators discussed above can be directly applied to the case of $Z_2$ insulators exhibiting helical edge states \footnote{ Helical edge states are counter-propagating edge states with opposite spins, which lead to the quantum spin Hall (QSH) effect \cite{Kane:2005}}. For instance, spin-resolved density measurements \cite{Weitenberg:2011} could be used to identify the propagation of the different spin species. Using cold atoms subjected to both synthetic spin-orbit couplings and magnetic fields, one could then identify the transition between QH and QSH phases, by studying the nature of the propagating edge states \cite{Goldman:2012,Beugeling:2012}. While the Bragg spectroscopy scheme of Refs. \cite{Goldman:2012prl,Goldman:2012epj} could be generalized to identify the edge states of any 2D topological phase, the ``wall-removal" strategy of Ref. \cite{GoldmanDalibard:2012} could be applied to any topological phase exhibiting propagating states (in 2D, but also 3D systems). Moreover, the methods to directly measure the Chern number could be extended to detect the topologically invariant spin-Chern number of $Z_2$ insulators \cite{Goldman:2010}. Finally, 3D optical lattices emulating $Z_2$ topological insulators provide a versatile platform for detecting emerging axion electrodynamics: the unusual modifications of Maxwell's equations due to the topological axion term \cite{Hasan2010,Qi2011}. A protocol to detect the fractional magnetic capacitor \cite{Wilczek:1987} -- a signature of axion electrodynamics -- was described in Ref. \cite{Bermudez:2010}.\\

\subsubsection{Majorana fermions in atomic topological superconductors}

The beautiful universality of topological band insulators motivated theorists to seek for similar structures in different physical systems. For example, the Bogoliubov-de Gennes Hamiltonian,  describing the excitations of superconductors, can also describe topological phases: topological superconductors.  Some topological superconductors -- those that break time-reversal (TR) symmetry -- host topologically-protected zero modes: Majorana fermions with non-Abelian exchange statistics \cite{Qi2011} (akin to the quasiparticle excitations of the $\nu=5/2$ quantum Hall state \cite{Nayak:2008}). Several platforms have been envisaged in the quest for these properties, such as (a) the interface between a 3D topological insulator and a conventional $s$-wave superconductor \cite{Fu:2008}, and (b) semiconductors with Rashba spin-orbit coupling, $s$-wave pairing and a TR-breaking perturbation (e.g. a Zeeman coupling) \cite{Sau2010,Alicea2010,Alicea:2011}. This second route has been envisaged for 1D spin-orbit coupled semiconducting wires that map onto Kitaev's superconducting chain \cite{Alicea:2011}, but also for 2D spin-orbit coupled semiconductors \cite{Sau2010,Alicea2010}. 

Neutral atoms with synthetic spin-orbit coupling therefore gives a natural and experimentally complementary platform for realizing TR-breaking topological superconductors, where s-wave pairing and TR breaking terms can be easily controlled.  In addition, Raman coupling and laser-assisted tunneling methods can also give 2D topological superconductors \cite{Sato:2009,Sau:2011,Liu:2012} and Kitaev's superconducting chain with cold atoms trapped in 1D optical lattices \cite{Jiang:2011,Kraus:2012,KrausDalmonte:2013,Nascimbene:2012}. 

These proposals are all motivated by the desire to clearly detect and manipulate the properties of atomic Majorana zero modes, with an eye for methods that are not experimentally practical in solid-state systems \cite{Zhu:2011,Kraus:2012,Liu:2012,Nascimbene:2012,Kraus:2013bis}.  A first signature might be the anomalous density of an atomic topological superfluid.  A Majorana mode located inside a vortex core, should contribute to the total density in a detectable manner \cite{Liu:2012}, and TOF images would reveal the Majorana mode's non-local correlations. By considering a topological atomic chain,  Kraus et al. \cite{Kraus:2012} discussed how the the Majorana mode's long-range correlations give modulations in the TOF images, and also, that this signal is related to the number of topological modes present in the system.  As  stressed above when discussing the detection of Chern insulators, it is crucial to reduce the large background stemming from the many bulk states to emphasize the Majorana signal, which could be realized by local addressing \cite{Kraus:2012}. Spectroscopic measurements, similar to the Bragg probe discussed above, both gives access to the energy of the Majorana states and provides a proof of their localization \cite{Kraus:2012,Nascimbene:2012}.  Finally, the Chern number measurement introduced by Alba et al. \cite{Alba:2011} could also be generalized to demonstrate the existence of Majorana modes in optical lattices \cite{Pachos:2013}. 

Braiding operations, which reveal the anyonic nature of the Majorana modes, have been proposed for 2D topological superfluids \cite{Zhu:2011}, where braiding is realized by externally moving the vortices hosting the zero modes. More recently, a braiding protocol was proposed for atomic Kitaev wires \cite{Kraus:2013bis}, where the braiding operations are realized through local addressing by locally switching on/off potentials, hopping and pairing terms. 


\section{Interacting gauge theories and dynamical gauge fields}
\label{sect:dynamical}


\label{sect:dynamical}

So far we have considered the effects of gauge fields -- or strictly speaking gauge \emph{potentials} -- that are classical and static (in the sense that any time-dependence of the field is experimentally specified). These gauge potentials are externally controlled, and thus, they typically depend on atom-light coupling parameters, on rotation frequency or on other types of external driving features.  As we have seen, effective magnetic and electric fields arise and give rise to observable effects, however, these {\it applied} fields do not reproduce a complete field-theory picture: they are not dynamical degrees of freedom. In contrast to ``real" fields, the synthetic fields considered in the previous Sections are not influenced by the matter fields (i.e. the atoms).  To be concrete, the synthetic electric and magnetic fields produced in laser-coupled atomic gases (Sect.~\ref{sect:lightmatter}) need not obey Maxwell's equations.  This aspect of synthetic gauge potentials emphasizes their artificial origin, and in fact, it could be exploited constructively to simulate exotic situations where electromagnetism is no longer ruled by Maxwell's equations. \\

Generating \emph{interacting gauge theories} with cold atoms, where matter fields and synthetic gauge structures are dynamically related, importantly connects simple experiments with intractable problems in QED and QCD \cite{Zohar:2012,Banerjee2012,Zohar2013,Zohar:2013bis,Banerjee2013}.  Such quantum simulators are a novel tool for gaining physical insight in issues encountered in high-energy physics, such as the fundamental problem of \emph{confinement}, which precludes the observation of free quarks in Nature  \cite{Zohar:2011,Zohar:2012}.  A first step towards realizing dynamical gauge fields is introducing back-action where atoms affect a synthetic gauge potential $\bs A (\bs r ,t)$ locally and dynamically. One possibility is to create a gauge potential that explicitly depends on the atomic density $\bs A (\bs r ,t) \sim \rho (\bs r,t)$, as was proposed in Refs. \cite{Edmonds2013,Keilmann2011}. These density-dependent gauge structures, (described in Sect.~\ref{sect:dens-dependent}), do not fully reproduce a conventional gauge field theory in the sense that the emerging fields do not exist in the absence of matter (i.e. when $\rho = 0$). However, they  give rise to interesting properties, such as anyonic structures and chiral solitons, suggesting novel perspectives in quantum simulation. Schemes that fully realize quantum field theories \cite{Cirac:2010,Zohar:2011,Zohar:2012,Banerjee2012,Zohar2013,Zohar:2013bis,Banerjee2013,Zohar:2013pra} require (a) quantum matter (fermionic) fields $\hat \psi$ \emph{and}  gauge (bosonic) fields $\hat a$ represented by different atomic species interacting with each other, and (b) gauge-invariance conditions usually synthesized by imposing some constraints.  These proposals highlight the birth of a very new and exciting field of research possibly connecting cold-atom and high-energy physicists, and are discussed in Sect.~\ref{sect:QGT}. A recent review on the quantum simulation of lattice gauge theories has been recently written by Wiese \cite{Wiese:2013uh}. Finally, we mention the possibility to realize dynamical gauge fields of condensed matter models, such as spin-ice materials, using polar molecules \cite{Tewari:2006js} or Rydberg atoms \cite{Glaetzle:2014vp}.

\subsection{Density-dependent gauge potentials}\label{sect:dens-dependent}

We first consider a non-interacting two component Bose gas evolving in space-dependent coupling fields described by the single-particle Hamiltonian
\be
\hat H = \frac{\bs p ^2}{2m} \hat{1} + \hat U =  \frac{\bs p ^2}{2m} \hat{1} + \frac{\hbar \Omega}{2} \begin{pmatrix}  0 &  e^{-i \phi} \\  e^{i \phi} &   0 \end{pmatrix},
\ee
where $\Omega$ is the Raman Rabi frequency; $\phi$ is the coupling laser's phase; and for simplicity the detuning $\delta$ from the two-photon Raman resonance is set to zero.  According to Sect.~\ref{sect:lightmatter}, a non-trivial gauge structure is generated through the atom-light coupling, when the gradient of the detuning is non-zero $\bs \nabla \delta \ne 0$. The general idea behind the concept of density-dependent synthetic gauge potentials, is that collisions between the atoms can induce an effective \emph{density-dependent detuning} at the mean-field level. Reference \cite{Edmonds2013} considered the mean-field Hamiltonian
\be
\hat H = \frac{\bs p ^2}{2m} \hat{1} + \hat U + \hat V_{MF} = \frac{\bs p ^2}{2m} \hat{1} + \begin{pmatrix}  g_{11}|\Psi_1|^2 +g_{12}|\Psi_2|^2 & \frac{\hbar \Omega }{2}e^{-i \phi} \\ \frac{\hbar \Omega }{2}e^{i \phi} & g_{22}|\Psi_2|^2 +g_{12}|\Psi_1|^2 \end{pmatrix},
\ee
for an interacting two-level atomic system, where $\rho_{1,2}=\vert \Psi_{1,2} \vert ^2$ are the densities associated with the two atomic species, and $g_{\mu \nu}$ are the species-dependent contact-interaction parameters.  When the atom-light coupling energy $\hbar \Omega$ is much larger than the mean field terms $g_{\mu \nu} \rho_{\mu} \ll \hbar \Omega$, the corresponding gauge potential and dressed states
\begin{align}
&|\chi_\pm\rangle=|\chi_\pm^{(0)}\rangle\pm \frac{g_{11}-g_{22}}{8\hbar\Omega}\rho_\pm|\chi_\mp^{(0)}\rangle,\\
&{\bf A}_\pm={\bf A}^{(0)}\pm{\bf a}_1\rho_\pm({\bf r}) = -\frac{\hbar}{2} \bs \nabla \phi  \pm \frac{g_{11} - g_{22}}{8 \Omega} (\bs \nabla \phi) \rho_\pm({\bf r}) ,\label{nonlin}
\end{align}
can be obtained perturbatively.  $|\chi_\pm^{(0)}\rangle$ are the standard unperturbed dressed states for $\hat V=0$, and $\rho_\pm = \vert \Psi _{\pm} \vert^2$ denote the dressed state densities. Equation \eqref{nonlin} highlights the main result: species-dependent collisions, with $g_{11} \ne g_{22}$, can produce a density-dependent gauge potential $\bs A \sim \rho (\bs r)$. This interaction-induced detuning is the simplest scheme realizing a pseudo-dynamical gauge theory with back action between the matter field $\Psi (\bs r)$ and the gauge potential $\bs A (\bs r)$.  The key point is that the parameters in the resulting density-dependent gauge potential $\bs A$ are governed by the Rabi frequency $\Omega$, the gradient of the phase $\bs \nabla \phi$ and the scattering length difference $a_{11}-a_{22} \propto g_{11}-g_{22}$. All these  parameters are largely adjustable by tuning the coupling lasers and the scattering lengths, which can be achieved using Feshbach resonances \cite{Bloch2008a}.

The properties of this unusual Bose gas can be studied through a generalized Gross-Pitaevskii (GP) equation, which takes into account the presence of the density-dependent gauge potential \cite{Edmonds2013}
\begin{equation}
i\hbar \partial_t \Psi= \left[\frac{{(\bf{p}-\bf{A})}^2}{2m}+ {\mathbf a}_1\cdot {\mathbf j} (\Psi, \Psi^*)+W+g\rho\right]\Psi,\label{cgp}
\end{equation}
where $\bs A = \bs A_{+}$ is given by Eq. (\ref{nonlin}), $g=(g_{11}+g_{22}+2g_{12})/4$, $W={|{\mathbf{A}^{(0)}}|^2}/{2m}$ and a single dressed-state branch has been isolated in the dynamics (i.e. $\Psi=\Psi_{+}$, $\rho=\rho_+$).  Here, the nonlinearity of the GP equation manifests itself both through the standard term $\sim g \rho$, and also through the current
\begin{equation}
{\bf j} (\Psi, \Psi^*)=\frac{\hbar}{2mi}\left[\Psi\left(\nabla +\frac{i}{\hbar}{\bf A}\right)\Psi^*-\Psi^*\left(\nabla -\frac{i}{\hbar}{\bf A}\right)\Psi\right].\label{currentGP}
\end{equation}
Thus, this gives rise to rich nonlinear dynamics: the modified GP equation \eref{cgp}-\eqref{currentGP} already leads to exotic properties in 1D, including chiral soliton solutions, as was already pointed out in the context of one-dimensional anyons \cite{Aglietti1996}. Supposing a coupling laser phase in the form $\phi=k x$ and setting $\Psi(x,t)=\varphi(x,t)e^{-ikx/2}$, the dynamics of the 1D atomic system is described by
\be
i \hbar \partial_t \varphi = \left[ \frac{(p - a_1 \rho)^2}{2m} + a_1 j + W_1 + g \rho \right ] \varphi , \label{1dsoliton}
\ee
where $a_1 \propto k (g_{11}-g_{22})$ is the $k$-dependent nonlinearity strength and $W_1=\hbar ^2 k^2/8m$. This equation supports bright or dark chiral solitons, depending on the sign of $\mathfrak{g}= g - 2 a_1 u$, where $u$ is the speed of the soliton. At the critical value $\mathfrak{g}=0$, the chiral soliton is destroyed. Hence, in contrast to a conventional soliton, the chiral solitons are strongly altered when reflected from a hard wall \cite{Edmonds2013}. Other non-trivial effects stemming from the unusual nonlinear equation \eqref{1dsoliton} have been reported in Ref. \cite{Edmonds2013}, including an asymmetric free expansion of the cloud accompanied with a drift of the centre of mass, and the existence of critical particle numbers for the onset of persistent currents in a ring geometry \cite{Edmonds2013}. \\

\begin{figure}
\begin{center}
	\includegraphics[width=0.55\columnwidth]{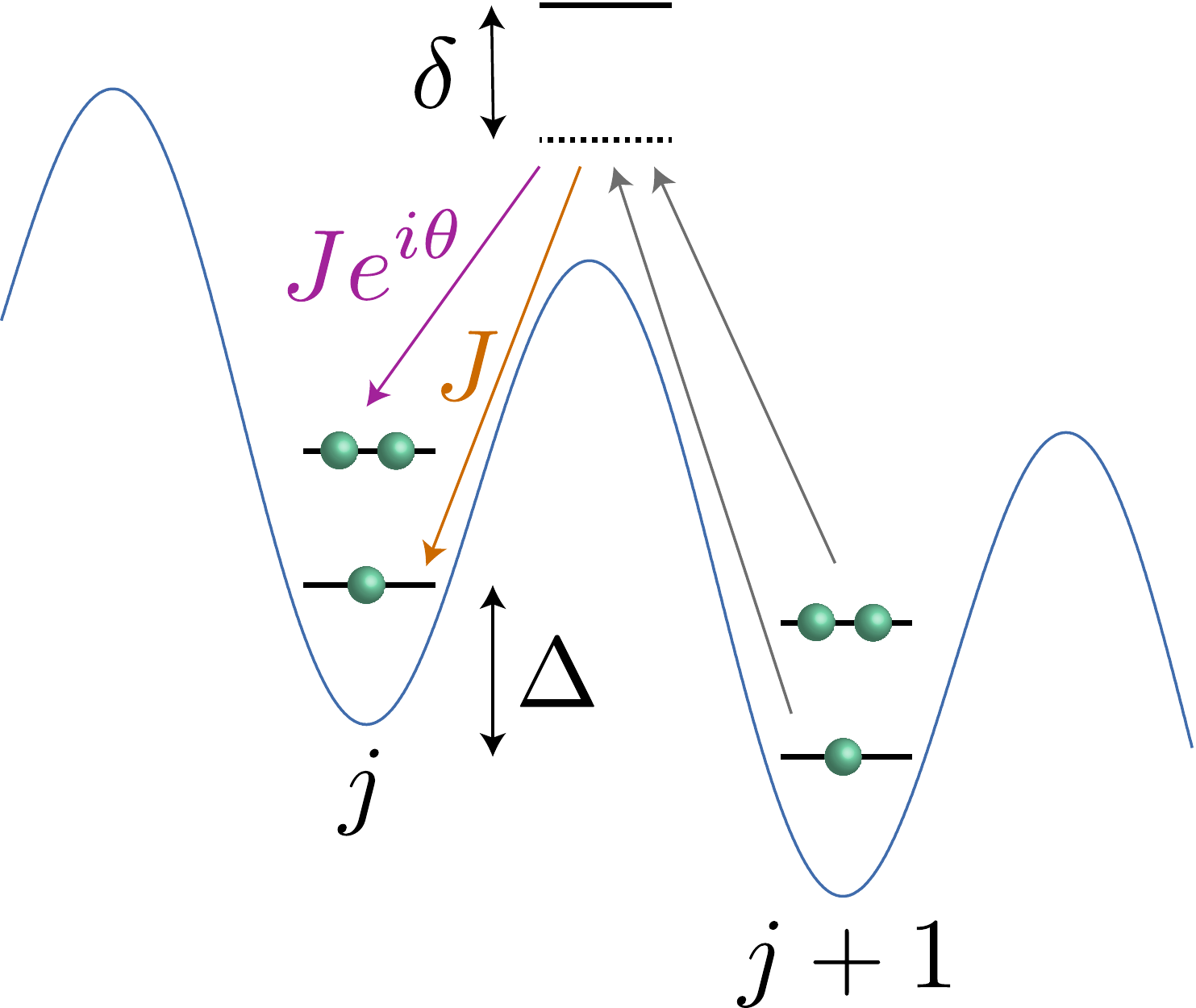}
	\caption{\label{keil} A schematic view of the proposal by Keilmann et al. for creating density dependent tunneling matrix elements $J e^{\phi_{jk}}= J e^{i \theta n_j}$, for the case $n_j \le 2$. Laser assisted tunneling using four detuned ($\delta$) lasers in combination with a tilted lattice given by the off-set $\Delta$ induces a phase which depends on the occupation number of the target site.}
	\end{center}
\end{figure}

A density-dependent gauge structure can also be created in optical-lattices. In the lattice framework (see Sect.~\ref{sect:lattices}), gauge potentials manifest themselves through the Peierls-modified hopping matrix element $J e^{i \phi_{jk}}$, describing hopping between lattice sites $j$ and $k$. Hence, the key idea is to engineer Peierls phases $\phi_{ij}$ that depend on the occupation number $ n_j$ at each lattice site $j$, e.g. $\phi_{ij} = \theta n_j$. Such a scenario was proposed by Keilmann et al. \cite{Keilmann2011}, giving the generalized Bose-Hubbard model
\begin{equation}
H=-J\sum_j\left(\hat b_j^\dagger \hat b_{j+1}e^{i\theta \hat n_j}+h.c.\right)+\frac{U}{2}\sum_j \hat n_j (\hat n_j-1). \label{bh}
\end{equation}
This is the lattice version of the Hamiltonian leading to Eq. \eqref{cgp}, contains the density operator $\hat n_j=\hat b^{\dagger}_j \hat b_j$ at site $j$, the onsite interaction strength $U$, and the  $\hat n_j$-dependent tunneling rate $J e^{i\theta \hat n_j}$. This unusual Peierls substitution can be realized using a generalization of the laser-assisted tunneling method presented in Sect.~\ref{experimentallattice}, as illustrated in Fig. \ref{keil} for the case $n_j \le 2$. In this picture, the two neighboring sites at $j$ and $j+1$ can either host $n=1$ or $n=2$ atoms (omitting the trivial case $n=0$), resulting in a four-dimensional ground state manifold $\{ \vert g_1 \rangle, \dots , \vert g_4 \rangle \}$. Coupling this subsystem to an excited state $\vert e \rangle$, with four different coupling fields, allows to individually address each hopping process, thanks to the energy offsets produced by the onsite interaction $U$ and by an additional lattice tilt $\Delta$, see Figure \ref{keil}. Finally, tuning the phases associated with each laser-induced tunneling process (see Sect.~\ref{experimentallattice}) results in the required $n_j$-dependent tunneling matrix elements between lattice sites $j$ and $j+1$ \cite{Keilmann2011}.

The Hamiltonian \eqref{bh} maps onto a theory of anyons in one-dimensional lattice.  The mapping between bosonic ($\hat b_j$) and anyonic ($\hat a_j$)  operators, $\hat a_j=\hat b_j \exp[i\theta\sum_j \hat n_j]$, gives rise  to the Anyon-Hubbard Hamiltonian
\begin{equation}
H=-J\sum_j\left(\hat a_j^\dagger \hat a_{j+1}+h.c.\right)+\frac{U}{2}\sum_j \hat n_j (\hat n_j-1),
\end{equation}
where the commutation relation for $\hat a_j^{(\dagger)}$ is anyonic:
\begin{equation}
\hat a_j\hat a_k^\dagger - e^{-i\theta \, \text{sign}(j-k)}\hat a_k^\dagger\hat a_j=\delta_{jk} .
\end{equation}
Since $\text{sign}(j-k)=0$ when $j=k$, particles behave as bosons on-site, and as anyons off-site with statistical angle $\theta$.  Thus, the anyonic nature of Eq. \eqref{bh} stems from the density-dependent Peierls phases $e^{i \theta \hat n_j}$. 

The Hamiltonian in Eq. \eref{bh} is a lattice version of the continuum Hamiltonian leading to the meanfield Gross-Pitaevskii equation in Eq. \eref{cgp}. As such it shows unconventional tunneling dynamics due to its nonlinearities (see Eqs. \eref{cgp}-\eqref{currentGP}),  which will affect any Josephson type dynamics in the lattice \cite{Edmonds2013b}. Reference \cite{Keilmann2011} showed that density-dependent tunneling gives rise to phase transitions between the standard superfluid phase and exotic Mott states, where the particle distributions show plateaus at fractional densities, due to the anionic statistics. 


\subsection{Simulating quantum gauge theories}\label{sect:QGT}

The simulation of Dirac fermions with atoms in optical lattices \cite{Lim2008,Lee:2009,Goldman:2009prl,Maraner:2009,Hou2009,Bermudez:2010NJP,Bermudez:2010,Tarruell2011,Goldman:2011,Mazza:2012,Zhang2012FP,Hauke:2012}, together with the possibility of generating background Abelian and non-Abelian gauge fields \cite{Osterloh:2005,Goldman:2009prl,Hauke:2012,Goldman:2013Haldane}, suggest that cold atoms could be exploited to deepen our understanding of quantum electrodynamics (QED) and quantum chromodynamics (QCD): a powerful alternative to numerical lattice-gauge-theory (LGT).  For instance, cold atoms in optical lattices of various spatial dimensions ($D=1,2,3$ \cite{Bloch2008a} and beyond $D>3$ \cite{Boada:2012}) may reveal the mechanisms by which confined phases emerge in various configurations: in Abelian and non-Abelian gauge theories; or in $D+1=3$ and $D+1=4$ space-time dimensions.  These mechanisms could be investigated in a setup where different atomic species encode the matter and the gauge degrees of freedom, and exploit the fact that the interactions between the particles can be tuned (e.g. by Feshbach resonances). Such cold-atom quantum simulators are a physical platform where the coupling strength of the gauge theory is externally controllable, and where phase transitions between confined and unconfined phases might be externally driven \cite{Zohar:2011,Zohar:2012}. 

This quantum-simulation scenario requires that both the matter and the gauge fields be dynamical, and also, that the simulated theory be gauge invariant.  For instance, realizing QED with cold atoms requires that the Gauss's law be imposed by a constraint, which generally requires a precise control over the simulated Hamiltonian \cite{Zohar:2011,Zohar:2012,Banerjee2012,Zohar2013,Zohar:2013bis,Banerjee2013}. The versatility of cold-atom systems also allows for the simulation of simple toy models, such as the Gross-Neveu model, which played a major role in the exploration of QCD-like effects \cite{Merkl2010,Cirac:2010}. 

\begin{figure}
\begin{center}
	\includegraphics[width=1\columnwidth]{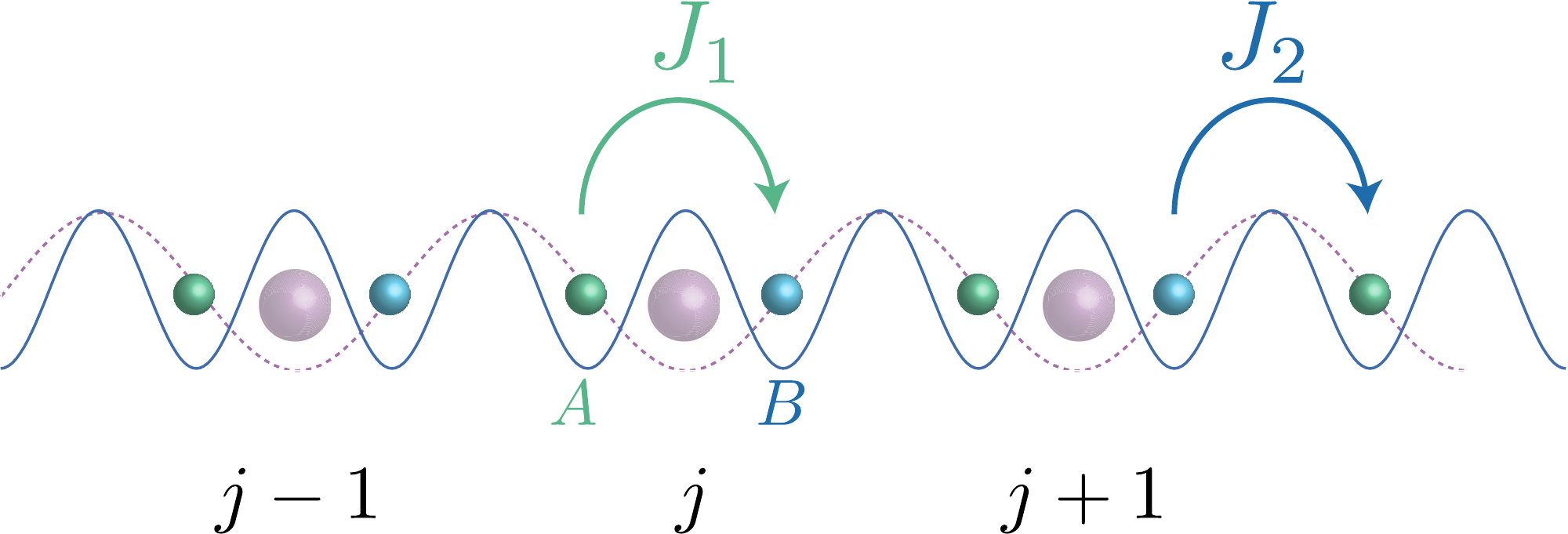}
	\caption{\label{grossfig} Optical-lattice implementation of Eq. \eqref{GrossOptical}. A primary optical lattice (blue) traps fermions in the two inequivalent sites $A$ and $B$, which are distinguished by the alternating tunneling rates $J_{1,2}$. A superimposed optical lattice (purple) traps a BEC between the $A$ and $B$ sites of each unit cell, labeled by the index $j$. The interaction between fermions and bosons within each unit cell $j$ results in a boson-mediated tunneling process for the fermions, giving rise to an effective quantum field theory at low energy \cite{Cirac:2010}.}
	\end{center}
\end{figure}

The quantum simulation of the Gross-Neveu model~\cite{Cirac:2010} already captures the general strategy and the main ingredients needed to simulate more elaborate field theories, such as QED and QCD.  The Gross-Neveu model describes the interaction between a massless Dirac fermion $\hat \Psi$ and a massive quantized scalar field $\hat \Phi$ in $D=1$ spatial dimension, with the Hamiltonian
\be
\hat H= \hbar \int dx \, \left [ c \, \hat \Psi^{\dagger} \gamma_1 \hat p \hat \Psi + g m \, \hat \Phi \hat \Psi^{\dagger} \hat \Psi + \frac{m ^2}{2} \hat \Phi^2 \right ],\label{GrossNeveu}
\ee
where $c$ is a velocity, $g$ is the coupling strength, $m$ is the mass of the scalar field and $\gamma_1$ is a Dirac matrix (the color quantum numbers $\sigma=1, \dots, N$ of the Dirac field $\hat \Psi_{\sigma}$ are implicit in Eq. \eqref{GrossNeveu}).  In this simple field theory the scalar field $\hat \Phi$ has no kinetic term, and the coupling between the Dirac $\hat \Psi$ and the scalar field $\hat \Phi$ can be formally traced back to the original Gross-Neveu model (which describes Dirac fermions interacting through the term $\sim g^2 (\hat \Psi^{\dagger} \hat \Psi)^2$ in the absence of the scalar field) \cite{Cirac:2010}. This interacting field theory could be implemented with a $N$-component Fermi gas in an optical lattice featuring spatially-modulated tunneling amplitudes $J=J_{1,2}$, see Fig. \ref{grossfig}, and along with an independent BEC loaded into a separate optical lattice with double spacing, as represented in Fig. \ref{grossfig}. The resulting Bose-Hubbard Hamiltonian is
\be
\hat H= - \sum_j \left [ J_1 \, \hat f_{j,A}^{\dagger} \hat f_{j,B} + J_2 \, \hat f_{j,B}^{\dagger} \hat f_{j+1,A} + \delta \, \hat b^{\dagger}_j \hat b_j \hat f^{\dagger}_{j,A} \hat f_{j,B} + \text{h.c.} \right ] - U \, (\hat b_j^{\dagger})^2 (\hat b_j)^2 - \mu \, \hat b_j^{\dagger} \hat b_j, \label{GrossOptical}
\ee  
where $\hat f_{j,A}^{\dagger}$ (resp. $\hat f_{j,B}^{\dagger}$) creates a fermion at lattice site $(j,A)$ (resp. $(j,B)$);  $\hat b_j^{\dagger}$ creates a boson at lattice site $j$; $U$ is the onsite interaction strength for bosons; $\mu$ is the chemical potential; and $\delta$ denotes the strength of the fermion-boson ``on-site" interaction, see Fig. \ref{grossfig}. According to Ref. \cite{Cirac:2010}, when the Fermi energy is at half-filling, the low-energy excitations exactly reproduce the dynamics dictated by \eqref{GrossNeveu} for specific values of the optical-lattice parameters. Since the matter-gauge field coupling strength $g$ of the simulated field theory results from a boson-mediated tunneling process of fermions that is proportional to the inter-species interaction strength $g \sim \delta / \sqrt{U}$ \cite{Cirac:2010}, the coupling constant $g$ is tunable. \\

Realizing more involved field theories, such as lattice QED or non-Abelian Yang-Mills lattice gauge theories, requires additional ingredients and control over inter-species interactions to  retain gauge invariance \cite{Zohar:2011,Tagliacozzo2011,Kapit:2011,Zohar:2012,Banerjee2012,Zohar2013,Zohar:2013bis,Banerjee2013,Kasamatsu:2012}.  These more complex proposals are still based on the strategy outlined above: different atomic species, fermions and bosons, are trapped in interpenetrating optical lattices, forming a many-body lattice system, whose effective (low-energy) description is that of an interacting quantum field theory.  These elaborate cold-atom schemes are largely based on available technologies -- taken to the extreme -- suggesting that experimental realizations of lattice gauge theories are within reach.   Very recently, a scheme has been proposed where the Gauss's law and gauge invariance emerge naturally (i.e. they are not imposed by an external constraint that modifies the effective low-energy description), based on the fundamental symmetries of the atomic system \cite{Zohar:2013pra}.  Optical-lattice implementations of lattice gauge theories offer unique measurement tools.  Experimenters might observe the mass dynamically acquired by fermions in the Gross-Neveu model \cite{Cirac:2010},  the ``electric flux tube" at the origin of confinement in $3+1$ Abelian LGT \cite{Zohar:2011}, the Wilson-loop area law \cite{Zohar2013}, cosmological inflation processes \cite{Kasamatsu:2012} and the nonperturbative string breaking effect present in quantum link models and QCD \cite{Banerjee2012}.


\section{Conclusions}
\label{sect:conclusion}


\label{sect:conclusion}

Artificial gauge potentials in cold gases are a new laboratory tool that allow access to new physical phenomena.  A handful of techniques for inducing the gauge potentials have been proven in the lab; these range from optically addressing the internal level structure of the atoms in combination with orbital motion, to mechanical shaking of optical lattices.  In particular, synthetic uniform magnetic fields were realized leading to quantized vortices in BEC~\cite{Lin2009b}, and non-trivial lattice gauge structures have been produced in optical lattices, leading to staggered ~\cite{Aidelsburger:2011,Aidelsburger:2013b,Struck2012,Jotzu} and uniform ~\cite{Aidelsburger:2013,Ketterle:2013,Aidelsburger:2014} magnetic flux configurations. Additionally, the (one-dimensional) spin-orbit coupling has now been created and its manifestations have been explored for atomic bosons and fermions~\cite{Lin2011,Wang2012,Cheuk2012,Zhang2012PRL,Fu2013PRA,Fu2013,Zhang2013PRA,Qu2013,LeBlanc2013,a9}. Examples of recent applications of artificial gauge potentials also include thermometry of cold atoms in optical lattices \cite{a1}, mapping the Berry curvature and Chern numbers of Bloch bands \cite{Jotzu,Aidelsburger:2014,a2}, Efimov states with SOC \cite{a3,a4} and gauge potentials in Rydberg gases \cite{a5,a6,a7,a8}.

Artificial electromagnetism has begun to mature.  It can be applied with or without an optical lattice present, leading to a broad spectrum of applications.  At the single particle level, a surge of theoretical work using optical lattices propose to emulate exotic condensed matter systems.  Prominent examples include topological insulators, quasi-relativistic systems, integer quantum Hall physics, and non-trivial spin dynamics from spin-orbit coupling.   Including collisional interactions enables a whole new field of research.  Interacting bosonic and fermionic gases show intriguing new effects when gauge fields are present. In the Abelian case, fractional quantum Hall (FQH) physics could be realized. To generate and stabilize such FQH liquids, it is crucial to develop schemes leading to topological flat bands \cite{CooperDalibard:2012,Baur:2014ux,Grushin:2014gt}, namely, dispersionless bands characterized by non-zero Chern numbers, which are well separated with respect to higher energy bands [see Fig. \ref{fig_edgestates} for such a configuration offered by the Hofstadter model at flux $\Phi=1/5$]. Note that such a topological-flat-band configuration has been realized experimentally in Munich \cite{Aidelsburger:2014}, where a Chern number $\nu_{\text{exp}}=0.99 (5)$ has been measured. Moreover, the rich set of spin-orbit coupled systems enable a range of novel effects such as unconventional superfluid properties in BCS-BEC crossover regimes, and finite momentum ground states for Fermi gases. Interacting spin-orbit coupled gases also offer an interesting route towards interacting topological phases in 3D, namely, the high-dimensional cousins of fractional quantum Hall states.

The prospect of creating non-trivial topological states of matter also opens up the possibility to address fundamental questions in quantum information. The availability of orbital magnetism provides a novel tool for atomtronics situations, i.e., the atomic version of spintronics.  The non-trivial topological states together with the concept of non-Abelian dynamics may also provide the route towards the mercurial and highly anticipated topological quantum computer~\cite{Nayak:2008}, which promises to be the holy grail of fault tolerant quantum computing.

Quantum simulators would in general benefit from artificial orbital magnetism. The gauge fields might enable the simulation of more exotic gauge theories that are computationally intractable.  A prominent example -- linked to the ability to create dynamical gauge fields -- is the quantum chromodynamics description of the strong force between elementary particles.  While dynamical gauge fields are still in their infancy and unrealized in the lab, recent advances and theoretical proposals using optical lattices show that it is in principle possible to engineer fully dynamical gauge fields.

It is intriguing to speculate on the future direction for artificial gauge fields. First of all there are challenges both theoretically and experimentally which need to be addressed. From a theoretical point of view perhaps the most pressing question is to what extent the gauge fields can be combined with strong collisional interactions and whether there are any fundamental or practical  limitations for doing so and still retain the dynamics governed by the effective magnetic fields. An interesting and promising direction in this respect is the inclusion of cavity QED where also strong interaction between the light and matter can be achieved which may provide a mechanism for creating a back-action between the matter field and the effective gauge fields \cite{Larson2010a,Larson2010b,Mivehvar-2014,Dong-2014,Deng-2014}, and by doing so also address the question of creating dynamical gauge fields but now possibly in the quantum regime.

Experimentally the challenges are also many. The level of controllability of key parameters such as stable laser frequencies, collisional properties, maintaining ultralow temperatures, avoiding numerous sources of heating and decay of the prepared quantum superpositions, are certainly taxing, and will require a constant development and refining of  experimental techniques and technologies.  This will have implications for reaching strongly correlated regimes in experiments.
All techniques for creating artificial magnetic fields to date could in principle reach the strongly correlated regime such as quantum Hall liquids. Each one of them have however their own experimental drawbacks. For instance the Raman based proposals including also flux lattices, will have problems with spontaneous emission which will cause heating. The mechanical shaking of a lattice will also induce heating eventually. Finally the rotational techniques have the disadvantage that there one would have to work with a very small number of particles. These limitations are all of a technical character, and we are likely to see a development of all them in parallel in the future.  

On a more speculative note artificial electromagnetism may provide a route to completely new physical scenarios which cannot be be found in conventional solid state systems. Emulating solid-state phenomena with the hope to shed some light on new and not so well understood phenomena, is the most natural application of the artificial gauge fields.  It is however not inconceivable to envisage a situation where the effective gauge field allows for the preparation of highly exotic quantum states which captures not only the solid state and condensed matter phenomena, but also addresses fundamental question regarding the fabric of space-time. For instance, gauge fields in the presence of analog gravity in Bose-Einstein condensates \cite{Boada:2011hd} could address important questions, such as the role of quantum entanglement and unitarity near black holes. The inclusion of gauge fields, and in particular dynamical gauge fields, into the picture of analog gravity, is largely uncharted territory and may provide some insight into the interplay between the quantum world and gravity.

\section*{Acknowledgments}

We acknowledge helpful discussions with 
M. Aidelsburger, 
B. Anderson,
E. Anisimovas, 
L. Aycock ,
M. A. Baranov,
R. Barnett, 
G. Barnich, 
A. Bermudez, 
J. Beugnon, 
I. Bloch, 
N. R. Cooper, 
J. Dalibard, 
M. Dalmonte, 
A. Dauphin, 
A. Eckardt,
M. Edmonds,
T. Esslinger, 
V. Galitski,
P. Gaspard, 
F. Gerbier, 
P. Hauke, 
W. Ketterle,
C. V. Kraus,
Z. Lan,
J. Larson, 
M. Lewenstein, 
D. Maldonado-Mundo, 
S. Nascimbene, 
M. Rechtsman, 
J. Ruostekoski,
J. Ruseckas,  
L. Santos, 
M. Valiente, 
H. Zhai and 
P. Zoller. 

N. G. thanks the FRS-FNRS and the ULB for financial support. G. J. acknowledges the financial support by the Lithuanian Research Council Project No. MIP-082/2012. P. \"O. acknowledges support from the UK EPSRC grant No. EP/J001392/1 and the Carnegie Trust for the Universities of Scotland. I. B. S. acknowledges the financial support by the NSF through the Physics Frontier Center at JQI, and the ARO with funds from both the Atomtronics MURI and DARPA's OLE Program.



\providecommand{\newblock}{}

\end{document}